\documentclass[aps,pre,reprint,amsmath,amssymb,longbibliography,floatfix]{revtex4-2}
\usepackage[T1]{fontenc}
\usepackage{graphicx,bm,booktabs,needspace}
\usepackage{tikz}
\usetikzlibrary{arrows.meta,positioning,fit}
\usepackage{xcolor}
\definecolor{figureblue}{HTML}{24518A}
\definecolor{figureorange}{HTML}{B85A32}
\usepackage{amsthm}
\usepackage{mathrsfs}
\usepackage[hidelinks]{hyperref}
\hypersetup{pdftitle={The resource cost of cooperative invasion},pdfauthor={Abhishek Chowdhury}}

\newtheorem{theorem}{Proposition}
\newtheorem{proposition}[theorem]{Proposition}

\newcommand{\E}{\mathbb E}
\newcommand{\one}{\bm 1}
\newcommand{\T}{\mathsf T}

\newcommand{\refl}{\mathrm{refl}}

\graphicspath{{figures/}}
\makeatletter
\newcommand{\labelalias}[2]{\@ifundefined{r@#2}{}{%
\protected@write\@auxout{}{\string\newlabel{#1}{\csname r@#2\endcsname}}}}
\makeatother
\begin{document}
\title{The resource cost of cooperative invasion}
\author{Abhishek Chowdhury}
\email{achowdhury@iitbbs.ac.in}
\affiliation{School of Basic Sciences, Indian Institute of Technology Bhubaneswar, India}
\begin{abstract}
A contact network can favor cooperation by sheltering a lineage until its descendants compete with defectors as a group. We determine the contact strengths, invasion times and measurement effort needed to realize this advantage at fixed population size. Individuals reproduce according to averaged cooperation payoffs, and offspring replace neighbors through contacts of equal weight in both directions. Contact contrast is the ratio of strongest to weakest contact. Under weak selection on a six-site path, the least contrast needed to favor a single cooperator over the reverse invasion by a defector grows as $656\epsilon^{-2}$ for uniformly placed mutants and $4050\epsilon^{-3}$ for mutations in newborn offspring, as the benefit--cost ratio $b/c=1+\epsilon$ approaches one. The extra power arises because rare replacement protects a site while suppressing mutation arrival there. Global bounds force two weakly coupled groups and exact mirror symmetry of the least-contrast contacts for sufficiently small $\epsilon$. Under uniform introduction, additional contrast shortens the leading mean time to extinction or fixation in the absence of selection, from $12/\epsilon$ toward $6/\epsilon$ reproduction attempts, and eventually breaks the mirror symmetry of the fastest design. Estimating the uniform-introduction advantage from fixation outcomes of independent completed invasions defines a third optimum. Within the six-site family with mirror-symmetric, strongly separated contacts, we optimize contacts and selection strength with reproduction weighted linearly or exponentially by payoff. At fixed relative precision, with contrast a fixed factor above its minimum, the leading summed trial duration grows as $\epsilon^{-3}$. Efficient measurement accepts a smaller advantage in exchange for shorter trials. The cheapest, fastest and most efficiently measured networks therefore differ.
\end{abstract}

\maketitle
\section{Introduction}\label{sec:intro}
A contact network can protect a cooperative lineage long enough for its descendants to change their surroundings. Realizing that advantage requires control of the contacts, time for the invasion to resolve, and enough observations to distinguish a small bias from fluctuations. These requirements are coupled. Strong protection can prolong a lineage's lifetime while making its founding site difficult for a new mutation to reach. We ask how the resulting competition between protection and access determines the resources needed for cooperation at fixed population size, and whether the same network minimizes contact cost, invasion time and measurement effort. On the six-site path, these objectives select distinct designs.

Cooperation benefits others at a direct cost to the individual. In the well-mixed donation game, that cost makes selection favor defection~\cite{Nowak2006}. Here each site holds one individual, and reciprocal contact weights, equal in both directions, govern both its averaged benefit encounters and the placement of its offspring. Cooperation costs $c$ and confers benefit $b$. An individual is chosen from the whole population to reproduce with a probability increasing with its payoff; its offspring then replaces a neighbor chosen with probability proportional to the weight of the connecting contact. This is source-first Birth--death updating. The payoff determines reproductive propensity, or fecundity, through a linear or exponential function. We compare a single cooperator taking over defectors with the reverse invasion, using the same distribution of founding sites. Eventual takeover is called fixation. At weak selection, where payoffs perturb reproduction to first order, the threshold is the benefit--cost ratio above which the cooperator has the larger fixation probability.

Reciprocity of contact weights permits strongly unequal copying probabilities. An individual is rarely replaced by a neighbor that directs most of its offspring along a stronger contact elsewhere, even though the individual itself often replaces that neighbor. Such a dynamically protected individual is called a pin; the nearby individuals it repeatedly replaces, and so aligns with its own trait, are its followers. The pin and its followers form a persistent domain. On the six-site path, two protected domains become almost uniform in trait before they compete. Their leading payoffs are $b-c$ for the cooperative domain and zero for the defector domain. This collective competition makes the threshold approach one. Six sites are the minimum needed for this limit on a path~\cite{ChowdhuryProtectedDomains}. We determine the relative contact strengths needed to approach this threshold and the dynamics of the designs that do so.

The introduction rule specifies where a new lineage begins. Uniform introduction samples every site equally. A mutation in a newborn offspring in a homogeneous population appears at the recipient of a copying event~\cite{AdlamChatterjeeNowak2015,McAvoyRaoHauert2021}, and we call this birth-associated introduction. The low replacement rate protecting a pin also makes newborn mutations arrive there rarely. Uniform introduction keeps access independent of the contact weights; birth-associated introduction links access to protection through the copying process. The required ratio of strongest to weakest contact therefore diverges as the inverse square or inverse cube of the target threshold gap above one, depending on the introduction rule.

Exact weak-selection response through coalescing ancestry, which follows lineages backward until they meet, makes the response of a heterogeneous network calculable~\cite{AllenEtAl2017,AllenEtAl2021,McAvoyAllen2021}. The location of a mutant's origin can materially affect its success~\cite{MaciejewskiFuHauert2014,AllenEtAl2015}. For unweighted regular graphs, the large-population, weak-selection approximation finds that source-first Birth--death updating disfavors cooperation~\cite{OhtsukiEtAl2006}. On weighted graphs, unequal replacement rates allow the protected domains considered here. A different route uses growing graph families with donation payoffs summed over neighbors. Such families can promote cooperation for prescribed positive selection strength and benefit--cost ratio above one, with quadratic fixation-time bounds~\cite{SvobodaChatterjee2025}. That process uses death-first updating, choosing the replacement site before its parent, and makes population size a resource. We keep the population finite and fixed, with averaged payoffs and source-first updating, and optimize the relative contact weights.

The distinction between successful invasion and rapid invasion is familiar in constant-selection amplifiers, networks that enhance fixation of a mutant with a constant fitness advantage~\cite{SlowTkadlec2019,SlowTkadlec2021}. Enumeration of unweighted graph topologies likewise separates fixation probability from the time taken by successful lineages~\cite{MollerHindersinTraulsen2019}. Constrained changes of directed weights have been optimized numerically to suppress a constant-fitness mutant~\cite{Bonneuil2026Resource}, and budgeted design has been studied by choosing the vertices at which a reproductive advantage or adoption bias acts~\cite{BrendborgEtAl2022,PetsinisEtAl2023}. Here the controls change the encounters and replacement probabilities that generate a payoff-dependent advantage. We optimize over all positive weights on a specified support, the set of retained contacts. Time counts global reproduction attempts, including copies that leave the population unchanged. This fixed clock makes waiting times comparable across designs.

Our contact resource is the contrast, the largest retained weight divided by the smallest. It measures the dynamic range of relative weights; an upper bound on it, the cap, acts as a contact budget. Common rescaling changes neither the copying probabilities nor the rate of reproduction attempts; converting contrast into an energetic or material cost would require a physical model of the contacts. For the target $b/c=1+\epsilon$, the least contrast on the six-site path is asymptotic to $656\epsilon^{-2}$ under uniform introduction and $4050\epsilon^{-3}$ under birth-associated introduction as $\epsilon\downarrow0$. The competing losses have a direct interpretation in the reflected path with successive weights $(Q,q,1,q,Q)$ and $Q\gg q\gg1$, whose outer pairs are the follower cores. Increasing $q$ suppresses replacement across the central contact, while increasing $Q/q$ suppresses replacement of a pin from its core. Under uniform introduction the leading losses are $6/q$ and $82q/(3Q)$. The least contrast allocates half of the allowed threshold gap $\epsilon$ to each loss. 

Under birth-associated introduction, frequent arrivals in followers are compared with rare direct arrivals at pins, and the second loss becomes $50q^2/(3Q)$. The optimal allocation then gives two thirds of the gap to the central loss and one third to the core loss. This weighted arithmetic--geometric mean balance predicts $q\sim12/\epsilon$ or $q\sim9/\epsilon$ and the two sharp contrast coefficients above. The global inequalities prove that no other positive weighting of the path improves them.

Proving global optimality requires control of every competing weighting, including designs with different or nested contact scales. Exact polynomial positivity bounds force every competitive six-site path design into the protected-domain hierarchy, in which the core contacts dominate the pin contacts and the pin contacts dominate the central contact. The small ratios of contact weights then describe the response regularly even as the weights diverge. Their constraints form a rank-one matrix relation, so the regular description also identifies which variations correspond to actual contacts. Uniform control of the response and its derivatives makes the unique leading minimum persist for all sufficiently small gaps. Path reversal then forces the exact optimizer to be mirror-symmetric. Exact symmetry thus follows from global localization and uniqueness.

The six-site comparison also extends continuously between introduction rules. A uniform fraction of order $\epsilon$ connects the two cost regimes. When the resource is the inverse smallest positive copying probability, the least required resource diverges with inverse-gap powers one and two. The extra inverse power for birth-associated introduction survives this change of resource, for copying probabilities generated by reciprocal weights.

A common convex balance organizes the resource comparisons. Positive monomials describe the separate losses of protection, and a variational identity determines their optimal balance and the cost of departures from it. Below a contrast cap that depends on the support and the introduction rule, no weighting favors cooperation at weak selection for any benefit--cost ratio; this first-promotion cap is the infimum of caps that admit a favorable weighting. For birth-associated introduction on the five-cycle, the variational identity expresses the first-promotion condition through relative entropy, measuring mismatch among terms in the response polynomial. On the five-site path, every competing minimum lies in a domain where a positively rescaled threshold constraint is convex in logarithmic contact coordinates. This proves a unique optimizer at every feasible cap. These constructions connect the first possibility of promotion to the approach to a specified threshold. The introduction rule reverses the order of first promotion between the five-cycle and the reflected six-site path, and an exact six-cycle weighting of contrast $51$ already promotes under uniform introduction. The four-cycle, five-cycle and five-site path have higher threshold infima and cubic inverse-gap costs under uniform introduction, with the gap measured from each support's own infimum.

Waiting time introduces a second optimization. We minimize neutral times subject to a weak-selection invasion target, so the dynamics without selection are evaluated on contacts constrained by the prescribed weak-selection threshold. Under uniform introduction on the six-site path, a fraction $1/3$ of introductions reaches a contest between opposite domains at leading order. Its mean duration is $3q$, so the overall mean time to extinction or fixation is $q$. The central loss $6/q$ uses half the allowed gap at least contrast and approaches the full gap as contrast increases. Thus extra contrast lowers the leading mean time from $12/\epsilon$ toward $6/\epsilon$, as the first-passage bounds in Sec.~\ref{sec:v3-tolerance} confirm. The fastest design is initially mirror-symmetric. With cores of unbounded strength, unequal pin contacts would shorten the waiting time, but they lengthen the delay caused by leakage from finite cores. Once the cap exceeds $1476\epsilon^{-4}$ at leading order, compared with $656\epsilon^{-2}$ for feasibility, this competition splits the fastest design into two asymmetric designs related by reflection, a supercritical pitchfork of the contact geometry. A coordinate invariant under exchange of the two pins follows these branches through every larger finite cap in the small-gap regime. Most of the time reduction occurs before this bifurcation; the later asymmetric gain is smaller and calculable. At small selection strength and contrasts of order $\epsilon^{-2}$, the unconditional and fixation-conditioned mean times share one mirror-symmetric minimizer under the selected invasion target (Proposition~\ref{prop:v7-selected-time}). On the five-cycle, a different change in the optimal contact ratios determines the complete leading time--contrast frontier and the limiting distribution of successful-fixation times.

Finite selection and imperfect contacts draw on the same small invasion margin. Near least contrast, deviations of the pin contacts enter quadratically, whereas shortfalls in the strongest contacts enter linearly. Nonlinear selection adds a further penalty. Because a singleton's descendants on a path remain a contiguous interval, a twenty-state chain on the six-site path determines these corrections exactly. A hierarchy of observables ordered by polynomial degree organizes successive selection orders without factorizing correlations, and for linear fecundity their sum is a finite matrix resolvent. This inverse operator combines neutral evolution with repeated selection perturbations. It distinguishes convergence of the response expansion from the smaller range of selection strengths that actually favors cooperation.

The same calculation makes measurement a design problem. For uniform introduction, we estimate the unknown fixation advantage from independent completed invasions, retaining the introduced type and site. Within the mirror-symmetric six-site hierarchy, the leading observation optimum jointly chooses contacts and selection strength for linear or exponential fecundity. At a fixed relative contrast surplus above feasibility, the advantage is of order $\epsilon$ and the pin-outcome variances remain of order one. Fixed relative precision therefore needs of order $\epsilon^{-2}$ trials, whose mean duration is of order $\epsilon^{-1}$, giving a summed duration of order $\epsilon^{-3}$. Near the signal maximum a small decrease of the pin contact loses signal only quadratically while shortening trials linearly. The observation optimum therefore accepts a smaller signal to obtain shorter trials. Near least contrast the largest advantage is of order $\epsilon$ times the $3/2$ power of the scaled contrast surplus, so a surplus of order $\epsilon$ raises the summed duration to order $\epsilon^{-6}$. Common preparation costs shift the balance toward signal maximization. This distinction between producing an advantage and resolving it statistically is consequential when a favorable invasion bias is small.

Changes of support test which costs belong to the protection mechanism and which belong to a particular contact geometry. Within a two-parameter family of two triangles joined by a weak link, sharing access between both sites of each follower core reduces the sharp contrast coefficients to $576$ and $3645$ for uniform and birth-associated introduction. The leading uniform-introduction waiting curve is unchanged when contrast is normalized by each family's own feasibility cost. Under uniform introduction on the four-cycle, a compulsory diagonal of fixed weight preserves the limiting threshold three within the hierarchy studied here and raises its leading contrast coefficient by a factor $7/4$ at the weakest permitted diagonal weight. Biological applications require measured encounter frequencies~\cite{AllenGoreNowak2013,NadellDrescherFoster2016}, an offspring-routing mechanism and a specified source of new traits. The statistical-physics object is a finite absorbing Markov process whose collective slow states have design-dependent accessibility, lifetimes and response. Analytical reductions and computer-assisted exact polynomial sign arguments prove the stated global bounds; Appendix~\ref{app:v10-sign-construction} gives the reconstruction and arithmetic. Symbolic calculations used SageMath, SymPy and Singular~\cite{SageMathSoftware,SymPySoftware,SingularSoftware}.

The argument proceeds from the invasion experiment to its resource requirements. Section~\ref{b:sec:experiment} fixes the game, introduction rules, clock and protected-domain inputs. Section~\ref{sec:cost} proves global contact bounds and identifies their optimal geometry, including changes of resource and introduction rule. Section~\ref{sec:finite-cap} connects these asymptotic results to first promotion and the complete five-site finite-cap problem. Section~\ref{sec:v3-tolerance} uses the remaining contact budget to optimize waiting times and quantify design tolerance. Section~\ref{sec:v3-selection-window} determines how finite selection converts that tolerance into a measurable signal and an observation cost. Section~\ref{sec:v7-robustness} then tests which conclusions transfer under changed games, encounters and support. Readers focused on protection and contact cost can follow Secs.~\ref{b:sec:experiment} and \ref{sec:cost}; the measurement problem in Sec.~\ref{sec:v3-selection-window} builds on the same design variables.

The appendices complete these arguments. Appendix~\ref{app:v10-response} develops the response and changed-support calculations, and Appendix~\ref{app:v10-onsets} proves the five-cycle and reflected six-site first-promotion onsets. Appendix~\ref{app:v10-geometry} constructs the compatible contact coordinates and convex resource bounds used in the global optimizations of Appendix~\ref{app:v10-cost}. Appendix~\ref{app:v10-times} follows physical waiting times through their changes of optimal geometry. Appendix~\ref{app:v10-selection} connects contact tolerance, finite selection and observation, including preparation costs and rare introductions.

\section{The invasion experiment and its design}
\label{b:sec:experiment}\label{sec:model}
A favorable invasion threshold leaves a physical design question. We can change the relative
contacts, but must specify where the trait enters, which advantage is
required and how long an invasion is allowed to take. This section fixes these choices; Table~\ref{tab:III-notation} collects the notation used across the calculations. The contact budget measures a range of interaction and
replacement weights; the observation cost later counts the time needed
to resolve the resulting fixation advantage.

\begin{table*}[tp]
\caption{Principal notation. Unmarked response and time quantities use
uniform introduction. The contact and error coordinates concern the six-site path with
consecutive weights $(Q_L,q_L,1,q_R,Q_R)$; its reflected specialization is $(Q,q,1,q,Q)$. A gap $\epsilon$ is always measured from the threshold
floor of the support and introduction rule under discussion.}
\label{tab:III-notation}
\begin{ruledtabular}
\begin{tabular}{p{.12\textwidth}p{.35\textwidth}p{.12\textwidth}p{.35\textwidth}}
$w_{ij},s_i,P_{ij}$&Reciprocal weights, strengths and normalized copying/encounter kernel&
$n_i,f_i,F_i$&Binary trait, averaged payoff and reproductive propensity\\[3pt]
$b,c,r,\delta,\kappa$&Benefit, cost, ratio $r=b/c$, selection strength $\delta$ and intensity $\kappa=\delta c$&
$\mu^{\rm U},\mu^{\rm b},\zeta$&Uniform and birth-associated introduction; uniform fraction in their mixture\\[3pt]
$\vartheta_i$&Incoming copying weight (temperature); $\vartheta_i/N$ is the neutral overwrite rate&
$\pi_i,Z$&Reproductive value (neutral singleton fixation probability) and inverse-strength normalizer\\[3pt]
$\rho_C^\mu,\rho_D^\mu$&Fixation probabilities in the paired introduction experiment&
$\Delta^\mu,\mathcal S^\mu$&Paired fixation advantage and weak susceptibility $\mathcal S^\mu=\partial_\delta\Delta^\mu|_0$\\[3pt]
$D_{ij}^\mu,\mathsf M$&Integrated neutral disagreement and killed ancestral-pair matrix&
$\gamma_{ij},\mathsf B$&Ancestral conductance and the signed benefit matrix\\[3pt]
$K^\mu,J^\mu,R^\mu$&Cost and benefit contractions; threshold $R^\mu=K^\mu/J^\mu$ on $J^\mu>0$&
$\begin{gathered}H(w),H,R_*^\mu\\C_P(w)\end{gathered}$&Actual contact contrast, cap (budget), threshold infimum on a fixed support, and inverse smallest positive copying probability\\[3pt]
$T_{\rm abs},T_{\rm fix}$&Mean absorption and success-conditioned time, in global attempts&
$\begin{gathered}M,T_{\rm total}\\ \widehat\Delta_{\rm str},r_{\rm prep}\end{gathered}$&Completed-trial count, summed run duration, stratified advantage estimate, and preparation time per trial\\[3pt]
$q,Q,u,v$&Reflected access/core weights; $u=1/q$, $v=q/Q$&
$p,y$&Pin proportion $p=q_R/(q_L+q_R)$ and imbalance $y=(q_L-q_R)/(q_L+q_R)$\\[3pt]
$H_*,q_*$&Uniform least contrast $H_*=H_{\min}^{\rm U}$ and its reflected pin contact at target $1+\epsilon$&
$d,\mathfrak r$&Scaled surplus $d=\epsilon^2(H-H_*)$ and relative surplus $\mathfrak r=(H-H_*)/H_*$\\[3pt]
$\alpha_L,\alpha_R$&Scaled nonnegative deficits of the two capped core weights&
$\xi_s,\xi_a$&Scaled common shift and imbalance of the pin contacts\\[3pt]
$\delta_c,\lambda,C_0$&Centered strength, intensity $\lambda=c\delta_c$, and the map-dependent quadratic margin penalty&
$\Lambda$&Large-budget coordinate $\Lambda=H\epsilon^4$ for the time-optimum bifurcation\\[3pt]
$\eta,p_H$&Scaled cap $\eta=H\epsilon^{p_H}$; $p_H=2$ on $P_6$ and $3$ on $C_4,C_5,P_5$&
$\theta,\theta_5$&Leading scaled neutral absorption time; $\theta_5(\eta)$ is the five-cycle frontier\\[3pt]
$\begin{gathered}H_{0,\mathsf G}^{\mu}\\H_{0,P_6,\refl}^{\mu}\end{gathered}$&First-promotion cap on support $\mathsf G$ for rule $\mu$; $\refl$ restricts to reflected paths&
$\omega_{\rm lay}$&Positive layer parameter, $d=\omega_{\rm lay}\epsilon$, or $H-H_*=\omega_{\rm lay}/\epsilon$
\end{tabular}
\end{ruledtabular}
\end{table*}

\subsection{Reciprocal contacts and the copying clock}
\label{b:sec:model}\label{sec:update}
Consider a finite connected, loopless undirected graph with $N\ge2$
sites. Each site contains one individual with inherited trait $n_i=1$
for cooperation or $n_i=0$ for defection. A retained edge has reciprocal
weight $w_{ij}=w_{ji}>0$; absent edges and self-contacts have weight zero.
The strengths and normalized contacts are
\begin{equation}
 s_i=\sum_jw_{ij},\qquad P_{ij}=\frac{w_{ij}}{s_i},\qquad
 \sum_jP_{ij}=1.
 \label{b:eq:kernel}\labelalias{eq:kernel}{b:eq:kernel}
\end{equation}
The same kernel averages benefit encounters and routes offspring.
A cooperator pays cost $c>0$ and gives benefit $b\ge0$ to its partner,
so its configuration-dependent payoff is
\begin{equation}
 f_i(n)=-cn_i+b(Pn)_i.
 \label{b:eq:payoff}\labelalias{eq:payoff}{b:eq:payoff}
\end{equation}
The term $(Pn)_i$ is the cooperative fraction among the encounters
sampled by site $i$. It describes a fixed encounter budget per recipient. A fixed quantity secreted by each cooperator and transported along its outgoing contacts would involve $P^{\mathsf T}n$ and define a different benefit model. With the other traits fixed, changing from
cooperation to defection gains $c$. Nevertheless, copying can let an
introduced lineage change its descendants' surroundings before the
population reaches consensus. The present question concerns the optimal realization of this mechanism under finite contact budgets.

Fecundity is the relative propensity to reproduce,
$F_i(n)=\phi(\delta f_i(n))>0$, where $\delta\ge0$ is selection strength
and $\phi(0)=\phi'(0)=1$. We use linear fecundity $\phi(x)=1+x$ and
exponential fecundity $\phi(x)=e^x$. Write $r=b/c$ and $\kappa=\delta c$.
The linear map requires $0\le\kappa<1$ to stay positive on every
configuration; the exponential map is positive for every finite
$\kappa\ge0$. Payoffs are averaged before applying $\phi$, with no
additional sampling of encounter rewards.

A Poisson process of rate one counts global reproduction attempts.
At an attempt the source is chosen with probability proportional to
its fecundity, and site $j$ copies source $i$ with probability $P_{ij}$.
For an observable $g$, the backward generator is
\begin{equation}
 (\mathcal A g)(n)=\sum_{i,j}
 \frac{F_i(n)P_{ij}}{\sum_kF_k(n)}
 \bigl[g(n^{j\leftarrow i})-g(n)\bigr],
 \label{b:eq:generator}\labelalias{eq:backward}{b:eq:generator}\labelalias{eq:rates}{b:eq:generator}
\end{equation}
where $n^{j\leftarrow i}$ is the configuration after the copy. Its neutral specialization is $\mathcal A_0=\mathcal A|_{\delta=0}$.
This source-first Birth--death rule is denoted Bd; the selected step
precedes replacement. Copies between identical traits still count as
attempts. Time thus compares designs at equal numbers of reproductive
opportunities. One sweep means $N$ global attempts, so at attempt-clock time $t$ the sweep time is $t/N$. Conversion to biological generations requires a calibration of the reproduction process.
The process is a finite Markov chain whose two consensus configurations are
absorbing.

The normalization in Eqs.~\eqref{b:eq:payoff} and \eqref{b:eq:generator} is part of the model. Accumulated payoffs would replace $f_i$ by $s_i f_i$, while strength-proportional reproduction would replace copying propensities $F_iP_{ij}$ by $F_iw_{ij}$. The distinction between averaged and accumulated payoffs is consequential on heterogeneous populations~\cite{MaciejewskiFuHauert2014}. Either change alters the selection response; the latter also makes neutral reproductive values uniform. The resource bounds below retain averaged payoffs and parent-normalized copying throughout. Section~\ref{sec:III-transfers} and Appendix~\ref{app:III-finite-comparisons} examine an accumulated-payoff example.

\subsection{Introduction, fixation and the measured advantage}
\label{b:sec:observables}\label{sec:observables}
Let $h_C(n)$ be the probability of eventual all-cooperator consensus
from configuration $n$. It is the fixation probability, or committor,
satisfying $\mathcal A h_C=0$ on mixed configurations with
$h_C(\mathbf0)=0$ and $h_C(\mathbf1)=1$. Connectedness and positive
fecundity make every mixed configuration transient.

We compare a cooperator introduced into defectors with a defector
introduced into cooperators. Both use the same selection-independent
introduction rule $\mu$, a probability distribution over sites. If $e_i$ is the configuration
with its sole cooperator at $i$, define
\begin{equation}
 \begin{aligned}
 \rho_C^\mu&=\sum_i\mu_i h_C(e_i),\\
 \rho_D^\mu&=\sum_i\mu_i[1-h_C(\mathbf1-e_i)],\\
 \Delta^\mu&=\rho_C^\mu-\rho_D^\mu,\qquad
 \mathcal S^\mu=\left.\partial_\delta\Delta^\mu\right|_{\delta=0}.
 \end{aligned}
 \label{b:eq:paired}\labelalias{eq:rho}{b:eq:paired}\labelalias{eq:response}{b:eq:paired}
\end{equation}
Positive $\Delta^\mu$ means that cooperation wins this paired invasion
comparison; $\mathcal S^\mu$ is its weak susceptibility. At finite selection this differs from comparing $\rho_C^\mu$
with its neutral baseline. For the common selection-independent singleton
introduction rule used here, type complementation gives
$\mathcal S^\mu=2\left.\partial_\delta\rho_C^\mu\right|_0$,
so the paired and baseline comparisons have the same first-order sign whenever $\mathcal S^\mu\ne0$. Section~\ref{b:sec:response-input} states the response conventions and cites their derivation.

Uniform introduction chooses sites equally. Birth-associated
introduction places a mutation at the recipient of an offspring in a
homogeneous population. Their distributions are
\begin{equation}
 \mu_i^{\rm U}=\frac1N,\qquad
 \mu_i^{\rm b}=\frac{\vartheta_i}{N},\qquad
 \vartheta_i=\sum_kP_{ki}.
 \label{b:eq:preparations}\labelalias{eq:v7-preparation}{b:eq:preparations}
\end{equation}
The incoming copying weight $\vartheta_i$ is called the site's
temperature in evolutionary graph theory~\cite{LiebermanHauertNowak2005};
it is unrelated to thermodynamic temperature. The second rule is also
called temperature initialization~\cite{AdlamChatterjeeNowak2015,AllenEtAl2021}.
In either homogeneous background all sources have the same fecundity,
so $\mu^{\rm b}$ is independent of selection strength even though it
depends on the contact design. We also consider
$\mu^\zeta=\zeta\mu^{\rm U}+(1-\zeta)\mu^{\rm b}$, where $0\le\zeta\le1$
is the uniform fraction. Unmarked observables use uniform introduction. Site-independent switching or controlled placement realizes the uniform rule; immigration does so only when its landing distribution is uniform. Mutation during reproduction realizes the birth-associated rule specified above.

For a uniformly introduced cooperator, let $\tau$ be the time to
either consensus state. The two time objectives are
\begin{equation}
 T_{\rm abs}=\mathbb E[\tau],\qquad
 T_{\rm fix}=\mathbb E[\tau\mid C\text{ fixes}].
 \label{b:eq:times}\labelalias{eq:times-definition}{b:eq:times}
\end{equation}
Failed introductions contribute to the first mean. The second concerns
successful invasions. Both definitions allow selection. Section~\ref{sec:v4-time} optimizes neutral times under a weak-selection target, and Proposition~\ref{prop:v7-selected-time} optimizes the selected six-site times under $\Delta\ge0$ at small selection and bounded scaled budgets.

\subsection{Exact response as the design objective}
\label{b:sec:response-input}
At neutrality the fixation weight of site $i$ is its reproductive value. The corresponding conserved mean, $\Phi$, is the fixation probability from any configuration,
\begin{equation}
 \begin{gathered}
 Z=\sum_i s_i^{-1},\qquad \pi_i=\frac{s_i^{-1}}Z,\\
 h_C^0(n)=\Phi(n)=\sum_i\pi_i n_i.
 \end{gathered}
 \label{b:eq:reproductive}\labelalias{eq:reproductive}{b:eq:reproductive}
\end{equation}
Thus $\rho_C^{0,\mu}=\rho_D^{0,\mu}=\rho_0^\mu=\sum_i\mu_i\pi_i$;
uniform introduction gives $1/N$. Slowly overwritten sites carry large
reproductive value in the protected geometries considered below.

The weak response follows from neutral ancestral pairs. Start with a
single cooperator at a site drawn from $\mu$. The integrated disagreement is
\begin{equation}
 D_{ij}^\mu=\int_0^\infty
 \mathbb E_0^\mu[(n_i(t)-n_j(t))^2]\,\mathrm dt,
 \qquad D_{ii}^\mu=0.
 \label{b:eq:disagreement}\labelalias{eq:disagreement}{b:eq:disagreement}
\end{equation}
Tracing a received copy backward to its source gives the exact pair
system
\begin{equation}
 \begin{aligned}
 &(\vartheta_i+\vartheta_j)D_{ij}^\mu
 -\sum_k(P_{ki}D_{kj}^\mu+P_{kj}D_{ik}^\mu)\\
 &\hspace{20mm}=N(\mu_i+\mu_j),\qquad i\ne j.
 \end{aligned}
 \label{b:eq:pairs}\labelalias{eq:pair-system}{b:eq:pairs}
\end{equation}
The pair matrix on the left, denoted $\mathsf M$, describes two ancestral lineages killed, or removed from the pair process, when they meet. Its inverse, the pair Green function, is finite on a connected positive-weight
support. The common factor $1/N$ in the physical ancestral clock has
been removed from $\mathsf M$ and retained on the right. The response
construction follows the coalescence approach of
Refs.~\cite{AllenEtAl2017,AllenEtAl2021,McAvoyAllen2021}. The source-first
normalization, introduction-dependent factors and type-complementation identity above are derived in
Ref.~\cite[Sec.~III\,C and Appendix~A; Sec.~V\,A and Appendix~B\,2]{ChowdhuryProtectedDomains},
which also develops the neutral operator representation used in Sec.~\ref{sec:III-filtered-observables}.

\begin{figure}[htbp]
\centering
\begin{tikzpicture}[x=1.1cm,y=1cm,font=\small,
 site/.style={circle,draw,minimum size=5.7mm,inner sep=0pt},
 pin/.style={site,fill=figureblue,draw=figureblue,text=white},
 core/.style={line width=1.8pt},
 access/.style={line width=1pt,draw=figureblue}]
\node[site] (p1) at (0,0) {$1$};
\node[site] (p2) at (1.15,0) {$2$};
\node[pin] (p3) at (2.30,0) {$3$};
\node[pin] (p4) at (3.45,0) {$4$};
\node[site] (p5) at (4.60,0) {$5$};
\node[site] (p6) at (5.75,0) {$6$};
\draw[core] (p1)--node[above=2pt] {$Q_L$}(p2);
\draw[access] (p2)--node[above=2pt] {$q_L$}(p3);
\draw (p3)--node[above=2pt] {$1$}(p4);
\draw[access] (p4)--node[above=2pt] {$q_R$}(p5);
\draw[core] (p5)--node[above=2pt] {$Q_R$}(p6);
\node[font=\footnotesize] at (.575,-.56) {follower core};
\node[font=\footnotesize] at (2.875,-.56) {dynamic pins};
\node[font=\footnotesize] at (5.175,-.56) {follower core};
\node[font=\footnotesize,align=center] at (2.875,-1.08)
 {Reflected hierarchy: $Q_L=Q_R=Q$, $q_L=q_R=q$, $1\ll q\ll Q$};
\end{tikzpicture}
\caption{The six-site contact design. Strong outer contacts suppress
copies from the follower cores into the two central pins, which continue
to send copies outward and align their followers. The unit central contact
couples the domains. Sites 1, 2 and 5, 6 are followers, aligned through copying from their nearest pin; each pin and its two followers form a domain. A pin is dynamically protected, and its trait can change like that of any other site.
Uniform introduction reaches each pin with probability $1/6$.
Birth-associated introduction reaches site $i$ with probability
$\vartheta_i/6$, where $\vartheta_i=\sum_jP_{ji}$ is its incoming copying
weight. The neutral overwrite rate is also $\vartheta_i/6$ in the global
attempt clock. The diagram separates access contacts $q_L,q_R$ from core
contacts $Q_L,Q_R$ before they are optimized.}
\label{fig:III-contact-geometry}
\end{figure}

Set the ancestral conductance $\gamma_{ij}=w_{ij}/(s_is_j)$ and let $L_\gamma$ be its symmetric
graph Laplacian, with off-diagonal entries $-\gamma_{ij}$ and diagonal
entries $\sum_j\gamma_{ij}$. The cost and benefit contractions are
\begin{equation}
 \begin{gathered}
 \mathsf B=\tfrac12(P^{\mathsf T}L_\gamma+L_\gamma P),\\
 K^\mu=\sum_{i<j}\gamma_{ij}D_{ij}^\mu,\qquad
 J^\mu=-\sum_{i<j}\mathsf B_{ij}D_{ij}^\mu,\\
 \mathcal S^\mu=\frac{2}{NZ}(bJ^\mu-cK^\mu).
 \end{gathered}
 \label{b:eq:response}\labelalias{eq:JK}{b:eq:response}\labelalias{eq:criterion}{b:eq:response}
\end{equation}
The positive $K^\mu$ measures the integrated cost of trait interfaces.
The signed $J^\mu$ measures the benefit of different cooperative
environments across them. On $J^\mu>0$, define
\begin{equation}
 R^\mu=\frac{K^\mu}{J^\mu},\qquad
 \mathcal S^\mu>0\quad\Longleftrightarrow\quad r>R^\mu.
 \label{b:eq:threshold}\labelalias{eq:threshold}{b:eq:threshold}
\end{equation}
A weighting with $J^\mu>0$ is a weak promoter. If $J^\mu\le0$,
nonnegative benefit cannot offset the positive cost to first order.
The response coefficients $\chi_b^\mu=2J^\mu/(NZ)$ and
$\chi_c^\mu=2K^\mu/(NZ)$ separate threshold from amplitude.

Changing the introduction rule changes the pair source vector, the right-hand side of Eq.~\eqref{b:eq:pairs}.
Uniform introduction weights ancestral meeting time; birth-associated
introduction weights the number of ancestral jumps, including the last
coalescing jump. These are different
averages of the same copying dynamics. Their contractions mix linearly under $\mu^\zeta$. Equation~\eqref{eq:v10-gap-mixture} below writes the normalized gap $(K^\zeta-J^\zeta)/K^\zeta$ of the mixture as the average of the two pure-rule gaps with weights $\zeta K^{\rm U}$ and $(1-\zeta)K^{\rm b}$, so $R^\zeta$ can depend nonlinearly on $\zeta$.
For finite positive contacts, one adjoint pair solve determines the
sensitivity to every contact, including the contact dependence of the birth-associated
introduction distribution (Appendix~\ref{app:v10-adjoint-contact}). For the six-site hierarchy,
Sec.~\ref{sec:v4-toric} constructs compatible contact ratios in which
the uniform-introduction threshold and its derivatives remain regular
as the original weights diverge.

\subsection{Contact resources and constrained targets}
\label{b:sec:resources}
Common rescaling of all weights leaves $P$, the introduction distributions and the
physical process unchanged. The available contact range is therefore
measured by
\begin{equation}
 H(w)=\frac{\max_{e\in\mathcal E}w_e}{\min_{e\in\mathcal E}w_e},
 \qquad H(w)\le H,
 \label{b:eq:contrast}\labelalias{eq:contrast}{b:eq:contrast}
\end{equation}
where $\mathcal E$ is the retained edge set and $H\ge1$ the imposed cap, also called the contact budget. Thus $H(w)$ is the contrast of a design, whereas $H$ bounds the admissible designs.
An absent edge is not included as a zero weight in this ratio. Contrast
is the dynamic range of relative encounter and replacement weights.
An energetic or material cost would require an additional relation
between these weights and the physical mechanism that maintains them. Experiments controlling offspring-routing probabilities directly lead to another resource, the inverse smallest positive entry of $P$. Section~\ref{sec:III-kernel-resource} derives its sharp small-gap costs and shows explicitly how the resource choice changes the absolute exponents.

Write $P_N$ and $C_N$ for paths and cycles on $N$ sites. For each support and introduction rule, $R_*^\mu$ is the infimum of $R^\mu$ over all positive-weight weak promoters, also called the threshold floor. Table~\ref{tab:III-input-floors} collects the threshold infima used in the resource comparisons.
\begin{table}[tbp]
\caption{Weak-selection threshold infima $R_*^\mu$ over all positive contact weights on each support, under uniform ($U$) and birth-associated ($\mathrm b$) introduction. The value $+\infty$ denotes absence of weak promotion. Each finite infimum is approached by positive weightings and is unattained at finite weights.}
\label{tab:III-input-floors}
\begin{ruledtabular}
\begin{tabular}{lcccc}
Support&$C_4$&$C_5$&$P_5$&$P_6$\\
\hline
$R_*^{\rm U}$&$3$&$3/2$&$7$&$1$\\
$R_*^{\rm b}$&$3$&$3/2$&$+\infty$&$1$
\end{tabular}
\end{ruledtabular}
\end{table}
Under uniform introduction, the diamond obtained by adding one diagonal to $C_4$ also has sharp floor three. These support bounds are derived in
Ref.~\cite[Propositions~4--6 and Appendix~D\,5]{ChowdhuryProtectedDomains}.

For a specified support and introduction rule with finite threshold infimum, the target is
$R^\mu\le R_*^\mu+\epsilon$, with $\epsilon>0$ measured from that
support's own threshold infimum. Let $H_{\min}^\mu(\epsilon)$ be the infimum of caps admitting a design that meets this target. On the six-site path under uniform introduction we also write $H_*(\epsilon)=H_{\min}^{\rm U}(\epsilon)$. We ask for this least contrast, for the shortest
time at a prescribed contrast, and for the tolerance and observable
advantage of the resulting designs. The first-promotion problem
locates the boundary of caps admitting any $J^\mu>0$. At that onset the benefit coefficient vanishes and the threshold diverges, so every finite threshold target needs a larger cap (Sec.~\ref{sec:finite-cap}). At fixed $r$ the closed weak target can be written
$rJ^\mu-K^\mu\ge0$, which automatically implies $J^\mu>0$.
The selected target uses the actual $\Delta^\mu$ of
Eq.~\eqref{b:eq:paired}. Its neutral extension is the limit of
$\Delta^\mu/\kappa$, since $\Delta^\mu$ itself is zero at neutrality.

\subsection{The protected-domain input}
\label{b:sec:domain-input}
The resource bounds below use the protected-domain reduction recalled here. The six-site path is the smallest path approaching threshold one. Interpreting a site as a deme would require a further reduction of within-deme dynamics to the same copying generator.
Its reflected hierarchy has consecutive contacts
\begin{equation}
 \begin{gathered}
 (Q,q,1,q,Q),\qquad u=q^{-1},\qquad v=q/Q,\\
 (u,v)\longrightarrow(0,0).
 \end{gathered}
 \label{b:eq:p6-input}\labelalias{eq:p6-weights}{b:eq:p6-input}
\end{equation}

Figure~\ref{fig:III-contact-geometry} identifies the contact roles. The exact copying probabilities into a pin across the central contact and from its follower core are $u/(1+u)$ and $v/(1+v)$, respectively. Thus $u$ and $v$ measure weak incoming replacement channels; leakage below refers to replacement through these channels.
The two central individuals rarely receive copies but continue to send
them to their followers. They are dynamically protected sites, or
pins. Each pin aligns its two followers into a triple before the two triples compete across
the unit bridge, the contact between the pins. A cooperative triple has leading payoff $b-c$,
while a defector triple has payoff zero. The protection emerges from
normalized reciprocal contacts; neither type is fixed externally.

Uniform singleton introduction reaches each mixed domain state with
probability $1/6$, while total probability $2/3$ disappears during
formation. Put $F_{\rm dom}=\phi(\delta(b-c))$, $d_0=\phi(0)$ and
$h_{\rm dom}=F_{\rm dom}/(F_{\rm dom}+d_0)$. In slow time $t/q$, each mixed state exits at
rate $1/3$ and cooperation wins with probability $h_{\rm dom}$.
Consequently
\begin{equation}
 \begin{gathered}
 \Delta=\frac{F_{\rm dom}-d_0}{3(F_{\rm dom}+d_0)}+O(u+v),\\
 T_{\rm abs}=q[1+O(u+v)],\quad
 T_{\rm fix}=3q[1+O(u+v)].
 \end{gathered}
 \label{b:eq:p6-domain-observables}\labelalias{eq:domain-probabilities}{b:eq:p6-domain-observables}\labelalias{eq:domain-times}{b:eq:p6-domain-observables}
\end{equation}
These are fixed-size hierarchy limits, uniform on compact parameter
sets with positive fecundities. At neutrality, an aligned contest ends at leading order when one pin copies across the central contact and the receiving domain rapidly realigns. The two central copying rates sum to $1/(3q)+o(q^{-1})$, so the contest's mean duration is $3q$ to leading order, while the total entrance probability $1/3$ gives the unconditional mean $q$. Follower-founded contests and reversals after a crossing enter the finite-contact corrections.
The probabilities of reaching the mixed domain states and the rate of leaving them are separate ingredients of the
time estimates. Near threshold one the finite-contact corrections
determine the sign even when the leading domain payoff difference is small.

The other supports test different collective contests. The four- and
five-cycles have two protected sites sharing a fringe of fluctuating followers;
the five-site path has a protected triple beside a dimer whose founder
must first win a selected establishment contest. The six-site reduction,
its site-resolved fixation probabilities and mean durations, and the controlled reductions for these smaller supports are derived in
Ref.~\cite[Secs.~III\,E--III\,F and Appendices~C\,1 (especially C\,1\,b), C\,3, C\,4, C\,7--C\,9]{ChowdhuryProtectedDomains}. The site-resolved reconstruction also controls follower introductions in Appendix~\ref{app:v7-terminal}. We retain the separate
threshold floors and entrance factors when comparing resources and
times. The next section asks whether every design close to the
six-site floor must realize the protected hierarchy, and what that
requirement costs.

\section{Geometry and cost of optimal contacts}
\label{sec:cost}
\begin{figure*}[tp]
\centering
\begin{tikzpicture}[x=1cm,y=1cm,>=Stealth,
 font=\fontsize{9}{11}\selectfont,
 box/.style={draw=black!40,line width=.5pt,rounded corners=2pt,
   fill=white,align=center,inner xsep=5pt,inner ysep=5pt},
 bluebox/.style={box,draw=figureblue,fill=figureblue!3},
 orangebox/.style={box,draw=figureorange,fill=figureorange!4},
 flow/.style={->,line width=.8pt,draw=black!60,rounded corners=3pt},
 uniformflow/.style={flow,draw=figureblue}]

\draw[black!15] (5.10,-.18)--(5.10,8.40);
\draw[black!15] (11.00,-.18)--(11.00,8.40);
\node[anchor=west,font=\bfseries] at (0,8.72)
  {(a) Introduction and resources};
\node[anchor=west,font=\bfseries] at (5.48,8.72)
  {(b) Geometry and margin};
\node[anchor=west,font=\bfseries] at (11.36,8.72)
  {(c) Observation time};

\node[bluebox,text width=1.60cm,minimum height=1.28cm] (unif) at (.94,7.16)
  {Uniform sites\\[2pt]$\mu_i^{\rm U}=1/6$};
\node[bluebox,text width=1.98cm,minimum height=1.28cm] (ucost) at (3.63,7.16)
  {$H_{\min}^{\rm U}$\\[3pt]$\sim656\epsilon^{-2}$};
\node[orangebox,text width=1.60cm,minimum height=1.28cm] (birth) at (.94,4.90)
  {Incoming births\\[2pt]$\mu_i^{\rm b}=\vartheta_i/6$};
\node[orangebox,text width=1.98cm,minimum height=1.28cm] (bcost) at (3.63,4.90)
  {$H_{\min}^{\rm b}$\\[3pt]$\sim4050\epsilon^{-3}$};
\draw[uniformflow] (unif)--(ucost);
\draw[flow,draw=figureorange] (birth)--(bcost);
\node[align=center,text width=4.50cm] at (2.40,2.70)
  {Same copying dynamics\\and target $R^\mu\le1+\epsilon$.};
\node[align=center,text width=4.50cm] at (2.40,1.34)
  {A uniform fraction $\zeta=O(\epsilon)$\\connects the two resource regimes.};

\node[bluebox,text width=4.50cm,minimum height=1.67cm] (geometry) at (8.03,7.16)
  {Compatible contact ratios\\[4pt]
   \mbox{\setlength{\arraycolsep}{1.9pt}$\displaystyle
   \operatorname{rank}\!\begin{pmatrix}
      q_L^{-1}&q_R/Q_L&q_R/Q_R\\q_R^{-1}&q_L/Q_L&q_L/Q_R
    \end{pmatrix}=1$}\\[4pt]
   Regular response};
\draw[uniformflow] (ucost.east)--(geometry.west);
\node[bluebox,text width=4.50cm,minimum height=1.43cm] (margin) at (8.03,4.90)
  {Available signal margin\\[3pt]
   $d-\mathcal L_{\rm con}-C_0\lambda^2$\\[3pt]
   $d=\epsilon^2[H-H_*(\epsilon)]>0$};
\draw[uniformflow] (geometry.south)--(margin.north);
\node[box,text width=4.50cm,minimum height=1.30cm] (clock) at (8.03,.60)
  {Trial duration\\[2pt]
   Slow domain exchange\\[2pt]
   $T_{\rm abs}\sim12/\epsilon$};
\draw[flow] (geometry.south west)--++(-.23,0)
  |-(clock.west);
\node[align=center,text width=3.65cm,text=black!75] at (8.03,2.58)
  {The same hierarchy fixes\\the physical clock.};

\node[align=center,text width=4.93cm,text=figureblue] at (14.08,7.16)
  {Uniform introduction\\near the least-contrast design\\[3pt]
   Joint limit $\epsilon,d\to0$, $d>0$};
\node[bluebox,text width=4.93cm,minimum height=1.43cm] (signal) at (14.08,4.90)
  {Best signal at zero contact error\\[3pt]
   $\displaystyle
    \Delta_{\rm peak}\sim\frac{\epsilon d^{3/2}}
                                  {11808\sqrt{3C_0}}$};
\draw[uniformflow] (margin.east)--(signal.west);
\node[box,text width=4.93cm,minimum height=1.45cm] (sampling) at (14.08,2.63)
  {Balanced completed trials\\[2pt]
   $\operatorname{Var}(\widehat\Delta_{\rm str})\sim1/(3M)$\\[2pt]
   Fixed signal-to-noise ratio\\
   $M\propto\Delta_{\rm peak}^{-2}$};
\draw[flow] (signal.south)--(sampling.north);
\node[box,text width=4.93cm,minimum height=1.30cm] (observation) at (14.08,.60)
  {Run count $\times$ mean duration\\[3pt]
   $\mathbb E[T_{\rm total}]\asymp\epsilon^{-3}d^{-3}$\\[3pt]
   $d=\omega_{\rm lay}\epsilon\ \Longrightarrow\ \mathbb E[T_{\rm total}]\asymp\epsilon^{-6}$};
\draw[flow] (sampling.south)--(observation.north);
\draw[flow] (clock.east)--(observation.west);
\path[use as bounding box] (-.10,-.21) rectangle (16.90,8.97);
\end{tikzpicture}
\caption{From contact resources to an observable invasion advantage on the
six-site path, at target benefit--cost ratio $b/c=1+\epsilon$ with
$\epsilon\downarrow0$. (a) Uniform and birth-associated introduction require
different least caps $H_{\min}^{\rm U}$ and $H_{\min}^{\rm b}$ on contrast $H(w)=\max w/\min w$, even though they share the
threshold infimum one. Here $\vartheta_i$ is the incoming copying weight,
and $\zeta$ is the uniform fraction in a mixture of the two introduction rules.
(b,c) The blue continuation concerns uniform introduction near its
least-contrast design, with consecutive weights $(Q_L,q_L,1,q_R,Q_R)$ and
$H_*(\epsilon)=H_{\min}^{\rm U}(\epsilon)=656\epsilon^{-2}+O(\epsilon^{-1})$. The scaled surplus is $d=\epsilon^2[H-H_*(\epsilon)]$. The matrix contains six ratios of these contacts; its three $2\times2$ minors enforce their compatibility. After a common monomial is removed, the response denominator stays nonzero at the small-ratio limit, so derivatives remain controlled.
At the reflected optimum,
contact errors reduce the leading margin by
$\mathcal L_{\rm con}=(\alpha_L+\alpha_R)/2+(41/9)(\xi_s^2+\xi_a^2)$, where the $\alpha$'s
are scaled core deficits and the $\xi$'s are scaled pin shifts.
The centered selection intensity is $\lambda$;
$C_0=24368/9$ for linear and $14296/9$ for exponential fecundity.
Optimizing selection at zero contact error gives the indicated peak.
Independent completed trials are balanced across the twelve introduction-type and site classes;
$M$ is their total number, $T_{\rm total}$ their summed duration and $\widehat\Delta_{\rm str}$ the stratified
estimate of the paired fixation advantage. The durations count global
copying attempts, including ineffective copies, with negligible reset overhead. The observation scaling
holds in the joint limit $\epsilon,d\to0$, $d>0$, at fixed signal-to-noise ratio;
the final line takes $d=\omega_{\rm lay}\epsilon$ with fixed $\omega_{\rm lay}>0$.
This scaling concerns estimation of unknown site-specific terminal probabilities from completed-trial outcomes. Trajectory data or a known microscopic parameter model change the information available, the likelihood and hence the applicable observation bound (Sec.~\ref{sec:discussion}).}
\label{fig:III-resource-to-signal}
\end{figure*}
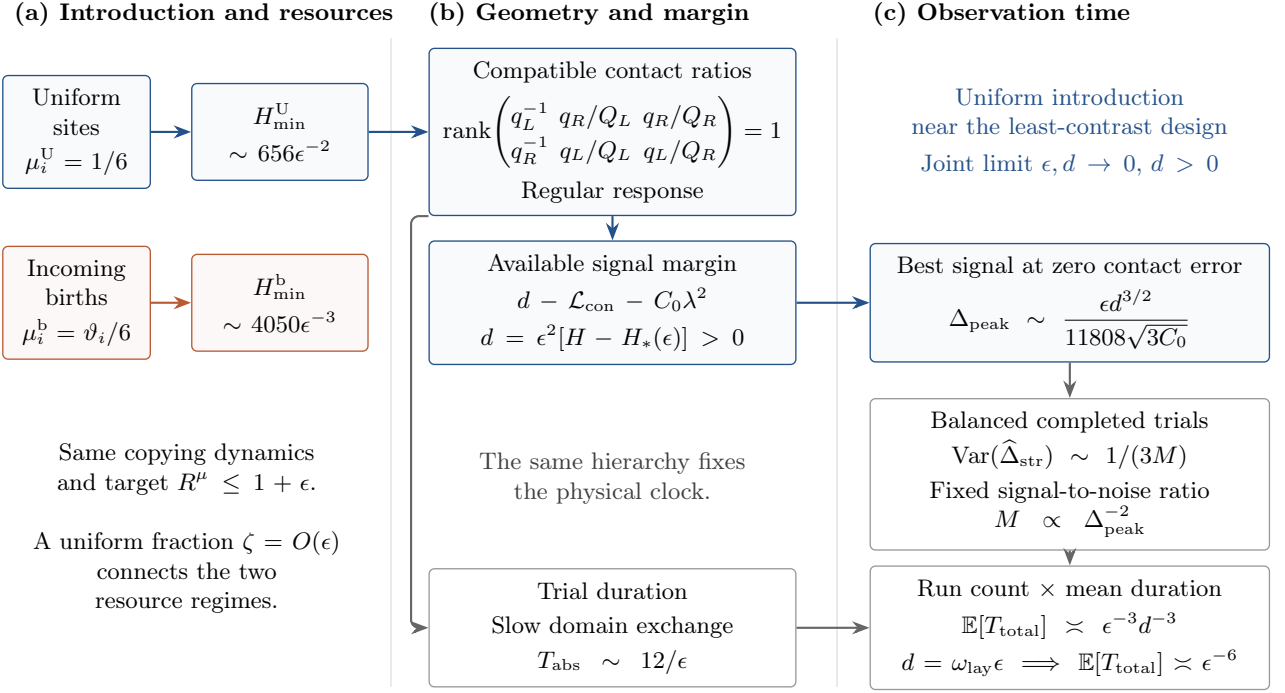

The sharp thresholds require singular ratios of contact strengths.
Common scaling leaves the kernel unchanged, so the physical design
space is $\mathbb R_{>0}^{|\mathcal E|}/\mathbb R_{>0}$. We constrain its contact contrast $H(w)$ from Eq.~\eqref{eq:contrast}.
Figure~\ref{fig:III-resource-to-signal} traces how this resource enters the invasion experiment. For the family $\mathcal W$ of positive weightings on a fixed support, define
\begin{equation}
 R_{\min}^{\mathcal W}(H)=
 \inf_{\substack{w\in\mathcal W,\ H(w)\le H\\J(w)>0}}R(w).
 \label{eq:constrained-threshold}
\end{equation}
In the stated limits, $\sim$ means that the ratio tends to one, whereas $\asymp$ means that it remains bounded above and below by positive constants.
Two later features, saturation of the strongest weights at the cap (Sec.~\ref{sec:v3-global-optimum}) and the split of contact errors into quadratic and linear directions (Sec.~\ref{sec:finite-budget-fold-v2}), reflect this choice of resource functional. We first force the six-site hierarchy globally, then compare the contact scales required by the different supports and introduction rules.

\subsection{A global bound forces the domain hierarchy}
\label{sec:contrast-design}\label{sec:asymmetric-entry-v2}
Normalize the six-site weights to $(Q_L,q_L,1,q_R,Q_R)$, without assuming
their order. The outer core weights $Q_L,Q_R$ join the two sites of each follower pair; the access weights $q_L,q_R$ connect these pairs to their respective pins. Clearing a common positive denominator in the cost and benefit contractions and removing common polynomial factors gives $R=U/V$, with $U>0$ and $\operatorname{sgn}V=\operatorname{sgn}J$. Put $\mathcal C=U-V$. Each exponent vector specifies the powers of $(Q_L,q_L,q_R,Q_R)$ in a monomial. Their convex hull is the Newton polytope; its boundary identifies terms that can dominate when contact ratios separate.
Exactly one monomial of $U$ is absent from the positive polynomial $\mathcal C$,
\begin{equation}
 \mathfrak m_\star=648Q_L^{10}q_L^8q_R^8Q_R^{10}.
 \label{eq:v4-newton-vertex}
\end{equation}
The six first-order exponent shifts from $\mathfrak m_\star$ give the terms of lowest degree in the small contact ratios. Dividing these terms of $\mathcal C$ by $\mathfrak m_\star$ gives
\begin{equation}
 F=\frac{3}{q_L}+\frac{3}{q_R}+
       \frac{53q_L+29q_R}{6Q_L}+\frac{29q_L+53q_R}{6Q_R}.
 \label{eq:v3-main-F}
\end{equation}
The exact coefficient inequalities 
\begin{equation}
 71\mathcal C-18(U-\mathfrak m_\star)\ge_{\rm coeff}0,\qquad \mathcal C-\mathfrak m_\star F\ge_{\rm coeff}0
 \label{eq:v4-global-remainders}
\end{equation}
hold with nonzero remainders; $\ge_{\rm coeff}0$ means every coefficient
is nonnegative. They imply $V/\mathcal C<53/18+1/F$, and on $J>0$,
\begin{equation}
 R-1>\frac{F}{1+53F/18}\qquad(J>0).
 \label{eq:v3-main-global-bound}
\end{equation}
The reconstruction from the $\binom62=15$ unordered ancestral pairs is detailed in
Appendix~\ref{v3-app-positive-p6}. Since every term of $F$ is
positive, $R-1\le\epsilon<18/53$ forces
$F<\epsilon/(1-53\epsilon/18)$ and hence
\begin{equation}
 q_L,q_R\longrightarrow\infty,\quad
 \frac{q_L+q_R}{Q_L},\frac{q_L+q_R}{Q_R}\longrightarrow0
       \quad\hbox{as }\epsilon\downarrow0.
 \label{eq:asymmetric-entry-v2}
\end{equation}
Thus the domain construction is forced by the global response, even
for sequences with arbitrarily imbalanced contacts. It is therefore the necessary setting for optimization near threshold one. Solving the bound for $F$ requires $\epsilon<18/53$, which keeps $1-53\epsilon/18$ positive. At larger gaps, such as the target $b/c=2$, the right-hand side stays below $18/53$ for every $F$, so the bound excludes no value of $F$; localizing the optimum there requires a finite-cap comparison beyond the localized corner (Appendix~\ref{v3-app-exact-global-reflection}).

For $p=q_R/(q_L+q_R)$, the contrast cap implies $Q_L,Q_R\le H$ and
\begin{equation}
 H>\frac{656}{4p(1-p)}
       \left(\epsilon^{-1}-\frac{53}{18}\right)^2.
 \label{eq:v3-main-finite-budget-bound}
\end{equation}
The independently controlled domain expansion also gives
\begin{equation}
 H\epsilon^2\ge\frac{656}{4p(1-p)}[1-O(\epsilon)],
 \quad \chi_b=\frac{2p(1-p)}3[1+O(\epsilon)].
 \label{eq:asymmetric-cost-amplitude-v2}
\end{equation}
Dividing out $p(1-p)$ before expanding keeps relative errors uniform
near $p=0,1$ (Appendix~\ref{app:asymmetric-entry-v2}). Imbalance therefore raises the leading cost and suppresses
the benefit susceptibility. In design terms, pin contacts in ratio $q_R/q_L=r_{\rm pin}$ multiply the
leading contrast by $(1+r_{\rm pin})^2/(4r_{\rm pin})$ and divide $\chi_b$
by the same factor.

\subsection{Analytic coordinates at the singular corner}
\label{sec:v4-toric}
The forced hierarchy has six small ratios but only four independent
weights. They are the entries of a rank-one matrix, which brings the
singular limit to the origin,
\begin{equation}
 \begin{gathered}
 \begin{pmatrix}q_L^{-1}&q_R/Q_L&q_R/Q_R\\q_R^{-1}&q_L/Q_L&q_L/Q_R\end{pmatrix}\\
 =\begin{pmatrix}q_L^{-1}\\q_R^{-1}\end{pmatrix}
   \begin{pmatrix}1&q_L q_R/Q_L&q_L q_R/Q_R\end{pmatrix}.
 \end{gathered}
 \label{eq:v4-segre}
\end{equation}

Write $z=(q_L^{-1},q_R^{-1},q_L/Q_L,q_R/Q_L,q_L/Q_R,q_R/Q_R)$. On the positive locus,
$q_L=1/z_1$, $q_R=1/z_2$, $Q_L=q_L/z_3=q_R/z_4$ and $Q_R=q_L/z_5=q_R/z_6$ recover the
physical weights, and every competitive design of
Sec.~\ref{sec:contrast-design} has $z$ near the origin.

Write $|z|=\sum_{i=1}^6|z_i|$. Every monomial of $U/\mathfrak m_\star$ and $\mathcal C/\mathfrak m_\star$ can be expressed with nonnegative
exponents in $z$. Their polynomial lifts satisfy
\begin{equation}
 \begin{gathered}
 U/\mathfrak m_\star=1+O(|z|),\quad \mathcal C/\mathfrak m_\star=F+O(|z|^2),\\
 V/\mathfrak m_\star=1+O(|z|).
 \end{gathered}
 \label{eq:v4-toric-unit}
\end{equation}
The denominator $V/\mathfrak m_\star$, divided by the same monomial as the numerator, remains nonzero at the hierarchy corner. This makes
$R-1=F+O(|z|^2)$ analytic, with derivatives in $z$ bounded uniformly near the corner,
although the weights themselves diverge. Appendix~\ref{app:v4-toric}
proves the nonnegativity of the exponents. Four suitable ratios parametrize the positive locus, but recovering the others can require division, as in $z_2=z_1z_3/z_4$. The six-ratio description keeps the lifted response polynomials regular at the hierarchy corner. The complete rank-one relations also matter at the boundary. The first two minors alone admit a spurious component at $z_1=z_2=0$; the third removes it (Appendix~\ref{v4:alg:toric-saturation}). The three minors generate the toric ideal, which encodes the relations between monomials~\cite[Chap.~4]{Sturmfels1996},
\begin{equation}
 I_{\rm tor}=(z_1z_3-z_2z_4,\ z_1z_5-z_2z_6,\ z_4z_5-z_6z_3).
 \label{eq:v10-toric-main-ideal}
\end{equation}
The coordinate algebra $\mathbb R[z_1,\ldots,z_6]/I_{\rm tor}$ identifies polynomials that differ by polynomial combinations of these constraints. Assigning degree one to each $z_i$, the Hilbert series
\begin{equation}
 \mathcal H_{\rm des}(\mathfrak t)=\frac{1+2\mathfrak t}{(1-\mathfrak t)^4}
 \label{eq:v10-hilbert-design-main}
\end{equation}
counts independent degree-$k$ polynomial functions of the contact ratios in the coefficient of $\mathfrak t^k$, where $\mathfrak t$ is a formal degree-counting variable. Six free variables would give 21 quadratic monomials; the three independent quadratic relations leave 18, the coefficient of $\mathfrak t^2$. The Hilbert series thus counts the independent correction terms available at each degree, and the response determines their coefficients. This graded count organizes higher corrections without treating the six ratios as independent. Appendix~\ref{v4:alg:toric-hilbert} derives the graded count and the relations among the generators. The global bound brings every competitive design into this chart, and the nonzero denominator in Eq.~\eqref{eq:v4-toric-unit} then controls the derivatives needed for stationarity and stability.

\subsection{Exact symmetry from unrestricted optimization}
\label{sec:v3-global-optimum}
We now minimize the threshold over all positive six-site path weights at fixed cap. Let $t_H=H^{-1/2}$ and put
$(Q_L,q_L,q_R,Q_R)=(a/t_H^2,X/t_H,Y/t_H,e/t_H^2)$. Then
$R=1+t_H\mathcal F+O(t_H^2)$, with
\begin{equation}
 \mathcal F=\frac3X+\frac3Y+
        \frac{53X+29Y}{6a}+\frac{29X+53Y}{6e}.
 \label{eq:v3-main-scaled-objective}
\end{equation}
The global bound confines competitive weightings to a compact positive
scaled set with $a,e\le1$. The leading minimum is unique at $a=e=1$,
$X=Y=3/\sqrt{41}$. Uniform derivative control forces the core weights
to saturate their cap; the positive pin Hessian gives a unique nearby minimum.
Reversal exchanges the pins, so uniqueness forces equality. For all
sufficiently large $H$,
\begin{equation}
 \begin{gathered}
 w_{\rm opt}=(H,q_H,1,q_H,H),\quad
 q_H=3\sqrt{H/41}+O(1),\\
 R_{\min}^{P_6}(H)=R_{\min}^{\refl}(H)
       =1+4\sqrt{41}\,H^{-1/2}+O(H^{-1}).
 \end{gathered}
 \label{eq:p6-optimum}
\end{equation}
Here $R_{\min}^{\refl}$ denotes the minimum restricted to reflected weights. Appendix~\ref{v3-app-exact-global-reflection} proves global optimality for every cap above a finite starting value. Its persistence argument, by compactness and uniform convergence, shows only that this value exists. Computing it requires explicit remainder bounds that keep competing designs in the localized region, preserve the signs of the core derivatives and keep the pin Hessian positive. 

Symmetry plays a different role in convex fastest-mixing design, which optimizes the relaxation of a reversible Markov chain toward stationarity~\cite{BoydDiaconisXiao2004}. There, averaging over graph symmetries preserves feasibility and cannot worsen the mixing objective~\cite[Theorem~2.1]{BoydEtAl2009Symmetry}. The present rational threshold requires global localization followed by uniqueness of the local minimum. Reflection maps optimizers to optimizers. Once uniqueness is proved, the optimizer must therefore be mirror-symmetric. This distinction becomes physical in Sec.~\ref{sec:v10-time-pitchfork}, where the fastest invasion breaks reflection symmetry.

\subsection{Resource exponents and their sharp constants}
\label{sec:cost-comparison}
\label{sec:positive-sums}
The hierarchy bound consists of positive contributions from competing contact scales. Their balance determines both an exponent and a coefficient. For any positive terms $\mathfrak p_l$, set $\mathcal Z=\sum_l\mathfrak p_l$ and $\omega_l=\mathfrak p_l/\mathcal Z$. Any reference probability weights $\widehat\omega_l$ obey
\begin{equation}
 \log\mathcal Z
 =\sum_l\widehat\omega_l\log\frac{\mathfrak p_l}{\widehat\omega_l}
   +D_{\rm KL}(\widehat\omega\Vert\omega),
 \label{eq:v10-positive-sum-identity}
\end{equation}
where $D_{\rm KL}(\widehat\omega\Vert\omega)=\sum_l\widehat\omega_l\log(\widehat\omega_l/\omega_l)$ is the Kullback--Leibler divergence, or relative entropy; zero reference weights are treated by continuity. Substituting $\omega_l=\mathfrak p_l/\mathcal Z$ proves the identity. Dropping the nonnegative last term and exponentiating gives the weighted arithmetic--geometric mean (AM--GM) inequality. When the terms are monomials in contact ratios, reference weights can cancel the free powers and determine a sharp resource bound. Retaining the last term measures the mismatch from a balanced design. Section~\ref{sec:v10-onset} uses this second interpretation at the birth-associated onset. Maximizing the first term over reference probability weights recovers $\log\mathcal Z$. This is the finite Gibbs variational identity, or convex conjugacy between log-sum-exp and negative entropy~\cite[Example~3.25]{MathBoydVandenberghe2004}. For positive monomial sums, logarithmic coordinates give the convex formulation used in geometric programming~\cite[Sec.~2.5]{MathBoydGP2007}. The probabilities in Eq.~\eqref{eq:v10-positive-sum-identity} weight terms of the response polynomial.

The four-cycle and five-site path obey analogous positive bounds.
For the four-cycle, take cyclic weights $(1,B,C,D)$ with minimum one
and opposite maximum $C\le H$. Then
\begin{equation}
 R-3>\frac8B+\frac8D+\frac{6BD}{C}\ge\frac8B+\frac8D+\frac{6BD}{H}
          \ge12(6/H)^{1/3}.
 \label{eq:c4-cost}
\end{equation}
For the five-site path with consecutive weights $(A,1,q,Q)$, oriented with $q\ge1$,
\begin{equation}
 R-7>\frac{83}{2q}+\frac{16q}{A}+\frac{83A}{3Q}
       \ge(496008/H)^{1/3}.
 \label{eq:p5-cost}
\end{equation}
The opposite-extrema condition for a cycle promoter and its global bound are proved in Appendix~\ref{app:resource-comparison}; Appendix~\ref{app:III-p5-resource} proves the path inequality.

The contested five-cycle obeys an inequality of the same type. Number sites cyclically as $1,\ldots,5$, with sites 1 and 2 the pins and sites 3, 4 and 5 their shared chain. Write
$\sigma=w_{45}/(w_{34}+w_{45})$ and $a(t)=(23+6t-4t^2)/8$. The coefficient $a(t)$ measures the leading threshold loss from finite protection of one pin when the intervening chain splits in fraction $t$; the opposite pin uses $a(1-t)$.

Rotate a weighting of $C_5$ so that a smallest weight is the bridge,
$w_{12}=1$. Every weak promoter with $R\le3$ obeys
\begin{equation}
 R-\frac32>\frac{a(\sigma)}{w_{23}}+\frac{a(1-\sigma)}{w_{51}}
 +\frac{11}{4}\,\frac{w_{23}w_{51}}{w_{34}+w_{45}},
 \label{eq:c5-global}
\end{equation}
its contrast satisfies $H>(185625/512)\,(R-3/2)^{-3}$, and its neutral
time satisfies $T_{\rm abs}>23/[4(R-3/2)]$. Every birth-associated weak promoter
with $R^{\mathrm b}-3/2=\epsilon<60/113$ has
$H>(12841875/2432)\,[(1-113\epsilon/60)/\epsilon]^3$.
Appendix~\ref{app:c5-global} proves the global sign comparisons for
both introduction rules and the neutral bound $T_{\rm abs}\ge2/(\vartheta_1+\vartheta_2)$ from the incoming overwrite rates of the two pins. Applying
the positive-sum identity, Eq.~\eqref{eq:v10-positive-sum-identity}, to
Eq.~\eqref{eq:c5-global} gives the contrast inequality; its sharp
coefficient occurs at the balanced chain $\sigma=1/2$.
The reflected design $(1,q,H,H,q)$ with $q=(25H/11)^{1/3}$ matches the
contrast bound to leading order,
\begin{equation}
 R_{\min}^{C_5}(H)=\frac32+\frac{15}{8}\,55^{1/3}H^{-1/3}+O(H^{-2/3}),
 \label{eq:c5-cost}
\end{equation}
and the family $(1,\varepsilon_{\rm cyc}^{-1},\varepsilon_{\rm cyc}^{-6},\varepsilon_{\rm cyc}^{-4},\varepsilon_{\rm cyc}^{-2})$, $\varepsilon_{\rm cyc}\to0$,
approaches the time bound with unequal pin contacts and an unbalanced chain.
The global onset in Sec.~\ref{sec:v10-onset} occurs at much smaller caps than this small-gap regime requires. It optimizes the sign of the benefit response; the present asymptotics optimize a prescribed finite threshold.

The five-cycle balance also determines the exact geometry. For either introduction rule $\mu\in\{\mathrm U,{\rm b}\}$ and all sufficiently large caps, the least-threshold weighting is unique up to scale and cycle symmetries and has the form $(1,q_H^{\mu},H,H,q_H^{\mu})$. Near the sharp cost, the three positive losses must approach equality, forcing the chain to balance and both stronger contacts to saturate. The remaining pin-contact minimum has positive curvature in both symmetric and antisymmetric directions. Uniqueness of the continued minimum, together with reversal symmetry, forces exact reflection, by the same organizing argument used on the six-site path. Appendix~\ref{app:III-c5-exact-resource} proves this conclusion and derives the first finite-gap corrections.

For the reflected six-site path,
\begin{equation}
 R-1=6u+\frac{82}{3}v+O((u+v)^2),
 \label{eq:p6-expansion}
\end{equation}
whose two linear terms also form the strict lower bound proved in Eq.~\eqref{eq:app-p6-strict-bound}. These leading terms are positive monomials, so their competing
contact scales can be balanced without cancellations. With $q\asymp H^{x_1}$, $A\asymp H^{x_2}$ and
$Q\asymp H$, the five-site exponents are $x_1,x_2-x_1,1-x_2$. Their
minimum is at most $1/3$, attained at $(x_1,x_2)=(1/3,2/3)$; the cycle has the
same three-way balance. The unrestricted six-site exponents are

\begin{equation}
 \begin{gathered}
 \bigl(x_{q_L},\,x_{q_R},\,x_{Q_L}-x_{q_L},\\
 x_{Q_L}-x_{q_R},\,x_{Q_R}-x_{q_L},\,x_{Q_R}-x_{q_R}\bigr),\\
 0\le x_{Q_L},x_{q_L},x_{q_R},x_{Q_R}\le1,
 \end{gathered}
 \label{eq:v4-six-exponents}
\end{equation}
for $(Q_L,q_L,q_R,Q_R)\asymp(H^{x_{Q_L}},H^{x_{q_L}},H^{x_{q_R}},H^{x_{Q_R}})$. The smallest of these
exponents is at most $1/2$, with equality only at
$x_{Q_L}=x_{Q_R}=1$, $x_{q_L}=x_{q_R}=1/2$.

The exponent comparison identifies the best power of the budget.
The same positive-sum identity, Eq.~\eqref{eq:v10-positive-sum-identity},
determines its coefficient. Choose the reference weights in Eq.~\eqref{eq:v10-positive-sum-identity} to cancel the free contact powers;
equality then fixes the relative sizes of the small copying probabilities into protected pins.
Table~\ref{tab:III-positive-balances} lists the resulting coefficient products.
\begin{table}[!htbp]
\caption{Positive-term balances for the sharp uniform-introduction contrast costs. The products are cubes of the leading threshold-gap coefficients on $C_4$, $C_5$ and $P_5$, and the square on $P_6$. The positive-sum identity, Eq.~\eqref{eq:v10-positive-sum-identity}, fixes their equality conditions; Table~\ref{tab:cost} compares the resulting resource and time costs.}
\label{tab:III-positive-balances}
\centering\small
\begin{ruledtabular}
\begin{tabular}{lcc}
Support & Product & Contrast coefficient\\
\hline
$C_4$ & $27(8\cdot8\cdot6)$ & $10368$\\[3pt]
$C_5$ & $27\left(\frac{25}{8}\,\frac{25}{8}\,\frac{11}{8}\right)$ & $\frac{185625}{512}$\\[3pt]
$P_5$ & $27\left(\frac{83}{2}\,16\,\frac{83}{3}\right)$ & $496008$\\[3pt]
$P_6$ & $4\cdot6\cdot\frac{82}{3}$ & $656$
\end{tabular}
\end{ruledtabular}
\end{table}

Appendix~\ref{app:v4-newton-dual} derives the exponent linear program
and its convex dual, with the entropy-dependent coefficient bound of geometric programming~\cite[Sec.~5.7.1, Example~5.5]{MathBoydVandenberghe2004}. Matching controlled response
expansions makes these bounds sharp without assuming convexity of the
full rational threshold. The same identity also gives a general degree rule. If
$R-R_*\ge\sum_ic_i\psi_i$ near the limit, with small contact ratios $\psi_i$ and
contrast $H=\prod_i\psi_i^{-k_i}$ and $c_i,k_i>0$, Eq.~\eqref{eq:v10-positive-sum-identity} gives
$H\epsilon^m\ge m^m\prod_i(c_i/k_i)^{k_i}$ with $m=\sum_ik_i$; the
exponent is the total degree of the contrast in these ratios.
Birth-associated introduction changes the six-site degree vector from
$(1,1)$ to $(2,1)$ because follower-originated introductions enter
relative to pin-originated ones as $q^2/Q$
(Sec.~\ref{sec:v7-preparation-cost}).

Table~\ref{tab:cost} compares the resulting contrast and neutral-time
costs, with $\epsilon$ measured above each support's own threshold infimum. Proposition~\ref{prop:v10-costs} states the exact scope of the contrast entries, and Secs.~\ref{sec:v4-time} and \ref{sec:v10-c5-time} derive the time bounds.
\begin{table*}[tp]

\caption{Leading contact and time costs as $\epsilon\downarrow0$ over all
positive weights, for a target $R\le R_*+\epsilon$ on the unit global
attempt clock. The contrast entries are the least $H$ under uniform and
birth-associated (temperature) introduction. The time entries are
neutral mean absorption times under uniform introduction, at the
least-contrast design and, in the time-only column, without a contrast
cap. Each entry is a sharp leading bound approached by an
explicit design. On the matching optimizing families $T_{\rm fix}/T_{\rm abs}\to N/2$, because the
limiting mass entering the mixed domain states is $2/N$. Sections~\ref{sec:v4-time} and \ref{sec:v10-c5-time} derive the time bounds; Appendix~\ref{app:III-c5-exact-resource} gives the five-cycle time at least contrast.}
\label{tab:cost}
\begin{ruledtabular}
\begin{tabular}{lcccccc}
Support&$R_*$&$H_{\min}^{\rm U}$&$H_{\min}^{\rm b}$&$T_{\rm abs}$ at $H_{\min}^{\rm U}$&Time-only $T_{\rm abs}$&$\lim T_{\rm fix}/T_{\rm abs}$\\
\hline
$C_4$&3&$10368\epsilon^{-3}$&$103680\epsilon^{-3}$&$24/\epsilon$&$16/\epsilon$&2\\
$C_5$&3/2&$\tfrac{185625}{512}\epsilon^{-3}$&$\tfrac{12841875}{2432}\epsilon^{-3}$&$75/(8\epsilon)$&$23/(4\epsilon)$&$5/2$\\
$P_5$&7&$496008\epsilon^{-3}$&no promoters&$498/\epsilon$&$166/\epsilon$&$5/2$\\
$P_6$&1&$656\epsilon^{-2}$&$4050\epsilon^{-3}$&$12/\epsilon$&$6/\epsilon$&3
\end{tabular}
\end{ruledtabular}

\end{table*}

\subsection{Contact cost under birth-associated introduction}
\label{sec:v7-preparation-cost}
The introduction rule changes which monomials of the numerators dominate
along a hierarchy, and with them the cost exponent.
On $P_6$, the two introduction rules have the same threshold infimum but
different requirements for approaching it. For a target $R\le1+\epsilon$
or $R^{\mathrm b}\le1+\epsilon$, respectively,
\begin{equation}
 \begin{aligned}
 H_{\min}^{\rm U}&=656\epsilon^{-2}+O(\epsilon^{-1}),\\
 H_{\min}^{\mathrm b}
 &=4050\epsilon^{-3}+6096\epsilon^{-2}+O(\epsilon^{-1}).
 \end{aligned}
 \label{eq:v7-preparation-costs}
\end{equation}
The birth-associated optimum is eventually exactly reflected, with
$q_*^{\mathrm b}=9/\epsilon+337/90+O(\epsilon)$ and both outer
weights equal to $H_{\min}^{\mathrm b}$. At this least-contrast design, the neutral fixation probability is $\rho_0^{\mathrm b}\sim\epsilon/54$, whereas uniform introduction gives $\rho_0^{\rm U}=1/6$. The total newborn arrival probability at the two pins is asymptotic to $1/(3q)$, and a pin-founded lineage fixes with probability tending to $1/2$. Their product is $1/(6q)\sim\epsilon/54$; follower introductions contribute at smaller order. Well-separated introductions at total rate $\nu$ per global attempt therefore produce successful neutral invasions at rate asymptotic to $\nu\epsilon/54$ under birth-associated introduction. Protection changes the rate at which successful lineages appear as well as the required contact range. Appendix~\ref{app:v7-birth-p6-cost} derives the source suppression.

The uniform balance in Eq.~\eqref{eq:p6-expansion} competes with a different leakage term under birth-associated introduction,
\begin{equation}
 R^{\mathrm b}-1=\frac6q+\frac{50q^2}{3Q}
       +O\!\left[(q^{-1}+q^2/Q)^2\right].
 \label{eq:v7-preparation-objectives}
\end{equation}
In this hierarchy, birth-associated arrival at a pin has probability of order $q^{-1}$. A follower is sampled with probability of order one, but its lineage reaches a protected pin during establishment with probability $O(q/Q)$. Relative to introduction at a pin, this contribution has scale $q^2/Q$. This explains the extra factor of $q$ in the leakage correction; the pair-response calculation fixes its positive coefficient. Balancing that correction against $q^{-1}$ requires $Q$ of order $q^3$. The weighted grading $(3,1,1,3)$ assigns degree three to $Q_L,Q_R$ and
degree one to $q_L,q_R$. It isolates this competition
before the optimization. An all-weight inequality in
Appendix~\ref{app:v7-preparation} proves that asymmetric designs cannot
improve its leading cost.

Under birth-associated introduction, the uniform-optimal hierarchy $Q=(41/9)q^2$ gives
$R^{\mathrm b}\to173/23$ (Appendix~\ref{app:v7-birth-p6-cost}). Reoptimizing the network is
therefore essential to comparing the two rules. On $C_4$ and $C_5$ the exponent remains three, with the source-dependent constants in Table~\ref{tab:cost}. For the reflected five-cycle $(1,q,Q,Q,q)$,
$R^{\mathrm b}-3/2=(25/4)q^{-1}+(761/38)q^2/Q
+O\bigl((q^{-1}+q^2/Q)^2\bigr)$, balanced at $Q=(3044/475)q^3$
(Appendix~\ref{app:c5-global}). Appendix~\ref{app:v7-birth-polynomials}
derives the four-cycle constant.

Table~\ref{tab:III-finite-gap-folds} evaluates the exact finite-state equations along the reflected branch at several target gaps. At $\epsilon=0.1$, the uniform and birth-associated contrasts are approximately $7.07\times10^4$ and $4.69\times10^6$. The leading terms underestimate them by about $7\%$ and $14\%$ of the finite values, respectively. These evaluations quantify corrections to the sharp resource asymptotics along the reflected branch.

The corresponding probabilities have a more direct copying interpretation. At the uniform-target design with $\epsilon=0.1$, a pin's offspring crosses to the opposite pin with probability about $0.0079$, while a follower next to a pin replaces it with probability about $0.0018$. At the birth-associated target the latter probability is about $2.0\times10^{-5}$. Reciprocal weights entail $s_iP_{ij}=s_jP_{ji}$, while the directed copying probabilities are generally unequal. These numbers specify an offspring-routing model; whether a physical implementation can control them depends on its replacement mechanism.

\begin{table*}[t]
\centering
\caption{Finite-gap values on the reflected six-site fold branch.
For each target $R^\mu=1+\epsilon$, the contacts $(H,q,1,q,H)$ solve
the exact response equation and $\partial_qR^\mu=0$ numerically at
65-digit precision. $H_{\rm lead}^{\rm U}=656\epsilon^{-2}$ and
$H_{\rm lead}^{\rm b}=4050\epsilon^{-3}$ are the leading asymptotic
contrasts. $T^{\rm U}$ denotes neutral absorption time from uniform
singleton introduction at the resulting design, in attempted updates.
The columns for birth-associated targets therefore use the same time
objective as the uniform-target columns. Independent full-configuration
and pair equations agree in the response to over forty decimal places;
the contiguous-interval and full-configuration time equations agree to
the same accuracy. Proposition~\ref{prop:v10-costs} identifies this branch with the unrestricted optimum for sufficiently small gaps. The $\epsilon=0.40$ row lies beyond the $\epsilon<18/53$ range of the localization bounds; for the smaller listed gaps, a conclusion about global optimality requires the quantitative localization and curvature estimates discussed in Sec.~\ref{sec:v3-global-optimum}.}
\begin{ruledtabular}
\begin{tabular}{c r r r r r r}
$\epsilon$ & $H^{\rm U}$ & $H^{\rm U}/H_{\rm lead}^{\rm U}$ & $T^{\rm U}[w^{\rm U}]$
& $H^{\rm b}$ & $H^{\rm b}/H_{\rm lead}^{\rm b}$ & $T^{\rm U}[w^{\rm b}]$\\
\hline
0.40 & 5436.234 & 1.32591 & 49.619 & 108776.911 & 1.71894 & 40.115\\
0.20 & 18998.024 & 1.15842 & 79.483 & 673096.961 & 1.32957 & 62.512\\
0.10 & 70720.222 & 1.07805 & 139.408 & 4688141.674 & 1.15757 & 107.461\\
0.05 & 272563.783 & 1.03873 & 259.370 & 34895125.898 & 1.07701 & 197.435\\
0.02 & 1665293.733 & 1.01542 & 619.346 & 521631273.729 & 1.03038 & 467.421\\
\end{tabular}
\end{ruledtabular}
\label{tab:III-finite-gap-folds}
\end{table*}

\subsection{Controlling copying probabilities}
\label{sec:III-kernel-resource}
The choice of controlled resource changes both the optimal contacts and
the absolute powers of the target gap. To separate this dependence from
the effect of introduction, consider the smallest nonzero copying
probability on the six-site path and set
\begin{equation}
 C_P(w)=\frac{1}{\min_{P_{ij}>0}P_{ij}}.
 \label{eq:III-kernel-resource}
\end{equation}
The endpoint rows contain a probability equal to one, so $C_P$ is also
the ratio of the largest to the smallest nonzero entry of $P$ on this
support. Reoptimizing over all positive path weights gives
\begin{equation}
 \inf_{R^{\rm U}\le1+\epsilon}C_P
 \sim\frac{100}{3}\epsilon^{-1},\qquad
 \inf_{R^{\rm b}\le1+\epsilon}C_P
 \sim400\epsilon^{-2}.
 \label{eq:III-kernel-costs}
\end{equation}
Thus the extra inverse power caused by birth-associated introduction
persists for this kernel resource, while the contact-contrast powers
reflect the reciprocal weights being controlled. The optimal contacts
also change. In central-contact normalization, the leading uniform
design has $q_L=q_R\sim100/(3\epsilon)$ and
$Q_L=Q_R\sim10000/(9\epsilon^2)$; the birth-associated design has
$q_L=q_R\sim12/\epsilon$ and $Q_L=Q_R\sim4800/\epsilon^3$.
Appendix~\ref{app:III-kernel-resource} derives these global leading
bounds from the same response inequalities.

\subsection{A vanishing uniform fraction changes the resource regime}
\label{sec:v10-crossover}
Interpolate between the experiments with the mixture $\mu^\zeta$ of Sec.~\ref{b:sec:observables}, keeping the generator fixed.
The contractions mix linearly; their ratio generally does not. In terms of $g_{\rm gap}^{\mu}=(K^\mu-J^\mu)/K^\mu$, the exact relation is
\begin{equation}
 g_{\rm gap}^{\zeta}=
 \frac{\zeta K^{\rm U} g_{\rm gap}^{\rm U}+(1-\zeta)K^{\rm b}g_{\rm gap}^{\rm b}}
 {\zeta K^{\rm U}+(1-\zeta)K^{\rm b}}.
 \label{eq:v10-gap-mixture}
\end{equation}
This identity remains meaningful if one benefit coefficient is negative. On the competitive hierarchy with bridge weight one, $K^{\rm U}$ is of order $\epsilon$ and $K^{\rm b}$ of order $\epsilon^2$. A uniform fraction of order $\epsilon$ can therefore compete with the entire birth-associated contribution.

\begin{figure}[tbp]
\centering
\includegraphics[width=\columnwidth]{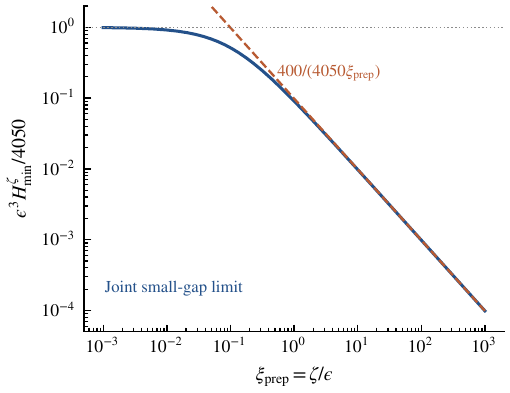}
\caption{Least contact contrast on the six-site path under the weak-selection target $R^\zeta\le1+\epsilon$. The uniform fraction $\zeta$ mixes uniform and birth-associated singleton introduction. In the joint limit $\epsilon\to0$ at fixed $\xi_{\rm prep}=\zeta/\epsilon$, the rescaled least contrast $\epsilon^3H_{\min}^\zeta/4050$ connects the birth-associated value one (dotted line) to the regime dominated by a small uniform fraction. The dashed curve is its large-$\xi_{\rm prep}$ asymptote, distinct from the fixed-$\zeta=1$ endpoint.}
\label{fig:III-introduction-crossover}
\labelalias{fig:v10-resource-geometry}{fig:III-introduction-crossover}
\end{figure}

Put $\xi_{\rm prep}=\zeta/\epsilon$, and let $x_\zeta\in[9,12]$ be the unique root there of
\begin{equation}
 \xi_{\rm prep}x_\zeta^2+
 [2(1-\zeta)-12\xi_{\rm prep}]x_\zeta-18(1-\zeta)=0.
 \label{eq:v10-crossover-root}
\end{equation}
The global optimum for $R^\zeta\le1+\epsilon$ satisfies
\begin{align}
 H_{\min}^{\zeta}
 &=\frac{(50+32\zeta)x_\zeta^3}
 {3\epsilon^3(x_\zeta-6)(1-\zeta+\xi_{\rm prep}x_\zeta)}
 [1+O(\epsilon)],\notag\\
 q_*^\zeta&=x_\zeta/\epsilon+O(1).
 \label{eq:v10-uniform-crossover}
\end{align}
The errors are uniform through both endpoints. Proposition~\ref{prop:v10-costs} states the exact optimizing geometry; Appendix~\ref{app:v10-crossover} proves global localization and uniform derivative control. At $\zeta=0$ this recovers $4050\epsilon^{-3}$; for each fixed $\zeta>0$ it gives $(256+400/\zeta)\epsilon^{-2}$. Equation~\eqref{eq:v10-uniform-crossover} also resolves $\zeta=O(\epsilon)$, where neither fixed-fraction limit suffices; Fig.~\ref{fig:III-introduction-crossover} shows this crossover.

\begin{proposition}[Global bounds and sharp contact costs]
\label{prop:v10-costs}\labelalias{prop:c5-global}{prop:v10-costs}
Normalize arbitrary positive $P_6$ contacts to $(Q_L,q_L,1,q_R,Q_R)$.
For $\mu\in\{\mathrm U,\mathrm b\}$, restrict to $J^\mu>0$.
The uniform and birth-associated gaps obey
Eqs.~\eqref{eq:v3-main-global-bound} and
\eqref{eq:v7-birth-p6-global-gap}, respectively.
Every sequence with $R^\mu\to1$ therefore enters the hierarchy
\eqref{eq:asymmetric-entry-v2} in this normalization.
For the target $R^\mu\le1+\epsilon$ with $0<\epsilon<18/53$,
the corresponding finite-gap contrast bounds are
Eqs.~\eqref{eq:v3-main-finite-budget-bound} and
\eqref{eq:v7-birth-finite-gap-bound}.
As $\epsilon\downarrow0$, the least contrasts satisfy
\begin{equation}
 H_{\min}^{\rm U}=656\epsilon^{-2}+O(\epsilon^{-1}),\quad
 H_{\min}^{\rm b}=4050\epsilon^{-3}+O(\epsilon^{-2}).
 \label{eq:v10-cost-summary}
\end{equation}
For sufficiently small positive $\epsilon$, each optimizer is unique up to common rescaling, exactly reflected, and has both outer contacts at the cap. At every sufficiently large fixed cap $H$, the least uniform-introduction threshold is likewise attained by the unique normalized design $(H,q_H,1,q_H,H)$ of Eq.~\eqref{eq:p6-optimum}. For the mixture $\mu^\zeta=\zeta\mu^{\rm U}+(1-\zeta)\mu^{\rm b}$, Eq.~\eqref{eq:v10-uniform-crossover} holds with errors uniform in $0\le\zeta\le1$. There is $\epsilon_0>0$, independent of $\zeta$, such that for $0<\epsilon<\epsilon_0$ the mixture optimizer is unique up to common rescaling, exactly reflected and has both outer contacts at the cap.

The uniform-introduction finite-weight bounds in Eqs.~\eqref{eq:c4-cost}, \eqref{eq:p5-cost} and \eqref{eq:c5-global}, together with the birth-associated bounds of Sec.~\ref{sec:v7-preparation-cost} and Appendices~\ref{app:c5-global} and \ref{app:v7-birth-polynomials}, imply the sharp leading contrast entries in Table~\ref{tab:cost}, with each gap measured above the corresponding support's threshold infimum. For either introduction rule, the five-cycle resource optimizer at sufficiently small positive gap is uniquely $(1,q_*^{\mu},H,H,q_*^{\mu})$, modulo scale and dihedral symmetries, where $q_*^\mu$ is the optimizing five-cycle access weight. Each finite leading cost is approached by positive-weight families.
\end{proposition}
\begin{proof}
The coefficient comparisons~\eqref{eq:v4-global-remainders}, reconstructed in Appendix~\ref{v3-app-positive-p6}, prove the uniform bound; Appendix~\ref{app:v7-birth-p6-cost} proves its birth-associated counterpart. The positive terms force the stated hierarchy, and the regular contact-ratio chart gives eventual global reflection in Appendix~\ref{v3-app-exact-global-reflection}. On that eventual branch the optimal threshold decreases strictly, connecting fixed-cap and fixed-gap optimization. Appendix~\ref{app:v10-crossover} makes the comparison uniform in the introduction fraction. The other support bounds and their matching limits are proved in Appendices~\ref{app:resource-comparison}, \ref{app:III-p5-resource} and \ref{app:c5-global}; Appendix~\ref{app:III-c5-exact-resource} proves the five-cycle optimizing geometry, and Appendix~\ref{app:v7-birth-polynomials} proves the birth-associated four-cycle bound.
\end{proof}

\section{First promotion and exact finite-cap design}
\label{sec:finite-cap}
Approaching a sharp floor requires caps that grow without bound as $\epsilon\downarrow0$, whereas first promotion asks for the infimum of caps admitting any promoter. We now determine the complete finite-cap optimum on the five-site path and first-promotion onsets on the five-cycle and reflected six-site path. Their sign polynomials retain the positive-sum structure that controlled the asymptotic contact costs. Comparing these onsets, together with an exact six-cycle example, separates the effects of contact geometry and introduction rule. We write $H_{0,\mathsf G}^{\mu}$ for the infimum of caps admitting $J^{\mu}>0$ on support $\mathsf G$ under introduction rule $\mu$. An additional $\refl$ index restricts this infimum to reflected weights.

\subsection{An exact optimizer at every feasible five-site cap}
\label{sec:v10-p5-allcap}
On the five-site path, increasing both outer weights by a common factor strictly lowers a promoting threshold. Minimizing the limiting threshold over their ratio gives its lower envelope at each fixed ratio of the two inner contacts. These facts follow from the polynomial calculations of Ref.~\cite[Sec.~IV\,C and Appendices~D\,2--D\,3]{ChowdhuryProtectedDomains}. We use two distinct polynomial inputs, the positive coprime numerator recalled in Appendix~\ref{app:III-p5-resource} and the signed outer degrees of the unreduced polynomials specified in Appendix~\ref{app:v10-p5-allcap}. A finite contrast cap restricts both the outer weights and the inner ratio. For uniform introduction, orient and normalize a candidate optimum as $(A,1,q,H)$. For this five-site path write $R=U/V$, with $U>0$ and $\operatorname{sgn}V=\operatorname{sgn}J$, using the polynomial normalization of Appendix~\ref{app:v10-p5-allcap}. At a trial level $r_{\rm lev}>7$, define
\begin{equation}
 L_{r_{\rm lev}}(\log A,\log q;H)
 =\frac{U-r_{\rm lev}V}{A^7q^6H^8}.
 \label{eq:v10-p5-level}
\end{equation}
Its sublevel $L_{r_{\rm lev}}\le0$ is precisely $V>0$, $R\le r_{\rm lev}$. A positive monomial changes the curvature of the defining function without changing the physical feasible set.

Exact coefficient grouping proves strict convexity in $(\log A,\log q)$ on two overlapping domains. Localization and a comparison design place all competing minimizers at each level in one of these convex domains. If two distinct designs minimized $R$ at the same level, their logarithmic midpoint would have $L_{r_{\rm lev}}<0$ and therefore a smaller threshold. This proves uniqueness without requiring the rational function $R$ itself to be convex.

The resulting optimizer $(A_H,1,q_{5,H},H)$ follows one analytic branch above the first-promotion cap $H_{0,P_5}^{\rm U}$. Numerical evaluation of this
analytically characterized root, enclosed by exact rational bounds in Eq.~\eqref{eq:III-p5-onset-bracket}, gives $H_{0,P_5}^{\rm U}\simeq1406.728807$,
with $(A_{\rm on},q_{\rm on})\simeq(133.918134,15.151087)$. Proposition~\ref{prop:v10-onsets} collects this conclusion with the five-cycle and reflected six-site onsets.

Appendix~\ref{app:v10-p5-allcap} proves the coefficient grouping and
complete cap coverage. Strict curvature fixes the stationary branch,
and outer-contact monotonicity determines its first onset.
Near onset the promoting region opens as an ellipse in logarithmic coordinates. Write $R_{\min}(H)$ for the optimal threshold at cap $H$. If $U_{\rm on}$ and $(\partial_HV)_{\rm on}>0$ are evaluated at onset, then
\begin{equation}
 R_{\min}(H)=\frac{U_{\rm on}}{(\partial_HV)_{\rm on}(H-H_{0,P_5}^{\rm U})}+O(1).
 \label{eq:v10-p5-onset-divergence}
\end{equation}
The divergent threshold and the large-cap limit seven describe opposite ends of the same analytic optimization branch.

\subsection{When promotion first becomes possible}
\label{sec:v10-onset}
Here the cap is raised until a positive benefit response first appears. With contact contrast as in Eq.~\eqref{eq:contrast}, both rules on $C_5$ have a finite first-promotion cap with a unique onset geometry. A positive monomial rescales the benefit numerator without changing
its sign. Consider a domain on which each normalized retained weight lies between one and a common upper bound exceeding the onset cap. There the rescaled
numerator increases when the two chain weights are raised to the cap
and, at fixed sum, when the two pin weights are equalized. Separate exact sign comparisons for the two
introduction rules therefore force a reflected onset geometry $(1,q,H,H,q)$,
giving
\begin{equation}
 H_{0,C_5}^{\rm U}=71.003074\ldots,\quad
 H_{0,C_5}^{\rm b}=316.241236884\ldots .
 \label{eq:v10-c5-onsets}
\end{equation}
At the unique onset geometry for either rule, $J=0$ and $K>0$. No finite benefit--cost ratio promotes exactly at onset. Appendix~\ref{app:v10-onsets} specifies each root by rational equations and isolating bounds, and proves the global reduction separately for the two introduction rules. Its exact polynomial inequalities control this full bounded domain of contacts.

The birth-associated calculation has a compact convex interpretation. Write $h$ for the capped core weight in this calculation. After positive factors are removed, its reflected benefit sign is
\begin{equation}
 F_{\rm b}(q,h)=304q^3h^7-\sum_{a,j}c_{aj}q^ah^j,
 \quad c_{aj}>0,\quad j\le7.
 \label{eq:v10-onset-polynomial}
\end{equation}
There are 57 negative monomials. Divide each by $304q^3h^7$ and call
the resulting positive terms $\mathfrak p_{aj}$, with
$\mathcal Z=\sum_{a,j}\mathfrak p_{aj}$ and
$\omega_{aj}=\mathfrak p_{aj}/\mathcal Z$.
The positive-sum identity \eqref{eq:v10-positive-sum-identity} applies to these monomials. The normalized weights $\omega_{aj}$ have a direct interpretation at the onset.

Let $q_{\rm on}^{\rm b}$ be the pin weight at the birth-associated onset $h=H_{0,C_5}^{\rm b}$. There $\mathcal Z(q_{\rm on}^{\rm b},H_{0,C_5}^{\rm b})=1$ and
$\partial_{\log q}\mathcal Z=0$. Choose
$\widehat\omega=\omega^*$, the onset weights. The mean exponent in
$q$ vanishes, leaving $m_h=\sum\omega^*_{aj}(j-7)<0$ and
\begin{equation}
 \frac{F_{\rm b}}{304q^3h^7}
 =1-\exp\!\left[m_h\log\frac h{H_{0,C_5}^{\rm b}}
       +D_{\rm KL}(\omega^*\Vert\omega)\right].
 \label{eq:v10-kl-onset}
\end{equation}
Increasing the cap compensates a nonnegative mismatch between the monomial weights. At fixed $h=H_{0,C_5}^{\rm b}$, every other $q$ therefore has strictly adverse benefit. At onset, the second derivative of $\log\mathcal Z$ in $\log q$ is the variance $\sigma_q^2=\sum\omega^*_{aj}(a-3)^2>0$, so the two promoting boundaries open with
\begin{equation}
 \left(\log\frac q{q_{\rm on}^{\rm b}}\right)^2
 =\frac{-2m_h}{\sigma_q^2}\log\frac h{H_{0,C_5}^{\rm b}}
 +o\!\left(\log\frac h{H_{0,C_5}^{\rm b}}\right).
 \label{eq:v10-onset-fold}
\end{equation}
This relative entropy is over monomials in an exact sign polynomial. It measures design mismatch among the monomials, a quantity distinct from population entropy production. The connection is the log-sum-exp geometry of positive monomial sums~\cite[Sec.~3.1.5]{MathBoydVandenberghe2004}. Appendix~\ref{app:v10-onset-entropy} refines the fold expansion. Appendix~\ref{app:v10-cycle-onsets} proves the uniform-introduction onset by a separate exact sign reduction.
\subsection{Six-site onset comparisons}
\label{sec:III-six-site-onsets}
On a reflected six-site path, positive normalization and concavity determine first promotion for both introduction rules,
\begin{equation}
 \begin{gathered}
 H_{0,P_6,\refl}^{\rm U}=69.160793\ldots,\\
 H_{0,P_6,\refl}^{\rm b}=320.082030\ldots .
 \end{gathered}
 \label{eq:III-p6-both-onsets}
\end{equation}
The respective pin contacts are $5.5006654696\ldots$ and $4.3881816022\ldots$.
For each rule the normalized benefit polynomial is strictly concave in
the pin contact and increases with the outer weight on the relevant
domain. Its unique zero maximum fixes the onset over every positive
reflected weighting. Appendix~\ref{app:III-p6-onsets} gives the
analytical argument and exact rational enclosures.

\begin{proposition}[Finite-contrast onset and optimization]
\label{prop:v10-onsets}
\labelalias{prop:v10-p5-allcap}{prop:v10-onsets}
Under uniform introduction on $P_5$, the first-promotion cap $H_{0,P_5}^{\rm U}$ lies in $(1406.728807,1406.728808)$, with exact rational endpoints as in Eq.~\eqref{eq:III-p5-onset-bracket}. At every $H>H_{0,P_5}^{\rm U}$ the least threshold is attained, up to reversal and common rescaling, by exactly one weighting $(A_H,1,q_{5,H},H)$. This optimizer follows an analytic branch; the promoting region is connected within the chart $(A,1,q,Q)$ with $q\ge1$ and its onset divergence is Eq.~\eqref{eq:v10-p5-onset-divergence}.

For $C_5$ and either introduction rule $\mu\in\{\mathrm U,{\rm b}\}$, the first-promotion cap is $H_{0,C_5}^{\mu}$ in Eq.~\eqref{eq:v10-c5-onsets}. Below it every weighting has $J^{\mu}<0$. At the cap, $J^{\mu}=0$ only at $(1,q_{\rm on}^{\mu},H_{0,C_5}^{\mu},H_{0,C_5}^{\mu},q_{\rm on}^{\mu})$, up to scale, cycle rotations and reflections, with $4<q_{\rm on}^{\mu}<5$. For every larger cap reflected promoters exist. Their boundary opens as a nondegenerate fold at onset. Since $K^{\mu}>0$, no finite benefit--cost ratio promotes exactly at either onset.

For arbitrary positive reflected $P_6$ weights $(Q,q,1,q,Q)$, the corresponding first-promotion caps are those of Eq.~\eqref{eq:III-p6-both-onsets}, with rational enclosures in Eqs.~\eqref{eq:III-p6-local-onset-bracket} and \eqref{eq:III-p6-birth-onset-bracket}. Below each cap every reflected weighting has $J^\mu<0$. At the cap there is exactly one zero up to scale, with the middle contact smallest and both outer contacts at the cap; above it reflected promoters exist. The ordering of the reflected six-site onset relative to the five-cycle onset reverses between introduction rules,
\begin{equation}
 H_{0,P_6,\refl}^{\rm U}<H_{0,C_5}^{\rm U},\qquad
 H_{0,C_5}^{\rm b}<H_{0,P_6,\refl}^{\rm b}.
 \label{eq:III-onset-ordering}
\end{equation}
\end{proposition}
\begin{proof}
Appendix~\ref{app:v10-p5-allcap} combines the five-site path's convex sublevels with global localization. Appendix~\ref{app:v10-cycle-onsets} proves the two cycle reductions and their rational isolating bounds; its nonzero cap derivative and negative pin curvature give the onset fold. Appendix~\ref{app:III-p6-onsets} proves the reflected six-site onsets by strict pin concavity and outer-contact monotonicity. The rational enclosures fix their ordering relative to the cycle.
\end{proof}

Proposition~\ref{prop:v10-onsets} shows how the introduction rule reverses the onset comparison between the unrestricted five-cycle and the reflected six-site family. Each reflected onset bounds the unrestricted six-site onset from above. Near onset the threshold diverges and leaves the range $R-1<18/53$ in which the global bounds~\eqref{eq:v3-main-global-bound} and \eqref{eq:v7-birth-p6-global-gap} constrain designs, so the unrestricted onsets require a finite-cap sign comparison of the benefit response with asymmetric paths. A different support can lower the cost further. The six-cycle with cyclic contacts
$(51,9/2,1,9/2,51,51)$ has $J^{\rm U}>0$, so its uniform onset lies below $51$.
Appendix~\ref{app:III-finite-comparisons} gives the exact sign. Finding the least cap on this support requires comparison with all positive cycle weights.

\begin{figure*}[t]
\centering
\begin{minipage}[c]{.49\textwidth}
\includegraphics[width=\linewidth]{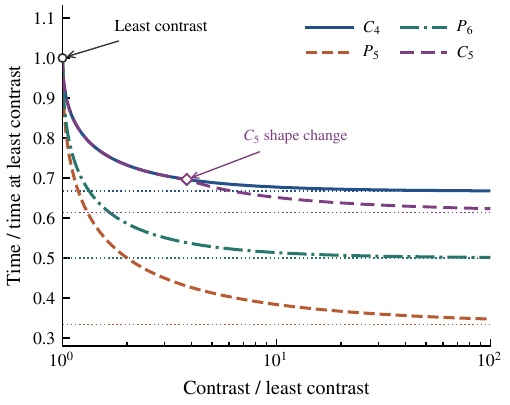}
\end{minipage}\hfill
\begin{minipage}[c]{.47\textwidth}
\centering
{\small Five-cycle contacts}\par\medskip
\begin{tikzpicture}[x=1cm,y=1cm,font=\footnotesize,
 site/.style={circle,draw,fill=white,thick,minimum size=4.5mm,inner sep=0pt},
 pin/.style={site,fill=figureblue,draw=figureblue,text=white}]
\node[pin] (c1) at (-1.15,0) {1};
\node[pin] (c2) at (1.15,0) {2};
\node[site] (c3) at (1.8,.85) {3};
\node[site] (c4) at (0,1.6) {4};
\node[site] (c5) at (-1.8,.85) {5};
\draw[thick] (c1)--node[below] {$1$}(c2);
\draw[figureblue,line width=1pt] (c2)--node[right] {$w_{23}$}(c3);
\draw[line width=1.8pt] (c3)--node[above,sloped] {$w_{34}$}(c4);
\draw[line width=1.8pt] (c4)--node[above,sloped] {$w_{45}$}(c5);
\draw[figureblue,line width=1pt] (c5)--node[left] {$w_{51}$}(c1);
\end{tikzpicture}
\end{minipage}
\caption{Leading neutral absorption-time frontiers under uniform singleton introduction on $C_4$, $C_5$, $P_5$ and $P_6$, normalized by each support's least-contrast design. The weak-selection target is $R\le R_*+\epsilon$, with $\epsilon\downarrow0$ measured from that support's threshold floor. The balanced $C_5$ curve lies exactly on $C_4$ until the marked change of limiting geometry at normalized contrast $3.80769\ldots$. The sketch identifies the five-cycle's unit bridge, blue access contacts and two chain contacts. Beyond the marked change, their optimizing ratio $w_{45}/(w_{34}+w_{45})$ tends to zero after a possible reflection, while all retained weights stay positive at finite gap. Dotted lines indicate the time-only limiting infima. All times use the global attempt clock. The frontiers are analytical small-gap limits at fixed scaled cap; locating the marked change at finite gap requires uniform control of the corrections as the chain ratio vanishes (Sec.~\ref{sec:v10-c5-time}).}
\label{fig:tradeoff}
\end{figure*}
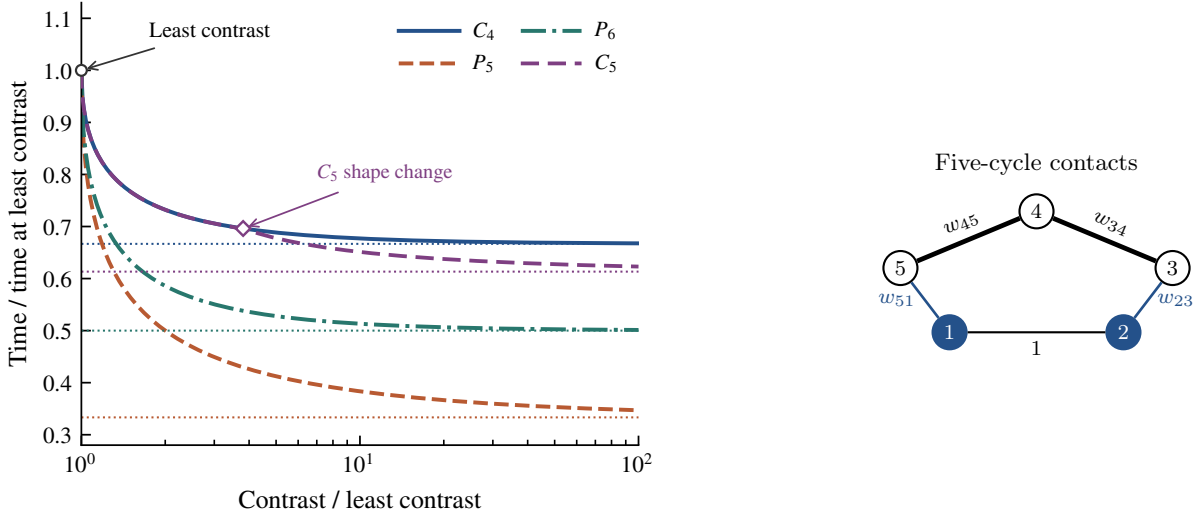

\section{Finite budgets, waiting time and design tolerance}
\label{sec:v3-tolerance}
Extra contrast can shorten an invasion by allowing faster exchange between domains. We first derive the time--contrast frontiers and the exact six-site optimizer near least contrast, then follow the five-cycle frontier through its change of limiting geometry. At larger six-site budgets the fastest design undergoes a pitchfork. Returning to the feasibility boundary, the contrast surplus above least contrast determines how much contact error the design can tolerate.

\subsection{The time--contrast frontier}
\label{sec:v4-time}
Let $\mathcal T_{\rm abs},\mathcal T_{\rm fix}$ be the separate infima of the
neutral times under $R\le R_*+\epsilon$ and contrast at most $H$.
Their lower envelope is the time--contrast Pareto frontier, where reducing one cost requires increasing the other. For fixed scaled budget $\eta=H\epsilon^{p_H}$ above its limiting
minimum, $\mathcal T_{\rm abs}=\theta/\epsilon+O_\eta(1)$, with the support-dependent relation between $\eta$ and $\theta$ in Table~\ref{tab:III-time-frontiers}.

Here $\theta$ is the leading coefficient of neutral absorption time in each support's hierarchy; its expression in contact coordinates is specified in the corresponding reduction. The success-conditioned leading times are larger by the factors $\lim T_{\rm fix}/T_{\rm abs}=N/2$ listed in Table~\ref{tab:cost}. The uncapped leading gains relative to least contrast are factors $3/2$, $3$ and $2$ on $C_4$, $P_5$ and $P_6$, respectively.
The $C_4$, $P_5$ and $P_6$ curves in Fig.~\ref{fig:tradeoff} are the faster branches of Table~\ref{tab:III-time-frontiers} and minimize time over all
positive weights; Appendices~\ref{app:v10-c4-time},
\ref{app:v10-p5-time} and \ref{app:v4-exact-time} prove the respective bounds.
The two branches correspond to slower and faster domain exchange at
the same threshold; the lower one resolves an invasion sooner. The stated errors hold at each fixed scaled cap above its minimum, where the lower threshold root is simple and an analytic correction meets the exact target. As the cap decreases to its minimum the two roots merge and this correction degenerates. On $C_4$ and $P_5$ a joint limit therefore needs higher-order feasibility and time estimates through the fold; the exact six-site construction below includes this regime. The five-cycle frontier in the same figure is derived in Sec.~\ref{sec:v10-c5-time}.

\begin{table}[!htbp]
\caption{Leading neutral absorption-time frontiers under uniform singleton introduction and the weak-selection target $R\le R_*+\epsilon$. At fixed $\eta=H\epsilon^{p_H}$ above its limiting minimum, $\mathcal T_{\rm abs}=\theta/\epsilon+O_\eta(1)$ in global attempts. The stated interval selects the faster branch. Its upper endpoint is the least-contrast limit; the lower endpoint is approached as the scaled cap diverges.}
\label{tab:III-time-frontiers}
\centering\small
\begin{ruledtabular}
\begin{tabular}{lccc}
Support & $p_H$ & $\eta(\theta)$ & Faster branch\\
\hline
$C_4$ & $3$ & $\dfrac{6\theta^3}{\theta-16}$ & $16<\theta<24$\\[7pt]
$P_5$ & $3$ & $\dfrac{1328\theta^3}{3(\theta-166)^2}$ & $166<\theta<498$\\[7pt]
$P_6$ & $2$ & $\dfrac{82\theta^2}{3(\theta-6)}$ & $6<\theta<12$
\end{tabular}
\end{ruledtabular}
\end{table}

\begin{figure*}[tp]
 \centering
 \includegraphics[width=0.94\textwidth]{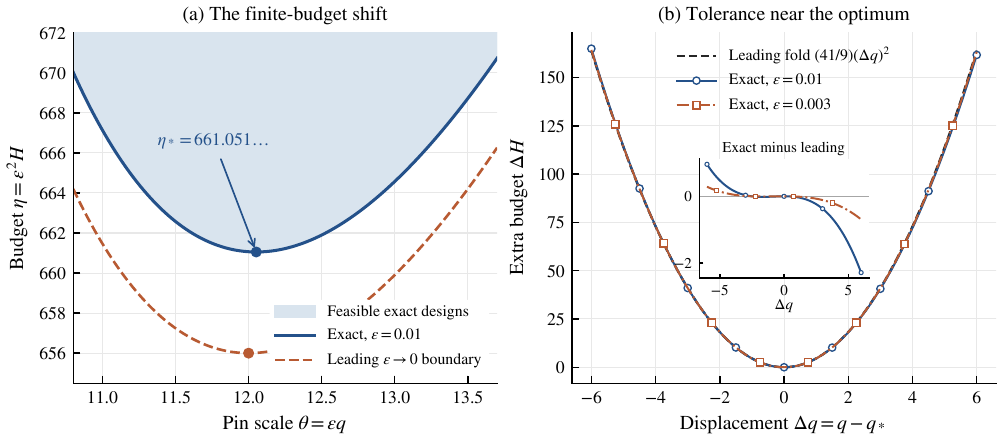}
 \caption{Finite-budget feasibility and parameter tolerance for the
 reflected six-site path with contacts $(H,q,1,q,H)$, uniform singleton
 introduction and target $R\le1+\epsilon$. The scaled coordinates are $\theta=\epsilon q$, $\eta=\epsilon^2H$, and the displacements are $\Delta q=q-q_*(\epsilon)$, $\Delta H=H-H_*(\epsilon)$. (a) The solid boundary is the exact rational
 weak-selection threshold at $\epsilon=0.01$; the shaded region is
 feasible. The dashed leading boundary underestimates the required
 budget. (b)  Exact curves, centered on their own finite-$\epsilon$
 minima, approach the quadratic fold relation. Distinct markers separate
 the nearly coincident curves; the inset shows the exact minus leading
 required budget. The curves are evaluations of the microscopic pair solution.}
 \label{fig:finite-budget-fold-v2}
\end{figure*}

For uniform introduction on the six-site path, let $q_*(\epsilon)$ be the reflected pin contact at the least contrast $H_*(\epsilon)=H_{\min}^{\rm U}(\epsilon)$. The exact analytic branch satisfies
$R=1+\epsilon$ and stationarity in $q$, with
\begin{equation}
 H_* =\frac{656}{\epsilon^2}+\frac{62032}{123\epsilon}+O(1),
 \quad q_* =\frac{12}{\epsilon}+\frac{26348}{5043}+O(\epsilon).
 \label{eq:finite-budget-leading-shift-v2}
\end{equation}
Further coefficients appear in Eq.~\eqref{eq:finite-budget-critical-v2}. At this design the neutral mean has the sharper expansion
\begin{equation}
 T_{\rm abs}=\frac{12}{\epsilon}+\frac{97483}{5043}+O(\epsilon).
 \label{eq:III-main-time-constant}
\end{equation}
The constant is about $19.33$ global attempts. It explains the appreciable time correction in Table~\ref{tab:III-finite-gap-folds}; Appendix~\ref{app:finite-budget-fold-v2} derives it directly from the interval generator.
Because the second term of $H_*$ in Eq.~\eqref{eq:finite-budget-leading-shift-v2} is positive, the leading budget $656/\epsilon^2$ is insufficient at
finite small gap, as Fig.~\ref{fig:finite-budget-fold-v2}(a) illustrates. Appendix~\ref{app:finite-budget-fold-v2} derives this finite-gap shift.

For sufficiently small positive gaps and caps $H\epsilon^2$ in a fixed bounded interval above feasibility, the exact time minimizers remain on the fast reflected branch. Proposition~\ref{prop:v10-allbudget} identifies their contacts and extends the classification through all larger finite caps.
Descendants of a singleton remain a contiguous interval on a path. The twenty-state transient resolvent therefore determines both exact means. Appendix~\ref{app:v4-exact-time} proves the shared minimum through the feasibility fold, using uniform derivative control and positive constrained curvature.

The target constraint $(1+\epsilon)J-K\ge0$ implies $J>0$ because $K>0$. The same contacts minimize both neutral times under the weak-selection target. Their ratio approaches three in the protected-domain limit, with finite-contact corrections set by the entrance probabilities and passage times. Proposition~\ref{prop:v10-allbudget} controls those corrections through the feasibility fold. We next determine the five-cycle frontier and conditional waiting distribution; Sec.~\ref{sec:v10-time-pitchfork} then follows the six-site optimum to the larger scale at which its geometry changes.

\subsection{The five-cycle frontier and successful fixation}
\label{sec:v10-c5-time}
The five-cycle has a second limiting geometry. At fixed
$\eta=H\epsilon^3>185625/512$, optimizing the neutral time compares
a balanced follower chain with a hierarchy in which one of its two
contacts is small relative to the other. Figure~\ref{fig:tradeoff} identifies the access and chain contacts. The balanced family has equal access contacts and equal chain contacts. On the boundary family, the chain fraction $w_{45}/(w_{34}+w_{45})$ tends to zero after a possible reflection; its optimizing access fraction is $p_{\rm cyc}=w_{51}/(w_{23}+w_{51})$. The two terms in the envelope below compare these families at a given leading time $\theta$.

Put $\mathfrak t_\theta=\theta-25/4$ and
\begin{gather}
 p_{\rm cyc}(\theta)=\frac{1-2\mathfrak t_\theta+\sqrt{4\mathfrak t_\theta^2+2\mathfrak t_\theta+1}}3,\notag\\
 \mathscr D_{\rm cyc}(\theta)=\max\left\{\frac{\mathfrak t_\theta}{2},
 p_{\rm cyc}(1-p_{\rm cyc})\left(\mathfrak t_\theta+\frac{p_{\rm cyc}}{2}\right)\right\},\notag\\
 \eta=\frac{11\theta_5^3}{16\mathscr D_{\rm cyc}(\theta_5)},
 \qquad \frac{23}{4}<\theta_5<\frac{75}{8}.
 \label{eq:III-c5-frontier}
\end{gather}
The last equation has a unique root in the indicated interval.
The first term in $\mathscr D_{\rm cyc}$ belongs to the balanced family and the second to the boundary family. Every retained contact remains positive at finite gap.
The two branches agree at
$\theta_c=6.5269531794\ldots$, corresponding to
$\eta_c=1380.4727394183\ldots$. The balanced limiting design minimizes
time below this scaled cap; the unequal-chain limiting design does so
above it. Appendix~\ref{app:III-c5-frontier} proves the global leading
frontier, including chain ratios tending to either endpoint.
At each fixed $\eta>185625/512$, the error is $o(\epsilon^{-1})$ on the full frontier; on the boundary family the matching trials take $\epsilon\to0$ at a fixed positive chain ratio and only then let that ratio approach zero. At the limiting crossing the two leading time objectives coincide, so their corrections can decide the preferred finite-gap geometry and any shift of the changeover. Resolving them needs higher-order response and time estimates that stay uniform as the chain ratio vanishes. After normalization by least-contrast values,
the balanced five-cycle and four-cycle frontiers coincide exactly,
$\widehat H=\widehat T^3/(3\widehat T-2)$. Their common three-term
balance explains this agreement despite their different threshold floors
and resource coefficients. The extra chain degree of freedom lets the
five-cycle leave that curve above $\eta_c$.

Under neutral copying and uniform singleton introduction, conditioning on successful fixation turns the absorption-time distribution into that of full ancestral coalescence. Once all present sites descend from one ancestor, the uniformly chosen introduced site agrees with that ancestor with probability $1/N$, independently of the realized ancestry. The success-conditioned probability of absorption by time $t$ is therefore the probability that all lineages have coalesced by $t$. This uses the graphical voter--coalescent duality described in Ref.~\cite[Sec.~14.3]{AldousFillReversible}, with the copying rates of Eq.~\eqref{b:eq:generator}. More generally, tracing copying arrows backward until their ancestral lineages merge gives $T_{\rm abs}\le T_{\rm fix}\le NT_{\rm abs}/2$ for neutral copying from uniformly placed singletons on any finite connected support (Appendix~\ref{app:v10-c5-time}). The upper bound compares means at the same weighting. The sharp conditional lower bound below follows from the coalescence-time distribution.

Full coalescence must wait for the first jump of the two lineages at the ends of a smallest cycle edge. The global threshold comparison bounds this jump rate and gives the following sharp constraints on successful invasion.
\begin{proposition}[Five-cycle waiting times]
\label{prop:III-c5-waiting}
Consider neutral copying on $C_5$, uniform singleton introduction and the rate-one Poisson clock of global attempts. Let $\tau$ be the absorption time. Every positive weighting with $J>0$ and $R\le3$ has $R>3/2$, satisfies $T_{\rm abs}>23/[4(R-3/2)]$, and obeys, for $t>0$ and real $m>0$,
\begin{gather}
 \Pr(\tau>t\mid C\text{ fixes})>
 \exp\!\left[-\frac{8(R-3/2)t}{115}\right],
 \label{eq:v10-c5-survival}\\
 \E[\tau^m\mid C\text{ fixes}]>
 \Gamma(m+1)\left[\frac{115}{8(R-3/2)}\right]^m,
 \label{eq:v10-c5-moments}
\end{gather}
where $\Gamma$ is Euler's gamma function. Without a contrast cap, these bounds are asymptotically sharp as $\epsilon\downarrow0$ for $J>0$ and $R\le3/2+\epsilon$. The time infima are $23/(4\epsilon)+O(1)$ and $115/(8\epsilon)+O(1)$ for absorption and successful fixation, respectively.

At each fixed $\eta>185625/512$, impose contrast at most $H=\eta\epsilon^{-3}$ and the same target. As $\epsilon\downarrow0$, the separate time infima obey
\begin{equation}
 \epsilon\mathcal T_{\rm abs}\longrightarrow\theta_5(\eta),
 \qquad
 \epsilon\mathcal T_{\rm fix}\longrightarrow\frac52\theta_5(\eta),
 \label{eq:III-c5-frontier-times}
\end{equation}
where $\theta_5$ is defined in Eq.~\eqref{eq:III-c5-frontier}. Along every feasible sequence with $\epsilon T_{\rm abs}\to\bar\theta$, the conditional distribution of $\epsilon\tau$ converges to the exponential distribution of mean $5\bar\theta/2$, together with all fixed positive moments. Consequently, for every real $m>0$,
\begin{equation}
 \lim_{\epsilon\downarrow0}\epsilon^m
 \inf\E[\tau^m\mid C\text{ fixes}]
 =\Gamma(m+1)\left[\frac52\theta_5(\eta)\right]^m,
 \label{eq:III-c5-moment-frontier}
\end{equation}
with the infimum over the same capped feasible set. Positive contacts approach these limits through the balanced and chain-boundary families described above. The matching chain-boundary construction takes $\epsilon\to0$ at fixed positive chain ratio before that ratio approaches zero.
\end{proposition}
\begin{proof}
The global comparison~\eqref{eq:c5-global} gives $R>3/2$, and Eqs.~\eqref{eq:c5-time-pins}--\eqref{eq:c5-time-only} prove the unconditional time bound. Appendix~\ref{app:v10-c5-time} derives the conditional bounds~\eqref{eq:v10-c5-survival-bound} and their matching uncapped limit from ancestral coalescence. Appendix~\ref{app:III-c5-frontier} proves the global capped frontier, the limiting distribution and the uniform tail estimate that controls all positive moments.
\end{proof}
For discrete global attempts the first-jump comparison is geometric; Appendix~\ref{app:v10-c5-time} gives its finite-gap form. Its means and leading scaled conditional moments agree with the Poisson-clock results.

The exponential limit reflects a separation between fast internal relaxation and slow exchange of the protected domains. Fixation-time distributions also depend on the limiting regime. At constant relative fitness, ring and complete-graph populations have different rescaled limiting distributions as their size grows~\cite[Secs.~III and IV]{HathcockStrogatz2019}. Here the population remains finite and increasing contact contrast separates its dynamical time scales.

\subsection{The fastest geometry at all neutral-time budgets}
\label{sec:v10-time-pitchfork}

We return to the six-site path, whose least-contrast scale is $H\asymp\epsilon^{-2}$. A second scale appears when the objective is time. Write $y=(q_L-q_R)/(q_L+q_R)$ and $\Lambda=H\epsilon^4$. After imposing the exact threshold target and saturating the cores, the imbalance-dependent terms at order $\epsilon$ are $f_0$ for $T_{\rm abs}$ and $3f_0$ for $T_{\rm fix}$, up to terms independent of $y$, where
\begin{equation}
 f_0(y;\Lambda)=-\frac23y^2+
 \frac{984}{\Lambda(1-y^2)}.
 \label{eq:v10-time-potential}
\end{equation}

\begin{figure}[!htbp]
\centering
\includegraphics[width=\columnwidth]{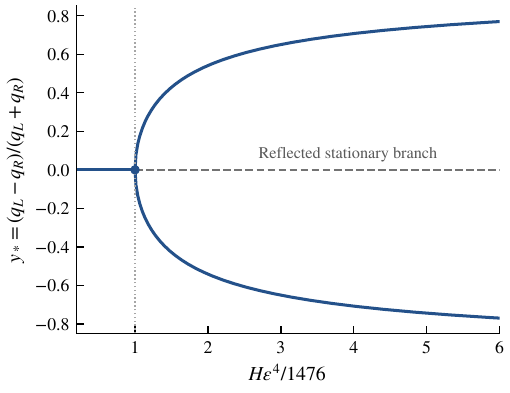}
\caption{Leading optimal pin-contact imbalance on the six-site path for neutral absorption and successful-fixation times under uniform singleton introduction and $R\le1+\epsilon$. The contacts $q_L,q_R$ connect the pins to their cores. Solid curves minimize Eq.~\eqref{eq:v10-time-potential}; the dashed reflected branch ceases to minimize above $H\epsilon^4=1476$. This leading bifurcation scale is common to the two objectives; their exact critical caps may differ at higher order (Proposition~\ref{prop:v10-allbudget}). Target feasibility occurs earlier, at $H\sim656\epsilon^{-2}$. Times use the global attempt clock.}
\label{fig:III-time-pitchfork}
\end{figure}

On the active threshold target, increasing pin imbalance reduces the ideal-core correction but increases the finite-core leakage delay. The first contribution is proportional to $g_p=(1-y^2)/4$ and the second to $g_p^{-1}$; their competition gives this potential. The reflected design loses its minimum at $\Lambda=1476$. Above that value the two minimizing leading imbalances obey
\begin{equation}
 y_*^2=1-\sqrt{1476/\Lambda}.
 \label{eq:v10-time-branches}
\end{equation}
The quartic term is positive at onset, giving a supercritical pitchfork of the optimizing contact geometry, shown in Fig.~\ref{fig:III-time-pitchfork}.
Writing $z=y^2\in[0,1)$ makes the minimization transparent. The derivative with respect to $z$ is $-2/3+984/[\Lambda(1-z)^2]$, which is strictly increasing and changes sign in the interval precisely when $\Lambda>1476$. Hence the minimized correction is
\begin{equation}
 f_{0,\min}(\Lambda)=
 \begin{cases}
 984/\Lambda,&0<\Lambda\le1476,\\
 -2/3+8\sqrt{41/\Lambda},&\Lambda\ge1476.
 \end{cases}
 \label{eq:III-pitchfork-value}
\end{equation}
At $\Lambda=1476$ the two branches and their first derivatives agree; the second derivative drops by $1/(3\cdot1476^2)$. Equation~\eqref{eq:v10-time-branches} also gives $y_*^2=(\Lambda-1476)/2952+O((\Lambda-1476)^2)$ on the asymmetric side. These are the Landau features of the minimized, imbalance-dependent order-$\epsilon$ time correction, with $y$ as order parameter. The singularity arises from optimizing over contacts. At fixed positive weights, neutral fixation probabilities and mean passage times are rational and smooth functions of those weights. The conditional correction carries the factor three stated above.

For fixed $\Lambda>1476$, the change of geometry has a calculable dynamical benefit. Comparing the optimum with the fast reflected branch at the same cap and active target $R=1+\epsilon$, Eq.~\eqref{eq:v10-time-potential} gives
\begin{equation}
 T_{\rm abs}^{\refl}-\mathcal T_{\rm abs}
 =\frac{2\epsilon}{3}
 \left(1-\sqrt{\frac{1476}{\Lambda}}\right)^2
 +O_\Lambda(\epsilon^2).
 \label{eq:v10-time-gain}
\end{equation}
The conditional gain has three times this leading coefficient. Both gains are of relative order $\epsilon^2$ against the leading times $6/\epsilon$ and $18/\epsilon$; the larger reduction from $12/\epsilon$ to $6/\epsilon$ occurs along the earlier time--contrast frontier. Equation~\eqref{eq:asymmetric-cost-amplitude-v2} also gives $\chi_b=\sqrt{1476/\Lambda}/6+O_\Lambda(\epsilon)$ on either asymmetric branch for fixed $\Lambda>1476$; below the bifurcation the reflected branch has $\chi_b=1/6+O_\Lambda(\epsilon)$. Greater imbalance therefore decreases the weak benefit susceptibility. This coefficient measures response to benefit; the active target itself has $\mathcal S=0$ at $b/c=1+\epsilon$.

\begin{proposition}[All-budget neutral-time branches]
\label{prop:v10-allbudget}
Consider uniform singleton introduction on $P_6$ and neutral times in the rate-one Poisson clock of global attempts. Minimize each mean over positive weights of contrast at most $H$ under the closed weak target $(1+\epsilon)J-K\ge0$, equivalently $J>0$ and $R\le1+\epsilon$. There is one $\epsilon_0>0$ such that, for $0<\epsilon<\epsilon_0$, each of $T_{\rm abs}$ and $T_{\rm fix}$ has a unique normalized global minimizer at every feasible cap below its own critical cap,
\begin{equation}
 (H,q_-,1,q_-,H),\qquad R(H,q_-,1,q_-,H)=1+\epsilon.
 \label{eq:v4-exact-time}
\end{equation}
Here $q_-$ is the lower exact threshold root, continued from the lower leading branch; the roots coincide at $H=H_*(\epsilon)$. At the critical cap the reflected minimizer is unique and quartically stable. Above it there are exactly two reversal-related global minimizers. At every feasible finite cap both cores equal $H$ and the target is active. The critical caps satisfy
\begin{equation}
 H_{{\rm c},\rm abs},H_{{\rm c},\rm fix}
 =\epsilon^{-4}[1476+O(\epsilon)],
 \label{eq:v10-critical-caps}
\end{equation}
and no further branch change occurs at larger finite cap. For the strict target $R<1+\epsilon$, the same values are infima at every $H>H_*(\epsilon)$; this target is infeasible at $H=H_*(\epsilon)$.

For $0<\epsilon<\epsilon_0$ and without a cap, the infima are the exact functions in Eqs.~\eqref{eq:v10-uncapped-abs} and \eqref{eq:v10-uncapped-exact}, with leading values $I_{\rm abs}(\epsilon)=6/\epsilon+O(1)$ and $I_{\rm fix}(\epsilon)=18/\epsilon+O(1)$. Positive finite contacts approach but do not attain them.
\end{proposition}
\begin{proof}
Appendix~\ref{app:v4-exact-time} proves uniqueness through the feasibility fold. Appendix~\ref{app:v10-allbudget} joins this bounded-budget result to the uniform large-budget reduction summarized below. Appendix~\ref{app:v10-uncapped} gives the exact uncapped infima and their nonattainment.
\end{proof}
The common leading critical cap follows because the two time objectives have proportional leading derivatives. Their critical caps may differ by $O(\epsilon^{-3})$; the next coefficients of the separate objectives determine that difference.
Let $p=q_R/(q_L+q_R)$ and $g_p=p(1-p)=(1-y^2)/4$, the coordinate introduced above, which is invariant under exchanging the pins. Put $\eta_H=(H\epsilon^2)^{-1}$, the reciprocal of the scaled cap $\eta$ of Table~\ref{tab:III-time-frontiers}. For either time objective write $L=\epsilon T$ on the exact target, with both cores at the cap. Its derivative factors as $\partial_{g_p}L=\epsilon^2a_{\rm time}-\eta_Hd_{\rm time}/g_p^2$. The positive analytic functions $a_{\rm time}$ and $d_{\rm time}$ are the coefficients of the ideal-core contribution and finite-core leakage, respectively, in this derivative; Appendix~\ref{app:v10-allbudget} constructs them in the localized chart. For absorption they approach $8/3$ and $246$; for successful fixation they approach $8$ and $738$. The stationary condition and its monotonicity become
\begin{equation}
 \begin{gathered}
 \frac{\eta_H}{\epsilon^2}=\Xi_{\epsilon,\eta_H}(g_p),\qquad
 \Xi_{\epsilon,\eta_H}(g_p)=\frac{g_p^2a_{\rm time}}{d_{\rm time}},\\
 \frac{\partial\log\Xi}{\partial\log g_p}
 =2+O(\epsilon+\eta_H/g_p)>0.
 \end{gathered}
 \label{eq:v10-time-invariant}
\end{equation}
Every candidate in this regime lies in the proved small-$\eta_H/g_p$ chart. Monotonicity of this stationarity function permits at most one interior value of $g_p$, hence exactly one pair of asymmetric optima above the critical cap. The symmetric point is its endpoint $g_p=1/4$. This control extends the local pitchfork through arbitrarily large finite budgets. As the cap grows without bound, these asymmetric branches approach the ideal-core endpoints studied in Appendix~\ref{app:v10-uncapped}.

\subsection{The geometry of feasible errors}
\label{sec:finite-budget-fold-v2}
Near the least-contrast six-site design under uniform introduction, admissible contact errors include signed shifts of the pins and one-sided reductions of the capped core weights. Perturbation theory for death-first updating quantifies the response to individual edge changes~\cite{MengMasuda2023}. Here we determine the coupled feasible region for source-first updating. Define
 \begin{align}
 d&=\epsilon^2(H-H_*), \notag \\
 (\alpha_L,\alpha_R)&=\epsilon^2(H-Q_L,H-Q_R), \label{eq:v3-main-error-coordinates} 
\\
 \xi_s&=\epsilon[(q_L+q_R)/2-q_*],\quad
 \xi_a=\epsilon(q_L-q_R)/2.\notag
 \end{align}
These are the contact-error coordinates used in Fig.~\ref{fig:III-resource-to-signal}(b). The coordinates $\xi_s,\xi_a$ measure a common pin shift and an
imbalance; $\alpha_L,\alpha_R\ge0$ are the scaled core deficits below the cap.
Expansion about the exact critical design gives the leading feasible region
\begin{equation}
 \frac{41}{9}(\xi_s^2+\xi_a^2)
       +\frac{\alpha_L+\alpha_R}{2}\le d.
 \label{eq:v3-main-tolerance}
\end{equation}
The positive pin Hessian makes the signed errors quadratic, whereas
nonzero inward core derivatives make the deficits linear. An analytic
change of coordinates absorbs the higher-order terms while keeping the
two one-sided core constraints; it is the Morse-lemma normal form of a
function restricted to a corner. In the new pin coordinates $X_s,X_a$
and core deficits $\widetilde\alpha_L,\widetilde\alpha_R$, the exact nearby region is
\begin{equation}
 X_s^2+X_a^2+\widetilde\alpha_L+\widetilde\alpha_R\le r_{\rm tol,\epsilon}(d),\qquad
 \widetilde\alpha_L,\widetilde\alpha_R\ge0,
 \label{eq:v4-morse-corner}
\end{equation}
with $r_{\rm tol,\epsilon}(0)=0$, $r_{\rm tol,\epsilon}'(0)>0$.
Appendix~\ref{v3-app-full-tolerance} constructs the change and proves
that the chart contains all feasible designs sufficiently near the
critical contrast. The exact reflected center and its fold are derived
in Appendix~\ref{app:finite-budget-fold-v2}.

On the reflected, core-saturated section, the roots open as
\begin{equation}
 \begin{aligned}
 q_\pm&=q_*\pm[3/\sqrt{41}+O(\epsilon)]\sqrt{H-H_*}\\
 &\quad+O\!\left(\epsilon(H-H_*)\right),
 \qquad \epsilon^2(H-H_*)\ll1.
 \end{aligned}
 \label{eq:finite-budget-width-v2}
\end{equation}
Figure~\ref{fig:finite-budget-fold-v2}(b) compares this width with the exact roots. The square-root dependence is a fold of the design constraint
\cite{Kuznetsov2004,PriorityKuznetsov2006}. This fold belongs to the set of feasible contacts; every finite positive-weight stochastic system retains the same two absorbing consensus states.

For multiplicative errors, use the logarithmic measure
$dV_{\log}=\mathrm dQ_L/Q_L\,\mathrm dq_L/q_L\,\mathrm dq_R/q_R\,\mathrm dQ_R/Q_R$ with the bridge fixed. If
$\mathfrak r=(H-H_*)/H_*$, the feasible volume is
\begin{equation}
 \mathcal V_{\log}=\frac{4\pi}{3}[1+O(\epsilon)]
             \mathfrak r^3[1+O(d)],\quad d=656\mathfrak r[1+O(\epsilon)].
 \label{eq:v3-main-volume}
\end{equation}
Appendix~\ref{v3-app-full-tolerance} derives this volume and its Jacobian. Two quadratic widths and two linear widths give the exponent
$2/2+2=3$. More generally, $m$ nondegenerate quadratic directions
and $k_{\rm lin}$ active linear constraints give power $m/2+k_{\rm lin}$ under a smooth
positive density; the Jacobian fixes the prefactor. Holding a core weight exactly at its cap removes its linear direction and lowers the power by one. The volume becomes a probability only when an error distribution is specified, as in the Gaussian example below.

For example, independent Gaussian logarithmic errors in the two pin
contacts, with the core weights fixed at their cap, lead to an
explicit confidence--contrast relation, Eq.~\eqref{eq:v7-gaussian-confidence} in Sec.~\ref{sec:III-nonlinear-margin}. Appendix~\ref{app:v7-gaussian-tolerance}
derives it from the same quadratic pin geometry and includes the
selection penalty.

The same fold width fixes the time gain just above least contrast. In the layer $H=H_*+\omega_{\rm lay}/\epsilon+O(1)$, fixed $\omega_{\rm lay}>0$,
the lower root also gives
\begin{equation}
 \begin{aligned}
 \mathcal T_{\rm abs}&=\frac{12}{\epsilon}
        -\frac{3\sqrt{\omega_{\rm lay}}}{\sqrt{41\epsilon}}+O_{\omega_{\rm lay}}(1),\\
 \mathcal T_{\rm fix}&=\frac{36}{\epsilon}
        -\frac{9\sqrt{\omega_{\rm lay}}}{\sqrt{41\epsilon}}+O_{\omega_{\rm lay}}(1).
 \end{aligned}
 \label{eq:finite-budget-time-layer-v2}
\end{equation}
At fixed $\omega_{\rm lay}>0$, the $\epsilon^{-1/2}$ correction follows from the
fold width. When $\omega_{\rm lay}=O(\epsilon)$, this shift becomes $O(1)$,
at the order omitted here; the exact-root statement resolves that
endpoint regime, including the fold itself. Appendix~\ref{app:finite-budget-fold-v2} derives this time layer from the same critical branch.

\section{Finite selection and observable advantage}
\label{sec:v3-selection-window}

The weak threshold is a derivative at neutrality. A finite advantage also depends on how selection changes establishment and collective competition. The explicit calculations use the six-site path with uniform introduction; the response algebra of Sec.~\ref{sec:III-filtered-observables} holds on any support. The joint small-gap and small-selection analysis must retain the finite-contact corrections that disappear in the fixed-selection hierarchy limit. Organizing the exact correlation response first makes those corrections reproducible without factorizing fluctuations. Proposition~\ref{prop:v7-selected-time} also determines the accompanying selected time optimum before we turn to signal and observation.
\subsection{Filtered observables and higher response}
\label{sec:III-filtered-observables}
\label{sec:v4-observable-algebra}
The threshold uses a pair response, but the observable advantage at
finite selection also depends on higher orders. We organize these
orders before calculating the shrinking selection window. With $\sigma_i=2n_i-1$,
polynomials agreeing on all binary configurations are identified in
\begin{equation}
 \mathscr A_N=\mathbb R[n_1,\ldots,n_N]/(n_i^2-n_i)
 =\mathbb R[\sigma_1,\ldots,\sigma_N]/(\sigma_i^2-1).
 \label{eq:v4-boolean}
\end{equation}
The quotient imposes the relations $n_i^2=n_i$, or $\sigma_i^2=1$.
Squarefree spin monomials $\sigma_S=\prod_{i\in S}\sigma_i$, with distinct site indices, form a
basis and multiply as $\sigma_S\sigma_T=\sigma_{S\triangle T}$. The symmetric difference $S\triangle T$ contains the indices belonging to exactly one of the two sets.
Repeated indices can lower the degree of a product. Degree therefore defines a filtration, the nested spaces $\mathscr F_k$ of polynomials of degree at most $k$, and degree is only subadditive under multiplication.

Polynomial degree organizes the first-order weak-selection equations~\cite{McAvoyAllen2021}. Here the binary algebra controls each successive order.
Neutral copying preserves $\mathscr F_k$. For fixation, multiplying $\mathcal A h_C=0$ by the positive total fecundity removes the normalizing denominator without changing its solutions. In this row-rescaled operator for hitting probabilities, an order-$m$ donation perturbation raises degree by at most $m$, and centered linear fecundity has only the order-one term. Timed observables retain the denominator, as detailed in Appendix~\ref{app:v3-sylvester-time}. Centering the payoff as
$\bar f_i=f_i-(b-c)/2=[-c\sigma_i+b(P\sigma)_i]/2$ removes a common fecundity
factor. In the coordinate $\delta_c=\delta/[1+\delta(b-c)/2]$ for linear
fecundity and $\delta_c=\delta$ for exponential fecundity,
$h_C=\Phi+\sum_{m\ge1}\delta_c^m h_m$ obeys
\begin{equation}
 \begin{gathered}
 \deg h_m\le\min(N,m+1),\\
 h_m(-\sigma)=(-1)^{m+1}h_m(\sigma),\quad m\ge1.
 \end{gathered}
 \label{eq:v4-degree-response}
\end{equation}
The neutral inverse with zero values at both consensus states preserves these spaces; it inverts the generator restricted to mixed configurations.
Appendix~\ref{app:v3-degree-closure} proves the induction; Appendix~\ref{app:v3-sylvester-time} retains the additional generator terms required by timed response.

These two properties organize the finite-selection response. Each order of selection raises
the degree of the response by at most one, so the hierarchy of
correlation equations closes exactly at every order, with a number of
unknowns that grows polynomially in $N$ at fixed order, compared with
$2^N$ configurations. The first-order correction uses pair observables, while the cubic signal also needs the intervening degree-three response and degree-four observables. The closure follows from this degree bound alone, without factorizing higher correlations into lower ones, as a moment-closure approximation would, and products of the pair disagreements
$d_{ij}=(1-\sigma_i\sigma_j)/2$ reduce through the binary relations
\begin{equation}
 d_{ij}^2=d_{ij},\qquad
 2d_{ij}d_{jk}=d_{ij}+d_{jk}-d_{ik}.
 \label{eq:v4-pair-relations}
\end{equation}
The parity under type complementation makes the advantage $\Delta$ exactly odd in the
centered selection intensity $\lambda$ of Eq.~\eqref{eq:v3-main-centered-selection} (Appendix~\ref{par:v3-joint-selection-coordinates}). On
the reflected six-site problem the degree bound and one consensus
condition per parity sector leave $9$, $12$ and $18$ response
coefficients at orders one, two and three, compared with $34$ mixed-configuration orbits after reflection alone (Appendix~\ref{app:v3-degree-closure}, counting of the reflected response sectors). For singleton introduction on a path, a separate geometric reduction to contiguous cooperative intervals permits the twenty-state calculation used below. The associated graded algebra keeps the highest-degree part of each relation, so its Hilbert series is
\begin{equation}
 \mathcal H_{\rm obs}(\mathfrak t)=(1+\mathfrak t)^N.
 \label{eq:v10-hilbert-observable}
\end{equation}
The coefficient of $\mathfrak t^k$ counts independent degree-$k$ observables modulo lower degrees. The neutral operator construction cited in Sec.~\ref{b:sec:response-input} develops this binary algebra and the one-lineage exterior representation of neutral propagation on paths and cycles; here the filtration controls successive selection orders. An exterior algebra has the same binomial count. In the present use the count enumerates commuting correlation observables of each degree modulo lower degrees; the one-lineage exterior representation applies the same count separately to neutral spectral propagation on paths and cycles.

For centered linear fecundity the hierarchy can also be summed exactly.
Write the denominator-removed harmonic operator as
$\mathcal G_0+\delta_c\mathcal G_1$. On functions vanishing at consensus, let
$\mathcal G_{0,D}$ and $\mathcal G_{1,D}$ be its Dirichlet restrictions,
where the subscript $D$ denotes mixed configurations with zero consensus
boundary values. Define the response insertion
$\mathsf R_{\rm ins}=-\mathcal G_{0,D}^{-1}\mathcal G_{1,D}$.
The first correction is
$h_1=-\mathcal G_{0,D}^{-1}(\mathcal G_1\Phi)|_D$, and the exact committor is
\begin{equation}
 h_C=\Phi+\delta_c(I-\delta_c\mathsf R_{\rm ins})^{-1}h_1.
 \label{eq:III-resummed-response}
\end{equation}
Each insertion combines a selection perturbation with neutral propagation
until absorption. Type complementation pairs these insertions in the
advantage, as derived in Appendix~\ref{app:III-response-resolvent}.
For fixed finite weights the advantage is rational in $\delta_c$. Its
uncancelled complex poles set the convergence radius of its response
series. On a reflected six-site path, parity bounds the denominator degree of $\Delta/\delta_c$ by fifteen in $\delta_c^2$. The interval restriction for singleton introduction sharpens it to eleven, as used in Appendix~\ref{par:v3-joint-radius-window}. These finite-dimensional reductions connect the correlation expansion to its convergence radius.

\subsection{Formation corrections and global control}
\label{sec:v3-formation}

At finite selection, the formation losses that set the weak threshold acquire selection-dependent coefficients. For the reflected six-site path with uniform introduction, retain $u=1/q$ and $v=q/Q$ and write $r=b/c=1+\epsilon$. At fixed positive $\kappa=\delta c$ in a compact admissible interval, correcting both entrance and domain competition gives
\begin{align}
 \Delta={}&\frac{\kappa}{6}
       [(r-1)-A_\kappa u-B_\kappa v]\notag\\
 &+O\!\left((r-1)^2+|r-1|(u+v)+(u+v)^2\right),\notag\\
 r_{\rm crit}-1={}&A_\kappa u+B_\kappa v+O((u+v)^2).
 \label{eq:selected-boundary-v2}
\end{align}
Here $r_{\rm crit}$ is the local benefit--cost ratio where $\Delta=0$. The positive coefficients quantify distinct formation losses. For linear fecundity,
\begin{equation}
 A_\kappa=\frac{2(\kappa^4+3)}{(1-\kappa^2)^2},\qquad
 B_\kappa=\frac{2(3\kappa^4+2\kappa^2+123)}
                 {(9-\kappa^2)(1-\kappa^2)^2}.
 \label{eq:selected-linear-coefficients-v2}
\end{equation}
Both fecundity maps have $A_0=6$ and $B_0=82/3$. Appendix~\ref{app:selected-boundary-v2} derives these coefficients and the exponential expressions. At $b=c$, the leading hierarchy dynamics makes aligned domains neutral.
Finite contacts retain an adverse payoff difference across the bridge
and a selected establishment transient, whose combined corrections
shift the invasion boundary.

The entries of the absorbing Green function are the expected times spent in mixed configurations before consensus. Its positive rooted-forest expansion uses the graph of configuration transitions, with copying rates as edge weights. This representation controls which designs can be favorable near neutrality. Up to a positive
normalization, the drift integral separates into favorable and
adverse time-integrated contributions, each a positive sum,
\begin{equation}
 \frac{\Delta}{\kappa}
 =\epsilon\,\mathcal U_\kappa-(1+\epsilon)\mathcal V_\kappa,
 \qquad
 \frac{\mathcal V_0}{\mathcal U_0}=1-\frac JK,
 \label{eq:v4-positive-rewards}
\end{equation}
for linear fecundity. For exponential fecundity the same decomposition
holds after each contribution of an edge joining unlike traits, in either reward, is
multiplied by the positive secant factor
$\Omega_{ij}=\int_0^1\exp\{\kappa[(1-t)f_D+tf_C]/c\}\,dt$,
where $f_C,f_D$ are the payoffs of the cooperator and defector on that
edge. These factors depend on the edge and configuration and remain
inside the occupation averages. The all-minors matrix-tree theorem~\cite{Chaiken1982} expresses cofactors through rooted forests. For this killed-generator inverse, the relevant forest sums have positive terms~\cite[Proposition~2]{AGNamGunawardena2025}.
Their common denominator cancels in the reward ratio, so bounded
changes of microscopic rates compare selected and neutral responses
uniformly over weights. Complement pairing improves the error to
$O(\kappa^2)$, even when the susceptibility is small.
Appendix~\ref{v3-app-selected-global-localization} proves this bound.

The forest bound forces every selected promoter, a weighting with $\Delta>0$ at the given
selection strength and benefit--cost ratio, into the same domain hierarchy for
sufficiently small positive $(\epsilon,\kappa)$. Designs with contrast
no greater than a reflected selected trial lie in a compact scaled
region. There, core saturation and pin uniqueness make the selected
critical branch globally reflected. This control covers all positive weights and is local in selection strength. As $\epsilon\downarrow0$, the forest comparison also forces entry into the hierarchy for selection in any fixed compact admissible interval. Extending core saturation and exact reflection to that interval requires the core-derivative and pin-curvature signs of the selected objective; the present proof controls those signs near neutrality (Appendix~\ref{v3-app-selected-global-localization}).

\begin{proposition}[Shared optimizer with selection]
\label{prop:v7-selected-time}
For each fixed finite $M_{\rm b}>656$, there are $\kappa_0,\epsilon_0>0$
such that the following holds for either linear or exponential
fecundity. Measure all four mean times in the rate-one global-attempt clock. Let $0\le\kappa\le\kappa_0$, $0<\epsilon<\epsilon_0$
and
\begin{equation}
 H_*(\epsilon,\kappa)\le H\le M_{\rm b}\epsilon^{-2},
 \label{eq:v7-selected-time-domain}
\end{equation}
where $H_*(\epsilon,\kappa)$ is the least contrast for the closed
selected target $\Delta\ge0$ at $b/c=1+\epsilon$.
At $\kappa=0$, the closed and strict targets and the root equation
below use the continuous limit of $\Delta/\kappa$, so $H_*(\epsilon,0)=H_*(\epsilon)$. The unconditional absorption time and the time conditioned
on fixation of the introduced type, for either uniform cooperator
or uniform defector introduction, minimized over positive $P_6$ weights with contrast at most $H$ and $\Delta\ge0$, have the same unique normalized
minimizer
\begin{equation}
 (H,q_-^{\rm sel},1,q_-^{\rm sel},H),\quad
 \Delta(H,q_-^{\rm sel},1,q_-^{\rm sel},H;\epsilon,\kappa)=0.
 \label{eq:v7-selected-time-root}
\end{equation}
The lower root continues the lower threshold branch and coincides
with the upper root at $H=H_*$. For strict positive advantage these
boundary values are infima when $H>H_*$; that strict target is
infeasible at the fold.
\end{proposition}
\begin{proof}
The interval resolvent
retains the physical attempt clock, and the strict core and
pin-curvature inequalities in the neutral proof persist at small
selection, including through the moving fold.
Appendix~\ref{app:v7-selected-time} proves this uniform
extension.
\end{proof}

The four mean times share their minimizing design, and their values there may differ at finite parameters.
Under the hypotheses of Proposition~\ref{prop:v7-selected-time},
with $\eta=\epsilon^2H$ fixed above $4A_\kappa B_\kappa$,
the common leading unconditional time is
$\theta_-^{\rm sel}/\epsilon$, and each success-conditioned time is
$3\theta_-^{\rm sel}/\epsilon$, where
\begin{equation}
 \theta_-^{\rm sel}(\eta,\kappa)
 =\frac{2A_\kappa}
 {1+\sqrt{1-4A_\kappa B_\kappa/\eta}}.
 \label{eq:v7-selected-time-frontier}
\end{equation}
The $O(1)$ time estimate is uniform when the scaled budget stays in a compact interval separated from the leading fold; the exact-root result follows the two roots through their merger. For every fixed finite $M_{\rm b}$, the proof controls sufficiently small gaps and selection strengths because the neutral core-derivative signs and pin curvature persist under the selected perturbation. The allowed $\epsilon_0$ and $\kappa_0$ remain implicit. Explicit bounds on the perturbation remainders and constrained curvatures would determine which specified parameter values are covered.

\subsection{Nonlinear selection and the tolerance margin}
\label{sec:III-nonlinear-margin}
Near least contrast, selection and contact errors draw on the same margin. Centering the payoffs at $b/c=1+\epsilon$, $\bar f_i/c=f_i/c-\epsilon/2$,
makes type complementation reverse their sign. The centered intensity is
\begin{equation}
 \begin{aligned}
 \lambda&=\frac{\kappa}{1+\epsilon\kappa/2},\quad
 \kappa=\frac{\lambda}{1-\epsilon\lambda/2}
       \quad \text{(linear)},\\
 \lambda&=\kappa\quad \text{(exponential)}.
 \end{aligned}
 \label{eq:v3-main-centered-selection}
\end{equation}
Thus $\lambda=c \delta_c$, with $\delta_c$ the centered coordinate introduced
in Sec.~\ref{sec:v4-observable-algebra}.
The normalized centered payoff $\bar f_i/c$ lies between $-(1+\epsilon/2)$ and
$1+\epsilon/2$. For the analytic continuation to signed selection
$\delta\in\mathbb R$, linear fecundity remains positive when
$|\lambda|<(1+\epsilon/2)^{-1}$, and exponential fecundity permits
every real $\lambda$. Complementing the two
introduction experiments makes $\Delta$ exactly odd on this common
domain. Near the weak critical design, the following result connects
that symmetry to the available contact margin.

\begin{proposition}[Selected margin and signal peak]
\label{prop:III-selected-margin}
Consider uniform singleton introduction on $P_6$ at $b/c=1+\epsilon$
with linear or exponential fecundity. Use the centered intensity of
Eq.~\eqref{eq:v3-main-centered-selection} and the contact coordinates of
Eq.~\eqref{eq:v3-main-error-coordinates}, centered on the exact weak
critical design $(H_*(\epsilon),q_*(\epsilon))$. In a common neighborhood
of that design in these contact coordinates, for sufficiently small
positive $\epsilon$ and $|\lambda|<\lambda_0$ with some fixed $\lambda_0>0$,
the quotient $6\Delta/(\epsilon\lambda)$ extends analytically to
$\epsilon=\lambda=0$ and is even in $\lambda$.

As $\epsilon\to0$ and a common scale $\upsilon\to0$, with
$d,\alpha_L,\alpha_R=O(\upsilon^2)$ and
$\xi_s,\xi_a,\lambda=O(\upsilon)$, the advantage obeys
\begin{equation}
 \begin{aligned}
 \Delta={}&\frac{\epsilon\lambda}{7872}
 \biggl[d-\frac{\alpha_L+\alpha_R}{2}
       -\frac{41}{9}(\xi_s^2+\xi_a^2)-C_0\lambda^2\biggr]\\
 &\quad+\epsilon\lambda\,O(\epsilon\upsilon^2+\upsilon^3),
 \end{aligned}
 \label{eq:v3-main-selected-signal}
\end{equation}
with
\begin{equation}
 C_0^{\rm lin}=24368/9,\qquad C_0^{\exp}=14296/9.
 \label{eq:v3-main-selection-cost}
\end{equation}
For sufficiently small $d>0$, optimize the contacts under the cap in this neighborhood.
After reducing $\lambda_0$ if necessary, the optimized signal has a
unique zero in $0<\lambda<\lambda_0$, denoted by
$\lambda_{\rm edge}$ and satisfying
\begin{equation}
 \lambda_{\rm edge}=\sqrt{d/C_0}[1+O(\epsilon+d)].
 \label{eq:v3-main-window}
\end{equation}
For every fixed $M>0$, its sign on $0<\lambda\le M\sqrt d$ is positive
precisely when $\lambda<\lambda_{\rm edge}$, for sufficiently small
$\epsilon,d$. Its maximum within that positive window satisfies
\begin{equation}
 \begin{gathered}
 \lambda_{\rm peak}=\frac{\lambda_{\rm edge}}{\sqrt3}
                       [1+O(\epsilon+d)],\\
 \Delta_{\rm peak}=\frac{\epsilon \,d^{3/2}}{11808\sqrt{3C_0}}
                       [1+O(\epsilon+d)].
 \end{gathered}
 \label{eq:v3-main-peak}
\end{equation}
\end{proposition}
\begin{proof}
Complement symmetry and the analytic elimination of fast excursions are
proved in Appendices~\ref{par:v3-joint-selection-coordinates} and
\ref{par:v3-joint-selected-analyticity}. The coefficient calculation and
selected boundary are in Appendices~\ref{par:v3-joint-selected-jet} and
\ref{par:v3-joint-joint-tolerance}.
Appendix~\ref{v3-app-selected-global-localization} controls contact
optimization in this neighborhood.
Appendix~\ref{par:v3-joint-signal-maximum} controls the differentiated
remainder after contact optimization and derives the peak.
\end{proof}

The analytic structure follows by dividing the effective operator for
aligned domains by its common escape-rate factor. Its limiting matrix
is invertible, as is the fast-excursion block. The cubic coefficient
can then be evaluated on the invariant twenty-state interval chain,
using the degree and parity constraints of
Sec.~\ref{sec:v4-observable-algebra}. Core deficits, pin errors and
nonlinear selection consume the same margin in
Eq.~\eqref{eq:v3-main-selected-signal}. For $\lambda>0$, the leading term determines the sign of $\Delta$
when its magnitude exceeds the additive remainder bound; the
analytic boundary resolves the zero when those terms compete.

The same coefficient measures the curvature of the finite-selection
resource boundary. Write
$A_\kappa=A_0+A_2\kappa^2+O(\kappa^4)$ and
$B_\kappa=B_0+B_2\kappa^2+O(\kappa^4)$, with
$A_0=6$ and $B_0=82/3$. At $\epsilon=0$, where $\kappa=\lambda$,
\begin{equation}
 4A_\kappa B_\kappa=656+C_0\kappa^2+O(\kappa^4),
 \quad C_0=4(A_2B_0+A_0B_2).
\label{eq:III-selected-fold-curvature}
\end{equation}
Thus the nonlinear penalty in the local signal and the increase in
the least scaled contact resource are two aspects of the same
selected establishment correction.

\begin{figure*}[tp]
\centering
\includegraphics[width=.94\textwidth]{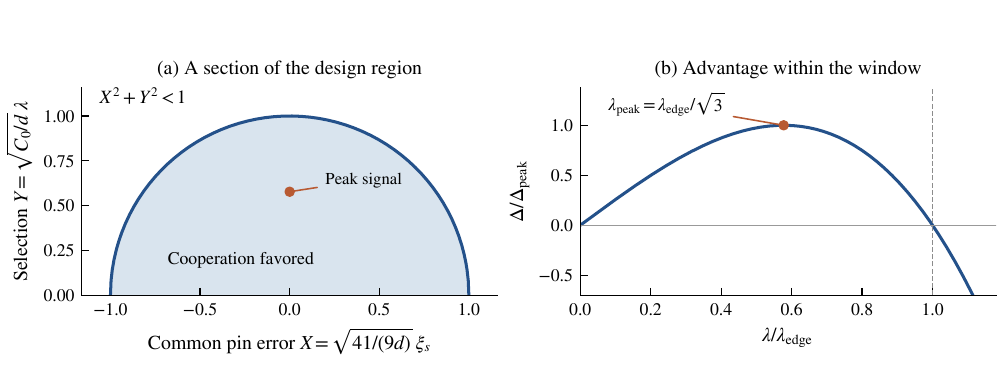}
\caption{Leading contact tolerance and paired fixation advantage $\Delta$ on the six-site path $P_6$ under uniform singleton introduction, at $b/c=1+\epsilon$ and small centered selection intensity $\lambda$. The surplus $d=\epsilon^2[H-H_*(\epsilon)]>0$ is measured from the least contrast for the weak target $R\le1+\epsilon$. The selection penalty $C_0$ equals $24368/9$ for linear fecundity and $14296/9$ for exponential fecundity.
(a) With saturated cores and no pin imbalance,
$X=\sqrt{41/(9d)}\,\xi_s$ rescales the common pin shift and $Y=\sqrt{C_0/d}\,\lambda$ rescales selection. The positive leading advantage lies inside $X^2+Y^2<1$, with $Y>0$. The full geometry additionally contains a second quadratic pin direction and two nonnegative core deficits.
(b) At optimized contacts, the normalized signal is $\Delta/\Delta_{\rm peak}=(3\sqrt3/2)Y(1-Y^2)$ to leading order, with its maximum
inside the positive interval. Here $\Delta_{\rm peak}$ is the maximum over selection at optimized contacts. 
The curves describe the joint local limit $\epsilon,d\to0$ with selection and pin errors of order $\sqrt d$. Their leading coefficients require small $d$, equivalently $656\mathfrak r\ll1$ for relative surplus $\mathfrak r=[H-H_*(\epsilon)]/H_*(\epsilon)$.}
\label{fig:v3-tolerance}
\end{figure*}

Nonmonotonic fixation at intermediate selection is known in cooperative
games~\cite{McAvoyRaoHauert2021}. Here the scaled surplus $d$ fixes the
selection strength at which the optimized advantage reverses. Selection
of order $\sqrt d$ multiplies a residual margin of order $d$, explaining
the signal exponent $3/2$. In particular,
$H-H_*=\omega_{\rm lay}/\epsilon$ permits selection of order
$\sqrt\epsilon$ and a peak advantage of order $\epsilon^{5/2}$.
This margin can also be expressed as a control probability. Keep both cores at their cap and give the two pin contacts independent Gaussian logarithmic errors of variance $s_{\rm e}^2$, centered on the optimized design. To leading order, confidence $1-p_{\rm fail}$ requires
\begin{equation}
 \mathfrak r\ge s_{\rm e}^2\log(1/p_{\rm fail})
                  +\frac{C_0\lambda^2}{656},
 \qquad \mathfrak r=\frac{H-H_*}{H_*}.
 \label{eq:v7-gaussian-confidence}
\end{equation}
Here $0<p_{\rm fail}<1$ is fixed, $\epsilon,d,s_{\rm e},\lambda\to0$, $s_{\rm e}^2/d$ stays in a compact positive interval and $\lambda^2/d$ is bounded. Appendix~\ref{app:v7-gaussian-tolerance} derives the Gaussian radial tail. This Gaussian model samples the two quadratic pin directions; random core deficits consume the margin linearly and would enter through an average over their own specified one-sided distribution. The same quadratic inequality measures pin errors
and nonlinear selection loss, as illustrated in Fig.~\ref{fig:v3-tolerance}.

On the reflected linear-fecundity family the radius of convergence of the
expansion in the centered selection intensity tends to one as
$\epsilon\to0$, while the favorable window shrinks as the scaled
surplus $d\to0$
(Appendix~\ref{par:v3-joint-radius-window}). The adverse nonlinear
response therefore reverses the sign within the convergent expansion.

\subsection{Signal and observation time}
\label{sec:signal-information}
The peak in Proposition~\ref{prop:III-selected-margin} fixes the
advantage that a terminal experiment must resolve. In terms of the relative surplus $\mathfrak r=(H-H_*)/H_*$, the scaled
surplus is $d=656\,\mathfrak r[1+O(\epsilon)]$. Substitution into Eq.~\eqref{eq:v3-main-peak} gives the small-surplus coefficients approximately $0.016\,\epsilon\,\mathfrak r^{3/2}$ for linear and $0.021\,\epsilon\,\mathfrak r^{3/2}$ for exponential fecundity, retaining relative error $O(\epsilon+d)$. These are leading-order coefficients for $d\ll1$, equivalently $656\mathfrak r\ll1$; an error bar at specified $\epsilon$ and surplus requires an explicit bound on the constant in the $O(\epsilon+d)$ remainder. At a fixed, sufficiently small positive $\mathfrak r$, the limit $\epsilon\to0$ leaves $d$ asymptotic to $656\mathfrak r$; the advantage is then of order $\epsilon$, and its relative error estimate retains an $O(d)$ contribution. The finite-surplus expression~\eqref{eq:III-main-finite-surplus} below keeps the full dependence on the scaled budget at leading order in $\epsilon$. The layer $d=\omega_{\rm lay}\epsilon$
below, a relative surplus of order $\epsilon$, is the extreme
near-critical regime.

To measure the advantage in the layer $d=\omega_{\rm lay}\epsilon$,
introduce each type equally often
and retain its introduction site. Equal numbers of independent
completed runs in the twelve type-and-site classes give an unbiased estimate of the uniform advantage. Let $h_i^+$ and $h_i^-$ be the probabilities of all-cooperator consensus after, respectively, cooperator and defector introduction at site $i$, and let $\widehat h_i^\pm$ be their observed fractions. For $M/12$ runs in each class, let $T_{\rm total}$ be their summed duration in global attempts. The estimator is
\begin{equation}
 \widehat\Delta_{\rm str}=\frac16\sum_{i=1}^6
       (\widehat h_i^++\widehat h_i^-)-1.
 \label{eq:III-observation-estimator}
\end{equation}
The class variances and mean durations imply
\begin{equation}
 \operatorname{Var}(\widehat\Delta_{\rm str})\sim\frac1{3M},
 \qquad
 \E[T_{\rm total}]\sim\frac{12M}{\epsilon}.
 \label{eq:v3-main-variance}
\end{equation}
Figure~\ref{fig:III-resource-to-signal}(c) summarizes how the signal, run count and duration combine. The pins dominate both uncertainty and duration. Their fixation
probabilities approach one half and their mean run time approaches
$36/\epsilon$, whereas follower outcomes have variance
$O(\epsilon)$ and bounded mean duration. Balancing the site counts in advance, or stratifying independently
randomized introductions by their retained labels, reduces the leading
duration needed for a fixed signal-to-noise ratio by a factor $3/5$
compared with randomized uniform sites whose labels are discarded. The required number of completed runs still scales as
$\epsilon^{-5}$, and their aggregate duration as $\epsilon^{-6}$
global attempts. At a sufficiently small fixed positive relative surplus the same estimator needs
$O(\epsilon^{-2}\mathfrak r^{-3})$ runs and $O(\epsilon^{-3}\mathfrak r^{-3})$
attempts, with the inverse-cube surplus dependence understood as $\mathfrak r\downarrow0$.

For positive class counts $m_i^\pm$, write $\varsigma_{i,\pm}^2=h_i^\pm(1-h_i^\pm)$ for the Bernoulli (binary-outcome) variances and $t_i^\pm$ for the mean completed-run durations, and put $\mathcal T=\sum_{i,\pm}m_i^\pm t_i^\pm$. Cauchy--Schwarz gives the exact variance--duration bound
\begin{equation}
 \operatorname{Var}(\widehat\Delta_{\rm str})\mathcal T
 \ge\frac1{N^2}\left(\sum_{i,\pm}\sqrt{\varsigma_{i,\pm}^2 t_i^\pm}\right)^2,
 \qquad N=6.
 \label{eq:v7-stratified-time-bound}
\end{equation}
Equality at continuous class counts requires $m_i^\pm\propto\sqrt{\varsigma_{i,\pm}^2/t_i^\pm}$. This is the unequal-cost version~\cite[Sec.~5.5, Theorem~5.6]{Cochran1977} of Neyman's stratified allocation~\cite{Neyman1934}. Here the four pin strata make the right-hand side asymptotic to $4/\epsilon$, which equal class counts already attain to leading order.

The same leading bound applies to regular terminal-outcome estimates that treat the site-specific probabilities as unknown and use mean duration as cost. Regularity means that the limiting error distribution is stable under local changes of those probabilities on the scale of their statistical uncertainty. At each fixed positive design the product Bernoulli experiment has a Gaussian local limit, called local asymptotic normality~\cite[Chap.~7]{VanDerVaart1998}. The H\'ajek--Le Cam convolution theorem then bounds its asymptotic estimation error~\cite{Hajek1970}, with a textbook treatment in Ref.~\cite[Sec.~8.5]{VanDerVaart1998}. In a joint small-$\epsilon$ limit, take diverging sample counts in all four pin classes and require regularity under their inverse-square-root count perturbations. Their probabilities tend to one half, so their Bernoulli variances remain uniformly positive and the Gaussian local limit holds along these sequences. Varying these four probabilities already enforces the leading bound; the exact variance calculation above needs no Gaussian approximation for follower outcomes. Appendix~\ref{app:v7-terminal} derives both statements and separates this measurement problem from estimating selection with known game and weights.

For a fixed-sample one-sided test in this Gaussian local regime, the signal-to-noise target also fixes limiting power. If $z_p$ is the $p$-quantile of a unit normal distribution, size $\alpha_{\rm err}$ and power $1-\beta_{\rm err}$ require $\mathfrak s=z_{1-\alpha_{\rm err}}+z_{1-\beta_{\rm err}}$. This translation assumes fixed errors with $\alpha_{\rm err}+\beta_{\rm err}<1$ and a consistent estimate of the terminal variance. The quoted observation asymptotics count completed-run durations with negligible reset overhead. Known reset durations can be added to each class cost. Uniformly bounded reset times leave the leading results unchanged. The calculation below also optimizes common preparation costs that grow with the hierarchy.

The establishment coefficients also resolve finite contrast surpluses.
On the reflected family $(H,q,1,q,H)$, fix
$\eta=H\epsilon^2$ and an admissible positive $\kappa$, and put
$\theta=\epsilon q$, which is also the leading scaled neutral absorption time in this reflected hierarchy. Equations~\eqref{eq:selected-boundary-v2} and
\eqref{eq:selected-linear-coefficients-v2}, with the exponential
counterpart in Appendix~\ref{app:selected-boundary-v2}, give
\begin{equation}
 \Delta=\frac{\epsilon\kappa}{6}G_\kappa^{(0)}+O(\epsilon^2),
 \qquad G_\kappa^{(0)}=1-\frac{A_\kappa}{\theta}
                           -\frac{B_\kappa\theta}{\eta}.
 \label{eq:III-main-finite-surplus}
\end{equation}
This profile keeps the finite-surplus dependence without expanding about
$\eta=656$. Its positive interval in $\theta$ is nonempty exactly when
$\eta>4A_\kappa B_\kappa$. The resulting observation problem compares
independent terminal experiments on the reflected family.

\begin{proposition}[Observation cost and design]
\label{prop:III-observation-design}
Use the terminal experiment defined above for uniform singleton
introduction on $P_6$ at $b/c=1+\epsilon$, with reflected contacts
$(H,q,1,q,H)$, $H=\eta/\epsilon^2$, $q=\theta/\epsilon$, and
unknown type-and-site absorption probabilities. Cost is expected
aggregate duration in global attempts. Let $0<\kappa<1$ for linear
fecundity and $\kappa>0$ for exponential fecundity, with
$G_\kappa^{(0)}(\theta,\eta)>0$. As $\epsilon\downarrow0$,
\begin{equation}
 \inf_{\{m_i^\pm\}}\operatorname{Var}(\widehat\Delta_{\rm str})\mathcal T
 =\frac{\theta}{3\epsilon}[1+O(\epsilon)],
 \label{eq:III-main-observation-coefficient}
\end{equation}
where the infimum is over positive continuous class allocations.
Rounding an optimal allocation to integer counts preserves this
leading value when the counts grow; equal class counts also attain it.
Consequently the least duration over these allocations for a fixed
signal-to-noise target $\mathfrak s>0$ obeys
\begin{equation}
 \mathcal T_{\mathfrak s}
 \sim\frac{12\mathfrak s^2}{\epsilon^3}
       \frac{\theta}{\kappa^2[G_\kappa^{(0)}]^2}.
 \label{eq:III-main-observation-design}
\end{equation}
The relative error is $O(\epsilon)$, uniformly on compact admissible
parameter sets with $G_\kappa^{(0)}$ bounded away from zero.

At fixed $(\eta,\kappa)$ with $\eta>4A_\kappa B_\kappa$, the leading
signal and observation duration have unique pin optima
\begin{equation}
 \theta_{\rm sig}=\sqrt{\frac{A_\kappa\eta}{B_\kappa}},\qquad
 \theta_{\rm obs}=\frac{6A_\kappa}
 {1+\sqrt{1+12A_\kappa B_\kappa/\eta}}.
 \label{eq:III-main-observation-pin}
\end{equation}
They satisfy
$\theta_-^{\rm sel}<\theta_{\rm obs}<\theta_{\rm sig}$, where
$\theta_-^{\rm sel}$ is the lower zero of $G_\kappa^{(0)}$.
For every fixed $\eta>656$, the leading observation objective
$\theta/\{\kappa^2[G_\kappa^{(0)}]^2\}$ has a unique joint minimum over
admissible $(\kappa,\theta)$ with positive $G_\kappa^{(0)}$.
\end{proposition}
\begin{proof}
Appendix~\ref{app:III-finite-surplus} derives the class variances and
physical durations, minimizes the leading pin objective and proves
joint uniqueness from strict log-convexity of the establishment
coefficients. The allocation and statistical bounds are derived in
Appendix~\ref{app:v7-terminal}.
\end{proof}

The observation optimum accepts a smaller advantage in exchange for
faster completed trials. Its uniqueness is a consequence of the
establishment coefficients. With $x=\log\theta$, strict log-convexity
of $A_\kappa$ and $B_\kappa$ makes
$\log\kappa-x/2+\log G_\kappa^{(0)}$ strictly concave in $(\kappa,x)$
on the convex region $G_\kappa^{(0)}>0$. This converts the coupled choice of selection and pin contact
into one concave optimization problem. At fixed scaled parameters the
advantage vanishes linearly with $\epsilon$, while the duration needed
to resolve it grows as $\epsilon^{-3}$.

These design optima concern the reflected regular hierarchy.
Appendix~\ref{app:III-finite-surplus} identifies the additional
quantities that a comparison with arbitrary contacts must control,
including independent variations of both pins and cores in the signal,
terminal variances and mean durations. At each fixed $\eta>656$, the
leading optimal $\kappa$ is positive and finite, and tends to zero as
$\eta\downarrow656$. Finite selection intensity is therefore part of
the observation design at fixed surplus.
Proposition~\ref{prop:v7-selected-time} treats
the duration of a single invasion over all positive path weights;
its continuation from neutrality holds at small selection and bounded
scaled budgets.

Preparation changes this balance continuously. Suppose each completed trial
also incurs a common known preparation duration $r_{\rm prep}$, measured in
the same units as the run and independent of the contacts and selection intensity. With
$\mathcal R_{\rm prep}=\epsilon r_{\rm prep}$ fixed, optimizing the allocation
among introduction classes gives the leading aggregate cost
\begin{equation}
 \mathcal T_{\mathfrak s}^{\rm prep}
 \sim\frac{4\mathfrak s^2}{\epsilon^3}
 \frac{3\theta+\mathcal R_{\rm prep}}
 {\kappa^2[G_\kappa^{(0)}]^2}.
 \label{eq:III-preparation-cost}
\end{equation}
At fixed $(\eta,\kappa)$ the minimizing pin contact increases monotonically
from $\theta_{\rm obs}$ to $\theta_{\rm sig}$ as preparation becomes dominant.
The exact allocation formula also allows class-dependent costs, whose
effect depends on the specified preparation protocol. The common-preparation-cost analysis of Appendix~\ref{app:III-finite-surplus}
derives the common-cost interpolation and the unique joint optimum when selection is also adjustable.

The surplus controls the size of the statistical task as well as the optimal contacts. For linear fecundity at $\epsilon=0.1$, direct finite-state evaluations at the limiting signal-optimal designs give advantages $2.82\times10^{-4}$ at $\eta=1000$ and $2.92\times10^{-3}$ at $\eta=5000$. Equal counts in the twelve classes require about $1.69\times10^7$ and $1.57\times10^5$ completed trials for signal-to-noise ratio two. Table~\ref{tab:III-observation-scales} compares these finite evaluations with both maps' limiting profiles. Combining the local cubic signal with the leading class variance gives $1.51\times10^{17}$ trials at $\epsilon=0.1$ and the much smaller scaled surplus $d=0.1$. A quantitative error bar for this leading-order count requires bounds on the signal remainder in Eq.~\eqref{eq:v3-main-peak} and on the finite-contact corrections to the class variances. Resolving the advantage becomes much harder close to least contrast. The finite-surplus calculation retains the establishment corrections needed farther from that boundary.

\begin{table*}[t]
\caption{Observation scales at $\epsilon=0.1$ for the reflected six-site
path under uniform introduction. For each $\eta$, the selection intensity
$\kappa$ and pin coordinate $\theta$ maximize the limiting signal using the
full coefficient functions $A_\kappa,B_\kappa$, solving Eq.~\eqref{eq:III-signal-optimal-selection} with $\theta=\theta_{\rm sig}$ from Eq.~\eqref{eq:III-main-observation-pin}.
The contacts are $(\eta/\epsilon^2,\theta/\epsilon,1,
\theta/\epsilon,\eta/\epsilon^2)$. The listed design parameters are rounded; the evaluations use the unrounded roots of these defining equations.
The columns labeled ``finite'' evaluate the full configuration chain at these same designs, without reoptimizing the contacts or selection strength at finite $\epsilon$.
Both run counts use equal numbers of independent completed trials in all
twelve type-and-site classes and target signal-to-noise ratio two.
The leading count is $4/[3(\Delta^{(0)})^2]$; the finite count uses the
computed class variances. The finite evaluations use 70-digit arithmetic,
with harmonic and mean-time residuals below $3\times10^{-70}$.
The residuals measure agreement with the finite-state equations; the entries are numerical evaluations without validated enclosures or a finite-sample power analysis.}
\label{tab:III-observation-scales}
\begin{ruledtabular}
\begin{tabular}{llrrrrrr}
Fecundity & $\eta$ & $\kappa$ & $\theta$ & $\Delta^{(0)}$ & $\Delta$ (finite)
 & $M^{(0)}$ & $M$ (finite)\\
\hline
Linear & 1000 & 0.1857 & 14.79 & $3.995\times10^{-4}$ & $2.819\times10^{-4}$
 & $8.355\times10^6$ & $1.689\times10^7$\\
Linear & 5000 & 0.4134 & 32.92 & $3.200\times10^{-3}$ & $2.920\times10^{-3}$
 & $1.302\times10^5$ & $1.572\times10^5$\\
Exponential & 1000 & 0.2479 & 14.73 & $5.287\times10^{-4}$ & $3.760\times10^{-4}$
 & $4.769\times10^6$ & $9.503\times10^6$\\
Exponential & 5000 & 0.6033 & 32.34 & $4.499\times10^{-3}$ & $4.126\times10^{-3}$
 & $6.588\times10^4$ & $7.893\times10^4$
\end{tabular}
\end{ruledtabular}
\end{table*}

Simulation can exploit a different observation protocol. At neutrality,
common directed copying events keep cooperator-at-$i$ and defector-at-$i$
runs exactly complementary, so their paired terminal contribution has zero
variance. At fixed positive contacts this coupling has variance
$O(|\kappa|)$ near neutrality. The estimate bounds the probability that
coupled source choices separate before absorption, so its coefficient
can grow with the mean lifetime along the protected hierarchy.
Determining the small-gap observation cost therefore requires control
of the coupled process through the slow domain-exchange time. The correlated-terminal-pair analysis of Appendix~\ref{app:III-finite-surplus} derives the coupling identity. The independent-run bounds above refer to the stated terminal experiment.

For the stationary mean cooperative fraction, the small signal also fixes how well successive introductions must be separated. Add independent symmetric spontaneous flips at total rate $\nu$, with each site chosen uniformly. At fixed positive contacts, the rare-introduction limit of that mean is $X_C^{\rm rare}=\rho_C/(\rho_C+\rho_D)$, so its bias above one half is $\Delta/[2(\rho_C+\rho_D)]$. In the optimized layer $d=\omega_{\rm lay}\epsilon$, with fixed $\omega_{\rm lay}>0$ and $q\asymp\epsilon^{-1}$, the
condition $\nu=o(\epsilon^2)$ suffices to preserve the positive
$O(\epsilon^{5/2})$ rare-introduction bias.
Appendix~\ref{app:v7-rare-signal} derives this joint limit by bounding
the correction relative to the positive bias by $O(\nu q/d)$.
This gives a sufficient rate condition tied to the signal margin.
More generally, within the local signal optimum, $\nu=o(\epsilon d)$
resolves the positive margin. The required separation of introductions
therefore changes with the available contrast surplus.

\section{Designs with changed games and contacts}
\label{sec:v7-robustness}
The preceding optima depend on the game, encounter rule and retained contacts. Response identities transfer some changes of game and encounters to the same design problem. Larger protected cores give achievable resource bounds. Two explicit changes of support then allow sharp cost comparisons within specified families.

\subsection{Games, encounters and larger cores}
\label{sec:III-transfers}
The game and encounter identities and the controlled larger-core reduction used below are derived in Ref.~\cite[Appendices~E\,1--E\,3]{ChowdhuryProtectedDomains}. The first two preserve the stated feasible sets; the core reduction bounds the response error on broader supports.

\paragraph*{Changing the game.}
For a symmetric two-strategy game, let $a_{XY}$ be the row player's payoff for type $X$ against type $Y$. With the same averaged contacts and common singleton introduction rule, the paired weak-response criterion is $\sigma_{\rm str}(a_{CC}-a_{DD})+a_{CD}-a_{DC}>0$, where the structure coefficient~\cite{TarnitaEtAl2009} is $\sigma_{\rm str}=(K+J)/(K-J)$. On $J>0$ it equals $(R+1)/(R-1)$ and decreases with $R>1$. If $a_{CC}>a_{DD}$ and $t_{\rm game}=(a_{DC}-a_{CD})/(a_{CC}-a_{DD})>1$, the design target is therefore
\begin{equation}
 R<\frac{t_{\rm game}+1}{t_{\rm game}-1}.
 \label{eq:III-game-target}
\end{equation}
The threshold-constrained contact and neutral-time results apply whenever this mapped target satisfies their hypotheses. In particular, a small-gap result requires proximity to the relevant support floor; the five-site path all-cap result of Proposition~\ref{prop:v10-onsets} retains its broader domain. If $a_{CC}>a_{DD}$ and $t_{\rm game}\le1$, every donation-promoting design satisfies the criterion, although other designs may do so too. General games keep a payoff term quadratic in the traits, which the donation game lacks. For payoff degree two the proof of Eq.~\eqref{eq:v4-degree-response} gives $\deg h_m\le\min(N,1+2m)$ (Appendix~\ref{app:v3-degree-closure}), so their higher-order response requires a new calculation in these larger observable spaces.

\paragraph*{Changing payoff normalization.}
With accumulated payoffs as defined in Sec.~\ref{b:sec:model}, a different geometry can promote cooperation. Under uniform introduction, the six-site path with contacts $(1,3203/100,\allowbreak4096/5,\allowbreak26837,\allowbreak10^6)$ has threshold $R\simeq13.9667$. Exact rational bounds prove positive weak advantage at $b/c=14$ (Appendix~\ref{app:III-finite-comparisons}). This example relies on a large gradient of contact strengths, and its threshold stays far above the protected-domain threshold one. A common rescaling of the accumulated-payoff weights changes both response amplitudes proportionally and leaves their ratio unchanged.

\paragraph*{Changing encounters.}
To separate encounters from replacement~\cite{OhtsukiNowakPacheco2007}, let encounters use $Q_{\rm int}^{\alpha}=(1-\alpha)P+\alpha W$, where $W_{ij}=(1-\delta_{ij})/(N-1)$ and $0\le\alpha\le1$, while offspring still use $P$. Write $R_\alpha$ for the resulting weak threshold. For a positive target $r_{\rm tar}$ and $0\le\alpha<1$, the exact feasible-set correspondence on the positive-benefit domain is
\begin{equation}
 R_\alpha\le r_{\rm tar}\quad\Longleftrightarrow\quad
 R\le\frac{(1-\alpha)r_{\rm tar}}{1+\alpha r_{\rm tar}/(N-1)}.
 \label{eq:III-encounter-target}
\end{equation}
This substitution transfers weak-threshold contrast, neutral-time and contact-tolerance optima within their proved target domains. Neutral replacement and its introduction distribution are unchanged. A support floor $R_*$ becomes $R_*/[1-\alpha-\alpha R_*/(N-1)]$ while its denominator is positive; at and above $\alpha=(N-1)/(N-1+R_*)$, no weak promoter exists. For exponential fecundity the full selected generator and physical clock reduce exactly to those with $b_{\rm eff}=(1-\alpha)b$ and $c_{\rm eff}=c+\alpha b/(N-1)$. Selected design conclusions transfer wherever these effective parameters satisfy their hypotheses. Linear fecundity retains the exact weak-response transfer but its configuration-dependent common payoff offset changes the full selected dynamics.

\paragraph*{Larger protected cores.}
For a broader fixed support $\mathsf G$, suppose the vertices partition into two adjacent pins and two disjoint connected follower cores, each containing at least two sites and attached to its own pin. Give core edges weight $Q$, selected pin-to-own-core edges fixed positive proportions of total weight $q$, and all remaining edges weight one. With $u=1/q$ and $v=q/Q$, the uniform weak response $\mathcal S$ has leading term $(b-c)/N$ and error $O_{\mathsf G}((u+v)(|b|+|c|))$. The birth-associated weak response $\mathcal S^{\rm b}$ has leading term $u(b-c)/N$, with error $O_{\mathsf G}((u^2+v)(|b|+|c|))$. At target $b/c=1+\epsilon$, choose $q$ a sufficiently large graph-dependent multiple of $\epsilon^{-1}$. Taking $Q=q^2$ under uniform introduction, or $Q=q^3$ under birth-associated introduction, then gives
\begin{equation}
 H_{\min}^{\mathsf G,{\rm U}}(\epsilon)=O_{\mathsf G}(\epsilon^{-2}),\qquad
 H_{\min}^{\mathsf G,\rm b}(\epsilon)=O_{\mathsf G}(\epsilon^{-3}).
 \label{eq:III-broader-cost}
\end{equation}
The response estimates above give these achievable upper bounds, with constants depending on the fixed graph and core proportions. Proving either power sharp on a given support requires a lower bound that excludes other positive-weight hierarchies with slower resource divergence. The access--protection competition can thus survive extra contacts, while the sharp constants in Table~\ref{tab:cost} remain specific to its paths and cycles.

\subsection{Resource costs of added contacts}
\label{sec:III-added-contacts}
Contact contrast measures the range of retained weights without charging separately for their number. An additional access route and a compulsory contact can therefore change the leading resource coefficient in opposite directions, as the following two families show.

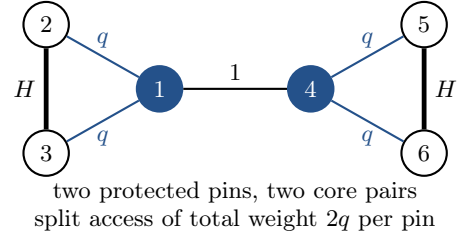
\begin{figure}[!htbp]
\centering
\begin{tikzpicture}[x=1cm,y=1cm,font=\small,
 pin/.style={circle,draw=figureblue,fill=figureblue,text=white,thick,minimum size=6mm,inner sep=0pt},
 core/.style={circle,draw,fill=white,thick,minimum size=6mm,inner sep=0pt}]
\node[pin] (p1) at (-1,0) {1};
\node[pin] (p4) at (1,0) {4};
\node[core] (c2) at (-2.5,.85) {2};
\node[core] (c3) at (-2.5,-.85) {3};
\node[core] (c5) at (2.5,.85) {5};
\node[core] (c6) at (2.5,-.85) {6};
\draw[thick] (p1)--node[above]{$1$}(p4);
\draw[thick,figureblue] (p1)--node[above]{$q$}(c2);
\draw[thick,figureblue] (p1)--node[below]{$q$}(c3);
\draw[thick,figureblue] (p4)--node[above]{$q$}(c5);
\draw[thick,figureblue] (p4)--node[below]{$q$}(c6);
\draw[line width=1.8pt] (c2)--node[left]{$H$}(c3);
\draw[line width=1.8pt] (c5)--node[right]{$H$}(c6);
\node[align=center] at (0,-1.55) {two protected pins, two core pairs\\split access of total weight $2q$ per pin};
\end{tikzpicture}
\caption{Two triangles joined by a unit bridge, with $1\le q\le H$ and contact contrast $H$. Each protected pin contacts both sites of its own core pair. Distributing a fixed total access weight over these two contacts reduces the leading leakage penalty relative to the six-site path. Equation~\eqref{eq:split-access-cost} gives the resulting sharp resource constants within this two-parameter family.}
\label{fig:III-split-access}
\end{figure}

\paragraph*{Split access lowers the contact cost.}
The constants depend on how each pin contacts its protected core.
Consider two triangles joined by a bridge of weight one. In each triangle,
the pin contacts both core sites with weight $q$, and the core edge has
weight $H$, where $1\le q\le H$ (Fig.~\ref{fig:III-split-access}). Write
$H_{\rm split}^\mu(\epsilon)$ for the least contrast within this
two-parameter family reaching $R^\mu\le1+\epsilon$.
The exact pair system gives
\begin{equation}
 \begin{aligned}
 R^{\rm U}-1&=\frac3q+\frac{48q}{H}
   +O\!\left[\left(q^{-1}+q/H\right)^2\right],\\
 H_{\rm split}^{\rm U}(\epsilon)&\sim576\epsilon^{-2},\\
 R^{\rm b}-1&=\frac3q+\frac{60q^2}{H}
   +O\!\left[\left(q^{-1}+q^2/H\right)^2\right],\\
 H_{\rm split}^{\rm b}(\epsilon)&\sim3645\epsilon^{-3}.
 \end{aligned}
 \label{eq:split-access-cost}
\end{equation}
Both remainders are uniform in a neighborhood of their respective
positive hierarchy corner. Appendix~\ref{app:split-access} proves
sharpness over the stated two-parameter family, including exclusion of
other limiting regimes. The constants are smaller than their path
values $656$ and $4050$, by the fractions $5/41$ and $1/10$, respectively. In terms of the total access weight
$q_{\rm eff}=2q$, the balances are
$6/q_{\rm eff}+24q_{\rm eff}/H$ and
$6/q_{\rm eff}+15q_{\rm eff}^2/H$; the leakage coefficients fall
from the path values $82/3$ and $50/3$, respectively. At fixed total access weight, the central loss is unchanged. The optimized contrast coefficient is proportional to the core-leak coefficient in each balance, so its fractional reduction is exactly $1-24/(82/3)=5/41$ or $1-15/(50/3)=1/10$. Since $576<656$ and $3645<4050$, for all sufficiently small positive $\epsilon$ the two-triangle family reaches $R^\mu\le1+\epsilon$ under either introduction rule with less contrast than every positive six-site path weighting that reaches the same target. The family uses a common access weight and a common core weight; a full two-triangle optimum would vary these contacts independently, and an optimum over all six-site supports would compare different contact patterns.

Split access lowers contrast without increasing the leading waiting
time. At its uniform least-contrast design, the neutral absorption and
successful-fixation means are $12/\epsilon+O(1)$ and
$36/\epsilon+O(1)$, respectively, as on the six-site path. More
generally, fix a cap $H=\eta\epsilon^{-2}$ with $\eta>576$, and let
$\mathcal T_{{\rm abs},{\rm sp}}$ and $\mathcal T_{{\rm fix},{\rm sp}}$
be the separate minimum neutral times within this family under uniform
introduction and the target $R^{\rm U}\le1+\epsilon$. Then
\begin{equation}
 \begin{aligned}
 \theta_{\rm sp}(\eta)&=\frac{12}{1+\sqrt{1-576/\eta}},\\
 \mathcal T_{{\rm abs},{\rm sp}}&=\frac{\theta_{\rm sp}(\eta)}\epsilon+O_\eta(1),\\
 \mathcal T_{{\rm fix},{\rm sp}}&=\frac{3\theta_{\rm sp}(\eta)}\epsilon+O_\eta(1).
 \end{aligned}
 \label{eq:split-time-frontier}
\end{equation}
Here $6<\theta_{\rm sp}<12$. The normalized curve agrees with the
path's when each contrast is divided by its own leading feasibility
constant. Appendix~\ref{app:split-access} derives the entrance probabilities, physical clock and uniform optimization. For each fixed $\eta>576$ the lower root is simple and the implicit-function theorem corrects it to the exact target. As $\eta\downarrow576$ the two leading roots merge, so a joint limit with $\epsilon\to0$ needs estimates uniform through this fold, as the exact-root construction of Sec.~\ref{sec:v4-time} gives for the six-site path.

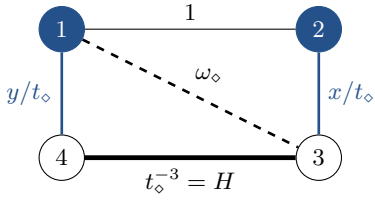
\begin{figure}[htbp]
\centering
\begin{tikzpicture}[x=1cm,y=1cm,font=\small,
 site/.style={circle,draw,minimum size=6mm,inner sep=0pt},
 pin/.style={site,fill=figureblue,draw=figureblue,text=white}]
\node[pin] (d1) at (0,1.7) {$1$};
\node[pin] (d2) at (3.4,1.7) {$2$};
\node[site] (d3) at (3.4,0) {$3$};
\node[site] (d4) at (0,0) {$4$};
\draw (d1)--node[above] {$1$}(d2);
\draw[figureblue,line width=1pt] (d2)--node[right] {$x/t_\diamond$}(d3);
\draw[line width=1.8pt] (d3)--node[below] {$t_\diamond^{-3}=H$}(d4);
\draw[figureblue,line width=1pt] (d4)--node[left] {$y/t_\diamond$}(d1);
\draw[dashed,line width=1pt] (d1)--node[above right,pos=.48] {$\omega_\diamond$}(d3);
\end{tikzpicture}
\caption{A compulsory diagonal separates the number of contacts from
contact contrast. The two dark pins share two followers; the dashed
edge has fixed retained weight $\omega_\diamond\ge1$ while $t_\diamond=H^{-1/3}\to0$.
For uniform introduction, the best balance within this family has
$y/x=1+3\omega_\diamond/4$. At $\omega_\diamond=1$, its leading contrast cost is
$18144\epsilon^{-3}$ for threshold $3+\epsilon$, compared with
$10368\epsilon^{-3}$ for the globally optimized four-cycle.
The comparison is within the indicated diamond family.}
\label{fig:III-diamond}
\end{figure}

\paragraph*{The cost of an additional contact.}
A support change can preserve a limiting threshold while changing the
finite resources needed to approach it. Here retain
source-first Bd, uniform introduction and the common contrast cap. The diamond and four-cycle share the sharp floor three recalled in Sec.~\ref{b:sec:resources}.
Add the diagonal $13$ to a four-cycle with cyclic contacts
$(1,x/t_\diamond,t_\diamond^{-3},y/t_\diamond)$, where $t_\diamond=H^{-1/3}$ and $x,y$ are local
positive contact coordinates. Give the diagonal a fixed weight
$\omega_\diamond\ge1$, included in the same cap (Fig.~\ref{fig:III-diamond}).

The exact pair expansion gives
\begin{equation}
 R_\diamond=3+t_\diamond\left[\frac8x+\frac{8+6\omega_\diamond}{y}+6xy\right]
                 +O(t_\diamond^2).
 \label{eq:III-diamond-response}
\end{equation}
The three positive terms have fixed product $384+288\omega_\diamond$.
Their sum is least when the three terms are equal, with
$y/x=1+3\omega_\diamond/4$. Optimizing this balance, with $x,y$ in a fixed positive compact
neighborhood of the minimum, gives the family contrast at target
$3+\epsilon$
\begin{equation}
 H_{\rm fam}(\epsilon,\omega_\diamond)
 =(10368+7776\omega_\diamond)\epsilon^{-3}[1+O(\epsilon)].
 \label{eq:III-diamond-cost}
\end{equation}
The weakest permitted diagonal, $\omega_\diamond=1$, multiplies the cycle's
sharp leading cost by $7/4$. Its extra copying route and the changed
normalization at pin 1 add the penalty $6\omega_\diamond/y$, which favors an
unequal pair of access contacts. Appendix~\ref{app:III-diamond} derives the expansion with uniform errors on fixed positive compact sets of $(x,y,\omega_\diamond)$. It determines the resource penalty when the diagonal weight stays fixed and the access ratios remain in this scaling sector. Allowing the diagonal to grow with the cap or the access ratios to leave these compact sets can change the leading balance; comparing those regimes is the remaining step toward an unrestricted diamond optimum.

\section{Discussion}\label{sec:discussion}
The contact pattern that protects a lineage also determines how a new lineage reaches it. Uniform introduction reaches every site equally, including the protected pins. Under birth-associated introduction, the low replacement rate preserving a pin suppresses mutation arrival at that same site. The six-site path gives a sharp measure of this conflict. Contacts optimized for uniform introduction have a birth-associated weak-selection threshold tending to $173/23$. Reoptimizing restores approach to the infimum one at cubic inverse-gap contrast cost, compared with the quadratic cost under uniform introduction. At the birth-associated least-contrast design for $b/c=1+\epsilon$, the neutral fixation probability is asymptotic to $\epsilon/54$, whereas uniform introduction gives exactly $1/6$. At leading order, a pin-founded lineage fixes with probability $1/2$, and the two pins receive total introduction probability $1/3$ under uniform placement but only $1/(3q)$ under newborn mutation. Their product explains the suppression to $1/(6q)\sim\epsilon/54$ at the birth-associated optimum $q\sim9/\epsilon$. Protection therefore changes both the contact resources and the frequency of successful introductions. A uniform fraction comparable to $\epsilon$ changes the cubic coefficient; any fixed positive fraction restores the quadratic regime.

The resource powers express a balance between replacement mechanisms. Positive terms in the threshold loss quantify competing replacement channels. At the optimum, their scales are fixed by the allowed gap, and the inverse powers in the contact resource determine its divergence. Measuring the resource by the inverse smallest positive copying probability changes the six-site powers from two and three to one and two, and changes the optimal contacts. At the optimizing reflected designs that smallest probability is asymptotic to $q/Q$. Its inverse therefore scales as $Q/q$, removing one factor $q\asymp\epsilon^{-1}$ relative to the contrast $Q$. The additional inverse power for birth-associated introduction persists under both choices. Thus the absolute power depends on what can be controlled, while its change between introduction rules retains the protection--access mechanism. An energetic cost would require a relation between contact weights and their physical implementation.

The global proofs make the protected-domain description predictive beyond a chosen family of examples. Positivity first forces every competitive six-site path weighting into the hierarchy. Compatible contact ratios then keep the response and its derivatives regular at a limit where the original weights diverge. Their rank-one relations prevent impossible variations from entering the optimization. A unique leading minimum persists at small finite gap, and path reversal forces its exact reflection symmetry. The coordinate algebra also organizes higher corrections. Its Hilbert series counts independent polynomial terms at each degree, accounting for relations among the contact ratios (Appendices~\ref{v4:alg:toric-saturation} and \ref{v4:alg:toric-hilbert}). On the dynamical side, nested spaces of increasing correlation degree organize successive selection orders. Both constructions keep the constraints of the physical problem explicit while reducing the independent quantities that must be calculated.

The convex formulation connects approach to a threshold infimum with the first appearance of promotion. Relative entropy measures the mismatch among monomial contributions to the birth-associated five-cycle sign polynomial. The same positive-sum identity fixes asymptotic resource balances and quantifies departures from them. At the first-promotion boundary the benefit response reaches zero while the cost response stays positive, so the threshold diverges on approach from the promoting side. No finite benefit--cost ratio promotes exactly at onset. For the five-site path, the unique optimizer follows one analytic branch from this boundary to its large-contrast limit. The introduction rule reverses the onset ordering between the unrestricted five-cycle and the reflected six-site path, and an exact six-cycle example promotes at still lower uniform contrast. These comparisons identify differences between supports that are already consequential before the small-gap regime.

Minimizing waiting time tests a different part of the hierarchy. Under uniform introduction on the six-site path, extra contrast first lowers the leading neutral mean time to extinction or fixation from $12/\epsilon$ toward $6/\epsilon$. A larger budget produces a pitchfork of the optimizing geometry, with two asymmetric designs related by reflection. For a fixed scaled cap $H\epsilon^4>1476$, where $H$ is the contact-contrast cap, their gain over the reflected design is $O(\epsilon)$ in absolute time and $O(\epsilon^2)$ relative to the mean. Most of the speed gain therefore precedes symmetry breaking. Increasing the asymmetry also suppresses the linear response to benefit. The minimized order-$\epsilon$ time correction has the Landau form derived in Eq.~\eqref{eq:III-pitchfork-value}. Just above the critical cap, the optimal pin imbalance grows as the square root of the excess scaled cap; the minimized correction has a continuous first derivative and a jump in its second derivative. These singularities arise from minimization over contacts. The neutral finite-chain hitting probabilities and mean times remain smooth at fixed positive weights, so this design bifurcation is separate from a thermodynamic phase transition. Reversal symmetry, a transverse curvature crossing and a positive quartic term together control the exact small-gap pitchfork in Proposition~\ref{prop:v10-allbudget}. Probability--time trade-offs in constant-selection amplifiers~\cite{SlowTkadlec2019,SlowTkadlec2021} and the different rankings obtained from long-run mean fitness under recurrent fitness mutations~\cite{SharmaTraulsen2022} already distinguish several meanings of a favorable network. Here the payoff-dependent target, introduction distribution and physical clock remain explicit throughout that comparison.

On the five-cycle, optimal speed changes the relative scales of two follower-chain contacts. Excursion rates from a pin and the probabilities of reaching the opposite pin compensate one another, leaving a controlled slow exchange rate even when one chain ratio vanishes. Proposition~\ref{prop:III-c5-waiting} uses this control to determine the leading time frontier, the limiting conditional exponential waiting distribution and every fixed positive conditional moment. Before the change of geometry, the normalized frontier agrees exactly with that of the four-cycle. A common positive-monomial balance explains the agreement, and the additional chain coordinate determines where it ends. Neutral death-first calculations on small unweighted graphs also associate long absorption with weakly connected groups~\cite{GaoLiuWu2024}. The present hierarchy fixes the slow rate under an explicit invasion target and contact constraint. All quoted times count reproduction attempts, including unchanged copies. Unconditional means include failed invasions, while conditional fixation means retain successful ones. Division by population size converts attempts into sweeps; conversion to biological time requires calibration of reproduction.

Changes of support separate protection from the cost of realizing it. Motif-controlled rewiring can alter conditional fixation times without changing fixation probability on regular constant-fitness graphs~\cite{KuoCarja2024}. In the weighted game studied here, a compulsory diagonal retains a cycle's limiting threshold but raises its contrast coefficient within the specified family. Sharing access between both sites of a protected core reduces the threshold penalty from leakage. At fixed total access weight, the two-triangle family lowers the core-leak coefficients from $82/3$ to $24$ under uniform introduction and from $50/3$ to $15$ under birth-associated introduction. The least contrast is proportional to this coefficient in both balances, so the sharp contrast constants fall by the same fractions, $5/41$ and $1/10$. The leading uniform waiting curve is unchanged after contrast is normalized by each family's feasibility constant. The improvement is a reduction in leakage cost at a common leading waiting time. Larger connected cores admit the same achievable inverse-gap powers on broader supports. Accumulated payoffs can promote cooperation through a different strength gradient, as the exact finite example shows; their optimal contacts must be determined with that changed payoff accounting.

Finite selection reveals why a favorable threshold alone does not determine a useful design. Near least contrast, nonlinear establishment corrections and contact errors draw on the same margin. Pin-contact deviations enter quadratically because the optimal pin contacts minimize the threshold. Core-weight shortfalls enter linearly because the strongest contacts lie on an active cap, and selection adds a quadratic penalty to the margin multiplying the leading response. The linear-fecundity resolvent separates this loss of favorable response from failure of perturbation theory. On reflected contacts, the centered response expansion retains a nonzero limiting convergence radius while the positive selection window shrinks with the scaled contrast surplus. The reversal of advantage therefore occurs inside a convergent expansion. Proposition~\ref{prop:v7-selected-time} follows the selected waiting-time optimum at small selection and bounded scaled budgets. Proposition~\ref{prop:v10-allbudget} follows neutral timing through all feasible caps in a common small-gap domain. Their different scopes reflect the additional establishment corrections that selected dynamics must retain.

Observation adds a third optimum. For uniform introduction, independent completed invasions estimate the unknown paired fixation advantage with the introduced type and site retained. Protected-site classes dominate both variance and duration, making stratification effective. Within the reflected six-site hierarchy, Proposition~\ref{prop:III-observation-design} identifies a unique leading optimum of contacts and selection strength for summed observation time. Its signal is smaller than the largest achievable signal at the same contrast, while shorter trials reduce the total time needed to resolve it. At fixed relative precision and fixed scaled contrast above feasibility, that time grows as $\epsilon^{-3}$. Near least contrast the maximal advantage is of order $\epsilon d^{3/2}$, where $d=\epsilon^2(H-H_*)$ is the scaled surplus above the exact weak-selection least contrast $H_*$. The required trial count is proportional to the inverse square of that advantage, which explains why the observation-time prefactor grows as the inverse cube of the surplus as the surplus vanishes. A controlled joint layer with surplus of order $\epsilon$ therefore has an $\epsilon^{-6}$ observation cost. Common preparation costs shift the optimum continuously toward signal maximization when preparation dominates.

Finite-state evaluations illustrate how strongly additional contrast can affect measurement. At $b/c=1.1$, increasing contrast from $10^5$ to $5\times10^5$ reduces the required number of completed trials from about $1.7\times10^7$ to $1.6\times10^5$ for signal-to-noise ratio two. These numbers evaluate the limiting signal-optimal designs with linear fecundity and equal counts in the twelve introduced-type and site classes (Table~\ref{tab:III-observation-scales}). The observation result concerns unknown terminal probabilities and summed occupied-run time. Parallel elapsed time, correlated copying events in simulations, incomplete trials and trajectory measurements each define a different statistical objective. Known microscopic parameters can also change which probabilities need to be inferred. These choices determine the information available and the corresponding likelihood, and therefore which observation bound applies.

The same margin controls the separation between spontaneous introductions. For independent symmetric trait switches at uniformly chosen sites, the sufficient condition $\nu=o(\epsilon d)$ preserves the positive stationary-abundance bias in the local signal-optimal regime. Here $\nu$ is the total introduction rate. This compares the correction from overlapping invasions with the shrinking bias itself. Requiring invasions to be well separated without reference to that bias would miss the stronger scale needed to resolve a positive abundance difference.

Neutral observables offer a route toward biological tests before resolving the small selective signal. Site-resolved fixation weights, suppression of successful birth-associated introductions, and conditional-to-unconditional waiting-time ratios probe the same protected-domain mechanism. Migration topology has changed the spread of beneficial mutations in networked microbial metapopulations~\cite{ChakrabortyNemzerKassen2023}. Their nodes contain populations with demographic rules different from the single-individual copying process here. Using the present coefficients in such a setting requires a reduction of within-node dynamics and a match between encounter and replacement mechanisms. The exact encounter-kernel comparison in Sec.~\ref{sec:III-transfers} identifies a controlled way to change one of these ingredients.

The remaining design questions have identifiable analytical obstacles. The complete leading five-cycle frontier leaves an exact finite-gap comparison near the crossing of its two limiting geometries, where a chain contact ratio approaches zero. The reflected six-site onset leaves the competing asymmetric contacts to be controlled before its unrestricted value can be determined. For selection in a fixed compact admissible interval beyond the neutral neighborhood, the rooted-forest comparison of Sec.~\ref{sec:v3-formation} still forces the hierarchy as the gap vanishes; core saturation and exact reflection require the corresponding derivative and curvature signs. Away from the small-gap regime, optimization needs global comparisons beyond that localized hierarchy. Explicit remainder and curvature bounds would determine the gaps and selection strengths to which the global-optimality and waiting-time propositions apply, including whether they reach the finite-gap designs in Table~\ref{tab:III-finite-gap-folds}. Unequal access weights and other supports test the optimality of the two-triangle and compulsory-diagonal families, while larger-core constructions require lower bounds to decide whether their achievable powers are sharp. Correlated-run observation calls for control of the coupling through the slow domain-exchange time. Growing arrays of domains introduce a further scale through the number and relaxation of the collective competitors.

The least-contrast design, the fastest design at a larger contact budget and the most efficient measurement design balance protection, access and duration differently. Global bounds identify the contact scales that every competitive weighting on the specified support must realize; regular coordinates then determine how its geometry changes with the objective. The same mechanism that makes cooperation possible thus fixes quantitative costs for realizing and observing its invasion advantage.

\begin{acknowledgments}
A.C. acknowledges the support and hospitality of the International Centre for Theoretical Sciences (ICTS), Bengaluru, during participation in the programs \emph{Refinements in Enumerative Geometry and Physics 2026} (ICTS/REGP2026/06) and \emph{Monopole Moduli 2026} (ICTS/MM2026/09).

The author acknowledges the use of the free versions of ChatGPT (GPT-5 series) and Claude (Claude 4/5 series) for assistance with organizing the literature, secondary numerical cross-checking, and LaTeX formatting. All primary research, derivations, and conclusions remain the sole work of the author, who retains full responsibility for the manuscript.
\end{acknowledgments}

\section*{Data availability}
The programs and data supporting the findings of this study are available from the author upon reasonable request.

\appendix
\section{Exact inputs and response computations}
\label{app:v10-response}
This appendix supports Secs.~\ref{b:sec:response-input}, \ref{sec:III-six-site-onsets}, \ref{sec:III-filtered-observables} and \ref{sec:v7-robustness}. Appendix~\ref{app:v10-adjoint-contact} states the birth-associated normalization and differentiates the response with respect to physical contacts, including the contact dependence of the pair source. Appendix~\ref{app:v3-operator-response} proves the degree closure and the parity resolvent used in Sec.~\ref{sec:III-filtered-observables} and expands timed observables on the global attempt clock. Appendices~\ref{app:III-diamond} and \ref{app:split-access} prove the added-contact and split-access results of Sec.~\ref{sec:III-added-contacts}, and Appendix~\ref{app:III-finite-comparisons} proves the exact signs in the six-cycle and accumulated-payoff examples of Secs.~\ref{sec:III-six-site-onsets} and \ref{sec:III-transfers}. High-precision evaluations of Tables~\ref{tab:III-finite-gap-folds} and \ref{tab:III-observation-scales} used mpmath 1.3.0~\cite{MpmathSoftware}.

\subsection{Contact derivatives and source normalization}
\label{app:v10-adjoint-contact}
\label{app:local-inputs}
Changing a contact changes both the response operator and, for birth-associated introduction, the source and neutral baseline. The normalization is therefore part of the design derivative.
\paragraph*{Birth-associated normalization.}
For birth-associated introduction the neutral normalization
can be written directly in terms of the same conductances,
\begin{equation}
 \begin{aligned}
 \rho_0^{\mathrm b}
 &=\sum_i\mu_i^{\mathrm b}\pi_i
   =\frac{2}{NZ}\sum_{i<j}\gamma_{ij},\\
 \frac{\chi_c^{\mathrm b}}{\rho_0^{\mathrm b}}
 &=\frac{\sum_{i<j}\gamma_{ij}D_{ij}^{\mathrm b}}
         {\sum_{i<j}\gamma_{ij}}\ge1.
 \end{aligned}
 \label{b:eq:birth-normalization}
\end{equation}
In the jump-count interpretation of Sec.~\ref{b:sec:response-input}, distinct lineages require at least one jump to coalesce, so $D_{ij}^{\mathrm b}\ge1$.
Through $\mu_i^{\rm b}\pi_i$, this normalization separates two properties of a protected site, its large reproductive value and the small probability that birth-associated introduction places a mutant there.

\paragraph*{Differentiating physical contacts.}
The response to a changed contact can be found from the same pair
problem. Fix the introduction rule $\mu(w)$ used in both paired experiments; within this subsection we suppress its superscript on $D,K,J,R$ and their derivatives. Write $g_{ij}=N(\mu_i+\mu_j)$ for the pair-source vector, so that $\mathsf M D=g$. Let $x_a$ be smooth contact parameters, and assume that the
retained graph stays connected with positive strengths. An added edge
may have zero weight at the reference point if the remaining graph is
connected. The pair inverse is then regular in a neighborhood of that
point. For a scalar contraction
$F=\boldsymbol\ell^\T D$, where $\boldsymbol\ell$ is the column of
readout coefficients, inverse differentiation uses the sensitivity calculus for absorbing Markov chains~\cite{Caswell2011}. Retaining the contact dependence of the pair source gives
\begin{equation}
 \begin{aligned}
 D_a&=\mathsf M^{-1}(g_a-\mathsf M_aD),\\
 D_{ab}&=\mathsf M^{-1}
 (g_{ab}-\mathsf M_{ab}D-\mathsf M_aD_b-\mathsf M_bD_a).
 \end{aligned}
 \label{eq:v10-pair-contact-derivatives}
\end{equation}
Subscripts denote parameter derivatives. One adjoint solve,
$\mathsf M^\T\boldsymbol y_{\rm ad}=\boldsymbol\ell$, then determines all first derivatives
and, with the first-order pair responses, all second derivatives,
\begin{equation}
 \begin{aligned}
 F_a={}&\boldsymbol\ell_a^\T D
       +\boldsymbol y_{\rm ad}^\T(g_a-\mathsf M_aD),\\
 F_{ab}={}&\boldsymbol\ell_{ab}^\T D
       +\boldsymbol\ell_a^\T D_b+\boldsymbol\ell_b^\T D_a\\
 &+\boldsymbol y_{\rm ad}^\T(g_{ab}-\mathsf M_{ab}D
                   -\mathsf M_aD_b-\mathsf M_bD_a).
 \end{aligned}
 \label{eq:v10-adjoint-contact-derivatives}
\end{equation}
Thus changing the contacts changes the propagator, the measurement and,
for a contact-dependent introduction rule, the source. Uniform
introduction has $g_a=g_{ab}=0$. Birth-associated introduction requires
the derivatives of $\vartheta_i+\vartheta_j$.

The physically relevant derivative also retains normalization. With
$W=bJ-cK$ and fixed $b,c$, the normalized weak susceptibility is
$\mathcal S^\mu=2W/(NZ)$ for $\phi'(0)=1$. Hence
\begin{equation}
 \begin{aligned}
 (\mathcal S^\mu)_a
 &=\frac{2}{NZ}\left(W_a-\frac{Z_a}{Z}W\right),\\
 R_a&=\frac{K_a-RJ_a}{J},\qquad J>0.
 \end{aligned}
 \label{eq:v10-physical-contact-derivative}
\end{equation}
The corresponding second derivative of $R$ is
$(K_{ab}-RJ_{ab}-R_aJ_b-R_bJ_a)/J$.
Under a common rescaling $w\mapsto\lambda_w w$, the copying kernel,
source and pair response are unchanged, whereas $K,J,Z$ each scale as
$\lambda_w^{-1}$. Both $R$ and $\mathcal S^\mu$ are invariant.
A favorable contact change is therefore identified through $R$ or $\mathcal S^\mu$; an unnormalized contraction such as $2K-3J$ can decrease under a common rescaling alone.
Equations~\eqref{eq:v10-pair-contact-derivatives}--\eqref{eq:v10-physical-contact-derivative}
are finite-weight identities. Near a singular hierarchy, uniform derivative estimates require control of the singular inverse together with its source and response contractions. The support-specific divisions and regular charts used near singular hierarchies are developed in Appendices~\ref{app:III-diamond}, \ref{v3-app-positive-p6} and \ref{app:asymmetric-entry-v2}.

\subsection{Response at fixed selection order}
\label{app:v3-operator-response}
The binary observable algebra closes at every fixed selection order. The closure and physical-clock expansion are distinct statements, both needed for the nonlinear design margin.
\subsubsection{A closed correlation space at fixed selection order}
\label{app:v3-degree-closure}

In the filtered Boolean algebra of Eq.~\eqref{eq:v4-boolean}, neutral
copying replaces an index or contracts two equal indices, preserving
each $\mathscr F_k$ on any graph. Its restriction to degree-at-most-$k$ observables vanishing
at both consensus states is invertible.  A zero-boundary harmonic
function vanishes by uniqueness of the absorbing Dirichlet problem, and
injectivity on this finite invariant space implies surjectivity.

Use the centered payoff and selection variable $\delta_c$ of Sec.~\ref{sec:v4-observable-algebra}. After removing the total-fecundity denominator from the harmonic equation, write $\mathcal G(\delta_c)=\sum_{p\ge0}\delta_c^p\mathcal G_p$ and $h=h_0+\sum_{m\ge1}\delta_c^mh_m$. The coefficients obey
\begin{equation}
 \mathcal G_0h_m=-\sum_{p=1}^m\mathcal G_ph_{m-p}.
 \label{eq:v3-harmonic-filtration}
\end{equation}
Only $\mathcal G_1$ is present for centered linear fecundity. Each order-$p$
operator increases degree by at most $p$ and has complement parity
$(-1)^p$, so the degree and parity bounds in Eq.~\eqref{eq:v4-degree-response} follow by induction; every copying difference
annihilates the constant part of $h_0$. For payoff degree $d$, the same
proof gives $\deg h_m\le\min(N,1+md)$ for analytic fecundity.
At fixed payoff degree and response order these spaces have polynomial
dimension in $N$, and graph symmetries reduce them further.
Before reflection, the associated graded space of the complement-even degree filtration
has Hilbert series $1+15\mathfrak t^2+15\mathfrak t^4+\mathfrak t^6$ on six sites, with $\mathfrak t$ counting one spin index.
For reflection-symmetric weights, the invariant count in
Eq.~\eqref{eq:v4-reflected-hilbert} gives the response dimensions quoted
in Sec.~\ref{sec:v4-observable-algebra}.

\subsubsection{Counting the reflected response sectors}
\label{par:III-response-counts}
The degree filtration in Eq.~\eqref{eq:v4-boolean} has associated graded algebra
\begin{equation}
\begin{gathered}
 \operatorname{gr}\mathscr A_N\simeq\mathbb R[x_1,\ldots,x_N]/(x_1^2,\ldots,x_N^2),\\
 \mathcal H_{\rm obs}(\mathfrak t)=(1+\mathfrak t)^N.
\end{gathered}
 \label{eq:v4-hilbert-full}
\end{equation}
Averaging the identity and reflection actions on squarefree monomials gives
\begin{align}
 \mathcal H_{\rm refl}(\mathfrak t)
 &=\frac{(1+\mathfrak t)^6+(1+\mathfrak t^2)^3}{2}\notag\\
 &=1+3\mathfrak t+9\mathfrak t^2+10\mathfrak t^3
 +9\mathfrak t^4+3\mathfrak t^5+\mathfrak t^6.
 \label{eq:v4-reflected-hilbert}
\end{align}
Reflection-fixed subsets are unions of mirrored pairs. The degree bound and the single consensus condition in each relevant parity sector leave $1+9-1=9$, $3+10-1=12$ and $1+9+9-1=18$ unknowns at orders one, two and three. This counts commuting correlation observables and is separate from the one-lineage exterior representation used for neutral propagation on paths and cycles (Sec.~\ref{sec:v4-observable-algebra}).

\subsubsection{Summing the centered linear response}
\label{app:III-response-resolvent}
Let $h_{\rm corr}=h_C-\Phi$, which vanishes at both consensus configurations.
The harmonic equation gives
$h_{\rm corr}=\delta_c h_1+\delta_c\mathsf R_{\rm ins}h_{\rm corr}$, proving
Eq.~\eqref{eq:III-resummed-response} wherever the Dirichlet inverse exists.
Near $\delta_c=0$ this is the sum of its Neumann series; the rational identity
continues beyond that series' convergence disk. Positive fecundities on
a connected finite graph ensure the unique absorbing Dirichlet solution
throughout the physical interval.

For uniform introduction define the paired linear readout
$\ell_{\rm U} g=N^{-1}\sum_i[g(e_i)+g(\mathbf1-e_i)]$ on zero-boundary
functions, extended by the same formula to all configuration observables.
Since $\ell_{\rm U}\Phi=1$, the advantage is $\Delta=\ell_{\rm U} h_{\rm corr}$.
The first response $h_1$ is complement-even, $\ell_{\rm U}$ annihilates
complement-odd functions, and $\mathsf R_{\rm ins}$ exchanges the two
parities. Let $\mathsf R_{\rm eo}$ take odd to even and $\mathsf R_{\rm oe}$
take even to odd. Summing paired insertions therefore gives
\begin{equation}
 \frac{\Delta}{\delta_c}
 =\ell_{\rm U}(I-\delta_c^2\mathsf R_{\rm eo}\mathsf R_{\rm oe})^{-1}h_1.
 \label{eq:III-parity-resolvent}
\end{equation}
The value at $\delta_c=0$ is understood continuously. This formula retains
all finite-state fluctuations and makes the centered oddness explicit.
On the reflected six-site path the Hilbert count above leaves $19$
complement-even and $15$ complement-odd zero-boundary observables.
Sylvester's determinant identity,
$\det(I-z\mathsf R_{\rm eo}\mathsf R_{\rm oe})
=\det(I-z\mathsf R_{\rm oe}\mathsf R_{\rm eo})$, therefore bounds the denominator
degree of $\Delta/\delta_c$ by $15$ in $z=\delta_c^2$; contraction with the source
and readout can reduce it further. These poles lie in the selection
variable $z=\delta_c^2$; physical relaxation rates come from the
generator spectrum at fixed selection. Timed observables require the
physical normalization retained next.

\subsubsection{Physical-clock insertions}
\label{app:v3-sylvester-time}

For time dependence define the copying difference
$\mathcal D_{ij}g(n)=g(n^{j\leftarrow i})-g(n)$,
$S_f=\sum_i\bar f_i$, $\mathcal A_0=\mathcal A|_{\delta=0}$,
$\mathcal G_0=N\mathcal A_0$, and
$\mathcal G_1=\sum_{ij}\bar f_iP_{ij}\mathcal D_{ij}$.
For centered linear fecundity write
$\mathcal A(\delta_c)=\sum_{k\ge0}\delta_c^k\mathcal A_k$. Its coefficients are
\begin{equation}
 \begin{gathered}
 \mathcal A(\delta_c)=\frac{\mathcal G_0+\delta_c\mathcal G_1}{N+\delta_cS_f},\qquad
 \mathcal A_1=\frac{\mathcal G_1}{N}-\frac{S_f\mathcal G_0}{N^2},\\
 \mathcal A_k=\left(-\frac{S_f}{N}\right)^{k-1}\mathcal A_1\quad(k\ge1).
 \end{gathered}
 \label{eq:v3-physical-vertices}
\end{equation}
All powers of $S_f$ multiply on the left. The higher-order coefficients enforce
the global event clock and cannot be discarded for timed observables.
If $e^{t\mathcal A(\delta_c)}g=\sum_{m\ge0}\delta_c^mg_m(t)$, the first two corrections are
\begin{equation}
 \begin{aligned}
 g_1(t)&=\int_0^t \mathsf U_0(t-s)\mathcal A_1\mathsf U_0(s)g\,ds,\\
 g_2(t)&=\int_0^t \mathsf U_0(t-s)\mathcal A_2\mathsf U_0(s)g\,ds\\
 &\quad+\int_{0<s_1<s_2<t}\mathsf U_0(t-s_2)\mathcal A_1\\
 &\hspace{16mm}\times \mathsf U_0(s_2-s_1)\mathcal A_1\mathsf U_0(s_1)g\,ds_1ds_2,
 \end{aligned}
 \label{eq:v3-physical-duhamel}
\end{equation}
where $\mathsf U_0(t)=e^{t\mathcal A_0}$. The triangular backward equation proves
$\deg g_m\le\min(N,\deg g+m)$, with alternating complement parity when $g$ has definite complement parity.
For payoff degree $d$, analytic fecundity gives $\deg g+md$.
For a pair $g=\sigma_a\sigma_b$, putting $Y_1=\mathcal G_0g$, $Y_2=\mathcal G_1g$ gives
$\mathcal A_1g=Y_2/N-S_f Y_1/N^2$ and
$\mathcal A_2g=S_f^2Y_1/N^3-S_f Y_2/N^2$. Thus the second response already
contains $t\mathcal A_2g$ at short times. The first two pair corrections
require at most three- and four-spin correlations; neutral propagation between the insertions is the exact semigroup $\mathsf U_0(t)=e^{t\mathcal A_0}$. On a positive reciprocal path or cycle it can also be evaluated through the one-lineage exterior representation recalled in Sec.~\ref{sec:v4-observable-algebra}.
No Gaussian factorization is assumed.

Let $M_f=\max_{n,i}|\bar f_i(n)|\le(b+c)/2$ and
$\|g\|_\infty=\max_n|g(n)|$, with the induced norm for operators
on configuration observables. On a complex disk
$|\delta_c|\le r_{\rm disk}<1/M_f$,
$\|\mathcal A(\delta_c)\|_\infty\le C_{r_{\rm disk}}=2/(1-r_{\rm disk}M_f)$.
Cauchy's estimate gives, for $|\delta_c|\le r_c<r_{\rm disk}$ and $0\le t\le T$,
\begin{equation}
\begin{aligned}
 &\left\|e^{t\mathcal A(\delta_c)}g-\sum_{m=0}^{m_*}\delta_c^mg_m(t)\right\|_\infty\\
 &\quad\le\|g\|_\infty e^{TC_{r_{\rm disk}}}\frac{(r_c/r_{\rm disk})^{m_*+1}}{1-r_c/r_{\rm disk}}.
\end{aligned}
 \label{eq:v3-physical-remainder}
\end{equation}
For exponential fecundity, $r_{\rm disk}M_f<\pi/2$ and
$C_{r_{\rm disk}}=1+e^{2r_{\rm disk}M_f}/\cos(r_{\rm disk}M_f)$ suffice, since the normalized denominator
then has positive real part. These finite-time bounds require no
spectral gap and are uniform in graph weights and $N$ in the global
attempt clock. The factor $e^{TC_{r_{\rm disk}}}$ grows exponentially
with the time horizon. The estimate controls each fixed horizon $T$; on a diverging metastable time scale its usefulness also depends on the selection and truncation factors in Eq.~\eqref{eq:v3-physical-remainder}. The slow-state reductions of Appendices~\ref{app:asymmetric-entry-v2} and \ref{app:v10-times} control the metastable regime by retaining its escape rates and physical clock.

\subsection{An added contact and its resource penalty}
\label{app:III-diamond}
For uniform introduction in the diamond family of
Fig.~\ref{fig:III-diamond}, the response expansion
\eqref{eq:III-diamond-response} has uniform errors on fixed positive
compact sets of $(x,y,\omega_\diamond)$. At fixed $\omega_\diamond\ge1$,
minimization over a fixed positive compact neighborhood of its leading
minimum gives the contrast cost in Eq.~\eqref{eq:III-diamond-cost}.
\begin{proof}
Order the six unordered
pairs as $(12,13,14,23,24,34)$ and expand the matrix of
Eq.~\eqref{b:eq:pairs} as
$\mathsf M=\mathsf M_0+t_\diamond\mathsf M_1+t_\diamond^2\mathsf M_2+\cdots$.
Uniform introduction gives $\mathsf M D=2\times \mathbf1$. The limiting
matrix has a simple zero mode. Its right and left vectors can be chosen as
\begin{equation}
 \begin{gathered}
 v_0=(1,2/3,1/3,1/3,2/3,1/3)^{\mathsf T},\\
 \ell_0=(1,0,0,0,0,0),\qquad \ell_0v_0=1,\\
 \ell_0\mathsf M_1v_0=x^{-1}+y^{-1}>0.
 \end{gathered}
 \label{eq:III-diamond-null}
\end{equation}
Write $D=t_\diamond^{-1}D_{-1}+D_0+O(t_\diamond)$. The two solvability conditions are
\begin{equation}
 \begin{gathered}
 D_{-1}=\frac{2xy}{x+y}v_0,\qquad
 \mathsf M_0D_0+\mathsf M_1D_{-1}=2\times \mathbf1,\\
 \ell_0(\mathsf M_1D_0+\mathsf M_2D_{-1})=0.
 \end{gathered}
 \label{eq:III-diamond-solvability}
\end{equation}
The last equation fixes the multiple of $v_0$ in $D_0$. Substituting in
the cost and benefit contractions gives

 \begin{align}
 K={}&\frac{2t_\diamond}{x+y}
 +\frac{2[-3\omega_\diamond+2x^2y+2xy^2-6]}{3(x+y)^2}t_\diamond^2+O(t_\diamond^3),\notag \\
 J={}&\frac{2t_\diamond}{3(x+y)}
 -\frac{2\mathcal P_\diamond}{9xy(x+y)^2}t_\diamond^2+O(t_\diamond^3),  \label{eq:III-diamond-contractions}\\
 \mathcal P_\diamond={}&6\omega_\diamond x^2+9\omega_\diamond xy+4x^3y^2+4x^2y^3 \notag \\
 &+8x^2+22xy+8y^2. \notag
 \end{align}

Their ratio is Eq.~\eqref{eq:III-diamond-response}. The nonzero
limiting fast eigenvalues and positive slow coefficient in
Eq.~\eqref{eq:III-diamond-null} bound the divided inverse and the
remainders uniformly on fixed positive compact parameter sets.

Put $m_d=(384+288\omega_\diamond)^{1/3}$. Equality in the three-term
arithmetic--geometric mean bound gives the unique leading minimum
$x=8/m_d$, $y=(8+6\omega_\diamond)/m_d$, with coefficient $3m_d$.
For each fixed $\omega_\diamond$ this is an interior nondegenerate minimum;
its continuation on a fixed compact neighborhood gives the relative
$O(\epsilon)$ error in Eq.~\eqref{eq:III-diamond-cost}.
\end{proof}
Compactness enters this proof twice, in the uniform bound on the divided inverse and remainders and in the continuation of the leading minimum. The cost is therefore the least contrast of the fixed-diagonal hierarchy on a fixed positive compact neighborhood of that minimum. Growing diagonal weights, unbounded rescaled access contacts and other placements of the strongest edge lie outside this comparison. Comparing those regimes requires lower bounds on the diamond response that cover them, such as an all-weight bound of the type proved for paths and cycles in Appendix~\ref{app:v10-cost}.

\subsection{Two triangles with split pin access}
\label{app:split-access}
Label the pins $1,4$ and the core pairs $\{2,3\},\{5,6\}$.
The bridge $14$ has weight one, the four pin--core contacts have
weight $q$, and contacts $23,56$ have weight $H$.
Within this two-parameter family, the corner response expansions and
sharp least-contrast costs are those of Eq.~\eqref{eq:split-access-cost}
under the two introduction rules. The minimization includes all family
weights satisfying $1\le q\le H$.
\begin{proof}
Within each source rule the independent integrated disagreements are
\begin{equation}
 x=D_{14},\quad y=D_{23},\quad z=D_{12},\quad
 w=D_{15},\quad h=D_{25}.
\end{equation}
Swaps within the two core pairs and interchange of the triangles
identify all remaining pairs with these five. Define the copying probabilities and core conductance
 \begin{gather}
 \alpha_c=\frac q{H+q},\quad \beta_c=\frac H{H+q},\notag \\
 \alpha_p=\frac q{2q+1},\quad \beta_p=\frac1{2q+1},\\
 \vartheta_p=2\alpha_c+\beta_p,\quad \vartheta_c=\alpha_p+\beta_c,\quad
 \gamma_c=\frac H{(H+q)^2}. \notag
 \end{gather}
Here $\vartheta_p,\vartheta_c$ are the pin and core temperatures, and $\gamma_c$ is the conductance of either core edge. The reduced pair matrix is
\begin{equation}
 \mathsf M_{\rm sp}=
 \begin{pmatrix}
 2\vartheta_p&0&0&-4\alpha_c&0\\
 0&2\vartheta_c&-2\alpha_p&0&0\\
 0&-\alpha_c&\vartheta_p+\alpha_p&-\beta_p&0\\
 -\alpha_p&0&-\beta_p&\vartheta_p+\alpha_p&-2\alpha_c\\
 0&0&0&-2\alpha_p&2\alpha_p
 \end{pmatrix}.
 \label{eq:split-pair-system}
\end{equation}
The five equations are $\mathsf M_{\rm sp}(x,y,z,w,h)^{\mathsf T}=g^\mu$, with sources
\begin{equation}
 \begin{aligned}
 g^{\rm U}&=2\times\bm 1,\\
 g^{\rm b}&=2\vartheta_c\bm 1+(\vartheta_p-\vartheta_c)(2,0,1,1,0)^{\mathsf T}.
 \end{aligned}
\end{equation}
Here $\bm1$ has five entries.
Its contractions are
\begin{equation}
 \begin{aligned}
 K&=\beta_p^2x+2\gamma_c y+4\alpha_c\beta_p z,\\
 J&=-\beta_p^2\vartheta_p x+2[(\alpha_p-\beta_c)\alpha_c\beta_p-\beta_c\gamma_c]y\\
 &\quad-2\beta_p(\alpha_c^2+\alpha_p\vartheta_p-\alpha_c\beta_c)z\\
 &\quad+2\beta_p^2(\alpha_c+\alpha_p)w.
 \end{aligned}
 \label{eq:split-contractions}
\end{equation}
These five linear equations and two contractions determine the exact
rational response $R=K/J$. The pair inverse exists for positive
$q,H$, as it is the Green function of a finite ancestral process
killed on meeting.

For uniform introduction put $u=q^{-1}$ and $v=q/H$; for
birth-associated introduction put $u=q^{-1}$ and $v=q^2/H$.
In either case the reduced rational response has a denominator equal
to $6$ at $(u,v)=(0,0)$ after a common sign normalization. Hence it is
analytic in a neighborhood of that corner. Exact expansion gives
\begin{equation}
 \begin{aligned}
 R^{\rm U}-1&=3u+48v+\frac72u^2+\frac{488}{3}uv\\
       &\quad+328v^2+O((u+v)^3),\\
 R^{\rm b}-1&=3u+60v+\frac72u^2+292uv\\
       &\quad+240v^2+O((u+v)^3).
 \end{aligned}
 \label{eq:split-corner-expansion}
\end{equation}
Analyticity makes the remainder and its fixed-order derivatives
uniform on a sufficiently small closed positive rectangle.

To exclude alternative minimizing regimes, clear positive common
factors in the exact response and write $R^\mu=N_\mu/D_\mu$ in
coprime integer polynomials with no common integer factor.
Choose $N_\mu$ with positive coefficients, so that $D_\mu>0$ exactly
on promoters. Direct polynomial subtraction gives the coefficientwise
bounds
\begin{equation}
 \begin{aligned}
 D_{\rm U}&\le6H^4q^3+8H^3q^4,\\
 N_{\rm U}-D_{\rm U}&\ge18H^4q^2+288H^3q^4,\\
 D_{\rm b}&\le6H^5q^3,\\
 N_{\rm b}-D_{\rm b}&\ge18H^5q^2+360H^4q^5.
 \end{aligned}
 \label{eq:split-global-bounds}
\end{equation}
Subtracting each denominator from its stated upper bound, and each stated lower bound from $N_\mu-D_\mu$, leaves a polynomial with nonnegative coefficients.
Consequently, throughout the promoting part of the family,
\begin{equation}
 \begin{aligned}
 R^{\rm U}-1&\ge\frac{3/q+48q/H}{1+4q/(3H)},\\
 R^{\rm b}-1&\ge\frac3q+\frac{60q^2}{H}.
 \end{aligned}
\end{equation}
These bounds imply $R^\mu>1$ and force $u,v=O(\epsilon)$ whenever
$R^\mu\le1+\epsilon$. The uniform inequality implies
$3/q+(48-4\epsilon/3)q/H\le\epsilon$, whose arithmetic--geometric
mean bound is
$H\ge576\epsilon^{-2}-16\epsilon^{-1}$ for $\epsilon<36$.
Minimizing the birth-associated lower bound over $q$ gives
$H\ge3645\epsilon^{-3}$. Conversely, the analytic expansions with
$q\sim6/\epsilon$, $H\sim576/\epsilon^2$, and with
$q\sim9/(2\epsilon)$, $H\sim3645/\epsilon^3$, respectively,
attain the corresponding leading costs. Increasing $H$ by a
relative amount of order $\epsilon$ enforces the closed target.
This proves Eq.~\eqref{eq:split-access-cost} over the entire stated two-parameter family.
\end{proof}

The first corrections to these optimal contrasts and pin contacts are
\begin{equation}
 \begin{aligned}
 H_{\rm split}^{\rm U}&=576\epsilon^{-2}+\frac{1435}{3}\epsilon^{-1}+O(1),\\
 q_{\rm U}&=6\epsilon^{-1}+\frac{1861}{576}+O(\epsilon),\\
 H_{\rm split}^{\rm b}&=3645\epsilon^{-3}+5913\epsilon^{-2}+O(\epsilon^{-1}),\\
 q_{\rm b}&=\frac9{2\epsilon}+\frac73+O(\epsilon).
 \end{aligned}
\end{equation}
\begin{proof}
The global lower bounds and the matching trial designs confine
all minimizers to compact sets in $k=\epsilon q$ and
$L=\epsilon^pH$, with $p=2,3$ respectively. The limiting problem minimizes $L$ subject to $F(k,L)\le1$, where
$F=3/k+48k/L$ or $F=3/k+60k^2/L$. Its unique minimizer satisfies
$F=1$, $F_k=0$, $F_{kk}>0$ and $F_L<0$. The Jacobian of
$(F-1,F_k)$ in $(k,L)$ has determinant $-F_LF_{kk}>0$.
Thus the implicit equations $R=1+\epsilon$ and $\partial_qR=0$
determine the eventual family minimizer;
inserting Eq.~\eqref{eq:split-corner-expansion} gives these
coefficients.
\end{proof}

\paragraph*{Waiting time at fixed contrast.}
Use uniform singleton introduction and the rate-one Poisson clock of
copying attempts. The neutral mean times obey
Eq.~\eqref{eq:split-neutral-times}, with errors uniform as
$q^{-1},q/H\to0$. At target $R^{\rm U}\le1+\epsilon$ and fixed cap
$H=\eta\epsilon^{-2}$ with $\eta>576$, the separate minimum absorption
and successful-fixation times over this family obey
Eq.~\eqref{eq:split-time-frontier}.
\begin{proof}
With $u=q^{-1}$ and $v=q/H$, a pin sends a copy
across the bridge with probability $u/(2+u)$ and into each core site
with probability $1/(2+u)$; a core site sends a copy into its pin with
probability $v/(1+v)$ and into the other core site with probability
$1/(1+v)$. At $u=v=0$ the pins cannot be overwritten, and each core
aligns with its own pin. The four aligned triangle states are the only
closed classes. The other 60 states have a uniformly bounded killed
inverse near this corner, since pin-to-core copies align both triangles
in a fixed finite sequence whose probability stays positive.

Each row leaving a mixed aligned state contains an exact factor $u$.
Eliminating the fast states gives the positive killed operator
$u[I_2/6+O(u+v)]$ on the two mixed states. Each bridge direction has
leading physical rate $u/12$, and a changed pin establishes its type
in its receiving core with probability one at the corner. Uniform
singleton introduction enters each mixed state with limiting probability
$1/6$; the other four singleton introductions disappear during the
fast stage. A mixed state's mean residence is therefore asymptotic to
$6/u$, and its fixation probability tends to $1/2$. Using the exact
uniform neutral fixation probability $1/6$ gives
\begin{equation}
 \begin{aligned}
 T_{\rm abs}&=2q[1+O(q^{-1}+q/H)],\\
 T_{\rm fix}&=6q[1+O(q^{-1}+q/H)].
 \end{aligned}
 \label{eq:split-neutral-times}
\end{equation}
The block inverse and the conditional-reward identity of
Eq.~\eqref{eq:v4-interval-times}, applied on the full configuration
space here, control both remainders. After the exact factor $u$ is
removed, the mixed-state operator and the fast inverse have analytic
inverses at the corner. The errors are thus uniform in a positive
neighborhood, without fixing $v/u$. At the uniform least-contrast
design, $q=6/\epsilon+O(1)$ and $H\sim576\epsilon^{-2}$, proving the
two waiting constants stated in Sec.~\ref{sec:III-added-contacts}.

For the fixed cap $H=\eta\epsilon^{-2}$, allow the actual common core
weight $H'$ to vary with $1\le q\le H'\le H$. The global response
bound above implies, throughout the feasible set,
\begin{equation}
 \frac3q+\left(48-\frac{4\epsilon}{3}\right)\frac q{H'}\le\epsilon.
\end{equation}
It confines every design to $q\asymp_\eta\epsilon^{-1}$ and
$H'\asymp_\eta\epsilon^{-2}$, so the response and time expansions
are uniform over that set. Put $\theta=2\epsilon q$, where the factor two counts the access contacts from each pin, and $\eta'=\epsilon^2H'\le\eta$. The limiting constraint is
$6/\theta+24\theta/\eta'\le1$. Increasing $\eta'$ relaxes it,
whereas both leading times increase with $\theta$. The least feasible
$\theta$ is therefore the lower root at $\eta'=\eta$, giving
Eq.~\eqref{eq:split-time-frontier}. For each fixed $\eta>576$ this
root is simple. The exact response differs by uniformly
$O_\eta(\epsilon)$ after division by $\epsilon$, so it changes the
lower bound on $\theta$ by $O_\eta(\epsilon)$. Taking $H'=H$ and
adjusting the root by the implicit-function theorem meets the exact
target with $q$ shifted by $O_\eta(1)$. These matching bounds prove
the separate time minima with the stated errors and fix their common
leading geometry.
\end{proof}

\subsection{Finite comparisons beyond the optimized families}
\label{app:III-finite-comparisons}
The support and payoff changes in Secs.~\ref{sec:III-six-site-onsets} and \ref{sec:III-transfers} can be checked at exact rational contacts. For the six-cycle with cyclic weights $(51,9/2,1,9/2,51,51)$, the fifteen-pair system \eqref{b:eq:pairs} has uniform source $g_{ij}=2$. Solving and contracting with Eq.~\eqref{b:eq:response} gives
\begin{equation}
 J^{\rm U}_{C_6}=\frac{12156722850366976407}{30404101865254614517064}>0.
 \label{eq:III-onset-witnesses}
\end{equation}
The corresponding threshold is approximately $1175.98$. Positivity
persists on reducing the largest contacts slightly, making the onset
upper bound strict. The large threshold also emphasizes the difference
between first promotion and an invasion target close to the support's
infimum.

For accumulated payoffs, retain parent-normalized copying and uniform introduction. On the path $(1,3203/100,4096/5,26837,10^6)$, write the paired response as $b\chi_b-c\chi_c$. Exact rational comparison gives
\begin{equation}
 \begin{aligned}
 \frac{28385}{10^5}&<\chi_b<\frac{28386}{10^5},\\
 \frac{396453}{10^5}&<\chi_c<\frac{396454}{10^5}.
 \end{aligned}
 \label{eq:III-accumulated-witness}
\end{equation}
Thus $14\chi_b-\chi_c>936/10^5>0$, proving promotion at $b/c=14$ and sufficiently small positive selection. The values of $\chi_b$ and $\chi_c$ refer to the weights as written; a common rescaling of the weights multiplies both by the same factor and leaves their ratio unchanged (Sec.~\ref{sec:III-transfers}).

\begin{proof}
Use the differentiated configuration process directly. For a function $v$ on configurations define the neutral operator with its common attempt factor removed,
\begin{equation}
 (\mathsf Q_0v)(n)=\sum_{ij}P_{ij}
       [v(n^{j\leftarrow i})-v(n)].
\end{equation}
Here $n^{j\leftarrow i}$ is the configuration after site $j$ copies site $i$. Set $v_b,v_c$ to zero at both consensus states and solve, on the 62 mixed configurations,
\begin{equation}
 \begin{aligned}
 -\mathsf Q_0v_b(n)&=\sum_{ij}P_{ij}s_i(Pn)_i\pi_j(n_i-n_j),\\
 -\mathsf Q_0v_c(n)&=\sum_{ij}P_{ij}s_i n_i\pi_j(n_i-n_j).
 \end{aligned}
\end{equation}
The neutral reproductive values $\pi_j$ remain those in Eq.~\eqref{b:eq:reproductive}. The common clock factor cancels between the neutral and perturbing operators. The paired coefficients are
\begin{equation}
 \chi_a=\frac1N\sum_i[v_a(e_i)+v_a(\mathbf1-e_i)],
 \qquad a\in\{b,c\}.
\end{equation}
All matrix entries and boundary data are rational. Direct solution followed by rational comparison proves Eq.~\eqref{eq:III-accumulated-witness}. Evaluating both initial types avoids assuming a factor-of-two relation for an altered payoff convention.
\end{proof}

\section{Exact first-promotion onsets}\label{app:v10-onsets}
This appendix proves the five-cycle and reflected six-site parts of Proposition~\ref{prop:v10-onsets}. Its first subsection also summarizes, in Table~\ref{tab:III-exact-signs}, the exact algebra behind the global comparisons of Appendices~\ref{app:v10-geometry} and \ref{app:v10-cost}. The onset optimizes the sign of the benefit response over the contact cube. Each introduction rule on the five-cycle has its own global reduction; the birth-associated sign also has the relative-entropy interpretation of Sec.~\ref{sec:v10-onset}. We then determine both onsets over the entire reflected six-site path family. The five-site path onset belongs to its all-cap optimization in Appendix~\ref{app:v10-p5-allcap}.
\subsection{Exact polynomial definitions and finite sign expansions}

\label{app:v10-sign-construction}
The following construction fixes the five-cycle polynomial normalization and the exact rational sign expansions used in the finite-contact comparisons. The pair matrix $\mathsf M$ in Eq.~\eqref{b:eq:pairs}
is the negative generator of a pair of ancestral lineages killed
when they meet, with the common physical factor $1/N$ removed. Its inverse is nonnegative and $\det\mathsf M>0$
on a connected positive-weight support. For a contraction vector
$\mathsf c$ on unordered pairs and the source
$g^\mu_{ij}=N(\mu_i+\mu_j)$, the bordered determinant identity is
\begin{equation}
 \mathsf c^{\mathsf T}\mathsf M^{-1}g^\mu
 =-\frac{1}{\det\mathsf M}
 \det\begin{pmatrix}
 \mathsf M&g^\mu\\
 \mathsf c^{\mathsf T}&0
 \end{pmatrix}.
 \label{b:eq:bordered}
\end{equation}
Equation~\eqref{b:eq:bordered} follows by taking the Schur complement of $\mathsf M$ and fixes the local polynomial calculation. Use
$\mathsf c_K=(\gamma_{ij})$ and
$\mathsf c_J=(-\mathsf B_{ij})$, with source $g^{\rm U}_{ij}=2$
or $g^{\mathrm b}_{ij}=\vartheta_i+\vartheta_j$.
Clear both rational contractions over a common denominator positive
for positive contacts. Divide their numerator pair by its polynomial
greatest common divisor, clear rational coefficient denominators,
remove common integer content, and choose the common sign so that
the cost numerator has positive leading coefficient. This defines
the coprime integer pair $U^\mu,V^\mu$ by
\begin{equation}
 K^\mu V^\mu=J^\mu U^\mu,\qquad
 R^\mu=\frac{U^\mu}{V^\mu}\quad(J^\mu>0).
 \label{b:eq:c5-polynomials}
\end{equation}
For $C_5$ each polynomial is homogeneous of degree 39. Every
coefficient of $U^\mu$ is positive. Thus
$J^\mu=(K^\mu/U^\mu)V^\mu$ has the sign of $V^\mu$
throughout the positive orthant, including its zero set.
We suppress the source superscript when one source has been fixed.
This construction determines the polynomials without listing their expanded monomials. The global sign comparisons use computer-assisted exact algebra. Integer or rational coefficient extraction, with the stated changes of variables, checks each complete finite expansion. The sizes and proof locations of the principal comparisons are summarized in Table~\ref{tab:III-exact-signs}. These exact checks are separate from the high-precision numerical evaluations, such as those in Table~\ref{tab:III-finite-gap-folds}, that test the stochastic reductions.
\begin{table*}[tp]
\caption{Exact algebra behind the global comparisons. Counts refer to the named polynomial before any indicated chamber substitution; degrees are total degrees. Each proof specifies its coefficient signs and the implication for the positive-weight domain. The bordered pair solve~\eqref{b:eq:bordered} and the rational transformation~\eqref{eq:v10-positive-transform} determine the coefficients without interpolation from sampled response values.}
\label{tab:III-exact-signs}
\centering\small
\setlength{\tabcolsep}{1.5pt}
\begin{tabular}{llll}
\hline\hline
Family and source & Polynomial sizes & Exact sign test & Proof location\\
\hline
$P_6$, uniform & $U$: 15467; $\mathcal C$: 15466 & Two nonnegative remainders; monomial lift & Appendices~\ref{app:v4-toric}, \ref{v3-app-positive-p6}\\
$P_6$, birth-associated & $U$: 15346; $U-V$: 15344 & Two nonnegative graded remainders & Appendix~\ref{app:v7-birth-p6-cost}\\
$C_5$, uniform & $U,V$: 25765 each & Twelve order chambers; positive gap & Appendix~\ref{app:c5-global}\\
$C_5$, birth-associated & $2U-3V$: 25590 & Sum over five bridge choices & Appendix~\ref{app:c5-global}\\
Split access, uniform & $N_{\rm U},D_{\rm U}$: 26 each; degree 7 & Two nonnegative remainders & Appendix~\ref{app:split-access}\\
Split access, birth-associated & $N_{\rm b},D_{\rm b}$: 35 each; degree 9 & Two nonnegative remainders & Appendix~\ref{app:split-access}\\
Onsets and finite caps & Reduced polynomials & Rational box signs and isolated roots & Appendices~\ref{app:v10-cycle-onsets}, \ref{app:III-p6-onsets}, \ref{app:v10-p5-allcap}\\
\hline\hline
\end{tabular}
\end{table*}

We use finite positive polynomial representations to prove several
domain-wide signs. Their mathematical content is an exact identity. If
$P(x)=\sum_\alpha p_\alpha x^\alpha$ has multidegree $n$ and the box endpoints
are $L_i/D_i,U_i/D_i$, define
\begin{equation}
Q(z)=\prod_iD_i^{n_i}(1+z_i)^{n_i}
P\left(\frac{L_i+U_iz_i}{D_i(1+z_i)}\right)
=\sum_\beta q_\beta z^\beta .
\label{eq:v10-positive-transform}
\end{equation}
An explicit coefficient formula is
\begin{equation}
q_\beta=\sum_\alpha p_\alpha
\prod_i\left[ D_i^{n_i-\alpha_i}
\sum_{k=0}^{\alpha_i}
\binom{\alpha_i}{k}L_i^{\alpha_i-k}U_i^k
\binom{n_i-\alpha_i}{\beta_i-k}\right],
\label{eq:v10-transform-coefficients}
\end{equation}
where a binomial coefficient outside its range is zero and the product
contains one such inner sum for each $i$.
Under $z_i=t_i/(1-t_i)$ the identity is an expansion in
$\prod_i t_i^{\beta_i}(1-t_i)^{n_i-\beta_i}$.
Every basis function is nonnegative on the closed unit cube, and at least one
is positive at every point. Thus $q_\beta>0$ for all coefficients proves
$P>0$ on the entire closed box. Equivalently, divide each coefficient by
$\prod_iD_i^{n_i}\binom{n_i}{\beta_i}$ to use the Bernstein basis~\cite{Garloff1993} whose
elements sum to one. The proof uses exact rational coefficients, without sampling values of $P$.

\paragraph*{Rational isolation of tangencies.}
The onset calculations also need unique zeros of polynomial systems.
For a system $F:\mathbb R^m\to\mathbb R^m$ with rational coefficients,
choose a rational center $x_0$ and radius $r_{\rm box}>0$ such that
$X=x_0+[-r_{\rm box},r_{\rm box}]^m$ lies in the positive orthant.
Assume $DF(x_0)$ is invertible and set
$\mathsf W_{\rm box}=[DF(x_0)]^{-1}$. If every row obeys
\begin{equation}
 \frac{|(\mathsf W_{\rm box}F(x_0))_i|}{r_{\rm box}}
 +\sum_j\sup_{x\in X}|(I-\mathsf W_{\rm box}DF(x))_{ij}|<1,
 \label{eq:III-onset-contraction-bound}
\end{equation}
then $x\mapsto x-\mathsf W_{\rm box}F(x)$ maps the closed box
into its interior and is a strict contraction. Its unique fixed point
is the unique zero of $F$ in the box. This is a preconditioned interval contraction of the type introduced by Krawczyk~\cite{Krawczyk1969}. To check the bound, evaluate
each derivative polynomial by enclosing every monomial between its
lower- and upper-endpoint values, then add these intervals with their
coefficient signs. Positivity of the box makes these endpoint bounds
valid. All operations use rational endpoints. The applications below
state their polynomial systems, centers and resulting row bounds.

\subsection{Both global five-cycle onsets}
\label{app:v10-cycle-onsets}
Normalize a smallest cycle contact to one and denote the consecutive
contacts by $(1,b,c,d,e)$. The letters in this tuple are local edge variables.
For either introduction rule, the unrestricted first-promotion cap and
its unique onset geometry are those in Proposition~\ref{prop:v10-onsets}.
Below that cap every weighting has negative benefit response, while
above it reflected promoters exist. The onset is a nondegenerate fold,
with $J=0$ and $K>0$ at the cap.
\begin{proof}
Subscripts on $V$ and $G$, and subscripts after a comma, denote partial derivatives; all other subscripts are labels. For either introduction rule let $V$ be the benefit polynomial $V^\mu$ of Eq.~\eqref{b:eq:c5-polynomials} and set
\begin{equation}
\begin{gathered}
\mathcal W=V/(bcde)^{16},\\
R_c=cV_c-16V,\quad R_d=dV_d-16V.
\end{gathered}
\end{equation}
On the equal-core face write
$V(b,h,h,e)=h^8G(s,p,h)$, where $s=b+e$, $p=be$, and set
$S=pG_p-16G$.
This invariant polynomial can be generated without elimination. For each
symmetric pair of monomials $b^ie^j+b^je^i$ with $i>j$, replace it by
$p^jT_{i-j}(s,p)$, where
$T_0=2$, $T_1=s$ and $T_k=sT_{k-1}-pT_{k-2}$; replace $b^ie^i$ by $p^i$.

For uniform introduction take $H_{\rm cmp}=72$; for birth-associated introduction
take $H_{\rm cmp}=1265/4$. In each case the following signs hold on the stated cubes,
\begin{equation}
\begin{gathered}
R_c,R_d>0\quad\hbox{on }[1,H_{\rm cmp}]^4,\\
S(b+e,be,h)>0\quad\hbox{on }[1,H_{\rm cmp}]^3.
\end{gathered}
\label{eq:v10-cycle-signs}
\end{equation}
For reproducibility, use Eq.~\eqref{eq:v10-transform-coefficients} after
splitting each of $b,e$ into $[1,8]$ and $[8,H_{\rm cmp}]$ and leaving each core
coordinate in $[1,H_{\rm cmp}]$. The four boxes cover each cube. The multidegrees of
$R_c$ and $S(b+e,be,h)$ are respectively $(16,15,16,16)$ and $(16,16,16)$.
All $78608$ coefficients per core box and all $4913$ coefficients per
pin box are strictly positive, for both sources independently.
Reflection $V(b,c,d,e)=V(e,d,c,b)$ gives $R_d$ from $R_c$.

The derivative identities
\begin{equation}
\begin{gathered}
\partial_c\mathcal W=\frac{R_c}{c(bcde)^{16}},\quad
\partial_d\mathcal W=\frac{R_d}{d(bcde)^{16}},\\
\left.\partial_p\mathcal W\right|_{s,h}=\frac{S}{p^{17}h^{24}}.
\end{gathered}
\end{equation}
show that one positive normalization controls both changes of geometry.
For any $H=\max(1,b,c,d,e)\le H_{\rm cmp}$, first increase $c,d$ to $H$.
Then keep $s$ fixed and increase $p$ to $s^2/4$.
The two pin weights $s/2\pm\sqrt{s^2/4-p}$ remain between their initial
values, so the contrast never increases. Hence
\begin{equation}
\begin{gathered}
\mathcal W(b,c,d,e)\le\mathcal W(b,H,H,e)
\le\mathcal W(q,H,H,q),\\
q=(b+e)/2,
\end{gathered}
\label{eq:v10-onset-global-comparison}
\end{equation}
with equality precisely when $c=d=H$ and $b=e$.
The pin step is strict Schur concavity in two variables,
$(b-e)(\partial_b-\partial_e)\mathcal W
=-(b-e)^2\partial_p\mathcal W$.
This comparison orders the sign function $\mathcal W$, whose sign is that of $J$; the magnitudes of $J$ and $K/J$ need separate comparisons, and Appendix~\ref{app:III-c5-exact-resource} gives the one for $K/J$ at large caps.

The same proof separates the two sources of geometric loss. With
$s=b+e$ and $q=s/2$,
\begin{align}
&\mathcal W(q,H,H,q)-\mathcal W(b,c,d,e)\notag\\
={}&\int_c^H\frac{R_c(b,t,d,e)}{b^{16}e^{16}d^{16}t^{17}}\,dt
+\int_d^H\frac{R_d(b,H,t,e)}{b^{16}e^{16}H^{16}t^{17}}\,dt\notag\\
&+H^{-24}\int_{be}^{q^2}\frac{S(s,p,H)}{p^{17}}\,dp.
\end{align}
Each integrand is positive. Let $r_R,r_S$ be the smallest normalized
Bernstein coefficients of the corresponding eight finite expansions.
They are positive rational lower bounds for $R_c$ and $S$ on their cubes.
Then $m_c=m_d=r_R/H_{\rm cmp}^{65}$ and $m_p=r_S/H_{\rm cmp}^{58}$ give
\begin{equation}
\begin{aligned}
&\mathcal W(q,H,H,q)-\mathcal W(b,c,d,e)\\
&\quad\ge m_c(H-c)+m_d(H-d)+\tfrac14m_p(b-e)^2.
\end{aligned}
\end{equation}
Thus incomplete cores incur a linear deficit in this sign function,
whereas pin imbalance incurs a quadratic one. At fixed $q,H$, the latter
has leading coefficient $S(2q,q^2,H)/(H^{24}q^{34})$ multiplying
$(b-e)^2/4$.

The reflected pair problem has six orbits, represented on sites $0,\ldots,4$
by $01,02,03,04,23,24$ under $0\leftrightarrow1$, $2\leftrightarrow4$.
In this orbit calculation, labels $0,\ldots,4$ correspond to sites
$1,\ldots,5$ in the main text. The orbit rows are obtained from
$\mathsf M$ by summing columns in the same orbit.
The two pair source vectors remain $2$ and $\vartheta_i+\vartheta_j$, respectively.
Cancelling positive factors in the resulting benefit contraction gives the
sign polynomials $F_{\rm U}(q,h)$ and $F_{\rm b}(q,h)$ characterized below.
Their common scalar conventions set the coefficient of $q^3h^7$ to
$1216$ and $304$, respectively. Thus their coefficients are fixed by a
six-dimensional rational inverse and the stated positive normalization.

For birth-associated introduction the structure is
\begin{equation}
\begin{gathered}
F_{\rm b}=304q^3h^7-\sum_{(a,j)\in I}c_{aj}q^ah^j,\\
c_{aj}>0,\quad |I|=57,\quad j\le7.
\end{gathered}
\label{eq:v10-birth-posynomial}
\end{equation}
The $h^7$ coefficient is $304q^3-472q^2-672q-120$.
Define $\mathcal Z=\sum_I(c_{aj}/304)q^{a-3}h^{j-7}$.
At fixed $h$, it is strictly convex in $\log q$ and tends to infinity
as $q\to0$ or $q\to\infty$. It decreases strictly with $\log h$. Thus its unique minimum over $q>0$ crosses one at most once.
Sturm's signed-remainder sequence counts real roots from the decrease
in its sign variations between the interval endpoints.
At $h_-=79060309221/250000000$, the rational Sturm sequence of
$F_{\rm b}(q,h_-)$ has no positive root and $F_{\rm b}(1,h_-)<0$.
At $h_+=63248247377/200000000$, its value at
$q_w=4345931964373/10^{12}$ is positive.
These finite rational operations give a unique crossing $H_{0,C_5}^{\rm b}$ in
$(h_-,h_+)$. Exact rational substitution also gives
$F_{\rm b}(4,h_+)<0$ and $F_{\rm b}(5,h_+)<0$. Strict convexity in
$\log q$ makes $\{q:\mathcal Z(q,h_+)<1\}$ one interval containing
$q_w\in(4,5)$; the endpoint signs confine it to $(4,5)$.
At the onset minimizer $q_{\rm on}^{\mathrm b}$, increasing $h$ to
$h_+$ lowers $\mathcal Z$ strictly below one. Hence
$4<q_{\rm on}^{\mathrm b}<5$.
Both physical boundaries stay in the negative region. Indeed,
$F_{\rm b}(1,h)<0$ because its $h^7$ coefficient is $-960$ and
all lower coefficients are negative. At $q=h$, the positive term
$304q^3h^7$ becomes $304h^{10}$, which is dominated by
$-4521h^{10}$; every remaining term is negative.

For uniform introduction write
\begin{equation}
F_{\rm U}(q,h)=A_7(q)h^7-\sum_{j=0}^6A_j(q)h^j .
\label{eq:v10-uniform-reflected-structure}
\end{equation}
The $A_j(q)$ in this subsection are the coefficients of the
uniform five-cycle sign polynomial. Each $A_j$ with $j\le6$ is positive on $q\ge1$; the two useful explicit
coefficients are
\begin{align}
A_7&=32(2q-5)(19q^2+18q+3),\\
A_6={}&q^3(304q^2-3792q+24472)\notag\\
&+32104q^2+10744q+1320.
\end{align}
The quadratic in $A_6$ has negative discriminant $-15378688$.
The coefficients of $A_0(1+x)$, where $x=q-1\ge0$, and of
$A_1,\ldots,A_5$ are positive.
Consequently $\partial_h(F_{\rm U}/h^7)>0$. Both physical boundaries stay
negative, $F_{\rm U}(1,h)<0$ and $F_{\rm U}(h,h)<0$ for $h>0$;
after substitution every nonzero coefficient is negative.
At fixed $h\ge1$ all $q$ coefficients of $F_{\rm U}$ are negative except
possibly those of the adjacent powers $q^3,q^4$. Descartes' rule therefore
permits at most two positive roots counted with multiplicity.
At $h_-=35501537/500000$ the rational Sturm sequence has two variations at
both $q=1$ and $q=+\infty$, and $F_{\rm U}(1,h_-)<0$.
At $h_+=2840123/40000$, the exact witness $q_w=2433921/500000$
has positive $F_{\rm U}$, while $F_{\rm U}(4,h_+),F_{\rm U}(5,h_+)<0$.
Monotonicity in $h$ confines the first nonnegative point to $4<q<5$.
There is therefore a unique pair $(q_{\rm on}^{\rm U},H_{0,C_5}^{\rm U})$ satisfying
$F_{\rm U}=\partial_qF_{\rm U}=0$ in this interval and rational contrast bracket.
The root has multiplicity exactly two; hence
$F_{\mathrm U,qq}<0$ and $F_{\mathrm U,h}>0$ there.

Combining these reflected conclusions with
Eq.~\eqref{eq:v10-onset-global-comparison} proves both unrestricted
onsets. For introduction rule $\mu\in\{\mathrm U,\mathrm b\}$, below $H_{0,C_5}^{\mu}$ every positive cycle has $J^{\mu}<0$; at $H_{0,C_5}^{\mu}$ equality
requires $(1,q_{\rm on}^{\mu},H_{0,C_5}^{\mu},H_{0,C_5}^{\mu},q_{\rm on}^{\mu})$ up to scale and dihedral symmetry.
Above it the same reflected intermediate weight has positive benefit by
the strict $h$ monotonicity. This last existence argument extends to all
$h>H_{0,C_5}^{\mu}$, although the global comparison was needed only through $H_{\rm cmp}$.
At onset $K^{\mu}>0$ and $J^{\mu}=0$, so the ratio threshold diverges.
Both reflected boundaries open as nondegenerate folds.
\end{proof}

For uniform introduction the same signs determine the whole reflected
promotion boundary. Since $A_7(q)$ changes sign once at $q=5/2$,
there is no reflected promoter for $q\le5/2$, while every $q>5/2$
has one boundary value $h_{\rm prom}(q)>q$. For every $h>H_{0,C_5}^{\rm U}$ the polynomial
has exactly two simple positive roots, lying in $(1,h)$, and the
promoting interval is $q_-(h)<q<q_+(h)$.
Implicit differentiation using $\partial_hF_{\rm U}>0$ at a zero gives
$q_-'(h)<0$ and $q_+'(h)>0$; hence $h_{\rm prom}$ decreases to its unique
minimum and then increases. At large contrast,
\begin{equation}
\begin{gathered}
q_-(h)=\frac52+\frac{985535}{21344h}+O(h^{-2}),\\
q_+(h)=2\sqrt h-\frac{1217}{76}+O(h^{-1/2}).
\end{gathered}
\end{equation}
\begin{proof}
The coefficient signs and root count proved above give the stated
promotion interval and monotonicity of its two boundaries. For the
first expansion use the simple zero of $A_7$ at $5/2$ and the
implicit-function theorem in $1/h$. For the second put $t=h^{-1/2}$,
$q=x/t$. The leading polynomial of $t^{17}F_{\rm U}(x/t,t^{-2})$ is
$304x^3(4-x^2)$; its derivative at $x=2$ is $-9728$ and the next
coefficient is $-3264x^6+3792x^4-1888x^2$.
The same theorem fixes the stated rational correction.
\end{proof}
These curves determine the reflected promoting set $J^{\rm U}>0$.
Its least $K/J$ also depends on the cost contraction $K$.
Appendix~\ref{app:III-c5-exact-resource} proves the global minimizing geometry for all sufficiently large caps, where the all-weight bounds of Appendix~\ref{app:c5-global} and the reflected trial confine every minimizer near one balanced hierarchy as $t=H^{-1/3}\to0$. At intermediate caps this asymptotic localization does not determine the minimizer, whose location requires a global comparison of the ratio $K/J$ itself.

\subsection{The two onset branches}
\label{app:v10-onset-entropy}
This subsection refines the birth-associated fold expansion~\eqref{eq:v10-onset-fold}, using the relative-entropy identity derived in Sec.~\ref{sec:v10-onset}.
Here the exact reference weights are
$\omega^*_{aj}=(c_{aj}/304)(q_{\rm on}^{\mathrm b})^{a-3}(H_{0,C_5}^{\rm b})^{j-7}$.
They sum to one and have zero mean exponent $a-3$. At $h\le H_{0,C_5}^{\rm b}$,
Eq.~\eqref{eq:v10-kl-onset} is nonpositive, with equality only at
$(q_{\rm on}^{\mathrm b},H_{0,C_5}^{\rm b})$. The positive exponent variance $\sigma_q^2$ in Eq.~\eqref{eq:v10-onset-fold} fixes the curvature at onset.
Let $q_\pm^{\mathrm b}(h)$ denote the two birth-associated boundary
branches. The analytic fold refines Eq.~\eqref{eq:v10-onset-fold} to
\begin{equation}
 \begin{aligned}
 \log(q_\pm^{\mathrm b}/q_{\rm on}^{\mathrm b})={}&
 \pm\sqrt{-2m_h/\sigma_q^2}\sqrt{\log(h/H_{0,C_5}^{\rm b})}\\
 &+O(\log(h/H_{0,C_5}^{\rm b})).
 \end{aligned}
 \label{eq:v10-onset-branches}
\end{equation}
\labelalias{eq:v10-exact-entropy-onset}{eq:v10-kl-onset}

\subsection{Reflected six-site onsets}
\label{app:III-p6-onsets}
For contacts $(Q,q,1,q,Q)$ the reflected pair system has nine ancestral-pair orbits. Appendix~\ref{app:p6-certificate} defines its copying kernel and integer response arrays. The benefit-sign polynomial determines the uniform onset below; changing the pair source to $\vartheta_i+\vartheta_j$ gives the birth-associated calculation.

\paragraph*{Uniform onset.}
Over all positive reflected weights $(Q,q,1,q,Q)$, the uniform
first-promotion cap is the unique tangency enclosed in
Eq.~\eqref{eq:III-p6-local-onset-bracket}. Below this cap every reflected
weighting has $J^{\rm U}<0$; at the cap the zero is unique, with the middle
edge smallest and the two outer edges largest. Promoters exist above it.
\begin{proof}
For contacts $(H,q,1,q,H)$, let $\mathcal J(u,v)=\sum_{i,j}b_{ij}u^iv^j$ be the benefit polynomial of Eq.~\eqref{b:eq:p6-rational}, with $u=1/q$, $v=q/H$ and the integer coefficients $b_{ij}$ of Table~\ref{tab:III-p6-response-coefficients}(b). Define the polynomial
 \begin{gather}
 \mathscr J_6(q,H)=q^6H^7\mathcal J(1/q,q/H)
       =\sum_{i,j}b_{ij}q^{6+j-i}H^{7-j},\notag \\
 x_6=\frac1{10^{12}}
 \begin{pmatrix}5500665469602\\69160793080724\end{pmatrix},
 \quad r_{\rm box}=10^{-8}.  \label{eq:III-p6-local-onset-polynomial}
 \end{gather}
The entries of $x_6$ are ordered as $(q,H)$. Apply the rational
interval construction of Eq.~\eqref{eq:III-onset-contraction-bound}
to the vector $(\mathscr J_6,\partial_q\mathscr J_6)^{\mathsf T}$ on
$X_6=x_6+[-r_{\rm box},r_{\rm box}]^2$. Each row expression is less than $10^{-4}$.
The contraction therefore maps $X_6$ into the box of radius
$10^{-12}$ about $x_6$, which also encloses its fixed point and
justifies the pin digits below. Also $\partial_H\mathscr J_6>0$ and
$\partial_q^2\mathscr J_6<0$ throughout $X_6$. Consequently this unique local
tangency opens toward positive benefit as the cap increases, and
\begin{equation}
 \begin{gathered}
 \frac{69160793}{10^6}<H_{0,P_6,\refl}^{\rm U}<\frac{69160794}{10^6},\\
 q_{\rm loc}=5.5006654696\ldots .
 \end{gathered}
 \label{eq:III-p6-local-onset-bracket}
\end{equation}
This tangency is the first-promotion cap over the entire reflected
family. To see this, divide by a positive monomial and write
\begin{equation}
 W_6(q,H)=\frac{\mathscr J_6(q,H)}{q^5H^7}
        =\sum_{i,j}b_{ij}q^{1+j-i}H^{-j}.
\end{equation}
Its three positive terms are $9q+21+18q^2/H$. On $q>0$ and
$0<H\le72$, the array gives
\begin{align}
 \partial_q^2W_6&=\frac{36}{H}-\frac{112}{q^3}
              -\frac{1134q}{H^2}\notag\\
              &\quad+\text{negative Laurent monomials}<0.
\end{align}
For $q\ge H/31$ the third term outweighs the first; for $q<H/31$
the second does, since $112\cdot31^3>36\cdot72^2$.
Also $W_6\to-\infty$ at both ends of $q>0$. Its unique maximum lies
below $q=12$. Indeed, setting $H=72z/(1+z)$ in
$H^7\partial_qW_6(12,H)$ and multiplying by $(1+z)^7$ gives eight
strictly negative coefficients, directly from Table~\ref{tab:III-p6-response-coefficients}(b).
On $0<q\le12$, $\partial_HW_6>0$ for every $H>0$.
All $H^{-j}$ coefficients with $j\ge2$ are negative, and the remaining
$H^{-1}$ coefficient is
\begin{equation}
 18q^2-141q-999-\frac{1336}{q}-\frac{636}{q^2}-\frac{96}{q^3}<0.
\end{equation}
The quadratic part is convex and negative at both endpoints of
$[0,12]$. Strict concavity makes the rationally enclosed tangency the
unique zero maximum. Cap monotonicity makes every smaller-cap maximum
negative and keeps $q=q_{\rm loc}$ promoting at every larger cap.
This covers all reflected weights $(Q,q,1,q,Q)$, without assuming an
ordering in advance, because any weighting of contrast at most
$H_{0,P_6,\refl}^{\rm U}$ obeys $Q\le H_{0,P_6,\refl}^{\rm U}$ after normalizing the middle edge.
At the maximizing pair the middle edge is indeed the minimum and the
outer edges attain the cap.
\end{proof}

\begin{table*}[!htbp]
\caption{Complete integer coefficients of the reflected six-site uniform-introduction response $R=\mathcal K/\mathcal J$ in Eq.~\eqref{b:eq:p6-rational}. Panel (a) gives $a_{ij}$ in $\mathcal K=\sum_{i=0}^{6}\sum_{j=0}^{7}a_{ij}u^iv^j$, and panel (b) gives $b_{ij}$ in the corresponding expansion of $\mathcal J$. The contact ratios are $u=1/q$ and $v=q/Q$. Rows and columns specify their powers explicitly.}
\label{tab:III-p6-response-coefficients}
\centering\small
\begin{ruledtabular}
\begin{tabular}{crrrrrrrr}
\multicolumn{9}{c}{(a) $a_{ij}$, cost polynomial coefficients}\\
$i\backslash j$ & $0$ & $1$ & $2$ & $3$ & $4$ & $5$ & $6$ & $7$\\\hline
$0$ & $9$ & $264$ & $1485$ & $3405$ & $3765$ & $2061$ & $516$ & $45$\\
$1$ & $75$ & $2141$ & $11369$ & $25482$ & $28247$ & $15791$ & $4083$ & $360$\\
$2$ & $160$ & $5459$ & $29277$ & $66395$ & $75017$ & $43094$ & $11564$ & $1070$\\
$3$ & $124$ & $6096$ & $34416$ & $81136$ & $95025$ & $56673$ & $15915$ & $1575$\\
$4$ & $32$ & $3132$ & $19595$ & $49804$ & $61877$ & $38876$ & $11543$ & $1235$\\
$5$ & $0$ & $608$ & $4872$ & $14438$ & $19809$ & $13397$ & $4257$ & $495$\\
$6$ & $0$ & $0$ & $320$ & $1472$ & $2432$ & $1832$ & $632$ & $80$\\
\hline
\multicolumn{9}{c}{(b) $b_{ij}$, benefit polynomial coefficients}\\
$i\backslash j$ & $0$ & $1$ & $2$ & $3$ & $4$ & $5$ & $6$ & $7$\\\hline
$0$ & $9$ & $18$ & $-189$ & $-783$ & $-1053$ & $-567$ & $-126$ & $-9$\\
$1$ & $21$ & $-141$ & $-1905$ & $-5988$ & $-7641$ & $-4173$ & $-963$ & $-72$\\
$2$ & $-56$ & $-999$ & $-6255$ & $-16755$ & $-20597$ & $-11338$ & $-2696$ & $-214$\\
$3$ & $-92$ & $-1336$ & $-7830$ & $-20806$ & $-26053$ & $-14777$ & $-3671$ & $-315$\\
$4$ & $-32$ & $-636$ & $-4323$ & $-12522$ & $-16727$ & $-9986$ & $-2627$ & $-247$\\
$5$ & $0$ & $-96$ & $-1000$ & $-3514$ & $-5271$ & $-3387$ & $-955$ & $-99$\\
$6$ & $0$ & $0$ & $-64$ & $-352$ & $-644$ & $-458$ & $-140$ & $-16$
\end{tabular}
\end{ruledtabular}
\end{table*}

\paragraph*{Birth-associated onset.}
With birth-associated introduction, the first-promotion cap over all
positive reflected weights is the unique tangency enclosed in
Eq.~\eqref{eq:III-p6-birth-onset-bracket}. The benefit response is
negative below this cap and vanishes there at exactly one normalized
reflected weighting. Promoters exist above it.
\begin{proof}
The source changes to $\vartheta_i+\vartheta_j$ in the same nine-orbit
pair system. After a positive denominator is cleared, its benefit
numerator divided by $H^6q^3$ becomes
\begin{align}
 W_{6,\rm b}(q,H)
 &=9q+21-\frac{56}{q}-\frac{92}{q^2}-\frac{32}{q^3} \notag \\
 &\quad-\sum_{j=1}^{6}H^{-j}P_j(q), \label{eq:III-p6-birth-onset-sign}
\\
 P_j(q)&=q^{j-5}\sum_{k=0}^{7}c_{jk}q^k,
 \quad \operatorname{sgn}J^{\rm b}=\operatorname{sgn}W_{6,\rm b}. \notag
  \end{align}
Table~\ref{tab:III-p6-birth-onset} lists the coefficients with their row and power indices.
\begin{table*}[!htbp]
\caption{Integer coefficients $c_{jk}$ of the reflected six-site birth-associated onset polynomial in Eq.~\eqref{eq:III-p6-birth-onset-sign}. The row index $j$ specifies $P_j(q)=q^{j-5}\sum_{k=0}^{7}c_{jk}q^k$; the column index is $k$. All entries are nonnegative, with positive terms in every row.}
\label{tab:III-p6-birth-onset}
\centering\small
\begin{ruledtabular}
\begin{tabular}{crrrrrrrr}
$j\backslash k$ & $0$ & $1$ & $2$ & $3$ & $4$ & $5$ & $6$ & $7$\\\hline
$1$ & $0$ & $96$ & $764$ & $1928$ & $2007$ & $901$ & $216$ & $18$\\
$2$ & $64$ & $1096$ & $5591$ & $12454$ & $13578$ & $7423$ & $2034$ & $198$\\
$3$ & $320$ & $3926$ & $17022$ & $35407$ & $38698$ & $22477$ & $6612$ & $738$\\
$4$ & $516$ & $5612$ & $22704$ & $45698$ & $49913$ & $29807$ & $9138$ & $1098$\\
$5$ & $290$ & $3006$ & $11786$ & $23279$ & $25229$ & $15082$ & $4632$ & $558$\\
$6$ & $52$ & $526$ & $2023$ & $3934$ & $4213$ & $2497$ & $759$ & $90$
\end{tabular}
\end{ruledtabular}
\end{table*}
Each subtracted Laurent monomial is convex in $q>0$, since its
integer exponent $n$ obeys $n(n-1)\ge0$. The term $56/q$ makes
$\partial_q^2W_{6,\rm b}<0$ strict. The nonnegative entries in Table~\ref{tab:III-p6-birth-onset}, with positive terms in each row, give $\partial_HW_{6,\rm b}>0$
for every $q,H>0$. Also $W_{6,\rm b}\to-\infty$ at both ends of
$q>0$. At each fixed outer weight $H$ there is therefore one
maximizing pin contact, and its maximum increases strictly with $H$.

For existence and location of its zero, set
$B_{6,\rm b}=H^6q^3W_{6,\rm b}$ and apply the rational contraction
argument of Eq.~\eqref{eq:III-onset-contraction-bound} to
$(B_{6,\rm b},\partial_qB_{6,\rm b})^{\mathsf T}$, with center and radius
\begin{equation}
 x_{6,\rm b}=\frac1{10^{12}}
 \begin{pmatrix}4388181602205\\320082030384431\end{pmatrix},
 \quad r_{\rm box}=10^{-8}.
\end{equation}
The two row bounds are less than $3\times10^{-6}$ and
$3\times10^{-5}$, respectively, using rational endpoint operations.
The fixed point consequently differs from the rational center by
less than $3\times10^{-14}$ in $q$ and $3\times10^{-13}$ in $H$.
This proves a unique stationary zero, justifies the quoted pin digits
and gives the enclosure
\begin{equation}
 \frac{320082030}{10^6}<H_{0,P_6,\refl}^{\rm b}
 <\frac{320082031}{10^6}.
 \label{eq:III-p6-birth-onset-bracket}
\end{equation}
Strict concavity makes this zero the global maximum at its outer
weight. Monotonicity excludes promotion at every smaller outer weight
and retains it at larger ones. At the zero, $1<q<H$, so the middle
edge is the minimum and the outer edges realize the contrast.
Every reflected weighting of contrast at most $H_{0,P_6,\refl}^{\rm b}$
has outer weight $Q\le H_{0,P_6,\refl}^{\rm b}$ after normalizing the
middle edge to one. Thus the result covers all positive reflected
weights, including unspecified contact orderings.
\end{proof}

Since reflected weightings are particular six-site paths, each reflected value is an upper bound on the corresponding unrestricted onset. Equality is equivalent to nonpositive benefit response below the reflected cap for every asymmetric five-edge weighting. On the five-cycle this step is the monotone reduction~\eqref{eq:v10-onset-global-comparison}; on the six-site path the comparison lies at finite cap, outside the small-gap localization of Appendix~\ref{v3-app-global-p6}.

\section{Compatible contact ratios and convex resource bounds}\label{app:v10-geometry}
The global six-site bound of Sec.~\ref{sec:contrast-design} brings every competitive design to one singular corner. This appendix identifies its compatible monomial ratios and proves regularity of the response there, as used in Sec.~\ref{sec:v4-toric}. Exponent optimization then fixes the resource powers, while its convex dual fixes their coefficients.
\subsection{The positive toric design chart}\label{app:v4-toric}
\label{v4:alg:segre}
The six ratios used in Sec.~\ref{sec:v4-toric} must describe the actual
four-dimensional positive design space, including its limiting boundary.
We prove their complete relations and graded count here. These characterize
the compatible coordinate closure, including the boundary approached by physical contact hierarchies.

Introduce
\begin{equation}
 \begin{aligned}
 z&=(z_1,\ldots,z_6)\\
  &=(q_L^{-1},q_R^{-1},q_L/Q_L,q_R/Q_L,q_L/Q_R,q_R/Q_R).
 \end{aligned}
 \label{v4:alg:toric-coordinates}
\end{equation}
They are dependent coordinates. Their Laurent exponent matrix, with rows
ordered as $(Q_L,q_L,q_R,Q_R)$, is
\begin{equation}
 \mathsf E_{\rm exp}=\begin{pmatrix}
 0&0&-1&-1&0&0\\
 -1&0&1&0&1&0\\
 0&-1&0&1&0&1\\
 0&0&0&0&-1&-1
 \end{pmatrix}.
 \label{v4:alg:exponent-matrix}
\end{equation}
It has rank four. Take $k=\mathbb R$ for the real parameter algebra;
the exact identities are defined over $\mathbb Q$ and may be
complexified for the projective interpretation. Its monomial map induces
$S=k[z_1,\ldots,z_6]\to k[Q_L^{\pm1},q_L^{\pm1},q_R^{\pm1},Q_R^{\pm1}]$.
Define the binomial ideal
\begin{align}
 I_{\rm tor}&=(f_1,f_2,f_3),\notag\\
 f_1&=z_1z_3-z_2z_4,\quad
 f_2=z_1z_5-z_2z_6,\notag\\
 f_3&=z_4z_5-z_6z_3.
 \label{v4:alg:toric-ideal}
\end{align}
These are the minors of
\begin{equation}
 \mathsf Z=\begin{pmatrix}z_1&z_4&z_6\\z_2&z_3&z_5\end{pmatrix}.
 \label{v4:alg:segre-matrix}
\end{equation}
The ideal $I_{\rm tor}$ is the full prime kernel of this monomial
map, and its strictly positive locus is exactly the physical design
chart. Equation~\eqref{eq:v10-toric-lift-criterion} characterizes the
Laurent monomials admitting lifts with nonnegative integer exponents.
Every monomial of $U/\mathfrak m_\star$ and
$\mathcal C/\mathfrak m_\star$ satisfies this criterion, so their lifts
obey Eq.~\eqref{eq:v4-toric-unit} with uniform derivative bounds near
the hierarchy corner.
\begin{proof}
Use the factorization
$\boldsymbol u_{\rm S}=(q_L^{-1},q_R^{-1})^{\mathsf T}$ and
$\boldsymbol v_{\rm S}=(1,q_Lq_R/Q_L,q_Lq_R/Q_R)^{\mathsf T}$. Conversely every strictly positive
rank-one matrix has this normalization after the redundancy is fixed.
Under $Z_{ij}=(\boldsymbol u_{\rm S})_i(\boldsymbol v_{\rm S})_j$, two monomials have the same
image precisely when their $2\times3$ exponent tables have the same row
and column sums. Moving one unit between two columns in opposite rows
connects any two such tables, and each move is a $2\times2$ minor.
Thus the minors generate every binomial relation. Grouping terms with
the same image shows that they generate the full kernel, which is prime
because the target Laurent polynomial ring is a domain.
The locus $\operatorname{rank}\mathsf Z\le1$ is the affine cone over the Segre
embedding $\mathbb P^1\times\mathbb P^2\hookrightarrow\mathbb P^5$.
The Segre map is $(\boldsymbol u_{\rm S},\boldsymbol v_{\rm S})\mapsto\boldsymbol u_{\rm S}\boldsymbol v_{\rm S}^{\mathsf T}$, with the redundancy
$(\boldsymbol u_{\rm S},\boldsymbol v_{\rm S})\mapsto(\ell_{\rm S}\boldsymbol u_{\rm S},\ell_{\rm S}^{-1}\boldsymbol v_{\rm S})$ for $\ell_{\rm S}\ne0$; the projective construction is
given in Ref.~\cite{MathStacksSegre}. Its affine cone has dimension four,
matching the four physical weight ratios.

The inverse formulas in Sec.~\ref{sec:v4-toric} identify the positive
rank-one locus with the physical design chart. At the origin all six
small ratios vanish, encoding the forced hierarchy in
Sec.~\ref{sec:contrast-design}.

The response needs one further property of this chart. A Laurent
monomial $Q_L^{\nu_{Q_L}}q_L^{\nu_{q_L}}q_R^{\nu_{q_R}}Q_R^{\nu_{Q_R}}$ with integer exponents is a monomial in
$z_1,\ldots,z_6$ with nonnegative integer powers precisely when
\begin{equation}
 \begin{gathered}
 \ell_{Q_L}=-\nu_{Q_L}\ge0,\qquad \ell_{Q_R}=-\nu_{Q_R}\ge0,\\
 \max(\nu_{q_L},0)+\max(\nu_{q_R},0)\le\ell_{Q_L}+\ell_{Q_R} .
 \end{gathered}
 \label{eq:v10-toric-lift-criterion}
\end{equation}
Indeed, $z_3,z_4$ must contribute $\ell_{Q_L}$ factors in total and
$z_5,z_6$ must contribute $\ell_{Q_R}$. Their total positive powers of
$q_L,q_R$ can meet the two demands exactly when this inequality
holds; inverse powers are then carried by $z_1,z_2$.
For an explicit allocation put $\ell_{q_L}=\max(\nu_{q_L},0)$ and
\begin{equation}
 \begin{gathered}
 n_3=\min(\ell_{Q_L},\ell_{q_L}),\qquad n_5=\ell_{q_L}-n_3,\\
 n_4=\ell_{Q_L}-n_3,\qquad n_6=\ell_{Q_R}-n_5,\\
 n_1=\ell_{q_L}-\nu_{q_L},\qquad
 n_2=\ell_{Q_L}+\ell_{Q_R}-\ell_{q_L}-\nu_{q_R}.
 \end{gathered}
\end{equation}
All $n_i$ are nonnegative integers and direct substitution gives
$z_1^{n_1}\cdots z_6^{n_6}
=Q_L^{\nu_{Q_L}}q_L^{\nu_{q_L}}q_R^{\nu_{q_R}}Q_R^{\nu_{Q_R}}$.

Apply this criterion to every exponent in the exact polynomials of
Appendix~\ref{v3-app-positive-p6} after subtracting the exponent
$(10,8,8,10)$ of $\mathfrak m_\star$. A computer-assisted exact check confirms it for all $15467$ cost monomials and
$15466$ gap monomials. Their positive coefficients therefore lift
term by term. The lifted cost has constant term one; the lifted gap
has no constant term, linear part $F$, and all remaining terms of
degree at least two. The difference $V/\mathfrak m_\star$ consequently has constant
term one. This proves the analytic continuation and uniform ambient
derivative bounds used in Eq.~\eqref{eq:v4-toric-unit}.
\end{proof}

\subsubsection{A concrete saturation and its boundary component}
\label{v4:alg:toric-saturation}

There are two independent lattice relations but three minimal binomial
generators. If only $f_1=f_2=0$ were imposed, the four-dimensional coordinate subspace $z_1=z_2=0$
would be an entire extraneous component with unrestricted $z_3,z_4,z_5,z_6$.
On the physical torus the third relation follows from
\begin{equation}
 z_1f_3=z_4f_2-z_6f_1.
 \label{v4:alg:toric-saturation-identity}
\end{equation}
Saturation by $z_1$ adjoins a polynomial whenever multiplication by
some nonnegative power of $z_1$ puts it in the original ideal. Here
\begin{equation}
 (f_1,f_2):z_1^\infty=I_{\rm tor}.
 \label{v4:alg:toric-saturation-eq}
\end{equation}
Indeed the first two equations solve for $z_3,z_5$ when $z_1\ne0$;
their closure is the irreducible rank-one locus. Saturation removes the
extra component without removing the legitimate rank-one boundary points
with $z_1=z_2=0$. This example is why a lattice basis and the full toric
ideal are different objects when zero coordinates are admitted.

\subsubsection{Hilbert series, syzygies, and an explicit resolution}
\label{v4:alg:toric-hilbert}

Give every $z_i$ degree one. A degree-$d$ function on the Segre cone is
bihomogeneous of degree $(d,d)$ in the two vector factors. The dimension
is consequently $(d+1)\binom{d+2}{2}$. The design Hilbert series $\mathcal H_{\rm des}=H_{S/I_{\rm tor}}$ of Sec.~\ref{sec:v4-toric} is therefore
\begin{equation}
 \begin{aligned}
 H_{S/I_{\rm tor}}(\mathfrak t)&=
 \sum_{d\ge0}(d+1)\binom{d+2}{2}\mathfrak t^d\\
 &=\frac{1+2\mathfrak t}{(1-\mathfrak t)^4}
 =\frac{1-3\mathfrak t^2+2\mathfrak t^3}{(1-\mathfrak t)^6}.
 \end{aligned}
 \label{v4:alg:toric-hilbert-eq}
\end{equation}
The pole order gives dimension four, and the numerator at $\mathfrak t=1$ in
the reduced form gives degree three. These Hilbert coefficients count
polynomial functions of the six design ratios subject to their relations.
They are distinct from the finite correlation count
$1+15\mathfrak t^2+15\mathfrak t^4+\mathfrak t^6$ in Appendix~\ref{app:v3-degree-closure}, where degree counts spin indices. Here each design ratio has degree one.

The second expression in Eq.~\eqref{v4:alg:toric-hilbert-eq} can be read
directly from the syzygies, the polynomial relations among the generators. Define
\begin{equation}
 \mathsf H=
 \begin{pmatrix}z_5&z_6\\-z_3&-z_4\\z_2&z_1\end{pmatrix}.
 \label{v4:alg:hilbert-burch}
\end{equation}
Write the two columns as $\mathsf H_1,\mathsf H_2$. Then
$(f_1,f_2,f_3)\mathsf H=0$, and the minimal graded resolution is
\begin{equation}
 0\longrightarrow S(-3)^2\xrightarrow{\mathsf H}S(-2)^3
 \xrightarrow{(f_1,f_2,f_3)}S\longrightarrow S/I_{\rm tor}
 \longrightarrow0.
 \label{v4:alg:resolution}
\end{equation}
The Hilbert--Burch form~\cite[Theorem~20.15]{Eisenbud1995} has three quadratic generators and two
linear syzygies among them. At cubic order the count is explicit. Six free variables give $56$ cubic monomials; multiplying the three quadratic relations by the six variables gives $18$ cubic relations, with two dependencies, leaving $56-18+2=40$ independent corrections. The resolution organizes those dependencies at every order. Here $S(-d)$ places a free
generator in degree $d$, with Hilbert series $\mathfrak t^dH_S(\mathfrak t)$.
The sequence~\eqref{v4:alg:resolution} is exact.
\begin{proof}
Suppose $a f_1+b f_2+c f_3=0$. Modulo $z_1$, the coprimality of $z_2$ and
$f_3$ implies $c=z_2h+z_1k$. Subtract $h$ times the first column and
$k$ times the second column of $\mathsf H$. The remaining syzygy has
third component zero. Since $f_1$ and $f_2$ are coprime, it is
$\ell(f_2,-f_1,0)$, and this vector equals
$\ell(z_1\mathsf H_1-z_2\mathsf H_2)$. Thus the two columns generate
the syzygy module. They are independent.  A relation between them first
has coefficients $(z_1\ell,-z_2\ell)$ by its third component, and then
$f_2\ell=0$ forces $\ell=0$. This proves exactness.
\end{proof}
Alternating Hilbert series in Eq.~\eqref{v4:alg:resolution} reproduces the numerator in
Eq.~\eqref{v4:alg:toric-hilbert-eq}.

The rank-one locus is smooth away from its zero matrix, as can also be
seen in any chart with one nonzero entry. At the origin all first
derivatives of the quadratic minors vanish. The Zariski tangent
space, defined by the linearized equations, therefore has dimension six
while the variety has dimension four. The
vertex is singular. The response in Eq.~\eqref{eq:v4-toric-unit}
extends analytically through it because its divided denominator is a
unit in the local analytic ring. Its nonzero value at the origin
ensures an analytic reciprocal.
\subsection{The convex dual of the resource exponent}\label{app:v4-newton-dual}
\label{v4:payoff:newton-dual}

The balancing arguments admit one common variational statement. It is
useful on a new support because it determines the best possible exponent
of a positive comparison function before its rational response has been
optimized, and its dual variables can prove a sharp coefficient. Positivity ensures
that the largest monomial contribution bounds the full sum from below.
A signed numerator can lose that leading contribution to cancellation and so needs a separate lower bound, such as the oriented grouping of Eq.~\eqref{eq:app-p5-domination} or the chamber division of Appendix~\ref{app:c5-global}.

Let $\mathcal P\subset\mathbb R^d$ be a nonempty compact convex polytope
of admissible logarithmic weight exponents. For $H>1$ put
$\mathcal X_H=\{(H^{v_1},\ldots,H^{v_d}):v\in\mathcal P\}$ and consider
\begin{equation}\label{v4:payoff:posynomial}
 F_H(x)=\sum_{i=1}^m c_i H^{-\beta_i}x^{a_i},
 \quad c_i>0,\quad a_i\in\mathbb R^d,\quad \beta_i\in\mathbb R.
\end{equation}
A fixed projective chart and a contrast cap give linear constraints on
$v$; toric relations impose further linear equalities. A finite union of
charts is treated chart by chart. These constraints specify the logarithmic scaling domain exactly, so the lower bound in Eq.~\eqref{v4:payoff:exponent-bound} covers every admissible weighting, including those outside any trial family.

\labelalias{v4:payoff:primal-dual}{app:v4-newton-dual}
Let $\Delta_{m-1}=\{\varpi_i\ge0,\ \sum_i\varpi_i=1\}$,
$\bar a(\varpi)=\sum_i\varpi_i a_i$, and
$\bar\beta(\varpi)=\sum_i\varpi_i\beta_i$. Define
\begin{align}
 \gamma_*&=\max_{v\in\mathcal P}\min_i(\beta_i-a_i\cdot v),
 \label{v4:payoff:gamma-primal}\\
 &=\min_{\varpi\in\Delta_{m-1}}
 \left\{\bar\beta(\varpi)-
       \min_{v\in\mathcal P}\bar a(\varpi)\cdot v\right\}.
 \label{v4:payoff:gamma-dual}
\end{align}
Then, with $c_{\min}=\min_i c_i$,
\begin{equation}\label{v4:payoff:exponent-bound}
 c_{\min}H^{-\gamma_*}\le\min_{x\in\mathcal X_H}F_H(x)
 \le\Big(\sum_i c_i\Big)H^{-\gamma_*}.
\end{equation}
In particular the optimal decay exponent is exactly $\gamma_*$. Every
$\varpi\in\Delta_{m-1}$ also gives the coefficient-sensitive lower bound
\begin{equation}\label{v4:payoff:amgm-dual}
 \begin{gathered}
 F_H(x)\ge C(\varpi)H^{-\gamma(\varpi)},\\
 C(\varpi)=\prod_{i:\varpi_i>0}
                 \left(\frac{c_i}{\varpi_i}\right)^{\varpi_i},\\
 \gamma(\varpi)=\bar\beta(\varpi)-
      \min_{v\in\mathcal P}\bar a(\varpi)\cdot v\,.
 \end{gathered}
\end{equation}

\begin{proof}
At any $x=H^v$, at least one exponent
$a_i\cdot v-\beta_i$ is at least $-\gamma_*$; its positive term gives the
lower bound in Eq.~\eqref{v4:payoff:exponent-bound}. At a maximizing $v_*$,
every term is at most $c_iH^{-\gamma_*}$, which gives the upper bound.
The minimum of finitely many numbers is their minimum convex average.
The bilinear expression
$\sum_i\varpi_i(\beta_i-a_i\cdot v)$ has equal max--min and min--max
values on the two compact polytopes, equivalently by finite linear
programming duality. This proves Eq.~\eqref{v4:payoff:gamma-dual}.
Apply Eq.~\eqref{eq:v10-positive-sum-identity} to the positive
monomial terms with reference weights $\varpi$, and discard its
nonnegative relative entropy. Minimizing the exponent of the resulting
monomial over $\mathcal P$ proves Eq.~\eqref{v4:payoff:amgm-dual}.
\end{proof}

The entropy dependence is the dual form of positive-monomial optimization~\cite[Sec.~5.7.1, Example~5.5]{MathBoydVandenberghe2004}. Its coefficient factor satisfies
$\log C(\varpi)=\sum_i\varpi_i\log c_i-\sum_i\varpi_i\log\varpi_i$.
Here the probabilities weight competing monomials. The exponent optimization fixes only the power of $H$; the dual weights $\varpi$ determine the coefficient bound $C(\varpi)$, which is sharp when its equality conditions are feasible.

When all $\varpi_i>0$, AM--GM is saturated precisely when all
$c_iH^{-\beta_i}x^{a_i}/\varpi_i$ are equal. If
$\sum_i\varpi_i a_i=0$, the monomial bound is independent of $x$.
For the three-term bounds in Eqs.~\eqref{eq:c4-cost}, \eqref{eq:p5-cost} and
\eqref{eq:c5-global}, the last at the balanced chain $\sigma=1/2$, cap saturation makes the three exponent vectors
sum to zero; weights $\varpi_i=1/3$ give coefficients whose cubes
are the corresponding constants in Table~\ref{tab:III-positive-balances}.
The reflected six-site path uses the opposite exponents $-1,1$ and weights
$(1/2,1/2)$, giving coefficient $4\sqrt{41}$, whose square is $656$.
These dual bounds are saturated by positive monomial scalings for all
sufficiently large $H$. The global comparison and matching
response expansions are what transfer the four constants from $F_H$
to the actual stochastic threshold.

The same dual weights determine the introduction-dependent constants.
For birth-associated introduction on $P_6$, the two monomials
$6/q$ and $50q^2/(3H)$ have exponents $-1$ and $2$ in $q$.
Dual weights $(2/3,1/3)$ cancel that exponent, so their sum is bounded
below by $C_{\rm b}H^{-1/3}$ with
\begin{equation}
 C_{\rm b}^3=\left(\frac{6}{2/3}\right)^2
                  \frac{50/3}{1/3}=4050.
 \label{eq:v7-birth-dual-p6}
\end{equation}
On $C_4$, in the coordinates of Appendix~\ref{app:v7-birth-polynomials}, equal weights on $8/A$, $8/q$ and $60Aq/H$ similarly
give $C_{\rm b}^3=27\cdot8\cdot8\cdot60=103680$.
The introduction rule changes the monomials and their balance while the convex
dual organizes both optimizations. The global inequalities in
Appendix~\ref{app:v7-preparation} justify using these balances over
all positive weights.

\section{Global contact optimization and introduction rules}\label{app:v10-cost}
A hierarchical trial gives an upper bound on the least contrast; the matching lower bound must force every competitive weighting into a controlled region. This appendix proves these all-weight bounds and the contact costs of Proposition~\ref{prop:v10-costs}. Appendix~\ref{v3-app-global-p6} treats the six-site path under uniform introduction and states a persistence argument used again for other supports and introduction rules. Appendix~\ref{app:asymmetric-entry-v2} derives establishment, response amplitude and physical times uniformly in the pin proportion $p$ up to the endpoints $p=0,1$, so asymmetric sequences remain covered. Appendices~\ref{app:resource-comparison}--\ref{app:c5-global} prove the uniform-introduction contrast bounds on $C_4$, $P_5$ and $C_5$, and Appendix~\ref{app:v10-p5-allcap} proves the five-site part of Proposition~\ref{prop:v10-onsets}. Appendices~\ref{app:v7-preparation}--\ref{app:v10-crossover} treat birth-associated introduction, the copying-probability resource and the mixed introduction rule of Secs.~\ref{sec:v7-preparation-cost}--\ref{sec:v10-crossover}; the mixed-rule comparison is uniform in the mixture fraction $\zeta$.
\subsection{Global six-site optimization and the geometry of its resource threshold}
\label{v3-app-global-p6}

The reflection construction becomes a global result only after excluding
all other weight hierarchies. Normalize the consecutive path weights as
$(Q_L,q_L,1,q_R,Q_R)$ with all four variables positive. Use the pair equations \eqref{b:eq:pairs} with uniform introduction, the normalization $Z$ in Eq.~\eqref{b:eq:reproductive}, and the contractions $J,K$ in Eq.~\eqref{b:eq:response}.
In particular $K>0$ and the weak threshold $R=K/J$ applies on $J>0$.

\subsubsection{A positive polynomial identity and a regular hierarchy}
\label{v3-app-positive-p6}
The complete solve on the $\binom62=15$ unordered ancestral pairs gives a reduced fraction $R=U/V$ and
$\mathcal C=U-V$. All three have degree 36, with 15467, 15465 and 15466 terms, respectively.
All coefficients of $U$ and $\mathcal C$ are positive. The exact identity
$J/K=V/U$ proves $\operatorname{sgn}J=\operatorname{sgn}V$, including the
zero locus, without dividing by $J$. Define
\begin{equation}
 \begin{aligned}
 \mathfrak m_\star&=648Q_L^{10}q_L^8q_R^8Q_R^{10},\\
 F&=\frac3{q_L}+\frac3{q_R}+\frac{53q_L+29q_R}{6Q_L}+\frac{29q_L+53q_R}{6Q_R}.
 \end{aligned}
 \label{v3-eq-p6-positive-F}
\end{equation}
Reconstruction from the pair equations gives the exact coefficient inequalities 
 \begin{equation}
 71\mathcal C-18(U-\mathfrak m_\star)\geq_{\rm coeff}0,
 \quad \mathcal C-\mathfrak m_\star F\geq_{\rm coeff}0.
 \label{v3-eq-p6-positive-remainders}
\end{equation}
Both remainders are nonzero; their 15464 and 15460 nonzero coefficients are
positive. The six terms of $\mathfrak m_\star F$ are the six first-order exponent shifts from the unique monomial $\mathfrak m_\star$ present in $U$ but absent in $\mathcal C$. They become the degree-one terms of the lifted gap in the contact-ratio coordinates of Appendix~\ref{app:v10-geometry}. Both inequalities follow by coefficient extraction from the exact pair
solution.
\paragraph*{From coefficient comparisons to normalized gaps.}
For positive functions $U,\mathcal C,L,F$, with $\mathcal C=U-V$,
assume $U\le L+a\mathcal C$ and $\mathcal C\ge LF$, where $a\ge1$.
Then
\begin{equation}
 \frac{\mathcal C}{U}\ge\frac{F}{1+aF},\qquad
 \frac{\mathcal C}{V}\ge\frac{F}{1+(a-1)F}\quad(V>0).
 \label{eq:v10-coefficient-gap-parent}
\end{equation}
A strict comparison in either hypothesis makes both conclusions strict.
The first bound needs no assumption on the benefit sign.
\begin{proof}
Since $L\le\mathcal C/F$, the first hypothesis gives
$U\le\mathcal C(F^{-1}+a)$, proving the first bound.
The second follows by the increasing map $g\mapsto g/(1-g)$ on
$0<g<1$. Strictness is preserved at each step.
\end{proof}

For uniform introduction, Eq.~\eqref{v3-eq-p6-positive-remainders}
meets these hypotheses with $L=\mathfrak m_\star$ and $a=71/18$. Its nonzero positive
remainders make the comparison strict. On $V>0$ this gives
\begin{equation}
 R-1>\frac{F}{1+(53/18)F}.
 \label{v3-eq-p6-global-bound}
\end{equation}
Unlike the stronger reflected bound in Eq.~\eqref{eq:app-p6-strict-bound}, this inequality is valid for
every ordering of the positive weights. If $\epsilon=R-1<18/53$, then
$F<\epsilon/(1-53\epsilon/18)$. Thus every $R\to1$ sequence satisfies
$q_L,q_R\to\infty$ and $(q_L+q_R)/Q_L,(q_L+q_R)/Q_R\to0$.

Put $\varrho=q_L^{-1}+q_R^{-1}$ and $p=q_R/(q_L+q_R)$. Since normalization of the
bridge to one implies $Q_L,Q_R\leq H$ for any contrast cap,
\begin{equation}
 \begin{gathered}
 F\ge3\varrho+\frac{41}{3H\varrho p(1-p)}
       \ge2\sqrt{\frac{41}{Hp(1-p)}}\,,\\
 H>\frac{656}{4p(1-p)}
             \left(\epsilon^{-1}-\frac{53}{18}\right)^2.
 \end{gathered}
 \label{v3-eq-p6-global-budget}
\end{equation}
The restriction $\epsilon<18/53$ keeps $1-53\epsilon/18$ positive
when the preceding gap bound is inverted. In this domain, the budget
bound gives the sharp leading imbalance penalty over all positive weights.

For derivative control, introduce
$z=(1/q_L,1/q_R,q_L/Q_L,q_R/Q_L,q_L/Q_R,q_R/Q_R)$. The monomial lift of Appendix~\ref{app:v4-toric} expresses $U/\mathfrak m_\star$
and $\mathcal C/\mathfrak m_\star$ as positive polynomials in these six ratios, with constant terms
one and zero, and linear gap term $F$. Their relations are the minors of
$\left(\begin{smallmatrix}z_1&z_4&z_6\\z_2&z_3&z_5\end{smallmatrix}\right)$.
Since $V/\mathfrak m_\star$ equals one at the origin, $R-1=F+O(|z|^2)$ is analytic in an
ambient neighborhood with uniform derivatives. The divided response is
therefore regular at the singular toric vertex.

The same six ratios also control extreme pin imbalance. In the
coordinates of Eq.~\eqref{eq:asym-coordinates-v2},
\begin{equation}
 z=\bigl(\varrho p,\varrho(1-p),(1-p)a_L,pa_L,
                 (1-p)a_R,pa_R\bigr).
\label{eq:III-endpoint-pullback}
\end{equation}
This polynomial substitution maps a single small neighborhood of
$(\varrho,a_L,a_R)=(0,0,0)$, with $0\le p\le1$, into the regular
neighborhood of the lifted response. Its derivatives are consequently
uniform on the whole closed pin interval. Substitution into the linear
term gives Eq.~\eqref{eq:v10-time-response-chart}; the separate
division of the response amplitudes in
Appendix~\ref{app:asymmetric-entry-v2} fixes their vanishing factor
$p(1-p)$.

\subsubsection{Eventual exact global reflection}
\label{v3-app-exact-global-reflection}
The global bound has localized every competitive design. We first state the persistence argument that turns this localization into exact symmetry, then apply it to the six-site path.
\paragraph*{Persistence of a localized optimum.}
\label{app:v10-persistence}
The following argument will also be used for other supports and
introduction rules. Suppose a comparison confines all minimizers to a
fixed compact positive set in capped coordinates $a_i\le1$ and free
coordinates $x$. Let the normalized objective $F_t(a,x)$ converge uniformly with
two derivatives on that compact set to a function with a unique minimum at $(\mathbf1,x_0)$.
Assume its capped derivatives are strictly negative and its free Hessian
is positive definite there, and that increasing a capped coordinate to
one remains admissible locally. Then, for all sufficiently small positive
$t$, the global minimum is unique, lies on the face $a=\mathbf1$,
and is fixed by every symmetry preserving the objective and domain.
\begin{proof}
Compactness and uniform convergence place all exact minimizers near
$(\mathbf1,x_0)$. The derivative signs
then force $a_i=1$. On that face the positive Hessian gives strict local
convexity, and the implicit-function theorem continues the stationary
point to a unique minimum. Localization excludes any other global
minimum. Any symmetry preserving the objective and domain must fix this
minimum. Analytic data give an analytic branch. The argument is uniform
in auxiliary parameters on a compact set when its bounds and limiting
minima are uniform.
\end{proof}

The least-cap problem is the generalized inverse of the least-threshold
problem,
\begin{equation}
 H_{\min}(\epsilon)=\inf\{H:R_{\min}(H)\le R_*+\epsilon\}.
\end{equation}
This follows from the nested contrast domains. On an eventual analytic
branch with $R_{\min}(H)=R_*+c_{\rm br}H^{-1/m}+O(H^{-2/m})$, $c_{\rm br}>0$, the
controlled derivative is strictly negative for sufficiently large $H$.
The branch can therefore be inverted near its limiting threshold. In the
localized protected hierarchy the bridge is the smallest contact, so
increasing a capped core preserves the admissible normalized domain.

\paragraph*{Application to uniform introduction on $P_6$.}
At finite $H$ the normalized contrast domain is compact. A reflected trial
has $R<2$ for large $H$; the competitive sublevel therefore has $V\geq U/2$
and is closed away from the promoter boundary. A minimizer exists. With
$t=H^{-1/2}$ write $q_L=X/t,q_R=Y/t,Q_L=a/t^2,Q_R=e/t^2$. The positive global
bound confines every competitor to a compact positive set with $a,e\leq1$.
The uniform expansion is
\begin{equation}
 R=1+t\left[\frac3X+\frac3Y+
       \frac{53X+29Y}{6a}+\frac{29X+53Y}{6e}\right]+O(t^2).
\end{equation}
The leading objective has its unique minimum at
$a=e=1$, $X=Y=3/\sqrt{41}$. Its core derivatives are $-\sqrt{41}$ and
its pin Hessian is $(82\sqrt{41}/9)I_2$. These are the signs needed by
the persistence argument; it gives $Q_L=Q_R=H$ and $q_L=q_R$ exactly.
This fixes the unique global weak-threshold optimizer for all
sufficiently large $H$. As discussed after Eq.~\eqref{eq:p6-optimum}, the persistence argument leaves the starting cap of this regime implicit; smaller caps require a finite-$H$ comparison of the threshold beyond the localized corner.
The critical-branch coefficients in
Eq.~\eqref{eq:finite-budget-critical-v2} therefore describe optimization
over all positive weights for sufficiently small $\epsilon$.

\subsection{Asymmetric establishment and endpoint-uniform response}
\label{app:asymmetric-entry-v2}

The unrestricted bound in Appendix~\ref{v3-app-global-p6} places
near-threshold designs in a hierarchy of protected triples. Here we derive
their establishment, response amplitude and physical times uniformly in
the relative pin rates, using weak selection and uniform introduction.

\subsubsection{A uniformly controlled weak interface}

Normalize the consecutive weights to $(Q_L,q_L,1,q_R,Q_R)$ and define
\begin{equation}
 \begin{gathered}
 \varrho=q_L^{-1}+q_R^{-1},\qquad p=\frac{q_R}{q_L+q_R},\\
 \ell_L=q_L/Q_L,\qquad\ell_R=q_R/Q_R,\\
 a_L=\frac{\ell_L}{1-p}=\frac{q_L+q_R}{Q_L},\qquad
 a_R=\frac{\ell_R}{p}=\frac{q_L+q_R}{Q_R}.
 \end{gathered}
 \label{eq:asym-coordinates-v2}
\end{equation}
The variables $\ell_L,\ell_R$ are within-domain leakages;
$a_L,a_R$ compare each outer core with both pin contacts. They differ
from the reflection variables $u,v$ used earlier. We consider
$\varrho\to0$, $0\leq\ell_L,\ell_R\leq M$ for a fixed finite $M$,
and allow $p$ to approach either endpoint. The actual finite paths
have $0<p<1$. Their nonzero copying probabilities are
\begin{equation}
 \begin{gathered}
 P_{12}=P_{65}=1,\quad
 P_{21}=\frac1{1+\ell_L},\quad
 P_{23}=\frac{\ell_L}{1+\ell_L},\\
 P_{32}=\frac1{1+\varrho p},\quad
 P_{34}=\frac{\varrho p}{1+\varrho p},\\
 P_{43}=\frac{\varrho(1-p)}{1+\varrho(1-p)},\quad
 P_{45}=\frac1{1+\varrho(1-p)},\\
 P_{54}=\frac{\ell_R}{1+\ell_R},\quad
 P_{56}=\frac1{1+\ell_R}.
 \end{gathered}
 \label{eq:asym-kernel-v2}
\end{equation}
At $\varrho=0$ the two triples independently reach their own
consensus. The four aligned states, labeled $00,10,01,11$ by the left and
right domain traits, are the only closed classes.
Their 60-state complement has a uniformly bounded killed inverse
on the stated compact set. A finite sequence aligning each triple
with its interface site has a uniformly positive probability,
because the required copying directions have probabilities at least
$(1+M)^{-1}$ or one. Fecundities near neutrality preserve this
bound.

Split the generator into aligned-state blocks $S$ and transient
fast blocks $F$. Eliminating the fast unknowns gives the Schur
complement $\mathcal A_{SS}-\mathcal A_{SF}\mathcal A_{FF}^{-1}\mathcal A_{FS}$.
Every aligned row has a factor $\varrho$. Dividing this reduced
harmonic operator by that factor gives an analytic limit.
Define the neutral pin-establishment probability and mixed-state
escape coefficient by
\begin{equation}
 k(\ell)=\frac{1+\ell}{1+3\ell+\ell^2},\quad
 \nu_0=p k(\ell_R)+(1-p)k(\ell_L).
 \label{eq:asym-module-k-v2}
\end{equation}
In slow time $t_{\rm slow}=\varrho t$, the mixed state $10$ exits to $11$ and
$00$ at rates $p k(\ell_R)/6$ and $(1-p)k(\ell_L)/6$,
respectively; the rates are interchanged for $01$. Both mixed states
have total exit rate $\nu_0/6$. On the reflected path
$\varrho=2/q$, so $t_{\rm slow}=2t/q$ and the rate in time $t/q$ is
$\nu_0/3$, approaching $1/3$ in the protected limit. Since $k$ is bounded below on $[0,M]$,
the divided mixed-state inverse is uniform even at $p=0,1$.
Consequently the committor and its weak derivative extend analytically
in $(\varrho,\ell_L,\ell_R,p)$ to a neighborhood of this compact
set. In particular, writing $\mathcal S=b\chi_b-c\chi_c$,
the coefficients and the deficit
$\chi_c-\chi_b$ are bounded there.

\subsubsection{An exact three-site establishment calculation}

For a triple with relative edge weights $(1/\ell,1)$, label
the interface site 3. Let $j(\ell)$ be the singleton fixation susceptibility from site 3 divided by $k(\ell)$, and let $e(\ell)$ be the sum of the three singleton fixation susceptibilities. Here susceptibility means the derivative of the all-cooperator fixation probability with respect to $\delta$ at neutrality. Separate the payoff
parameters as $j=j_b b+j_c c$ and $e=e_b b+e_c c$. Solving the
six transient committor equations gives
 \begin{gather}
 j_b(\ell)=-\frac{\ell}{1+\ell}, \quad 
 j_c(\ell)=-\frac{\ell(3+2\ell)}{1+3\ell+\ell^2},  \label{eq:asym-module-response-v2}
\\
 e_b(\ell)=-\frac{3\ell}{1+3\ell+\ell^2},
 e_c(\ell)=-\frac{20\ell(1+\ell)^2}
                         {3(1+3\ell+\ell^2)^2}\,. \notag
 \end{gather}
The neutral committor is the inverse-strength
weighted occupation of this triple; each derivative solves
$\mathcal A_{0,\mathrm{mix}}h_1=-\mathcal A_1h_0$ with zero boundary values.
Here $\mathcal A_1=\partial_\delta\mathcal A|_0$ also equals the
first $\delta_c$ coefficient because
$\partial_\delta\delta_c|_0=1$.
These rational functions follow from the microscopic rates in
Eq.~\eqref{eq:rates}. All four
coefficients are strictly negative for $\ell>0$ and vanish at
$\ell=0$. The overall embedded clock changes residence times but
not these domain committors.

Put $h=p k(\ell_R)/\nu_0$. A rare copy across the central edge
must establish in the receiving domain before converting it.
The derivative of cooperative fixation from either mixed state is
$h(1-h)[b-c+j(\ell_L)+j(\ell_R)]$. Uniform introductions reach
the two mixed states with neutral probability $1/6$ each and derivatives
$e(\ell_L)/6$, $e(\ell_R)/6$. Hence, at $\varrho=0$,
\begin{equation}
 \begin{aligned}
 \chi_b^0={}&\frac{2h(1-h)}3[1+j_b(\ell_L)+j_b(\ell_R)]\\
 &+\frac{h e_b(\ell_L)+(1-h)e_b(\ell_R)}3,\\
 \chi_c^0={}&\frac{2h(1-h)}3[1-j_c(\ell_L)-j_c(\ell_R)]\\
 &-\frac{h e_c(\ell_L)+(1-h)e_c(\ell_R)}3.
 \end{aligned}
 \label{eq:asym-interface-response-v2}
\end{equation}
The actual coefficients differ by uniformly $O(\varrho)$. In the
protected coordinates of Eq.~\eqref{eq:asym-coordinates-v2}, $\ell_L=(1-p)a_L$ and
$\ell_R=pa_R$ make both responses vanish at the pin-dominance endpoints.
Dividing out their common entrance factor then controls the threshold
ratio uniformly.

\subsubsection{Uniform response, contrast and time consequences}
The all-weight inequality in Eq.~\eqref{v3-eq-p6-global-bound} forces the protected corner without an initial leakage assumption. In the present coordinates,
$F=3\varrho+(53-24p)a_L/6+(29+24p)a_R/6$.
For a promoter with $\epsilon=R-1<18/53$,
$F<\epsilon/(1-53\epsilon/18)$, uniformly in $0<p<1$.
Thus $\varrho,a_L,a_R=O(\epsilon)$, with
$\ell_L=(1-p)a_L\to0$ and $\ell_R=pa_R\to0$.
This recovers the hierarchy entry in Eq.~\eqref{eq:asymmetric-entry-v2}, now in the coordinates of Eq.~\eqref{eq:asym-coordinates-v2}.
The response amplitudes still require division by their vanishing entrance factor.

In the forced coordinates $(\varrho,a_L,a_R,p)$ the same bounded inverses apply. At either endpoint $p=0,1$ with $\varrho>0$, one pin cannot be overwritten and determines final consensus. Both response coefficients vanish there, and their analytic continuation retains this identity at $\varrho=0$. Analytic division by $p(1-p)$ before Taylor expansion
therefore preserves uniform errors. With
$m=\varrho+a_L+a_R$, the result is
 \begin{align}
 \chi_b&=\frac{2p(1-p)}3[1+O(m)],  \label{eq:asym-uniform-expansion-v2}
\\
 R-1&=3\varrho+\frac{53-24p}{6}a_L
                    +\frac{29+24p}{6}a_R+O(m^2). \notag
 \end{align}
The coefficients follow from the pair system
Eq.~\eqref{eq:pair-system}.  Expanding its matrix as
$\mathsf M(t)=\mathsf M_0+t\mathsf M_1+t^2\mathsf M_2+O(t^3)$ along
$(\varrho,a_L,a_R)=t(r,c_L,c_R)$, with fixed ray coefficients $r>0$ and $c_L,c_R\ge0$, the singular pair solution
starts with $2v_0/(tr)$, where $v_0$ is one on pairs crossing the
central edge and zero on within-domain pairs. The next coefficient
solves $\mathsf M_0d^{(0)}+\mathsf M_1(2v_0/r)=2\times\mathbf1$, with the next-order left-null
solvability condition fixing its remaining component.
Contracting these terms into $J,K$ gives
$3r+(53-24p)c_L/6+(29+24p)c_R/6$. The matrix coefficients and left null vector are determined directly
by Eq.~\eqref{eq:pair-system}; these two coefficient equations fix the
required contraction.

The global bound in Eq.~\eqref{v3-eq-p6-global-budget} already proves the leading contrast penalty of Eq.~\eqref{eq:asymmetric-cost-amplitude-v2}. Attaining the least coefficient $656$ requires $p\to1/2$, saturation of both outer weights and $\varrho\sim\epsilon/6$. A bounded $H\epsilon^2$ consequently keeps $p$ away from zero and one.

The physical neutral times can also be computed before the
leakages become small. Uniform introductions reach each mixed state
with mass $1/6$, and their common residence time is
$6/(\varrho \nu_0)$ to leading order. The fast-state inverse is uniformly bounded on the compact leakage
domain. After the slow factor $\varrho$ is removed, its divided
escape rate $\nu_0$ stays positive there, so $\varrho T$ is analytic
at $\varrho=0$. Taylor expansion of this reconstructed physical
time gives a uniformly bounded correction. Consequently
\begin{equation}
 T_{\rm abs}=\frac{2}{\varrho \nu_0}+O(1),\qquad
 T_{\rm fix}=\frac{6}{\varrho \nu_0}+O(1).
 \label{eq:asym-times-v2}
\end{equation}
Conditional normalization uses the exact neutral fixation
probability $1/6$. For each fixed $s>0$, the unconditional survival
probability obeys
$\Pr(\varrho\tau>s)=\frac13\exp(-\nu_0s/6)+o(1)$,
uniformly on the bounded-leakage domain. If $\nu_0$ converges, the
limiting distribution has mass $2/3$ at zero and mass $1/3$ in an
exponential distribution with the corresponding mean $6/\nu_0$.
In the protected corner,
$\nu_0=1+O(a_L+a_R)$, and the positive $3\varrho$ term in the gap
proves $\liminf\epsilon T_{\rm abs}\ge6$ and
$\liminf\epsilon T_{\rm fix}\ge18$. Reflected trials attain both time-only
bounds to leading order.

The construction assumes leakages at most $M$, which bounds the needed copying probabilities below by $(1+M)^{-1}$ and keeps $k(\ell)$ bounded below. An arbitrarily weak outer core, with $\ell\to\infty$, sends $k(\ell)$ to zero and can make the divided escape rate vanish. For promoters with $R-1<18/53$ the all-weight inequality~\eqref{v3-eq-p6-global-bound} forces $\ell_L,\ell_R=O(\epsilon)$, so the assumption holds before the local expansion is used.

\subsection{The global contrast bound on the four-cycle}
\label{app:resource-comparison}

We prove the four-cycle comparison in Eq.~\eqref{eq:c4-cost} and its sharp contrast cost in Table~\ref{tab:cost}, under uniform singleton introduction and the promoter condition \(J>0\). The prescribed tolerance \(\epsilon>0\) bounds the threshold gap of an admissible graph. An exact polynomial inequality controls all positive weights; balancing its leading positive terms identifies the asymptotically optimal geometry. Appendix~\ref{app:v10-c4-time} derives the associated waiting times with the physical attempt clock. The five-site path comparison follows in Appendix~\ref{app:III-p5-resource}.

\subsubsection{An all-weight polynomial identity for the four-cycle}

Normalize a smallest cyclic edge to one and write the weights as
\((1,B,C,D)\). Put \(S=B+D\) and \(T=BD\). The exact pair system
Eq.~\eqref{eq:pair-system}, equivalently its configuration-space Poisson
equation, gives
\begin{equation}
 R-3=\frac{4g(C,S,T)}{\beta(C,S,T)}.
 \label{eq:app-c4-ratio}
\end{equation}
Here \(g\) and \(\beta\) are integer polynomials of total degree 17
with 540 and 541 nonzero monomials, respectively. All coefficients of
\(g\) and the cost numerator \(3\beta+4g\) are positive. The exact
identity \(K\beta=J(3\beta+4g)\), with \(K>0\), therefore shows
that \(\beta\) has the sign of \(J\), including its zero locus.

For reconstruction, substitute the cycle kernel into the pair system,
contract with $\gamma$ and $\mathsf B$ as in Eq.~\eqref{eq:JK}, and reduce
$(K-3J)/J$ to the integer-polynomial fraction $4g/\beta$.

The useful polynomial identity concerns
\begin{equation}
 L(C,S,T)=4CTg-(8CS+6T^2)\beta.
 \label{eq:app-c4-remainder}
\end{equation}
After reflecting the cycle, there are three exhaustive order chambers.
In each, parameterize the largest, middle and smallest of \(B,C,D\)
by \(1+X+Y+Z,1+X+Y,1+X\), respectively, with \(X,Y,Z\ge0\).
Exact expansion determines the signs in Table~\ref{tab:c4-chamber-signs}.
\begin{table}[!htbp]
\caption{Exact coefficient signs in the three weight-order chambers of the four-cycle with contacts $(1,B,C,D)$. The benefit-sign polynomial $\beta$ is defined by Eq.~\eqref{eq:app-c4-ratio}, and $L$ is the comparison remainder in Eq.~\eqref{eq:app-c4-remainder}. The counts follow substitution of the three ordered weights by $1+X+Y+Z,1+X+Y,1+X$, with $X,Y,Z\ge0$.}
\label{tab:c4-chamber-signs}
\centering
\small
\vspace{0.5em}
\begin{tabular}{@{}l@{\hspace{1em}}l@{\hspace{1em}}l@{}}
\toprule
Chamber & Polynomial & Coefficients\\
\midrule
\(C\ge B\ge D\ge1\)&\(L(C,B+D,BD)\)&2952 positive\\
\(B\ge C\ge D\ge1\)&\(\beta(C,B+D,BD)\)&2041 negative\\
\(B\ge D\ge C\ge1\)&\(\beta(C,B+D,BD)\)&2301 negative\\
\bottomrule
\end{tabular}
\end{table}

The constant terms are \(60129542144\), \(-2147483648\), and
\(-2147483648\), respectively. Every sign is therefore strict even
on chamber boundaries. The last two chambers exclude adjacent extreme
edges from the promoter domain, including tied weightings that admit an adjacent-extrema labeling.
Every promoter consequently has an opposite-extrema representation
with \(C\ge B,D\ge1\). Dividing the first-chamber inequality \(L>0\) by
\(CT\beta>0\) proves
\begin{equation}
 R-3>8(B^{-1}+D^{-1})+6BD/C.
 \label{eq:app-c4-global-bound}
\end{equation}
In particular \(R>3\) for every finite positive weighting with \(J>0\).

With \(p=\sqrt{BD}\), the right side is at least
\(16/p+6p^2/C\). Its minimum occurs at \(p^3=4C/3\), giving
the lower bound in Eq.~\eqref{eq:c4-cost} because \(C\le H\).
The reflection family \((1,q,H,q)\) attains the matching upper
construction. In the independent small variables \(u=q^{-1}\) and
\(v_4=q^2/H\), its exact threshold has the regular expansion
\begin{equation}
 R=3+16u+6v_4+64u^2+99uv_4+6v_4^2+O((u+v_4)^3).
 \label{eq:app-c4-expansion}
\end{equation}
Taking \(q=(4H/3)^{1/3}\) gives
\begin{equation}
 R_{\min}^{C_4}(H)=3+12\sqrt[3]6\,H^{-1/3}+O(H^{-2/3}).
 \label{eq:c4-optimum}
\end{equation}
The necessary contrast \(H>10368\epsilon^{-3}\) is strict for
every graph satisfying \(R-3\le\epsilon\), and its coefficient is
asymptotically attainable.

The global inequality also prevents a competing optimum from escaping
into another weight hierarchy. Write \(t=H^{-1/3}\) and
\((B,C,D)=(x/t,h/t^3,y/t)\). For a fixed $C_{\rm loc}>0$, if \(R-3\le C_{\rm loc}t\), then
\begin{equation}
 \frac8{C_{\rm loc}}\le x,y\le\frac{C_{\rm loc}^2}{48},\qquad
 \frac{384}{C_{\rm loc}^3}\le h\le1.
 \label{eq:app-c4-localization}
\end{equation}
The exact scaled objective converges with derivatives on this compact
positive set to \(8/x+8/y+6xy/h\). Its unique minimum is
\(h=1\), \(x=y=(4/3)^{1/3}\). Its Hessian in \(x,y\) is
\(\left(\begin{smallmatrix}12&6\\6&12\end{smallmatrix}\right)\).
The capped derivative in $h$ is negative. The persistence argument of
Appendix~\ref{app:v10-persistence} therefore gives uniqueness
up to scale and cycle symmetry for all sufficiently large $H$, with
$B=D$ exactly.

\subsection{The global contrast bound on the five-site path}
\label{app:III-p5-resource}

We prove the five-site path bound~\eqref{eq:p5-cost} and its sharp contrast cost in Table~\ref{tab:cost}. Reversing and renormalizing \((A,1,q,Q)\) gives
\((Q/q,1,1/q,A/q)\). We may therefore impose \(q\ge1\)
without losing any positive weighting or changing the uniform-introduction
observables. Introduce
 \begin{gather}
 u=q^{-1}\in(0,1],\qquad e=q/A>0,\qquad w=A/Q>0, \notag \\
 uew=Q^{-1}\ge H^{-1}.  \label{eq:app-p5-coordinates}
\end{gather}
The letter \(e\) here is a dimensionless ratio, not an edge label.
For these contacts the nonzero entries of the five-site kernel are
 \begin{gather}
 P_{12}=P_{54}=1,\qquad
 P_{21}=\frac{A}{A+1},\qquad P_{23}=\frac1{A+1},\notag \\
 P_{32}=\frac1{q+1},\qquad P_{34}=\frac q{q+1}, \label{b:eq:p5-kernel} \\
 P_{43}=\frac q{Q+q},\qquad P_{45}=\frac Q{Q+q}. \notag
 \end{gather}
Use this kernel in the ten-pair system~\eqref{b:eq:pairs}
with source $2\times \mathbf1$, and form the contractions
in Eq.~\eqref{b:eq:response}. After substituting
$A=(ue)^{-1}$, $q=u^{-1}$ and $Q=(uew)^{-1}$, reduce
their ratio to coprime polynomials and choose the common constant
so that
\begin{equation}
 R=\frac{\mathscr P(u,e,w)}{\mathscr Q(u,e,w)},\qquad
 \mathscr Q(0,0,0)=4356.
 \label{b:eq:p5-rational}
\end{equation}
Exact expansion gives nonnegative coefficients for the nonzero polynomial $\mathscr P$, so $\mathscr P>0$ on positive contacts.
Consequently $\mathscr Q$ is positive precisely on the promoter
domain. The quantitative identity below sharpens that support result.

At the origin,
\begin{equation}
 R=7+\tfrac{83}{2}u+16e+\tfrac{83}{3}w
       +O((u+e+w)^2).
 \label{eq:app-p5-local}
\end{equation}
For \(\mathscr E=\mathscr P-
[7+(83/2)u+16e+(83/3)w]\mathscr Q\), exact expansion gives
\begin{equation}
\begin{aligned}
 \mathscr E={}&\mathscr E_+
 +16434u^7ew(1-u)+180774u^6w(1-u)\\
 &+120516u^4w^2(1-u^2)+85822u^4w^2(1-u).
\end{aligned}
 \label{eq:app-p5-domination}
 \end{equation}
Every nonzero coefficient of \(\mathscr E_+\) is positive, including
the coefficient \(361548\) of \(w^2\). This proves strict positivity
on the complete oriented domain, even at \(u=1\). Without orientation,
the four negative monomials in Eq.~\eqref{eq:app-p5-domination}
would prevent a coefficientwise positivity argument. Division by \(\mathscr Q>0\)
proves Eq.~\eqref{eq:p5-cost}. AM--GM applied to its three
terms gives \((R-7)^3>496008/Q\ge496008/H\).

Write \(a_5=83/2\), \(b_5=16\), \(c_5=83/3\), and
\(\ell=(a_5b_5c_5)^{1/3}\). With
\(Q=H=t^{-3}\), \(q=x/t\), \(A=\alpha/t^2\), the leading
gap is \(t[a_5/x+b_5x/\alpha+c_5\alpha]\).
Its three terms are equal at
\(x=a_5/\ell\), \(\alpha=\ell/c_5\), giving
\begin{equation}
 R_{\min}^{P_5}(H)=7+\sqrt[3]{496008}\,H^{-1/3}
                         +O(H^{-2/3}).
 \label{eq:app-p5-optimum}
\end{equation}
This upper construction matches the all-weight lower bound. The latter
also gives \(u,e,w=O(t)\) for every competitive graph; their product
is at least \(t^3\), so each is bounded below by a positive multiple
of \(t\). Thus the expansion is uniform over a single compact positive
scaling sector, and \(1<q<A<Q\) eventually. This ordering follows from
the global bound after reversal to \(q\ge1\).

\subsection{All-weight bounds and the contrast cost on the five-cycle}
\label{app:c5-global}
The hierarchy input for the five-cycle has two pins, sites 1 and 2,
and a three-site fringe between them, whose sites form a chain. With equal pin contacts,
write
\begin{equation}
 \begin{gathered}
 (w_{12},w_{23},w_{34},w_{45},w_{51})=(1,q,Q_1,Q_2,q),\\
 u=q^{-1},\qquad y_{\rm cyc}=\frac{q^2}{Q_1+Q_2},\\
 \sigma=\frac{Q_2}{Q_1+Q_2},\qquad
 \xi_{\rm cyc}=\frac{Q_2}{Q_1}=\frac{\sigma}{1-\sigma}.
 \end{gathered}
 \label{b:eq:c5-hierarchy}
\end{equation}
The controlled class reduction for response and physical time cited in Sec.~\ref{b:sec:domain-input}
uses $\beta=1-\sigma$ and $y=y_{\rm cyc}$ in its notation. It is regular as $u,y_{\rm cyc}\to0$ with $\sigma$ in a compact subset
of $(0,1)$. Its uniform and birth-associated thresholds have
the local forms

 \begin{align}
 R^{\rm U}&=\frac32+[6+\sigma(1-\sigma)]u+\frac{11}{4}y_{\rm cyc}
             +O((u+y_{\rm cyc})^2), \notag  \\
 R^{\mathrm b}&=\frac32+[6+\sigma(1-\sigma)]u
             +c^{\mathrm b}(\xi_{\rm cyc})y_{\rm cyc}  \label{b:eq:c5-first-response}\\
             &\quad+O((u+y_{\rm cyc})^2), \notag
 \end{align}

where the needed birth coefficient is
\begin{equation}
\begin{gathered}
 c^{\mathrm b}(\xi_{\rm cyc})=
 \frac{\Pi_6(\xi_{\rm cyc})}
 {2\xi_{\rm cyc}(1+\xi_{\rm cyc})^2(32\xi_{\rm cyc}^2+69\xi_{\rm cyc}+32)},\\
 \begin{aligned}
 \Pi_6(\xi_{\rm cyc})&=186(1+\xi_{\rm cyc}^6)+2895(\xi_{\rm cyc}+\xi_{\rm cyc}^5)\\
 &\quad+10453(\xi_{\rm cyc}^2+\xi_{\rm cyc}^4)+15548\xi_{\rm cyc}^3.
 \end{aligned}
\end{gathered}
\label{b:eq:c5-birth-coefficient}
\end{equation}
In particular, $c^{\mathrm b}(1)=761/19$.
The coefficients quantify finite pin protection and leakage
through the chain. The global argument below determines when
these local losses bound every competitive design.

Write the cycle weights as $(w_{12},w_{23},w_{34},w_{45},w_{51})$. The
ten-pair solution gives $R=U/V$ in lowest terms, with $U$ and $V$
homogeneous of degree $39$ and $25765$ monomials each, invariant under the
dihedral symmetries of the cycle. Every coefficient of $U$ is positive, so
$V$ has the sign of $J$, including its zero locus. The polynomial $2U-3V$
has $25720$ monomials, all with positive coefficients (the smallest is
$564$). On $V>0$ this gives $R>3/2$ for all positive weights, without an
ordering of the weights. The birth-associated numerators behave in the
same way. $U^{\mathrm b}$ is positive, and $2U^{\mathrm b}-3V^{\mathrm b}$
has $25590$ positive coefficients (the smallest is $372$). A second exact
elimination, by a Schur complement on the five adjacent pairs and a
fraction-free Cramer solution of the remaining $5\times5$ system,
reproduces $K/J=U/V$ and $K^{\mathrm b}/J^{\mathrm b}=U^{\mathrm b}/V^{\mathrm b}$
as polynomial identities.

For Eq.~\eqref{eq:c5-global}, rotate a smallest weight to $w_{12}=1$ and
divide the orderings of the other four weights into twelve chambers modulo
reflection, parameterized as in Table~\ref{tab:c4-chamber-signs} by
cumulative nonnegative increments. In four chambers $V$ has only negative
coefficients, so they contain no promoters. In these four and in four
further chambers the numerator of $R-3/2-L$, with $L$ the right side of
Eq.~\eqref{eq:c5-global}, has only positive coefficients. In the remaining
four chambers that comparison fails, because the chain term of $L$ is
unbounded far from the hierarchy; there $U-3V$ has only positive
coefficients, so every promoter in them has $R>3$. With $w_{12}=1$ the
contrast satisfies $w_{34}+w_{45}\le H/\max(\sigma,1-\sigma)$, and weighted
AM--GM applied to the three terms of $L$ gives
\begin{equation}
 L^3\ge\frac{27}{H}\,a(\sigma)\,a(1-\sigma)\,\frac{11}{4}\,
 \max(\sigma,1-\sigma)\ge\frac{27}{H}\,\frac{6875}{512}.
\end{equation}
The product $a(\sigma)a(1-\sigma)\max(\sigma,1-\sigma)$ is minimal at $\sigma=1/2$, as Eq.~\eqref{eq:III-c5-chain-positive} below shows, and $a(1/2)=25/8$. Hence
$H>(185625/512)(R-3/2)^{-3}$ for every promoter with $R\le3$, which proves
the uniform-introduction five-cycle contrast bound in Proposition~\ref{prop:v10-costs}.

On the reflected family $(1,q,Q,Q,q)$, with $u=q^{-1}$ and $v_{\rm cyc}=q^2/Q=2y_{\rm cyc}$, the
exact expansions are
\begin{equation}
 \begin{aligned}
 R={}&\tfrac32+\tfrac{25}{4}u+\tfrac{11}{8}v_{\rm cyc}+\tfrac{125}{8}u^2\\
 &+\tfrac{2269}{76}uv_{\rm cyc}+\tfrac{11}{32}v_{\rm cyc}^2+O((u+v_{\rm cyc})^3),\\
 R^{\mathrm b}={}&\tfrac32+\tfrac{25}{4}u+\tfrac{761}{38}v_{\rm cyc}
 +\tfrac{125}{8}u^2+\tfrac{857725}{5776}uv_{\rm cyc}\\
 &+\tfrac{376621}{5776}v_{\rm cyc}^2+O((u+v_{\rm cyc})^3).
 \end{aligned}
 \label{eq:c5-reflected-expansions}
\end{equation}
Balancing the two linear terms at $Q=H$ gives Eq.~\eqref{eq:c5-cost} and
the birth-associated constant $27\,(25/8)^2\,(761/38)=12841875/2432$.

Birth-associated introduction admits a global bound of the same type once
all five choices of the bridge are combined. Grade the monomials of
$U^{\mathrm b}$ and $V^{\mathrm b}$ by the weights $(0,1,3,3,1)$ of
$(w_{12},w_{23},w_{34},w_{45},w_{51})$, which follow the hierarchy of
Eq.~\eqref{b:eq:c5-hierarchy}. The top graded part of $V^{\mathrm b}$ is
 \begin{align}
 W_0={}&6\,w_{12}^2(w_{23}w_{51})^6(w_{34}w_{45})^8(w_{23}+w_{51})\\
 &\times(w_{34}+w_{45})^6(32w_{34}^2+69w_{34}w_{45}+32w_{45}^2), \notag
 \end{align}
the top part of $U^{\mathrm b}$ is $\tfrac32W_0$, and the top part of
$\mathcal C_5^{\mathrm b}=2U^{\mathrm b}-3V^{\mathrm b}$ is $2L^{\mathrm b}W_0$, with
\begin{equation}
 \begin{aligned}
 L^{\mathrm b}={}&a^{\mathrm b}(\sigma)\frac{w_{12}}{w_{23}}
 +a^{\mathrm b}(1-\sigma)\frac{w_{12}}{w_{51}}\\
 &+\sigma(1-\sigma)\,c^{\mathrm b}\,
 \frac{w_{23}w_{51}(w_{34}+w_{45})}{w_{12}w_{34}w_{45}},
 \end{aligned}
 \label{eq:c5-birth-functional}
\end{equation}
where $a^{\mathrm b}(t)=(5+3t-t^2)/2$ and $c^{\mathrm b}$ is
Eq.~\eqref{b:eq:c5-birth-coefficient} at $\xi_{\rm cyc}=\sigma/(1-\sigma)$. Let a
superscript $(r)$ denote the relabeling $i\mapsto i+r$ of the sites. Among
the monomials of $V^{\mathrm b}$, exactly $90$ have a vanishing coefficient
in $\mathcal C_5^{\mathrm b}$. They form the supports of the five rotations
$W_0^{(r)}$, which are disjoint because the exponent pattern of each $W_0^{(r)}$ identifies its bridge; every edge can serve as the
bridge. With $W_{\rm all}=\sum_{r=0}^4W_0^{(r)}$,
both
\begin{equation}
 W_{\rm all}+\tfrac{113}{120}\,\mathcal C_5^{\mathrm b}-V^{\mathrm b},\qquad
 \mathcal C_5^{\mathrm b}-2\sum_{r=0}^{4}\bigl(L^{\mathrm b}W_0\bigr)^{(r)}
 \label{eq:c5-birth-positivity}
\end{equation}
have only nonnegative coefficients, $25580$ and $25355$ of them nonzero,
and $113/120$ is the smallest constant for which the first statement
holds. Let $\underline L^{\mathrm b}$ be the least of the five values
$L^{\mathrm b(r)}$. On $V^{\mathrm b}>0$ the second polynomial gives
$\mathcal C_5^{\mathrm b}\ge2\underline L^{\mathrm b}W_{\rm all}$, the first then
gives $V^{\mathrm b}\le\mathcal C_5^{\mathrm b}[1/(2\underline L^{\mathrm b})+113/120]$,
and
\begin{equation}
 R^{\mathrm b}-\frac32>\frac{\underline L^{\mathrm b}}
 {1+113\,\underline L^{\mathrm b}/60}
 \label{eq:c5-birth-global}
\end{equation}
for every birth-associated promoter, without a chamber division. For any
bridge the opposite chain weights obey
$w_{34}w_{45}/(w_{34}+w_{45})\le H\,w_{12}\min(\sigma,1-\sigma)$, and AM--GM
applied to the three terms of each $L^{\mathrm b(r)}$ gives
$(\underline L^{\mathrm b})^3\ge27\,\Theta^{\mathrm b}/H$, where
$\Theta^{\mathrm b}=\min_{0<\sigma<1}\Theta_{\rm b}(\sigma)$, with
$\Theta_{\rm b}(\sigma)=a^{\mathrm b}(\sigma)a^{\mathrm b}(1-\sigma)\max(\sigma,1-\sigma)\,c^{\mathrm b}$.
The positive derivative comparison below puts this minimum uniquely at
$\sigma=1/2$, so $\Theta^{\mathrm b}=(25/8)^2(1/2)(761/19)=475625/2432$. Every
birth-associated promoter with $R^{\mathrm b}=3/2+\epsilon$ and
$\epsilon<60/113$ therefore has
\begin{equation}
 H>\frac{12841875}{2432}\left(\frac{1-113\epsilon/60}{\epsilon}\right)^3 .
 \label{eq:c5-birth-cost}
\end{equation}
\begin{table}[!htbp]
\caption{Numerator coefficients of the derivative of the birth-associated five-cycle chain factor $\Theta_{\rm b}$ with respect to $\xi_{\rm cyc}$, in powers $z^k$ with $z=\xi_{\rm cyc}-1$. The derivative denominator is $8(\xi_{\rm cyc}+1)^8(32\xi_{\rm cyc}^2+69\xi_{\rm cyc}+32)^2$. Positivity of every coefficient makes this derivative strictly positive for $z\ge0$.}
\label{tab:III-c5-chain-numerator}
\centering\small
\begin{ruledtabular}
\begin{tabular}{cr}
Power $k$ & Coefficient of $z^k$\\\hline
$0$ & $3542455000$\\
$1$ & $18269034252$\\
$2$ & $42709017323$\\
$3$ & $60009556894$\\
$4$ & $56685702169$\\
$5$ & $38162173398$\\
$6$ & $18929820574$\\
$7$ & $7035654790$\\
$8$ & $1960687409$\\
$9$ & $401358706$\\
$10$ & $57215507$\\
$11$ & $5064780$\\
$12$ & $208320$
\end{tabular}
\end{ruledtabular}
\end{table}

The reflected family with $Q=(3044/475)\,q^3$ gives
$R^{\mathrm b}=3/2+(75/8)q^{-1}+(113758442875/2816844544)q^{-2}+O(q^{-3})$,
so the least birth-associated contrast is
$(12841875/2432)\,\epsilon^{-3}[1+O(\epsilon)]$. Two
separately written programs checked both positivity statements of
Eq.~\eqref{eq:c5-birth-positivity} over all monomials.

The minimum over chain shape needs no root isolation. By reflection take
$\xi_{\rm cyc}=\sigma/(1-\sigma)\ge1$ and put $z=\xi_{\rm cyc}-1$.
For the uniform chain factor
$\Theta_{\rm U}=(11/4)a(\sigma)a(1-\sigma)\max(\sigma,1-\sigma)$,
\begin{equation}
\begin{gathered}
 \frac{d\Theta_{\rm U}}{d\xi_{\rm cyc}}
 =\frac{P_{\rm U}(z)}{256(\xi_{\rm cyc}+1)^6}>0,\\
 P_{\rm U}(z)=110000+211024z+144804z^2\\
 \hspace{14mm}+41624z^3+4169z^4.
\end{gathered}
 \label{eq:III-c5-chain-positive}
\end{equation}
The birth-associated derivative has positive denominator
$8(\xi_{\rm cyc}+1)^8(32\xi_{\rm cyc}^2+69\xi_{\rm cyc}+32)^2$.
Table~\ref{tab:III-c5-chain-numerator} lists its numerator coefficients by power of $z$.

The positive coefficients in Eq.~\eqref{eq:III-c5-chain-positive} and Table~\ref{tab:III-c5-chain-numerator} prove strict increase away from the balanced chain for both rules, including the endpoint directions.

\subsubsection{Exact five-cycle resource geometry near the floor}
\label{app:III-c5-exact-resource}
For either introduction rule and all sufficiently large contrast caps,
the exact global threshold minimum is unique after a smallest edge is
chosen as the unit bridge. Its weights are $(1,q,H,H,q)$, with $q$ on
an analytic branch. The corresponding small-gap corrections are
Eq.~\eqref{eq:III-c5-resource-corrections}.
\begin{proof}
Let $\mu\in\{\mathrm U,{\rm b}\}$ name the introduction rule, and set
$\mathfrak b_{\rm U}=11/8$, $\mathfrak b_{\rm b}=761/38$.
With $t=H^{-1/3}$, write the positive weight chart as
\begin{equation}
 (w_{12},w_{23},w_{34},w_{45},w_{51})
 =(1,X/t,a_c/t^3,e_c/t^3,Y/t).
 \label{eq:III-c5-resource-chart}
\end{equation}
Here $a_c,e_c$ are fractions of the cap, and $X,Y$ are scaled pin
contacts. On compact positive sets the exact ratio is
$R^{\mu}=3/2+tF_{\mu}+O(t^2)$, where
\begin{equation}
\begin{aligned}
 F_{\rm U}&=\frac{a(\sigma)}X+\frac{a(1-\sigma)}Y
             +\frac{11XY}{4(a_c+e_c)},\\
 F_{\rm b}&=\frac{a^{\rm b}(\sigma)}X
       +\frac{a^{\rm b}(1-\sigma)}Y
       +\frac{c^{\rm b}(e_c/a_c)XY}{a_c+e_c},\\
 \sigma&=\frac{e_c}{a_c+e_c}.
\end{aligned}
\label{eq:III-c5-resource-objectives}
\end{equation}
This expansion controls derivatives as well as values. After the
respective factors $t^{-86}$ and $t^{-85}$ are removed from the exact
polynomial denominators, their nonzero limits are
\begin{equation}
\begin{aligned}
 V_{\mathrm U,0}&=12X^7Y^7a_c^8e_c^8(a_c+e_c)^6\mathcal P_c,\\
 V_{{\rm b},0}&=6X^6Y^6a_c^8e_c^8(X+Y)(a_c+e_c)^6\mathcal P_c,\\
 \mathcal P_c&=32a_c^2+69a_ce_c+32e_c^2.
\end{aligned}
\end{equation}
The top numerator is $3V_{\mu,0}/2$ and the next divided difference is
$F_{\mu}V_{\mu,0}$. These polynomial identities prove analyticity at $t=0$.

Put $d_{\mu}=[25/(8\mathfrak b_{\mu})]^{1/3}$ and
$C_{\mu}=3[(25/8)^2\mathfrak b_{\mu}]^{1/3}$. The reflected trial has gap
$C_{\mu}t+O(t^2)$. For uniform introduction, rotating an actual minimum
edge to the bridge gives
$R-3/2>L_{\rm U}\ge C_{\rm U}t$. Every minimizer must consequently approach
equality in the chain-shape and three-term comparisons above.
It follows that $(X,Y,a_c,e_c)\to(d_{\rm U},d_{\rm U},1,1)$.
For birth-associated introduction, choose the bridge that minimizes
$L^{\rm b(r)}$. Equation~\eqref{eq:c5-birth-global} and the same trial give
$C_{\rm b}t\le\underline L^{\rm b}\le C_{\rm b}t+O(t^2)$.
The cap bounds every edge ratio by $H$, even before the chosen bridge
is known to be smallest. Equality in the chain comparison therefore
forces $(X,Y,a_c,e_c)\to(d_{\rm b},d_{\rm b},1,1)$.
All four other contacts then exceed the normalized bridge, which is
indeed the minimum. The five possible rotations exhaust every edge
placement. Thus every global minimizer lies in one compact positive
neighborhood of the corresponding balance. Existence follows from
the normalized closed cap cube and divergence of $K/J$ as $J\downarrow0$.

At the balance, each core derivative is
$\partial_{a_c}F_{\mu}=\partial_{e_c}F_{\mu}=-\mathfrak b_{\mu}d_{\mu}^2/2<0$.
These signs persist for the exact scaled response, forcing
$a_c=e_c=1$. On that face,
\begin{equation}
 \begin{gathered}
 F_{\mu}=\frac{25}{8}\left(\frac1X+\frac1Y\right)+\mathfrak b_{\mu}XY,\\
 D^2F_{\mu}\big|_{X=Y=d_{\mu}}
 =\mathfrak b_{\mu}\begin{pmatrix}2&1\\1&2\end{pmatrix}.
 \end{gathered}
\end{equation}
Its eigenvalues are $\mathfrak b_{\mu}$ and $3\mathfrak b_{\mu}$.
Together with the compact localization, analytic denominator and negative
core derivatives above, these data satisfy the persistence conditions of
Appendix~\ref{app:v10-persistence}. The exact global minimum therefore
saturates both core contacts and follows one analytic pin branch.
Reflection fixes that unique branch at $X=Y$ for both introduction rules.

For the finite-gap correction set $q=k/\epsilon$, $H=L/\epsilon^3$ in
Eq.~\eqref{eq:c5-reflected-expansions}, so that
$R^{\mu}-3/2=\epsilon f_{\mu}(k,L)+\epsilon^2g_{\mu}(k,L)+O(\epsilon^3)$, with
\begin{equation}
\begin{aligned}
 f_{\mu}&=\frac{25}{4k}+\mathfrak b_{\mu}\frac{k^2}{L},\\
 g_{\mu}&=\frac{125}{8k^2}+\mathfrak c_{\mu}\frac{k}{L}
                         +\mathfrak d_{\mu}\frac{k^4}{L^2},\\
 (\mathfrak c_{\rm U},\mathfrak d_{\rm U})&=(2269/76,11/32),\\
 (\mathfrak c_{\rm b},\mathfrak d_{\rm b})&=(857725/5776,376621/5776).
\end{aligned}
\end{equation}
The leading equations $f_{\mu}=1$, $\partial_kf_{\mu}=0$ give $k_0=75/8$ and
$L_{\mu,0}=27(25/4)^2\mathfrak b_{\mu}/4$. At the next order,
$f_LL_1+g_{\mu}=0$ and $f_{kk}k_1+f_{kL}L_1+\partial_kg_{\mu}=0$.
The nonsingular Jacobian fixes
\begin{equation}
\begin{aligned}
 H_{\min}^{C_5,{\rm U}}
 &=\frac{185625}{512\epsilon^3}
   +\frac{5131575}{4864\epsilon^2}+O(\epsilon^{-1}),\\
 H_{\min}^{C_5,{\rm b}}
 &=\frac{12841875}{2432\epsilon^3}
   +\frac{1023825985875}{140657152\epsilon^2}+O(\epsilon^{-1}),\\
 q_*^{\rm U}&=\frac{75}{8\epsilon}+\frac{16961}{2508}+O(\epsilon),\\
 q_*^{\rm b}&=\frac{75}{8\epsilon}
               +\frac{4647618155}{1056316704}+O(\epsilon).
\end{aligned}
\label{eq:III-c5-resource-corrections}
\end{equation}
The optimal threshold decreases strictly with $H$ in this domain,
justifying inversion from a fixed cap to a fixed gap.
\end{proof}

\paragraph*{Neutral time at the resource optimum.}

Retain the normalization $w_{12}=1$. The neutral time at least contrast is forced asymptotically even
without imposing reflection. For any sequence with $R\le3/2+\epsilon$
and $H=(185625/512)\epsilon^{-3}[1+o(1)]$, every step of the
three-term bound above must approach equality. The unique minimum in
$\sigma$ forces $\sigma\to1/2$ and $w_{34}/H,w_{45}/H\to1$.
Equality in AM--GM then gives $w_{23},w_{51}\sim75/(8\epsilon)$.
Choose any fixed $\eta_+>185625/512$. Every such least-resource sequence eventually has contrast at most $\eta_+\epsilon^{-3}$, so the uniform estimates of Eq.~\eqref{eq:III-c5-uniform-times} apply with bounded $O_{\eta_+}(1)$ errors. That proof retains the physical attempt clock and is uniform even as a chain ratio approaches an endpoint. The pin weights forced above consequently give
\begin{equation}
 T_{\rm abs}\sim\frac{2}{w_{23}^{-1}+w_{51}^{-1}}
       \sim\frac{75}{8\epsilon},\quad
 T_{\rm fix}\sim\frac{5}{w_{23}^{-1}+w_{51}^{-1}}.
 \label{b:eq:c5-balanced-time}
\end{equation}
Thus $T_{\rm fix}/T_{\rm abs}\to5/2$ on these least-resource
sequences. The factor $5/2$ comes from the uniform entrance mass $2/5$ into the contested pin states and the common slow residence time in the ancestral reduction.

For uniform introduction at neutrality, every weak promoter on the five-cycle with
$3/2<R\le3$ obeys the time bound~\eqref{eq:c5-time-only}.
Its leading coefficient is asymptotically sharp as $R\downarrow3/2$.
Without a contrast cap, unequal pin contacts and an unbalanced chain approach
this coefficient below the uncapped reflected value $25/(4\epsilon)$.
At neutrality a copy into site $i$ occurs at rate $\vartheta_i/N$ on the
attempt clock, independently of the configuration. A singleton introduced
at pin $1$ is absorbed only after a copy into pin $1$ (extinction) or into
pin $2$ (fixation), and symmetrically for pin $2$. Uniform introduction
therefore gives, for every weighting,
\begin{equation}
 T_{\rm abs}\ge\frac2N\cdot\frac{N}{\vartheta_1+\vartheta_2}=\frac{2}{\vartheta_1+\vartheta_2}.
 \label{eq:c5-time-pins}
\end{equation}
For the five-cycle, write consecutive contacts $(A,B,C,D,E)$ with $A$
smallest, and let sites 1 and 2 be its endpoints. Put
\begin{equation}
\begin{gathered}
\Pi=(A+B)(D+E)(A+E)(B+C),\\
N_T=\Pi\left[\frac A{A+B}+\frac E{D+E}
+\frac A{A+E}+\frac B{B+C}\right].
\end{gathered}
\end{equation}
Thus $\vartheta_1+\vartheta_2=N_T/\Pi$ exactly.
With the uniform polynomial pair of Eq.~\eqref{b:eq:c5-polynomials},
define $\mathcal C_5=2U-3V$ and
\begin{equation}
Q_T=4\mathcal C_5\Pi-23VN_T.
\end{equation}
The four order chambers relevant to $3/2<R\le3$, after reflection, are
\begin{equation}
\begin{gathered}
B\le C\le E\le D,\qquad B\le E\le C\le D,\\
B\le E\le D\le C,\qquad C\le B\le E\le D.
\end{gathered}
\end{equation}
Set $A=1$ and substitute ordered contacts
$1+x_1$, $1+x_1+x_2$, $1+x_1+x_2+x_3$,
$1+x_1+x_2+x_3+x_4$. The resulting $Q_T$ polynomials have respectively
$122052$, $121737$, $121736$, $136624$ strictly positive
coefficients. The other chambers have either $V<0$ or $U-3V>0$ by the
chamber signs proved for Eq.~\eqref{eq:c5-global}.
Thus every positive cycle in this threshold range obeys
\begin{equation}
\epsilon_5=R-\tfrac32>\tfrac{23}{8}(\vartheta_1+\vartheta_2).
\label{eq:v10-cycle-temperature-time}
\end{equation}

Combining this inequality with Eq.~\eqref{eq:c5-time-pins} gives
\begin{equation}
 T_{\rm abs}>\frac{23}{4\,(R-3/2)} .
 \label{eq:c5-time-only}
\end{equation}
The family $(w_{12},w_{23},w_{34},w_{45},w_{51})=(1,\varepsilon_{\rm cyc}^{-1},\varepsilon_{\rm cyc}^{-6},\varepsilon_{\rm cyc}^{-4},\varepsilon_{\rm cyc}^{-2})$
approaches this bound. Exact solution of the chain of contiguous mutant arcs on the cycle
and of the ancestral-pair system
over $\mathbb Q(\varepsilon_{\rm cyc})$ gives $R-3/2=\tfrac{23}{8}\varepsilon_{\rm cyc}+\tfrac{587}{64}\varepsilon_{\rm cyc}^2+O(\varepsilon_{\rm cyc}^3)$,
$T_{\rm abs}=2\varepsilon_{\rm cyc}^{-1}+\tfrac{593}{96}+O(\varepsilon_{\rm cyc})$, $T_{\rm fix}=5\varepsilon_{\rm cyc}^{-1}+\tfrac{25}{3}+O(\varepsilon_{\rm cyc})$, and
$(R-3/2)\,T_{\rm abs}=\tfrac{23}{4}+O(\varepsilon_{\rm cyc})$. The time-only infimum is therefore
$\tfrac{23}{4}\epsilon^{-1}[1+O(\epsilon)]$, with $T_{\rm fix}/T_{\rm abs}\to5/2$. It is
approached with $w_{51}\gg w_{23}$ and $w_{45}\ll w_{34}$, where the pin
coefficient $a(\sigma)$ of Eq.~\eqref{eq:c5-global} reaches its least value
$a(0)=23/8$.

\subsection{The five-site path at an arbitrary feasible contrast}
\label{app:v10-p5-allcap}
\label{app:v10-p5-all-cap}
This subsection concerns the uniform-introduction weak threshold on $P_5$,
with no large-contrast approximation. Reverse the path if necessary
and orient its contacts as $(A,1,q,Q)$ with $q\ge1$.
For every $H>H_{0,P_5}^{\rm U}$, the global optimizer is unique in this
oriented chart, saturates its strongest contact and follows a
real-analytic branch with strictly decreasing threshold. The promoting
region is connected in this chart. Its unique first onset and the
threshold divergence are those of Proposition~\ref{prop:v10-onsets}
and Eq.~\eqref{eq:v10-p5-onset-divergence}; the rational enclosure is
Eq.~\eqref{eq:III-p5-onset-bracket}.
\begin{proof}
Using Eq.~\eqref{b:eq:p5-kernel} and its ten-pair matrix,
define
 \begin{gather}
 T_5=(A+1)(q+1)(Q+q),\qquad
 \mathscr D_5=T_5^8\det\mathsf M/q,\notag \\
 U=\mathscr D_5K,\qquad V=\mathscr D_5J.  \label{eq:v10-p5-normalization}
\end{gather}
These polynomial representatives are kept unreduced throughout this
subsection. A common positive factor in $U,V$ leaves $R$ unchanged but can
change the logarithmic curvature used in the concavity and convexity steps
below.

Exact expansion of the unreduced polynomials gives $U>0$, with all 736 coefficients positive. Under simultaneous scaling of $A,Q$, the signed grades are

 \begin{gather}
 U=\sum_{m=0}^{15}U_m,\qquad V=\sum_{m=0}^{15}V_m, \label{b:eq:p5-outer-input}\\
 U_m\ge_{\rm coeff}0,\quad V_m<0\ \text{for positive weights}\quad(m<15),\notag\\
 V_{15}=A^7Q^7(2q+3)(3q+2)
                  [Q\Pi_5(q)-Af(q)],\notag\\
 \Pi_5(q)=11q^4-16q^3-124q^2-128q-33, \notag\\
 f(q)=33q^4+128q^3+124q^2+16q-11. \notag
 \end{gather}
 
Here $\Pi_5(q)$ is a polynomial in the inner ratio and is distinct
from the positive normalization $\mathscr D_5$.
For $m\le13$, every coefficient of $V_m$ is negative. The two positive
coefficients at grade 14 combine with negative terms to give

\begin{align}
 -\frac{V_{14}}{A^6Q^6}
 ={}&3(1+q^7)(A-Q)^2+A^2p_{14}(q) \label{eq:III-p5-grade14-positive}
\\
 &+Q^2q^7p_{14}(q^{-1})\notag \\
 &+AQ\,q(1+q)\bigl[4671(1+q^4)\notag \\
 &\hspace{23mm}+21680(q+q^3)+34329q^2\bigr],\notag \\
 p_{14}(q)={}&129+829q+4967q^2+17415q^3\notag \\
 &+30466q^4+26838q^5+11328q^6+1797q^7.\notag
\end{align}

Every term on the right is nonnegative, and $p_{14}(q)>0$ for $q>0$.
Thus $V_{14}<0$ throughout the positive-weight domain, despite its mixed
coefficient signs.
The unique positive zero $q_c$ of $\Pi_5$ exceeds four.
Since $f(q)>0$ for $q\ge1$, every finite promoter requires
$q>q_c$ and $Q/A>f(q)/\Pi_5(q)>3$.
Along every promoting outer ray, increasing both outer contacts
strictly decreases $R$.

At cap $H$, joint scaling of $A,Q$ along such a ray lowers $R$ without increasing the contrast while $A<1$, so an optimizer has $A\ge1$. The sign comparisons below then
make the strongest contact saturate the cap. Thus every optimizer
has
\begin{equation}
 \begin{gathered}
 (A,1,q,H),\qquad q_c<q\le H,\\
 1\le A< H \Pi_5(q)/f(q),\qquad V(A,q,H)>0.
 \end{gathered}
 \label{b:eq:p5-cap-domain}
\end{equation}
The ratio condition is necessary at a finite point, and together
with $q>q_c$ it characterizes the rays that promote at sufficiently large
outer scale. A finite point on such a ray promotes only where $V>0$, so
Eq.~\eqref{b:eq:p5-cap-domain} keeps this condition separately.
In the rest of this subsection $Q=H$ in
Eq.~\eqref{eq:v10-p5-normalization}.

Two further sign expansions sharpen those bounds.
First, $V(A,A+x,H)$ has 1289 nonzero coefficients, all negative, for
$A,H>0$ and $x\ge0$. Hence every promoter has $q<A$.
Together with the outer-degree bounds this gives
\begin{equation}
\begin{gathered}
q_c<q<A<H/3,\qquad q_c>4,\\
11q_c^4-16q_c^3-124q_c^2-128q_c-33=0.
\end{gathered}
\end{equation}
Second, write $V$ as a polynomial in $H$ and substitute
$q=1+s$, $A=q+a$. Its coefficients at $H$-degrees zero through seven
are strictly negative for $s,a\ge0$; Table~\ref{tab:III-p5-positive-coefficients}(b) lists the nonzero term counts at each degree.
The degree-eight coefficient may have either sign.
\begin{table}[tbp]
\caption{Coefficient data for the five-site path polynomials $U,V$. (a) The six monomials with positive coefficients in $V$, grouped by their coefficient ratio in $U$ and $V$; exponents are ordered as $(\deg_A,\deg_q,\deg_H)$. (b) Nonzero coefficient counts in each $H$-degree of $V(q+a,q,H)$ after $q=1+s$. Every counted coefficient is negative for $s,a\ge0$; the degree-eight part can have either sign.}
\label{tab:III-p5-positive-coefficients}
\centering\small
\begin{ruledtabular}
\begin{tabular}{lc}
\multicolumn{2}{c}{(a) Positive monomials in $V$}\\
Exponent triples & Coefficient ratio\\
\hline
$(7,6,8),(8,0,7)$ & $7$\\
$(7,5,8),(8,1,7)$ & $3068/47$\\
$(7,7,7),(7,0,7)$ & $3167/3$\\
\hline
\multicolumn{2}{c}{(b) Negative terms after substitution}\\
$H$-degree & Nonzero coefficient count\\
\hline
$0$ & $162$\\
$1$ & $153$\\
$2$ & $144$\\
$3$ & $135$\\
$4$ & $126$\\
$5$ & $117$\\
$6$ & $108$\\
$7$ & $99$
\end{tabular}
\end{ruledtabular}
\end{table}

Consequently $V/H^8$ increases strictly with $H$ on the ordered domain.
The positive coefficients of $U$, its degree bound of eight in $H$,
and its nonzero lower-degree terms make $U/H^8$ decrease strictly.
Their ratio therefore decreases strictly wherever $V>0$, so the
strongest contact saturates an optimizing cap. These sign expansions
also exclude the ordering faces from $V=0$ and give $V_H>0$ there.

The dependence on the other outer contact will also be needed.
At fixed $q\ge1$ and $H>0$, the signed outer degrees give
\begin{gather}
V(A,q,H)=c_7(q,H)A^7-
\sum_{i\in\{0,1,\ldots,6,8\}}b_i(q,H)A^i,\notag\\
b_i>0.
\end{gather}
Thus $v(x)=V(e^x,q,H)e^{-7x}$ is strictly concave and tends to
$-\infty$ as $x\to\pm\infty$. Its unique maximum is determined by
$AV_A-7V=0$. Promotion at fixed $q,H$, when possible, occupies one interval in $A$.
At a collapsed endpoint $V=V_A=0$, strict concavity gives $V_{AA}<0$.
This is the second nonzero derivative needed below for the
analytic endpoints of the projection onto $q$.
This concavity is a property of the representative $V$ fixed in
Eq.~\eqref{eq:v10-p5-normalization}.

For $r_{\rm lev}\ge7$, set $x=\log A$, $y=\log q$ and
\begin{equation}
L_{r_{\rm lev}}(x,y;H)=\frac{U-r_{\rm lev} V}{A^7q^6H^8}.
\label{eq:v10-p5-level-proof}
\end{equation}
Since $U>0$, $L_{r_{\rm lev}}\le0$ is exactly the physical sublevel
$V>0$, $U/V\le r_{\rm lev}$.
Table~\ref{tab:III-p5-positive-coefficients}(a) identifies the six monomials with positive coefficients in $V$ and their coefficient ratios in $U$ and $V$.

Every other coefficient of $U-r_{\rm lev} V$ is positive. The three exceptional
$(A,H)$ grades are $A^8H^7P_d$, $A^7H^8P_h$, $A^7H^7P_m$, with
\begin{align}
P_d={}&(198r_{\rm lev}+440)q^6\notag\\
&+(1197r_{\rm lev}+2968)q^5\notag\\
&+(2606r_{\rm lev}+7982)q^4\notag\\
&+(2476r_{\rm lev}+10964)q^3\notag\\
&+(886r_{\rm lev}+8116)q^2\notag\\
&+(3068-47r_{\rm lev})q+462-66r_{\rm lev},\\
P_m={}&(6334-6r_{\rm lev})(q^7+1)\notag\\
&+(4671r_{\rm lev}+49176)(q^6+q)\notag\\
&+(26351r_{\rm lev}+157658)(q^5+q^2)\notag\\
&+(56009r_{\rm lev}+271432)(q^4+q^3),
\end{align}
and $P_h(q,r_{\rm lev})=q^6P_d(1/q,r_{\rm lev})$.
Let $D_q=q\partial_q$. For $q\ge4$, $r_{\rm lev}\ge7$,
\begin{equation}
P_d>0,\qquad P_dD_q^2P_d-(D_qP_d)^2>0.
\end{equation}
These inequalities follow by $q=4+t$, $r_{\rm lev}=7+s$, after which the two
polynomials have respectively 14 and 36 positive nonzero coefficients.
The contribution $e^{x-6y}P_d(e^y,r_{\rm lev})/H$ therefore has positive definite
Hessian. Its Hessian determinant divided by the square of the contribution
is $\partial_y^2\log P_d>0$.

Two convexity domains cover every optimizer. If
$7\le r_{\rm lev}\le r_{{\rm lev},0}=3068/47$, the normalized monomials outside the convex $P_d$ group all have
nonnegative coefficients except the constant $-66(r_{\rm lev}-7)$.
Hence $L_{r_{\rm lev}}$ is strictly convex for all $A>0,q>4$ in this level range.
For arbitrary $r_{\rm lev}\ge7$, the head and middle terms have no $x$ dependence.
Their $y$-curvature numerators are
\begin{equation}
C_h=(D_q-6)^2P_h,\qquad C_m=(D_q-6)^2P_m.
\end{equation}
With $q=(4+70t)/(1+t)$ and $r_{\rm lev}=7+s$, clearing the positive denominators
gives respectively 12 and 16 strictly positive coefficients.
Thus $L_{r_{\rm lev}}$ is strictly convex for $A>0$, $4<q<70$, at every level $r_{\rm lev}\ge7$.
Each finite sign assertion is reproducible from the explicit polynomials
and Eq.~\eqref{eq:v10-transform-coefficients}.

The coverage of these domains is exact. Every promoter satisfies $H>32q$.
Indeed, if $H\le32q$, monotonicity of $V/H^8$ would imply
$V(A,q,32q)>0$. Write $A=aq$. The ordering confines $1<a<32/3$.
For the polynomial $-V(aq,q,32q)$ use $q=4+s$ and split
$a$ into $[1,4]$, $[4,6]$, $[6,8]$, $[8,10]$, $[10,32/3]$.
The projective box transform in $a$, multiplied by $(1+t)^8$, has 198
strictly positive coefficients on each interval, contradicting promotion.
If $H<2200$, every promoter consequently has $q<68.75<70$.
If $H\ge2200$, the exact design $(A,q,H)=(150,15,2200)$ has
{\small
\begin{multline}
R=\frac{1920936003318510951592810514449244279307148864}
{33545297254534441962908949665701249893137799}\\
<\frac{3068}{47}.
\end{multline}
}
Strongest-edge monotonicity retains this bound at all larger caps.
Every global minimizer therefore falls within one of the two convexity domains.

After normalizing the weakest contact to one, the cap box $[1,H]^3$ is
compact, and
$R\to+\infty$ as $V\downarrow0$. The strict ordering excludes an
$A$ or $q$ cap-face minimum. An interior global minimum exists and obeys
\begin{equation}
U_AV-UV_A=0,\qquad U_qV-UV_q=0.
\label{eq:v10-p5-stationary}
\end{equation}
If two distinct minima had common value $r_{\rm lev}$, strict convexity would give
$L_{r_{\rm lev}}<0$ at their logarithmic midpoint, implying $R<r_{\rm lev}$ there.
This contradiction proves uniqueness. At the minimum,
\begin{equation}
\operatorname{Hess}_{x,y}R
=\frac{A^7q^6H^8}{V}\operatorname{Hess}_{x,y}L_{r_{\rm lev}}>0.
\end{equation}
The optimizer is a real-analytic branch in every feasible finite cap, and
its threshold decreases strictly with the cap. Stationary points of $R$ at threshold
values above $r_{{\rm lev},0}=3068/47$ with $q>70$, if any, lie outside
both convexity domains; classifying them would require a separate sign
analysis of Eq.~\eqref{eq:v10-p5-stationary} in that region.

Within the oriented chart the promoting region is connected. Define
$D_A=A\partial_A$ and
\begin{equation}
\mathcal P_{\rm conn}=\left[\frac{D_A-7}{10}+D_q-6\right]V.
\end{equation}
For $q\ge70$ it is strictly negative. A compact grouped verification writes
\begin{equation}
\mathcal P_{\rm conn}=A^6H^7[a_2(q)A^2+a_1(q)A+a_0(q)]+\mathcal P_{\rm conn,rem},
\end{equation}
where $a_j$ is the corresponding coefficient of $A^{6+j}H^7$ in
$\mathcal P_{\rm conn}$. After $q=70+s$, the polynomials
$-a_2,-a_0,4a_2a_0-a_1^2$ have respectively 7, 9 and 15 positive
coefficients; $-\mathcal P_{\rm conn,rem}$ has 907. The grouped quadratic is
negative definite. Along $A=A_i(q/q_i)^{1/10}$ at fixed $H$,
\begin{equation}
\frac{d}{d\log q}\frac{V}{A^7q^6}
=\frac{\mathcal P_{\rm conn}}{A^7q^6}<0.
\end{equation}
Decreasing $q$ from a promoter with $q>70$ preserves promotion and joins it
to $q=70$; $A$ decreases and $A/q$ increases, so the cap and ordering remain
admissible. For $4<q\le70$, the function
$W=-V/(A^7q^6H^8)$ is strictly convex in $(\log A,\log q)$.
This follows from the same group calculations at the leading order in $r_{\rm lev}$.
The dual group and its logarithmic curvature have respectively 7 and 12
positive coefficients after $q=4+t$; the head and middle curvature
transforms have respectively 6 and 8 on $4\le q\le70$.
Its negative sublevel is convex. The two constructions show that the
promoting set in this chart is path connected. Restoring labeled edge
positions gives the separate component obtained by reversal, since a
promoting inner-contact ratio cannot pass continuously through one.

There is a unique promotion onset $128<H_{0,P_5}^{\rm U}<2200$ and a unique geometry
$(A_{\rm on},q_{\rm on})$ satisfying $V=V_A=V_q=0$ there. To see this, first contact
with $V\ge0$ occurs in the interior by the strict ordering and $H>32q$.
It lies below 2200, where $W$ has positive definite Hessian. A zero minimum
of a strictly convex function is unique, and $V_H>0$ on $V=0$ follows
from strongest-edge monotonicity. The Jacobian of these three equations
is consequently nonsingular. Write $z=(\log A,\log q)$,
$z_{\rm on}=(\log A_{\rm on},\log q_{\rm on})$,
$U_{\rm on}=U(A_{\rm on},q_{\rm on},H_{0,P_5}^{\rm U})$, and
$\delta_H=H-H_{0,P_5}^{\rm U}$. With $B=-\operatorname{Hess}_zV|_{\rm on}$,
\begin{align}
V={}&V_H|_{\rm on}\delta_H-\tfrac12(z-z_{\rm on})^{\mathsf T}B(z-z_{\rm on})\\
&+O(\delta_H^2+\delta_H|z-z_{\rm on}|+|z-z_{\rm on}|^3),\qquad B>0. \notag
\end{align}
The promoting region opens elliptically with linear dimensions
$O(\sqrt{\delta_H})$. The stationary equations
\eqref{eq:v10-p5-stationary} have geometry Jacobian
$-U_{\rm on}\operatorname{Hess}V|_{\rm on}$, hence the optimal geometry moves by
$O(\delta_H)$ and
\begin{equation}
R_{\min}(H)=\frac{U_{\rm on}}{V_H|_{\rm on}(H-H_{0,P_5}^{\rm U})}+O(1).
\end{equation}
The projection of the promoting region onto $q$ is one interval.
Its endpoints satisfy $V=V_A=0$ and are nondegenerate. For $q\le70$ the strict Hessian excludes a second zero stationary point once a promoter exists; for $q\ge70$ the strict Euler inequality excludes $V_q=0$.
The implicit endpoints are analytic, with the lower one decreasing and
the upper one increasing as $H$ grows.

Two useful slice consequences follow from the same grading. Every promoter
in fact has $A>2q$. The coefficient of $H^8$ in $V$, divided by $A^7$,
increases with $A$ and has 13 negative coefficients at $A=2q$; lower
$H$ grades are already negative on $A>q$. At fixed $A>0$, $H\ge3A$,
all $q$ coefficients of $V$ are negative except the adjacent degrees
five and six. Thus
$V/q^5=c_5+c_6q-\sum_{j\notin\{5,6\}}b_jq^{j-5}$, with $b_j>0$,
is strictly concave on $q>0$ and tends to $-\infty$ at both ends.
The numerator $U/q^5$ is strictly convex. The promoting slice is one
interval, when nonempty, and $(U-r_{\rm lev} V)/q^5$ is strictly convex for $r_{\rm lev}>0$.
It follows that the threshold has one conditional $q$ minimum.
The slice statements hold at fixed $A$ with $H\ge3A$; uniqueness of the
joint optimizer rests on the logarithmic sublevel argument above.

A rational contraction argument locates this unique root more closely.
Let $F_5=(V,V_A,V_q)^{\mathsf T}$, with the polynomial $V$ fixed by
Eq.~\eqref{eq:v10-p5-normalization}, and set
\begin{equation}
 x_5=\frac1{10^{12}}
 \begin{pmatrix}133918133808002\\15151087136153\\1406728807115288\end{pmatrix},
 \quad r_{\rm box}=10^{-8}.
 \label{eq:III-p5-onset-box-center}
\end{equation}
The entries are ordered as $(A,q,H)$. On
$X_5=x_5+[-r_{\rm box},r_{\rm box}]^3$, each row expression in
Eq.~\eqref{eq:III-onset-contraction-bound} is less than $10^{-4}$,
using rational endpoint operations. The unique zero in this box is
the onset already identified above, so
\begin{equation}
 \frac{1406728807}{10^6}<H_{0,P_5}^{\rm U}<\frac{1406728808}{10^6}.
 \label{eq:III-p5-onset-bracket}
\end{equation}
\end{proof}

\subsection{Introduction-dependent thresholds and contact costs}
\label{app:v7-preparation}

\begin{table}[tbp]
\caption{Coefficient signs for birth-associated introduction on the four-cycle with contacts $(1,A,Q,q)$. The polynomial $V$ has the sign of the benefit response, and $\mathcal P_{\mathrm b}$ is the comparison remainder in Eq.~\eqref{eq:v7-birth-c4-remainder}. Counts refer to the nonzero coefficients after the ordered-weight substitution in the text.}
\label{tab:III-birth-c4-chambers}
\begin{ruledtabular}
\begin{tabular}{llr}
Chamber&Polynomial&Coefficients\\
\hline
$Q\ge A\ge q\ge1$&$\mathcal P_{\mathrm b}$&2928 positive\\
$A\ge Q\ge q\ge1$&$V$&2041 negative\\
$A\ge q\ge Q\ge1$&$V$&2301 negative
\end{tabular}
\end{ruledtabular}
\end{table}

The copying process fixes the ancestral-pair operator; the introduction rule
determines which histories its inverse weights. We use this separation
to derive the birth-associated classification and contact costs in
Sec.~\ref{sec:v7-preparation-cost}. The source interpretation leads to
all-weight bounds and the global six-site optimum. Appendix~\ref{app:III-kernel-resource} then changes the controlled resource to copying probabilities, and Appendix~\ref{app:v10-crossover} treats mixtures of the two introduction rules.

\subsubsection{Exact polynomial reconstruction on the small supports}
\label{app:v7-birth-polynomials}
The bordered determinant of Eq.~\eqref{b:eq:bordered} has the following equivalent adjugate form for birth-associated introduction.
Order the unordered pairs lexicographically, put
$g^{\mathrm b}_{ij}=\vartheta_i+\vartheta_j$, $k_{ij}=\gamma_{ij}$ and
$j_{ij}=-\mathsf B_{ij}$, and form
\begin{equation}
 \frac{U}{V}
 =\frac{k^{\mathsf T}\operatorname{adj}(\mathsf M)g^{\mathrm b}}
        {j^{\mathsf T}\operatorname{adj}(\mathsf M)g^{\mathrm b}}.
 \label{eq:v7-birth-reconstruction}
\end{equation}

Here $\operatorname{adj}(\mathsf M)$ is the adjugate, the transposed
matrix of cofactors. Clear rational
denominators, remove the common polynomial factor, and choose the
common sign so that $U$ is positive at unit weights. This defines
coprime integer polynomials up to a common positive multiplier.
The identity $UJ^{\mathrm b}=VK^{\mathrm b}$ is an exact consequence
of the reconstruction. In each case below, $U$ has positive
coefficients, so $V$ has the sign of $J^{\mathrm b}$ throughout
the positive domain.

For the four-cycle with cyclic weights $(1,A,Q,q)$, the
$6\times6$ reconstruction gives 996 monomials in each of $U,V$.
The cost numerator $U$ and the 988-term difference $U-3V$ have positive
coefficients, while $V$ has mixed signs. Thus
$R^{\mathrm b}>3$ on the positive-benefit domain. To obtain a
global contrast bound, rotate a smallest edge to the unit position
and reflect so that $A\ge q\ge1$. Define
\begin{equation}
 \mathcal P_{\mathrm b}
 =(U-3V)AqQ-[8Q(A+q)+60A^2q^2]V.
 \label{eq:v7-birth-c4-remainder}
\end{equation}
The three possible order chambers are parameterized by the ordered
weights $1+X$, $1+X+Y$, $1+X+Y+Z$, with $X,Y,Z\ge0$.
Table~\ref{tab:III-birth-c4-chambers} gives the exact coefficient signs after this substitution.

The constant terms have the indicated strict signs, so boundaries
are included. A promoter must therefore lie in the first chamber,
where division by $AqQV>0$ gives
\begin{equation}
 R^{\mathrm b}-3>
 \frac8A+\frac8q+\frac{60Aq}{Q}
 \ge\frac{16}{\sqrt{Aq}}+\frac{60Aq}{H}
 \ge\left(\frac{103680}{H}\right)^{1/3}.
 \label{eq:v7-birth-c4-global}
\end{equation}
The last minimum has $(Aq)^{3/2}=2H/15$. In the reflected chart
$\upsilon_{\rm b}=q^{-1}$, $\omega_{\rm b}=q^2/Q$, the exact ratio is regular at the origin
and equals $3+16\upsilon_{\rm b}+60\omega_{\rm b}+O((\upsilon_{\rm b}+\omega_{\rm b})^2)$. Taking
$q=24/\epsilon+O(1)$ and
$Q=103680\epsilon^{-3}+O(\epsilon^{-2})$ matches the global bound.
Every competitive sequence is confined to the scaled chart
$(A,q,Q)=(a/t,d/t,h/t^3)$, $t=H^{-1/3}$, with positive scaled
variables in a compact set. Its leading gap divided by $t$ is
$8/a+8/d+60ad/h$. This decreases in $h$, has its unique minimum at
$h=1$, $a=d=(2/15)^{1/3}$, and has pin Hessian
$\left(\begin{smallmatrix}120&60\\60&120\end{smallmatrix}\right)$
there, and the capped $h$ derivative is negative. Together with the
localization and regularity just checked, these data meet the
persistence conditions in Appendix~\ref{app:v10-persistence}. The
finite-weight optimum is therefore eventually exactly reflected.

\subsubsection{The six-site grading and sharp global cost}
\label{app:v7-birth-p6-cost}
Normalize the path weights to $(Q_L,q_L,1,q_R,Q_R)$. For its
$15\times15$ reconstruction, set $C=U-V$ and use the grading
$(3,1,1,3)$ on $(Q_L,q_L,q_R,Q_R)$. The degrees of $U,V,C$ are
$75,75,74$, respectively. Normalize the common multiplier so that
their leading parts obey
\begin{equation}
 \begin{gathered}
 \relax[U]_{75}=[V]_{75}=U_{\rm lead},\\
 U_{\rm lead}=324Q_L^{10}Q_R^{10}q_L^7q_R^7(q_L+q_R),\\
 [C]_{74}=U_{\rm lead}F_{\rm res}^{\mathrm b},\\
 F_{\rm res}^{\mathrm b}=\frac3{q_L}+\frac3{q_R}
  +\frac{25}{3}q_Lq_R\left(\frac1{Q_L}+\frac1{Q_R}\right).
 \end{gathered}
 \label{eq:v7-birth-p6-leading-grade}
\end{equation}
The notation $[U]_{75}$ denotes the component of weighted degree
75. The complete polynomials $U$ and $C$ have positive
coefficients. Two coefficient comparisons connect their leading
graded parts to all positive weights,
\begin{equation}
 71C-18(U-U_{\rm lead})\ge_{\rm coeff}0,\quad
 C-U_{\rm lead}F_{\rm res}^{\mathrm b}\ge_{\rm coeff}0.
 \label{eq:v7-birth-p6-coefficients}
\end{equation}
The coefficient order is the one defined in Sec.~\ref{sec:contrast-design}.
Both expressions are polynomials; the negative powers in
$F_{\rm res}^{\mathrm b}$ cancel against $U_{\rm lead}$. These finite comparisons,
obtained from Eq.~\eqref{eq:v7-birth-reconstruction}, involve the
complete 15346-term numerator and 15344-term gap polynomial.
Equation~\eqref{eq:v7-birth-p6-coefficients} meets the hypotheses of
Eq.~\eqref{eq:v10-coefficient-gap-parent}, now with $L=U_{\rm lead}$,
$F=F_{\rm res}^{\mathrm b}$ and $a=71/18$. On $V>0$ it gives
\begin{equation}
 R^{\mathrm b}-1\ge
 \frac{F_{\rm res}^{\mathrm b}}{1+53F_{\rm res}^{\mathrm b}/18}.
 \label{eq:v7-birth-p6-global-gap}
\end{equation}

The central-edge normalization ensures $Q_L,Q_R\le H$ even before the
smallest edge is identified. With $q_g=\sqrt{q_Lq_R}$,
\begin{equation}
 F_{\rm res}^{\mathrm b}\ge\frac6{q_g}+\frac{50q_g^2}{3H}
 \ge\left(\frac{4050}{H}\right)^{1/3}.
 \label{eq:v7-birth-p6-resource}
\end{equation}
The second inequality is saturated at $q_g^3=9H/50$. Hence every
target-feasible weighting with $0<\epsilon<18/53$ satisfies
\begin{equation}
 H\ge4050\left(\frac{1-53\epsilon/18}{\epsilon}\right)^3.
 \label{eq:v7-birth-finite-gap-bound}
\end{equation}
The reflected expansion in Eq.~\eqref{eq:v7-preparation-objectives}
attains its leading coefficient.

The grading also controls the local expansion required for an exact
optimizer. Put $t=H^{-1/3}$ and
$(Q_L,q_L,q_R,Q_R)=(a/t^3,X/t,Y/t,e/t^3)$. Equation~\eqref{eq:v7-birth-p6-global-gap} confines every minimizing sequence
to a compact positive scaled domain. After the leading powers are
removed, the denominator tends to a positive multiple of
$a^{10}e^{10}X^7Y^7(X+Y)$. Thus
$(R^{\mathrm b}-1)/t$ is analytic there, with limiting function
\begin{equation}
 \frac3X+\frac3Y+
 \frac{25}{3}XY\left(\frac1a+\frac1e\right).
 \label{eq:v7-birth-scaled-objective}
\end{equation}
The core derivatives are negative. On their capped face $a=e=1$,
the unique limiting minimum is $X=Y=(9/50)^{1/3}$, with Hessian
$\left(\begin{smallmatrix}100/3&50/3\\50/3&100/3\end{smallmatrix}\right)$
and positive eigenvalues $50$, $50/3$. The persistence argument of
Appendix~\ref{app:v10-persistence} now gives $Q_L=Q_R=H$, $q_L=q_R$ exactly.
Thus the reflected subleading coefficients are the unrestricted global
coefficients.

Their evaluation needs only the quadratic part of the regular reflected
chart,
\begin{equation}
 \begin{aligned}
 R^{\mathrm b}-1={}&6\upsilon_{\rm b}+\frac{50}{3}\omega_{\rm b}+10\upsilon_{\rm b}^2
       +\frac{1478}{9}\upsilon_{\rm b}\omega_{\rm b}+\frac{100}{3}\omega_{\rm b}^2  \\
       &+O((\upsilon_{\rm b}+\omega_{\rm b})^3),\quad \upsilon_{\rm b}=q^{-1},\quad \omega_{\rm b}=q^2/Q.
 \end{aligned}
  \label{eq:v7-birth-reflected-quadratic}
 \end{equation}
Write $q=k/\epsilon$, $Q=L/\epsilon^3$ and
$(R^{\mathrm b}-1)/\epsilon=f(k,L)+\epsilon g(k,L)+O(\epsilon^2)$.
Equation~\eqref{eq:v7-birth-reflected-quadratic} gives
\begin{equation}
 \begin{aligned}
 f&=\frac6k+\frac{50k^2}{3L},\\
 g&=\frac{10}{k^2}+\frac{1478k}{9L}+
                         \frac{100k^4}{3L^2}.
 \end{aligned}
 \label{eq:v7-birth-optimizer-expansion}
\end{equation}
The leading equations $f=1$, $f_k=0$ have the solution
$(k,L)=(9,4050)$. For
$k=9+\epsilon k_1+O(\epsilon^2)$ and
$L=4050+\epsilon L_1+O(\epsilon^2)$, the next equations are
$f_LL_1+g=0$ and
$f_{kk}k_1+f_{kL}L_1+g_k=0$, evaluated at that point.
They give $L_1=6096$ and $k_1=337/90$, as used in
Eq.~\eqref{eq:v7-preparation-costs}.

The coefficient $25/3$ has a local ancestral explanation. Within a
fast triple the leading jump counts satisfy
\begin{equation}
 \begin{aligned}
 D^{\mathrm b,[0]}_{12}&=1+D^{\mathrm b,[0]}_{13}/3,\\
 D^{\mathrm b,[0]}_{13}&=1+D^{\mathrm b,[0]}_{23},\\
 D^{\mathrm b,[0]}_{23}&=1+D^{\mathrm b,[0]}_{13}/2,
 \end{aligned}
 \label{eq:v7-birth-fast-pairs}
\end{equation}
so $(D^{\mathrm b,[0]}_{12},D^{\mathrm b,[0]}_{13},D^{\mathrm b,[0]}_{23})=(7/3,4,3)$. In the edge contribution
\begin{equation}
 \begin{aligned}
 \mathscr L_{ij}&=D^{\mathrm b}_{ij}
 -\frac12\sum_k(P_{ik}-P_{jk})(D^{\mathrm b}_{jk}-D^{\mathrm b}_{ik}),\\
 K^{\mathrm b}-J^{\mathrm b}&=\sum_{i<j}\gamma_{ij}\mathscr L_{ij},
 \end{aligned}
 \label{eq:v7-birth-edge-deficit}
\end{equation}
the two fast edges contribute $14/3$ and $11/3$ after extracting
their conductances. Their sum is $25/3$. The central edge has zero
leading deficit; its first correction is
$t(3/X+3/Y)+O(t^2)$. This accounts separately for the
pin waiting and follower establishment terms in
Eq.~\eqref{eq:v7-birth-scaled-objective}.

For comparison at a fixed design, substitution $Q=kq^2$ with fixed
$k>2$ gives
\begin{equation}
 R^{\mathrm b}(kq^2,q)=\frac{9k+132}{9k-18}+O(q^{-1}).
 \label{eq:v7-birth-uniform-design}
\end{equation}
The uniform optimum has $k=41/9$, giving $173/23$. At the
birth-associated optimum the two pin introduction masses are each
$\epsilon/54+O(\epsilon^2)$, their reproductive values tend to
$1/2$, and follower contributions to neutral fixation are
$O(\epsilon^2)$. Indeed, $q\asymp\epsilon^{-1}$ and
$Q\asymp\epsilon^{-3}$ give $Z=\sum_i s_i^{-1}\asymp\epsilon$;
each follower has reproductive value $s_i^{-1}/Z=O(\epsilon^2)$,
and its introduction probability is at most one.
Equation~\eqref{b:eq:birth-normalization}
therefore gives $\rho_0^{\mathrm b}\sim\epsilon/54$. For separated introductions of total rate $\nu$, the successful neutral invasion rate is consequently asymptotic to $\nu\epsilon/54$.

\subsection{Changing the controlled resource to copying probabilities}
\label{app:III-kernel-resource}
Normalize the path weights to $(Q_L,q_L,1,q_R,Q_R)$, impose
$C_P\le C$ and set $S=C-1$. For either introduction rule, the sharp
small-gap kernel cost is Eq.~\eqref{eq:III-kernel-costs}. For all
sufficiently large $C$, the exact least-threshold design is reflected,
and both outer-to-inner weight ratios saturate their allowed bounds.
\begin{proof}
Every interior row contains two probabilities,
so the cap is equivalent to bounding each adjacent ratio of edge weights
between $S^{-1}$ and $S$. In particular,
$q_L,q_R\le S$ and $Q_L\le Sq_L$, $Q_R\le Sq_R$.
The uniform all-weight loss in Eq.~\eqref{v3-eq-p6-positive-F} then obeys
\begin{equation}
 \begin{aligned}
 F&\ge\frac3{q_L}+\frac3{q_R}
 +\frac{106+29(q_L/q_R+q_R/q_L)}{6S}\\
 &\ge\frac{100}{3S}.
 \end{aligned}
 \label{eq:III-kernel-uniform-bound}
\end{equation}
The birth-associated loss in
Eq.~\eqref{eq:v7-birth-p6-leading-grade} satisfies
\begin{equation}
 F_{\rm res}^{\rm b}
 \ge\sum_{i=L,R}\left(\frac3{q_i}+\frac{25q_i}{3S}\right)
 \ge\frac{20}{\sqrt S}.
 \label{eq:III-kernel-birth-bound}
\end{equation}
Combining these inequalities with the exact global gap comparisons
gives, for $0<\epsilon<18/53$,
\begin{equation}
 \begin{aligned}
 C_P&\ge1+\frac{100}{3}\left(\epsilon^{-1}-\frac{53}{18}\right)
 &&({\rm uniform}),\\
 C_P&\ge1+400\left(\epsilon^{-1}-\frac{53}{18}\right)^2
 &&(\text{birth-associated}).
 \end{aligned}
 \label{eq:III-kernel-finite-bounds}
\end{equation}
The reflected trials $(Q,q,1,q,Q)$ with
$(q,Q)=(S,S^2)$ and
$(q,Q)=(3\sqrt S/5,3S^{3/2}/5)$ attain the respective leading
coefficients. The regular response expansions give gaps
$100/(3S)+O(S^{-2})$ and
$20/\sqrt S+O(S^{-1})$. They prove
Eq.~\eqref{eq:III-kernel-costs} for all positive path weights.
The strict target has the same leading infimum, approached from larger
kernel budgets.

The same calculation also identifies the eventual exact geometry.
For uniform introduction write $q_i=Sx_i$ and $Q_i=S^2x_i y_i$.
The cap imposes $x_i,y_i\le1$; the forcing inequality confines any
competitive sequence to a compact positive set of these variables.
The leading scaled gap decreases strictly in each $y_i$.
At $y_L=y_R=1$ it is
\begin{equation}
 \frac3{x_L}+\frac3{x_R}
 +\frac{106+29(x_L/x_R+x_R/x_L)}6,
\end{equation}
with unique minimum at $x_L=x_R=1$ and both outward $x_i$ derivatives
equal to $-3$. Uniform derivative control in the regular contact chart
therefore keeps all four cap faces active. For all sufficiently large
$C$, the exact least-threshold design is
$(S^2,S,1,S,S^2)$.
For birth-associated introduction use
$q_i=\sqrt S x_i$, $Q_i=S^{3/2}x_i y_i$.
The global gap comparison and the matching trial force
$F_{\rm res}^{\rm b}=O(S^{-1/2})$ for every competitive design.
The positive terms in Eq.~\eqref{eq:III-kernel-birth-bound} then bound
$x_L,x_R$ above and below. The positive terms $q_Lq_R/Q_i$ in
$F_{\rm res}^{\rm b}$ bound $Q_i/S^{3/2}$ below, while the cap bounds
them above. Thus $x_i,y_i$ lie in a compact positive set, where
the leading core derivatives are negative and the core faces
$y_i=1$ remain active. The leading gap on those
faces is $\sum_i(3/x_i+25x_i/3)$, with unique minimum
$x_L=x_R=3/5$ and positive Hessian
$\operatorname{diag}(250/9,250/9)$.
The regular birth-associated chart gives uniform derivatives and a
unique exact nearby minimum. Path reversal then forces $q_L=q_R$
exactly, with $Q_i=S q_i$ and
$q_i=3\sqrt S/5+O(1)$. These estimates retain the required derivative
signs for sufficiently large $C$. The size of $C$ needed here is left implicit, because the compact localization
and the regular-chart derivative bounds above carry unevaluated constants.
An explicit lower cap requires explicit versions of both.
\end{proof}

\subsection{A crossover between introduction rules with errors uniform in the mixture fraction}
\label{app:v10-crossover}
\label{app:v10-preparation-crossover}
Let $\mu^\zeta=\zeta\mu^{\rm U}+(1-\zeta)\mu^{\mathrm b}$, $0\le\zeta\le1$, and
set $v_\zeta=1-\zeta$. We optimize the weak target
$R^\zeta\le1+\epsilon$ over all positive contacts $(Q_L,q_L,1,q_R,Q_R)$.
Superscripts $U$ and $\mathrm b$ denote uniform and birth-associated introduction.
The global least contrast and optimal pin weight satisfy
Eq.~\eqref{eq:v10-uniform-crossover} as $\epsilon\downarrow0$, with errors
uniform for $0\le\zeta\le1$. For one sufficiently small-gap domain
independent of $\zeta$, the exact optimizer is reflected and saturates
both core contacts.
\begin{proof}
For each introduction rule define $\mathscr D_{\rm gap}=K-J$ and $g_{\rm gap}=\mathscr D_{\rm gap}/K$. Positivity of $K,\mathscr D_{\rm gap}$ gives
the exact weighted gap identity
\begin{equation}
g_{\rm gap}^\zeta=
\frac{\zeta K^{\rm U} g_{\rm gap}^{\rm U}+v_\zeta K^{\mathrm b}g_{\rm gap}^{\mathrm b}}
{\zeta K^{\rm U}+v_\zeta K^{\mathrm b}}.
\label{eq:v10-mixture-gap}
\end{equation}
The target implies $g_{\rm gap}^\zeta\le\epsilon/(1+\epsilon)<\epsilon$.
Feasibility constrains only the mixture, so one of $J^{\rm U}$ and $J^{\mathrm b}$
may be negative at a feasible design.

We first justify the hierarchy without assuming reflection. The normalized
deficits are $(U-V)/U$ in the respective polynomial reconstructions.
Equation~\eqref{eq:v10-coefficient-gap-parent} therefore gives, throughout
the positive-contact domain,
\begin{equation}
 g_{\rm gap}^{\mu}\ge\frac{F_{\rm res}^{\mu}}{1+(71/18)F_{\rm res}^{\mu}},\qquad \mu\in\{\mathrm U,\mathrm b\},
 \label{eq:v10-birth-gap-global}
\end{equation}
where $F_{\rm res}^{\rm U}=F$ is the function in Eq.~\eqref{v3-eq-p6-positive-F} and
$F_{\rm res}^{\mathrm b}$ is the function in Eq.~\eqref{eq:v7-birth-p6-leading-grade}.
The uniform inequality is strict, and both comparisons hold throughout the positive-contact domain, including where the corresponding benefit coefficient vanishes or is negative.
At least one participating component in Eq.~\eqref{eq:v10-mixture-gap}
has $g_{\rm gap}\le\epsilon$, so either bound forces
\begin{equation}
m=q_L^{-1}+q_R^{-1}+(q_L+q_R)(Q_L^{-1}+Q_R^{-1})=O(\epsilon)
\label{eq:v10-crossover-hierarchy}
\end{equation}
uniformly in $\zeta$. In the following bounds, $c,c'$ and $C$ are positive comparison constants, independent of $\epsilon$, $\zeta$ and the pin imbalance, and their values may change between bounds. On this hierarchy, with $S=q_L+q_R$,
\begin{equation}
\begin{gathered}
K^{\rm U}=\frac2S[1+O(m)],\qquad K^{\mathrm b}\le2K^{\rm U},\\
K^{\mathrm b}\ge c[(q_Lq_R)^{-1}+Q_L^{-1}+Q_R^{-1}],\\
\mathscr D_{\rm gap}^{\mathrm b}\ge c'(Q_L^{-1}+Q_R^{-1}).
\end{gathered}
\end{equation}
The lower bound on $K^{\mathrm b}$ uses $D^{\mathrm b}_{ij}\ge1$, since the birth pair response counts the ancestral jumps until coalescence, so $K^{\mathrm b}\ge\sum_{i<j}\gamma_{ij}$. The lower bound on $\mathscr D_{\rm gap}^{\mathrm b}$ follows by multiplying
$g_{\rm gap}^{\mathrm b}\ge c z/(1+z)$, $z=q_Lq_R(Q_L^{-1}+Q_R^{-1})$, by
$K^{\mathrm b}\ge c(1+z)/(q_Lq_R)$. The upper bound $K^{\mathrm b}\le2K^{\rm U}$ follows entrywise
from $\vartheta_i=\sum_kP_{ki}\le\deg(i)\le2$ on a path, so $\mu_i^{\mathrm b}\le2\mu_i^{\rm U}$. Positivity of the pair inverse gives $D^{\mathrm b}\le2D^{\rm U}$, and contraction with the positive conductances gives the claimed bound. These estimates remain uniform in pin imbalance.

To obtain differentiated expansions, put
\begin{equation}
\begin{gathered}
Q_L=\frac a{t^2\eta_{\rm mix}},\quad Q_R=\frac e{t^2\eta_{\rm mix}},\\
q_L=\frac X t,\quad q_R=\frac Y t,\qquad 0\le\eta_{\rm mix}\le2.
\end{gathered}
\end{equation}
Here $a,X,Y,e$ belong to compact positive intervals. Positive $t,\eta_{\rm mix}$ describe finite weights; the face $\eta_{\rm mix}=0$ is included by analytic continuation of the divided pair system. Divide the $(3,4)$
row of the 15-pair matrix by $t$. The divided inverse can be checked through its fast and slow blocks.
At $t=0$, the fourteen fast pair rows divide into two within-domain
three-pair blocks, two one-follower--one-pin blocks and one
two-follower block. Each within-domain block has determinant $3$.
With the follower sites ordered from the outer end toward their pin, the two-follower block is $\mathsf T\otimes I_2+I_2\otimes\mathsf T$, and
the one-follower block is
\begin{equation}
 \begin{gathered}
 \mathsf T=\begin{pmatrix}1&-1\\-1&2\end{pmatrix},\qquad \det\mathsf T=1,\\
 \det(\mathsf T\otimes I_2+I_2\otimes\mathsf T)=36.
 \end{gathered}
\end{equation}
The fast determinant is therefore $3^2\cdot36=324$. The limiting
divided $(3,4)$ row acts as
\begin{equation}
 \frac{\eta_{\rm mix}X}{a}(D_{34}-D_{24})
 +\frac{\eta_{\rm mix}Y}{e}(D_{34}-D_{35})
 +\left(\frac1X+\frac1Y\right)D_{34}.
\end{equation}
Fast harmonic extension of the slow value has
$D_{24}=D_{35}=D_{34}$. Its Schur scalar is thus
$X^{-1}+Y^{-1}$, independently of $\eta_{\rm mix}$, proving the
determinant $324(X+Y)/(XY)$. Compact positivity of $a,e,X,Y$
then controls the inverse and all its derivatives uniformly,
including $\eta_{\rm mix}=0$.
The inverse therefore has uniform Laurent expansions and differentiated remainders. Solving its zeroth and first coefficient equations gives
\begin{align}
K^{\rm U}&=\frac{2t}{X+Y}+O(t^2),\\
\mathscr D_{\rm gap}^{\rm U}&=\frac{6t^2}{XY}+\frac{t^2\eta_{\rm mix}}{3(X+Y)}\\
&\quad\times\left(\frac{53X+29Y}{a}+\frac{29X+53Y}{e}\right)+O(t^3),\notag \\
K^{\mathrm b}&=\frac{t^2}{XY}+\frac{t^2\eta_{\rm mix}}{3(X+Y)}\\
&\quad\times\left[\frac{28X+16Y}{a}+\frac{16X+28Y}{e}\right]+O(t^3),\notag \\
\mathscr D_{\rm gap}^{\mathrm b}&=\frac{25}{3}t^2\eta_{\rm mix}(a^{-1}+e^{-1}) \label{eq:v10-crossover-chart}\\
&\quad+\frac{3(X+Y)t^3}{(XY)^2}+O(t^3\eta_{\rm mix}+t^4).\notag
\end{align}
The coefficient equations are obtained by expanding $\mathsf M D=g$ and solving
$\mathsf M_0D_k=g_k-\sum_{j\ge1}\mathsf M_jD_{k-j}$ after the row division.
The birth source is analytic; the uniform source has a simple $t^{-1}$
term in the divided row. This finite recursion fixes all coefficients in
Eq.~\eqref{eq:v10-crossover-chart}.

The smaller birth remainder at the ideal-core face follows from a
short pair calculation. At $\eta_{\rm mix}=0$, put
$\beta_L=t/(X+t)$ and $\beta_R=t/(Y+t)$. The central pair has
$D_{34}^{\rm b}=1$, and the surviving conductance is
$\gamma_{34}=\beta_L\beta_R$. The fast equations give
\begin{equation}
 \begin{aligned}
 D_{24}^{\rm b}-D_{23}^{\rm b}
     &=1-4\beta_R+O((\beta_L+\beta_R)^2),\\
 D_{35}^{\rm b}-D_{45}^{\rm b}
     &=1-4\beta_L+O((\beta_L+\beta_R)^2).
 \end{aligned}
\end{equation}
Using Eq.~\eqref{eq:v7-birth-edge-deficit},
\begin{equation}
 \mathscr D_{\rm gap}^{\rm b}
 =3\beta_L\beta_R(\beta_L+\beta_R)+O(t^4)
 =\frac{3(X+Y)t^3}{(XY)^2}+O(t^4).
\end{equation}
Analyticity in $\eta_{\rm mix}$ makes the difference of the
$t^3$ coefficient from its ideal-core value $O(\eta_{\rm mix})$.
Together with the $t^2$ coefficient above, this proves the
$O(t^3\eta_{\rm mix}+t^4)$ remainder in
Eq.~\eqref{eq:v10-crossover-chart}.

With $t=\epsilon$, $\eta_{\rm mix}=\zeta+\epsilon$, the reflected trial
$q_L=q_R=12/\epsilon$, $Q_L=Q_R=20000/[\epsilon^2(\zeta+\epsilon)]$
has leading gap below $697\epsilon/1000$ and error $O(\epsilon^2)$.
Every competitive design thus has
$H\le C/[\epsilon^2(\zeta+\epsilon)]$.
Together with Eq.~\eqref{eq:v10-crossover-hierarchy} this proves the
uniform compact localization
\begin{equation}
q_L,q_R\asymp\epsilon^{-1},\qquad
Q_L,Q_R\asymp[\epsilon^2(\zeta+\epsilon)]^{-1}.
\label{eq:v10-crossover-compact}
\end{equation}
The endpoint-sensitive step has three cases. If $\zeta\ge1/2$,
$K^{\mathrm b}\le2K^{\rm U}$ gives $g_{\rm gap}^{\rm U}\le3\epsilon$, and the uniform global bound
and trial cap give Eq.~\eqref{eq:v10-crossover-compact}.
If $\zeta<1/2$ and $g_{\rm gap}^{\mathrm b}\le2\epsilon$, the birth bound gives
$q_Lq_R\le C\epsilon H$ and $Q_L,Q_R\ge cq_Lq_R/\epsilon$; the lower bounds
$q_L,q_R\ge c/\epsilon$ give the required upper bounds on both pins.
Finally, if $\zeta<1/2$ and $g_{\rm gap}^{\mathrm b}>2\epsilon$, feasibility implies
$v_\zeta \mathscr D_{\rm gap}^{\mathrm b}\le2\epsilon\zeta K^{\rm U}$, hence
$Q_L^{-1}+Q_R^{-1}\le C\epsilon\zeta/(q_L+q_R)$.
Combining this with $Q_L^{-1}+Q_R^{-1}\ge2/H$ bounds $q_L+q_R\le C/\epsilon$
and each outer contact below by $c/[\epsilon^2\zeta]$.
At $\zeta=0$ the mixture gap equals $g_{\rm gap}^{\mathrm b}<\epsilon$, so the third
case arises only for $\zeta>0$.

On the compact set just proved define
\begin{align}
K_0&=\frac{2\zeta}{q_L+q_R}+\frac{v_\zeta}{q_Lq_R},\\
\mathscr D_{{\rm gap},0}&=3(q_L^{-1}+q_R^{-1})K_0\notag\\
&\quad+\frac\zeta{3(q_L+q_R)}
\left(\frac{53q_L+29q_R}{Q_L}+\frac{29q_L+53q_R}{Q_R}\right)
\notag\\
&\quad+\frac{25v_\zeta}{3}(Q_L^{-1}+Q_R^{-1}).
\end{align}
Equation~\eqref{eq:v10-crossover-chart} gives
$K^\zeta=K_0[1+O(\epsilon)]$ and
$R^\zeta-1=\mathscr D_{{\rm gap},0}/K_0+O(\epsilon^2)$, including scaled derivatives,
uniformly in $\zeta$. In particular $K_0\asymp\epsilon(\zeta+\epsilon)$,
which is of order $\epsilon^2$ at the pure-birth endpoint. The smaller
birth remainder in Eq.~\eqref{eq:v10-crossover-chart} keeps the threshold
error of order $\epsilon^2$ after division by this vanishing denominator.

The leading objective decreases with both outer contacts. Set $Q_L=Q_R=H$
and $q_{\rm harm}=2q_Lq_R/(q_L+q_R)$. Then
\begin{equation}
\begin{gathered}
\mathscr D_{{\rm gap},0}/K_0=\frac6{q_{\rm harm}}+
\frac{C_\zeta q_Lq_R}{H(v_\zeta+\zeta q_{\rm harm})},\\
C_\zeta=(50+32\zeta)/3.
\end{gathered}
\end{equation}
The identity $q_Lq_R-q_{\rm harm}^2=q_Lq_R(q_L-q_R)^2/(q_L+q_R)^2\ge0$ forces equal pins at the
leading optimum. Write $x=\epsilon q_{\rm harm}$, $\xi_{\rm prep}=\zeta/\epsilon$.
The required contrast is minimized by
\begin{equation}
H_{\rm lead}(x)=\frac{C_\zeta x^3}
{\epsilon^3(x-6)(v_\zeta+\xi_{\rm prep}x)},\qquad x>6.
\end{equation}
Its derivative has the sign of
$\xi_{\rm prep}x^2+(2v_\zeta-12\xi_{\rm prep})x-18v_\zeta$.
There is one root $x_\zeta\in[9,12]$, since the quadratic is increasing
for $x>6$, negative at 6 and has values $-27\xi_{\rm prep}$ and
$6v_\zeta$ at 9 and 12.
Write $\mathcal H_\zeta(\epsilon)=\min_{x>6}H_{\rm lead}(x)$.
At fixed $\zeta$, stationarity in $x$ gives
\begin{equation}
\epsilon\,\partial_\epsilon\log\mathcal H_\zeta
=-3+\frac{\xi_{\rm prep}x_\zeta}{v_\zeta+\xi_{\rm prep}x_\zeta}
\in[-3,-2].
\end{equation}
Replacing the target by $\epsilon+O(\epsilon^2)$ therefore
changes its optimal contrast by a relative $O(\epsilon)$ uniformly.
This proves
\begin{equation}
H_{\min}^\zeta=
\frac{C_\zeta x_\zeta^3}
{\epsilon^3(x_\zeta-6)(v_\zeta+\xi_{\rm prep}x_\zeta)}
[1+O(\epsilon)].
\end{equation}

Exact eventual reflection requires the derivative bounds as well. Put
$a=Q_L/H$, $e=Q_R/H$ and $y=(q_L-q_R)/(q_L+q_R)$, and solve the exact target
$R^\zeta=1+\epsilon$ for the required cap
$H_{\rm req}(a,e,x,y;\epsilon,\zeta)$. Normalize
$v_\zeta+\xi_{\rm prep}$ by setting
$\tau_{\rm mix}=\xi_{\rm prep}/(v_\zeta+\xi_{\rm prep})\in[0,1]$, and define
\begin{align}
f_{\tau_{\rm mix}}(x)&=\frac{x^3}{(x-6)(1-\tau_{\rm mix}+\tau_{\rm mix} x)},\notag\\
W_\zeta(a,e,y)&=\frac1{50v_\zeta+82\zeta}
\left[\frac{25v_\zeta+\zeta(41+12y)}a\right.\notag\\
&\hspace{4em}\left.+\frac{25v_\zeta+\zeta(41-12y)}e\right].
\end{align}
On the compact minimizer neighborhood, the response estimate implies
\begin{equation}
\frac{\epsilon^3(v_\zeta+\xi_{\rm prep})}{C_\zeta}
H_{\rm req}
=\frac{f_{\tau_{\rm mix}}(x)W_\zeta(a,e,y)}{1-y^2}+O(\epsilon),
\label{eq:v10-required-cap}
\end{equation}
uniformly with two scaled derivatives. Indeed, $x-6$ stays uniformly
positive there, so the leading target derivative with respect to the
normalized cap stays away from zero. The implicit-function theorem
therefore transfers the response remainder to this solved objective.
Both capped derivatives are uniformly negative since
$25v_\zeta+\zeta(41\pm12y)\ge25$ for $|y|\le1$.
At $a=e=1$, $W_\zeta=1$, and the unique leading minimum has $y=0$,
$9\le x\le12$, a uniformly positive $x$ Hessian and $y$ Hessian
$2f_{\tau_{\rm mix}}(x)>0$. These signs and compact localization verify the
conditions of Appendix~\ref{app:v10-persistence}. Both cores saturate
and the pins are exactly equal. The $C^1$ error gives
$q_*^\zeta=x_\zeta/\epsilon+O(1)$.
The uniform derivative bounds give one small-$\epsilon$ domain for the
whole mixture interval. The localization and the expansions of
Eq.~\eqref{eq:v10-crossover-chart} use the small-gap bound $m=O(\epsilon)$ in
Eq.~\eqref{eq:v10-crossover-hierarchy}. At larger target gaps, determining the
optimal geometry requires a comparison that also covers designs outside this localized hierarchy.
\end{proof}

\section{First-passage distributions and optimal waiting}\label{app:v10-times}
The reflected response arrays used in Appendix~\ref{app:III-p6-onsets} are defined first, since they also determine the critical time branch. All times in this appendix use the rate-one Poisson attempt clock unless a discrete version is stated explicitly. Appendices~\ref{app:v10-c4-time}--\ref{app:v4-exact-time} derive the neutral time--contrast frontiers on $C_4$, $P_5$ and $P_6$. The six-site analysis continues through every feasible finite cap in Appendix~\ref{app:v10-allbudget} and to the uncapped infimum in Appendix~\ref{app:v10-uncapped}, for sufficiently small target gaps. Appendices~\ref{app:III-c5-frontier} and \ref{app:v10-c5-time} treat the five-cycle frontier and conditional distributions. The selected extension in Appendix~\ref{app:v7-selected-time} is proved for small selection, small positive target gaps and bounded scaled budgets $\epsilon^2H$.
\subsection{Reflected six-site response and the critical branch}
\label{app:p6-certificate}
The reflected contacts are $(Q,q,1,q,Q)$, with the pins at
sites 3 and 4. Set $u=1/q$ and $v=q/Q$. Their nonzero copying
probabilities are
\begin{equation}
 \begin{aligned}
 P_{12}=P_{65}&=1,&
 P_{21}=P_{56}&=\frac1{1+v},\\
 P_{23}=P_{54}&=\frac v{1+v},&
 P_{32}=P_{45}&=\frac1{1+u},\\
 P_{34}=P_{43}&=\frac u{1+u}.
 \end{aligned}
 \label{b:eq:p6-kernel}
\end{equation}
The protected-domain reduction and physical times for this
kernel are stated in Eq.~\eqref{b:eq:p6-domain-observables}.
Here we need the exact finite-contact response. Use
$\mathbf D$ for the integrated-disagreement vector in
Eq.~\eqref{b:eq:pairs}, with $N=6$ and uniform source
$2\times \mathbf1$. Reflection reduces the fifteen ancestral pairs
to nine orbits. The contractions and susceptibility normalization
are Eq.~\eqref{b:eq:response}; $R=K/J$ on $J>0$.
Substitute Eq.~\eqref{b:eq:p6-kernel} into this pair system
and clear the common rational factor in the ratio to obtain
\begin{equation}
 R=\frac{\mathcal K(u,v)}{\mathcal J(u,v)}.
 \label{b:eq:p6-rational}
\end{equation}
 
Write
$\mathcal K=\sum_{i=0}^6\sum_{j=0}^7a_{ij}u^iv^j$ and
$\mathcal J=\sum_{i=0}^6\sum_{j=0}^7b_{ij}u^iv^j$.
Table~\ref{tab:III-p6-response-coefficients} lists both coefficient arrays, indexed by the powers of $u$ and $v$.

The polynomial $\mathcal K$ is strictly positive for positive $u,v$.
The exact identity $K\mathcal J=J\mathcal K$, with $K>0$,
then gives the same sign for $J$ and $\mathcal J$, including their zero locus.
The coefficients in Table~\ref{tab:III-p6-response-coefficients} give
\begin{equation}
 \begin{aligned}
 \mathcal K-\mathcal J-\left(6u+\frac{82}{3}v\right)\mathcal J
 &=90u^2+1600uv+1182v^2\\
 &\quad+\text{positive higher monomials}.
 \end{aligned}
 \label{eq:app-p6-positive-certificate}
\end{equation}
Every nonzero coefficient is positive; the complete left side is
specified by Table~\ref{tab:III-p6-response-coefficients}, so this is an exact polynomial identity. On $\mathcal J>0$, division gives
\begin{equation}
 R-1>6u+\frac{82}{3}v.
 \label{eq:app-p6-strict-bound}
\end{equation}
Since $\mathcal J(0,0)=9$, Taylor division also gives the convergent expansion
\begin{equation}
 R=1+6u+\frac{82}{3}v+10u^2+
 \frac{1600}{9}uv+\frac{394}{3}v^2+O((u+v)^3).
 \label{eq:app-p6-taylor}
\end{equation}

In particular, the exact reflected bound gives
\begin{equation}
 R-1>\frac6q+\frac{82q}{3Q}
       \ge4\sqrt{41/Q}\ge4\sqrt{41/H}.
 \label{eq:p6-cost-bound}
\end{equation}
On this reflected subfamily it is sharper than the all-weight bound in
Eq.~\eqref{v3-eq-p6-global-bound}. Appendix~\ref{v3-app-global-p6} shows why the eventual
unrestricted minimum lies in this subfamily.

The trial $Q=H$, $q=3\sqrt{H/41}$ matches the leading lower bound in
Eq.~\eqref{eq:p6-cost-bound}. The eventual unique exact optimum has
$Q=H$ and $q=q_H=3\sqrt{H/41}+O(1)$, as proved over all positive
weights in Appendix~\ref{v3-app-exact-global-reflection}.
The time frontier follows from the exact optimizer in
Appendix~\ref{app:v4-exact-time}; the no-cap limit follows from
Appendix~\ref{app:v10-uncapped}. We next determine the exact
feasibility center needed by both the tolerance and time arguments.

\subsubsection{The finite-budget fold and its uniform time boundary layer}
\label{app:finite-budget-fold-v2}

The two roots in the leading frontier coalesce at
$\theta=12$, $\eta=656$. To resolve this endpoint, use the exact
ratio $\mathscr R=\mathcal K/\mathcal J$ with the coefficients in Table~\ref{tab:III-p6-response-coefficients}, and put
\begin{equation}
 \begin{gathered}
 \theta=\epsilon q,\qquad \eta=\epsilon^2 Q,\\
 g(\theta,\eta,\epsilon)
 =\frac{\mathscr R(\epsilon/\theta,\epsilon\theta/\eta)-1}{\epsilon}-1.
 \end{gathered}
 \label{eq:app-fold-g-v2}
\end{equation}
Since $\mathcal J(0,0)=9$, the apparent singularity at $\epsilon=0$
is removable. Taylor division gives $g=F_1-1+\epsilon F_2+
\epsilon^2F_3+O(\epsilon^3)$ uniformly near that point, with
\begin{align}
 F_1&=\frac6\theta+\frac{82\theta}{3\eta},\notag\\
 F_2&=\frac{10}{\theta^2}+\frac{1600}{9\eta}
                         +\frac{394\theta^2}{3\eta^2},\notag\\
 F_3&=\frac{38}{\theta^3}+\frac{14764}{27\eta\theta}
          +\frac{12304\theta}{9\eta^2}+
                           \frac{2330\theta^3}{3\eta^3}.
 \label{eq:app-fold-series-v2}
\end{align}
At $(12,656)$, $F_{1,\theta}=0$, $F_{1,\theta\theta}=1/144$,
$F_{1,\eta}=-1/1312$ and $F_2=3877/10086$. The Jacobian of
$(g,g_\theta)$ with respect to $(\eta,\theta)$ is invertible.
The implicit-function theorem therefore gives an analytic critical
branch $(\eta_*(\epsilon),\theta_*(\epsilon))$, whose finite-weight geometry appears in Fig.~\ref{fig:finite-budget-fold-v2}. Writing
$\eta_*=656+\gamma\epsilon+\zeta_{\rm fold}\epsilon^2+O(\epsilon^3)$ and
$\theta_*=12+a\epsilon+O(\epsilon^2)$, its coefficients follow from
\begin{equation}
 \begin{aligned}
 \gamma&=-F_2/F_{1,\eta},\qquad
 a=-\frac{\gamma F_{1,\theta\eta}+F_{2,\theta}}{F_{1,\theta\theta}},\\
 \zeta_{\rm fold}&=-\frac{1}{F_{1,\eta}}\biggl[
 \frac{a^2}{2}F_{1,\theta\theta}+a\gamma F_{1,\theta\eta}
       +\frac{\gamma^2}{2}F_{1,\eta\eta}\\
 &\hspace{20mm}+aF_{2,\theta}+\gamma F_{2,\eta}+F_3\biggr]_* .
 \end{aligned}
 \label{eq:app-fold-recursion-v2}
\end{equation}
All quantities on the right are evaluated at $(12,656)$.
Exact rational arithmetic gives
\begin{equation}
 \begin{aligned}
 H_*&=\frac{656}{\epsilon^2}+\frac{62032}{123\epsilon}
                  +\frac{48137852}{620289}+O(\epsilon),\\
 q_*&=\frac{12}{\epsilon}+\frac{26348}{5043}+O(\epsilon).
 \end{aligned}
 \label{eq:finite-budget-critical-v2}
\end{equation}
Appendix~\ref{v3-app-exact-global-reflection} proves that this exact
branch is the unrestricted critical design for sufficiently small
$\epsilon$. The strict target has the same critical budget as an
infimum and requires $H>H_*$.

The neutral uniform-introduction time also has a controlled constant
term. In the reflected kernel let $u=1/q$ and write $v=k_{\rm time}u$, where
$k_{\rm time}=q^2/Q$ remains in a fixed compact subset of $[0,\infty)$.
Then
\begin{equation}
 T_{\rm abs}^{\rm U}
 =q+\frac{41}{3}+\frac{2q^2}{Q}+O(q^{-1}).
 \label{eq:III-reflected-time-constant}
\end{equation}
To obtain it, eliminate the eighteen fast proper intervals from the
twenty-state neutral interval generator. With physical rates $P_{ij}/6$,
the remaining two aligned domains have the Schur-complement expansion
\begin{equation}
 \begin{aligned}
 \mathsf S(u)&=\frac u3 I_2-\frac{1+2k_{\rm time}}{3}u^2 I_2+O(u^3),\\
 b_{\rm eff}(u)&=(1+6u)\mathbf1+O(u^2).
 \end{aligned}
\end{equation}
Their mean remaining times are
$3/u+6k_{\rm time}+21+O(u)$. Substituting these times into the fast equations and
averaging over the six singleton states gives
Eq.~\eqref{eq:III-reflected-time-constant}. The fast killed block is
invertible at $u=0$, including $k_{\rm time}=0$, so the remainder is uniform on
the stated compact sets. At the two least contact-contrast designs this
becomes
\begin{equation}
 \begin{aligned}
 T_{\rm abs}^{\rm U}[w_*^{\rm U}]
 &=\frac{12}{\epsilon}+\frac{97483}{5043}+O(\epsilon),\\
 T_{\rm abs}^{\rm U}[w_*^{\rm b}]
 &=\frac9\epsilon+\frac{1567}{90}+O(\epsilon).
 \end{aligned}
 \label{eq:III-least-cost-time-constants}
\end{equation}
The superscripts on $w_*$ specify the introduction rule used to choose
the threshold-optimal contacts. Both times in this comparison start
from a uniformly introduced singleton.

Let $d=\eta-\eta_*(\epsilon)$. The parameter-dependent
quadratic normal form gives, uniformly for small $d\geq0$ and
$\epsilon\geq0$,
\begin{equation}
 \begin{gathered}
 \theta_\pm=\theta_*\pm c_f(\epsilon)\sqrt d+O(d),\\
 c_f(\epsilon)=
 \left[-\frac{2g_\eta}{g_{\theta\theta}}\right]_*^{1/2}
 =\frac3{\sqrt{41}}+O(\epsilon).
 \end{gathered}
 \label{eq:app-fold-root-v2}
\end{equation}
Division by $\epsilon$ gives Eq.~\eqref{eq:finite-budget-width-v2}.
The exact shared time optimizer follows the lower root by
Appendix~\ref{app:v4-exact-time}. In the full error coordinates of
Eq.~\eqref{eq:v3-main-error-coordinates}, the leading time is
controlled by the harmonic mean of the pin contacts,
\begin{equation}
 \frac{2q_Lq_R}{q_L+q_R}
 =\frac{\theta_*+\xi_s}{\epsilon}
  -\frac{\xi_a^2}{\epsilon(\theta_*+\xi_s)}.
\end{equation}
This explains why the symmetric fold displacement determines the
$\epsilon^{-1/2}$ time shift, while imbalance enters quadratically.
 Since the reflected lower-root design has physical times
$T_{\rm abs}=q+O(1)$ and $T_{\rm fix}=3q+O(1)$ uniformly in this neighborhood,
$d=\omega_{\rm lay}\epsilon+O(\epsilon^2)$ gives
Eq.~\eqref{eq:finite-budget-time-layer-v2}. At fixed positive $\omega_{\rm lay}$
the square-root time shift is of order $\epsilon^{-1/2}$. When $\omega_{\rm lay}=O(\epsilon)$ it
is $O(1)$, the order of the omitted remainder, and the exact lower root of
Appendix~\ref{app:v4-exact-time} determines the optimal times there, including at the fold itself. The positive constant shift in
Eq.~\eqref{eq:finite-budget-critical-v2} means even
$656/\epsilon^2+\gamma/\epsilon$ remains insufficient.

There is also a selected local version. Fix $\kappa>0$ in a compact
interval inside the positive-fecundity domain for either map in
Sec.~\ref{b:sec:model}. Analyticity of the fast and reduced
Poisson systems, together with the simple zero in
Eq.~\eqref{eq:selected-boundary-v2}, makes $r_{\rm crit}(u,v;\kappa)$
analytic near $(0,0)$. Replace $F_1$ by
$F_\kappa=A_\kappa/\theta+B_\kappa\theta/\eta$. Its critical point
is $(2A_\kappa,4A_\kappa B_\kappa)$ and
\begin{equation}
 (F_\kappa)_{\theta\theta,*}=\frac{1}{4A_\kappa^2},\qquad
 (F_\kappa)_{\eta,*}=-\frac{1}{8A_\kappa B_\kappa}.
 \label{eq:app-selected-fold-curvature-v2}
\end{equation}
Consequently an exact local selected fold exists, with
\begin{equation}
 \begin{aligned}
 H_{*,\kappa}&=4A_\kappa B_\kappa\epsilon^{-2}
                              +O_\kappa(\epsilon^{-1}),\\
 q_{*,\kappa}&=2A_\kappa\epsilon^{-1}+O_\kappa(1),\\
 q_\pm^{\rm sel}&=q_{*,\kappa}\pm
       [\sqrt{A_\kappa/B_\kappa}+O_\kappa(\epsilon)]
                 \sqrt{H-H_{*,\kappa}}\\
       &\hspace{10mm}+O_\kappa\bigl(\epsilon(H-H_{*,\kappa})\bigr).
 \end{aligned}
 \label{eq:app-selected-fold-v2}
\end{equation}
Here $H_{*,\kappa},q_{*,\kappa}$ locate the reflected selected fold, and $q_\pm^{\rm sel}$ are its two roots. The target is $r_{\rm crit}\leq1+\epsilon$, and the center is exact. This extends the tolerance calculation to finite
selection inside the controlled reflection hierarchy.
Appendix~\ref{v3-app-selected-global-localization} identifies this reflected fold with the least selected contrast $H_*(\epsilon,\kappa)$ at sufficiently small positive gap and intensity.
The $O_\kappa$ remainders in Eq.~\eqref{eq:app-selected-fold-v2} are uniform on compact admissible selection intervals. The endpoint discussion following the asymptotics of $A_\kappa$ and $B_\kappa$ in Appendix~\ref{app:selected-boundary-v2} explains the extra control required for a simultaneous hierarchy and selection-endpoint limit. The weak limit $A_0=6$, $B_0=82/3$
recovers the preceding curvatures and width.

\labelalias{v3-app-global-time-frontier}{app:v4-exact-time}
\labelalias{v3-eq-unrestricted-time-constraint}{eq:v4-time-constraint}
\labelalias{v3-eq-unrestricted-time-frontier}{eq:v4-time-lower-root}
\labelalias{v3-app-global-time-layer}{app:finite-budget-fold-v2}
\labelalias{v3-eq-global-time-layer}{eq:finite-budget-time-layer-v2}

\subsection{The four-cycle time reduction and joint optimum}\label{app:v10-c4-time}

This subsection derives the $C_4$ time entries of Table~\ref{tab:cost} and the $C_4$ row of Table~\ref{tab:III-time-frontiers}. For the four-cycle contacts $(w_{12},w_{23},w_{34},w_{41})=(1,x/t_4,h/t_4^3,y/t_4)$ with $t_4=H^{-1/3}$, sites 1 and 2 are pinned by the limiting
fast process. Its two mixed recurrent classes have stationary measures
uniform on binary configuration codes \(\{1,9,13\}\) and
\(\{14,6,2\}\), where a code is \(\sum_i2^{i-1}n_i\).
Let \(\mathsf A^{(4)}(t_4)=\mathsf A^{(4)}_0+t_4 \mathsf A^{(4)}_1+O(t_4^2)\) be the backward generator killed at
consensus, \(\mathsf E_4\) the two mixed-pinned-sector indicator columns, and
\(\mathsf V_4\) the two stationary rows. Direct evaluation gives
\begin{gather}
 \mathsf A^{(4)}_0\mathsf E_4=0,\quad \mathsf V_4\mathsf A^{(4)}_0=0,\quad \mathsf V_4\mathsf E_4=I_2,\notag\\
 \mathsf V_4\mathsf A^{(4)}_1\mathsf E_4=-\ell I_2,\quad
 \ell=\tfrac14(x^{-1}+y^{-1}),\notag\\
 (-\mathsf A^{(4)}(t_4))^{-1}=\frac{\mathsf E_4\mathsf V_4}{t_4\ell}+O(1).
 \label{eq:app-c4-resolvent}
\end{gather}
The fast nonzero eigenvalues are \(-1/4,-3/4,-1\), each with
multiplicity four; the two independent zero modes account for the
full multiplicity of the zero eigenvalue, which is therefore semisimple. Inverting
the fast block and taking its Schur complement proves the uniform
Laurent estimate on every compact positive scaling set
\cite[Sec.~3]{SlowAvrachenkovHaviv2004}. Physically, the weak pair aligns at
rates \(t_4/(4x)\) and \(t_4/(4y)\).

Uniform singleton introduction enters the two mixed sectors with masses
\(1/4,1/4\). The sources for total and successful first moments are
\(\one\) and the fixation committor, respectively. The neutral
fixation probability of this introduction rule is exactly \(1/4\).
Thus, with \(q_{4,\rm harm}=2xy/(x+y)\),
\begin{equation}
 T_{\rm abs}=q_{4,\rm harm}/t_4+O(1),\quad T_{\rm fix}=2q_{4,\rm harm}/t_4+O(1).
 \label{eq:app-c4-times}
\end{equation}
At the contrast optimum these give $T_{\rm abs}\sim24/\epsilon$ in
Table~\ref{tab:cost} and $T_{\rm fix}\sim48/\epsilon$ through the limiting ratio two.

For the fixed-budget time problem, set \(H=\eta\epsilon^{-3}\)
and require \(J>0\), \(R-3\le\epsilon\). Define
\(z^2=(x+y)^2/(4xy)\ge1\). The leading threshold coefficient is
\begin{equation}
 F=16/q_{4,\rm harm}+6z^2q_{4,\rm harm}^2/h\ge16/q_{4,\rm harm}+6q_{4,\rm harm}^2.
 \label{eq:app-c4-reduced-objective}
\end{equation}
Equation~\eqref{eq:app-c4-global-bound} localizes every feasible graph,
not only minimizers. It gives \(F<\eta^{1/3}\) exactly. The smallest
limiting \(q_{4,\rm harm}\) lies on the decreasing branch of
\(16/q_{4,\rm harm}+6q_{4,\rm harm}^2=\eta^{1/3}\). Setting \(\theta=\eta^{1/3}q_{4,\rm harm}\)
gives the $C_4$ row of Table~\ref{tab:III-time-frontiers}.
Equality in the limiting problem requires \(x=y\), \(h=1\).
For fixed \(\eta>10368\), let \(\theta_-\in(16,24)\) denote the lower root of \(16/\theta+6\theta^2/\eta=1\); the derivative there is
nonzero. An analytic correction of the reflection trial enforces the
exact target and proves separate optimal times
\(\mathcal T_{\rm abs}=\theta_-/\epsilon+O_\eta(1)\) and
\(\mathcal T_{\rm fix}=2\theta_-/\epsilon+O_\eta(1)\).

The time-only problem has no contrast cap, so its constant needs
estimates that hold at every scaled budget, including extreme pin asymmetry.
Put \(a=B^{-1}\), \(d=D^{-1}\), \(z=BD/C\), and
\(\varrho_4=a+d\). Equation~\eqref{eq:app-c4-global-bound} gives
\(R-3>8\varrho_4+6z\). The exact projection on the same fast classes is
 \begin{gather}
 \mathsf V_4\mathsf A^{(4)}\mathsf E_4=-\bar\ell I_2,  \label{eq:app-c4-exact-escape}\\
 \bar\ell=\frac14\left[
 \frac a{1+a}+\frac d{1+d}
 +\frac13\left(\frac{za}{1+za}+\frac{zd}{1+zd}\right)\right].\notag
 \end{gather}
The factor \(1/3\) is the probability of disagreement with a strong
neighbor in the fast stationary class. At fixed \(z\) in a bounded
interval, the generator is analytic in \(\varrho_4\), uniformly for
\(a/\varrho_4\in[0,1]\). Its fixed fast block is invertible and its
off-diagonal slow--fast blocks are \(O(\varrho_4)\). Since
\(\bar\ell\ge\varrho_4/[4(1+\varrho_4)]\), block inversion gives
\((-\mathsf A^{(4)})^{-1}=\mathsf E_4\mathsf V_4/\bar\ell+O(1)\), including extreme asymmetry.
The two sources then give \(T_{\rm abs}=(2\bar\ell)^{-1}+O(1)\) and
\(T_{\rm fix}=\bar\ell^{-1}+O(1)\). Using
\(\bar\ell\le\varrho_4(1+z/3)/4\) gives the lower constants
\(16/\epsilon\) and \(32/\epsilon\).
The trial \((1,q,q^4,q)\), with \(q=16/\epsilon+5\), has
\(R-3=\epsilon-5\epsilon^2/128+O(\epsilon^3)\), and attains
the matching upper bounds with \(O(1)\) remainders. Together they
determine the leading infima as \(\epsilon\to0\), with bounded
additive errors. This comparison fixes the times only up to bounded terms, and the
uncapped normalized weight domain is noncompact. Attainment of the infima at fixed
finite gap therefore remains open; settling it requires minimizing the exact
interval-resolvent times over that domain.

\subsection{Five-site time control and joint optimization}\label{app:v10-p5-time}

This subsection derives the $P_5$ time entries of Table~\ref{tab:cost} and the $P_5$ row of Table~\ref{tab:III-time-frontiers}, using the coordinates $u,e,w$ of Eq.~\eqref{eq:app-p5-coordinates} and the coefficients $a_5,b_5,c_5$ of Appendix~\ref{app:III-p5-resource}. The exact five-site kernel is regular in \((u,e,v_5)\), with
\(v_5=q/Q=ew\). At the origin its four fast closed classes are
\((l,l,r,r,r)\), \(l,r\in\{0,1\}\); the remaining 28 states
have a bounded transient inverse. The effective rows leaving either
mixed class have an exact factor \(u\). After division by \(u\),
both neutral total exit rates tend to \(1/10\).
Uniform singleton introduction reaches each mixed class with probability
\(1/5\). Either left-dimer singleton establishes with probability
\(1/2\), while only the protected right singleton survives the fast
layer. The remaining mass \(3/5\) rapidly reaches all defection.
Only the right-cooperative mixed class fixes cooperation at this corner.
These facts give mean absorption \((2/5)(10/u)\) and successful
first moment \((1/5)(10/u)\); dividing the latter by the exact
neutral fixation probability \(1/5\) gives
\begin{equation}
 T_{\rm abs}=q[4+O(u+e+v_5)],\quad
 T_{\rm fix}=q[10+O(u+e+v_5)].
 \label{eq:app-p5-times}
\end{equation}
The bounded fast inverse, the exact factor \(u\), and the positive
limiting mixed-class exit rates make the scaled Poisson solutions
analytic on a fixed neighborhood. The errors are therefore uniform
without fixing ratios of the small coordinates. In particular, the
global threshold constraint forces \(u,e,w=O(\epsilon)\),
\(v_5=O(\epsilon^2)\), throughout its admissible domain.

For \(H=\eta\epsilon^{-3}\), balancing the last two terms of
Eq.~\eqref{eq:p5-cost} at fixed \(q\) gives
\begin{equation}
 R-7>\frac{a_5}{q}+2\sqrt{b_5c_5q/H}.
 \label{eq:app-p5-fixed-q}
\end{equation}
Set \(\theta_q=4\epsilon q\). Every feasible graph obeys
\begin{equation}
 1>\frac{166}{\theta_q}
       +\sqrt{\frac{1328\theta_q}{3\eta}},\qquad
 \theta_q>166.
 \label{eq:app-p5-feasibility}
\end{equation}
Squaring with the positive remaining side produces the $P_5$ row of Table~\ref{tab:III-time-frontiers}. Its derivative is
\((1328/3)\theta^2(\theta-498)/(\theta-166)^3\), so its minimum
is \(496008\) at \(\theta=498\). For fixed larger \(\eta\),
let \(\theta_-\in(166,498)\) denote the lower root. The uniform
time estimate proves the global leading lower bounds. For the upper
bound take \(Q=H\), \(b_5e=c_5w\), and
\(q=\theta_-/(4\epsilon)+O(1)\). In the variable
\(x=\epsilon q\), the leading threshold function
\(a_5/x+2\sqrt{b_5c_5x/\eta}\) has nonzero derivative at this
root. Its analytic correction therefore enforces the exact target,
giving separate infima \(\theta_-/\epsilon+O_\eta(1)\) and
\(5\theta_-/(2\epsilon)+O_\eta(1)\).
At the least-contrast design the limiting root is 498, giving
$T_{\rm abs}\sim498/\epsilon$ in Table~\ref{tab:cost} and, through the ratio $5/2$, $T_{\rm fix}\sim1245/\epsilon$.

Without a contrast bound, Eq.~\eqref{eq:p5-cost} gives
\(q>a_5/\epsilon\). Equation~\eqref{eq:app-p5-times} then gives
the lower leading constants 166 and 415. A matching strictly feasible
trial is \(q=a_5/\epsilon+7\), \(e=w=\epsilon^3\), for which
\begin{equation}
 R=7+\epsilon-\frac{4537}{75779}\epsilon^2+O(\epsilon^3).
 \label{eq:app-p5-time-trial}
\end{equation}
Its time corrections are \(O(1)\), proving the time-only entries.
For both graphs the leading endpoint budgets
$10368\epsilon^{-3}$ and $496008\epsilon^{-3}$ are infeasible at
finite \(\epsilon\); the joint-frontier errors hold at each fixed
scaled budget above that endpoint. The analytic correction of each trial needs a nonzero derivative at the lower root, and this
derivative vanishes where the two roots merge at the endpoint. Uniform control up to the
endpoint requires higher-order feasibility and time estimates through this fold,
as for $P_6$ in Appendices~\ref{app:finite-budget-fold-v2} and \ref{app:v4-exact-time}. Selected-time optimization on $C_4$ and $P_5$ requires a
fecundity-dependent interval operator, the analogue of $L_{\rm int}^{\sigma}$ in
Eq.~\eqref{eq:v7-selected-interval-times}, which Appendix~\ref{app:v7-selected-time} analyzes for $P_6$.

\subsection{Exact finite-budget optimization of absorption times}
\label{app:v4-exact-time}

We first prove the neutral-time result and then extend the same constrained minimum to the four selected absorption objectives in Proposition~\ref{prop:v7-selected-time}.
The bounded-cap part of Proposition~\ref{prop:v10-allbudget} concerns the same exact finite
neutral chain for both objectives. Its proof needs derivative control
through the moving fold in addition to convergence of the leading times.
In the neutral proof, all weights are positive and the initial
distribution is a uniform cooperator singleton. The constraint is the closed weak target
$(1+\epsilon)J-K\ge0$, which implies $J>0$ because $K>0$.
At fixed contrast cap, normalize the minimum weight to one. The
resulting domain is compact, the target inequality is closed, and both
interval-resolvent times are continuous there. A feasible minimum
therefore exists before its structure is determined below.

\subsubsection{An interval resolvent with the physical clock}
On a nearest-neighbor path a singleton's descendants remain a contiguous
interval $[l,r]$. There are $N(N+1)/2-1$ proper nonempty intervals,
twenty for $P_6$. When $l>1$, expansion to $[l-1,r]$ occurs at rate
$w_{l-1,l}/(Ns_l)$ and contraction at rate
$w_{l-1,l}/(Ns_{l-1})$. The right rates are analogous, with source $r$
for expansion and source $r+1$ for contraction. Missing exterior
boundaries give no such event; singleton contractions both lead to
extinction and add. These rates keep the unit global attempt clock.

Let $L_{\rm int}=-\mathcal A_{{\rm int},{\rm mix}}$ be minus the
physical backward generator restricted to proper nonempty intervals.
Its inverse is entrywise nonnegative. With $\mu$ the uniform singleton
row and $h([l,r])=\sum_{i=l}^r\pi_i$, the exact times are
\begin{equation}
 T_{\rm abs}=\mu L_{\rm int}^{-1} \mathbf1,\qquad
 T_{\rm fix}=N\mu L_{\rm int}^{-1}h.
 \label{eq:v4-interval-times}
\end{equation}
For the conditional expression use
$\E[\tau\mathbf1_{\{C\text{ fixes}\}}]=\E\!\left[\int_0^\tau h_C(n(t))\,dt\right]$
and $\mu h=1/N$. Together, the distinguished algebraic threshold root and
this finite rational inverse give an exact specification of the optimum.

Fix $M_{\rm b}>656$ and $\eta=\epsilon^2H\le M_{\rm b}$. The global inequality in
Eq.~\eqref{v3-eq-p6-global-bound} forces every feasible design into a compact
positive chart
\begin{equation}
 \begin{gathered}
 \eta_L=\epsilon^2Q_L,\quad \eta_R=\epsilon^2Q_R,\\
 \theta=\frac{2\epsilon q_Lq_R}{q_L+q_R},\quad p=\frac {q_R}{q_L+q_R}.
 \end{gathered}
 \label{eq:v4-time-chart}
\end{equation}
The central edge becomes the unique minimum. At $\epsilon=0$ the
interval process has two slow aligned mixed states $[1,3]$ and $[4,6]$
and eighteen fast transient intervals. A uniformly positive sequence of
copies aligns the fast intervals with their pins, bounding the fast
inverse. The killed slow Schur complement is $\epsilon S$ with
$S(0)=I_2/(3\theta)$. The block inverse therefore makes
$\epsilon L_{\rm int}^{-1}$ analytic with two uniformly controlled
derivatives on the compact chart. Entrance mass is $1/3$ in total,
while the two success weights are $p$ and $1-p$. It follows that
 \begin{equation}
 \epsilon\,T_{\rm abs}\longrightarrow\theta,\qquad
 \epsilon\,T_{\rm fix}\longrightarrow3\theta.
 \label{eq:v4-time-leading}
\end{equation}
The limiting threshold constraint is $g_0\le0$, where
\begin{equation}
 g_0=\frac6\theta+
 \frac{\theta}{12p(1-p)}
 \left(\frac{53-24p}{\eta_L}+\frac{29+24p}{\eta_R}\right)-1.
 \label{eq:v4-time-constraint}
\end{equation}
 This derives the
needed $C^2$ control from the interval inverse, without differentiating an $O(1)$
remainder for a mean time.

\subsubsection{The lower branch away from the fold}
At fixed $\eta>656$, the limiting constraint in
Eq.~\eqref{eq:v4-time-constraint} decreases as either core increases, so
its smallest value at each $\theta$ has $\eta_L=\eta_R=\eta$. It then equals
$6/\theta+41\theta/[6\eta p(1-p)]-1$, minimized at $p=1/2$.
Because both leading times increase with $\theta$, their common global
minimum is the smaller root of this constraint,
\begin{equation}
 \theta_- =\frac{12}{1+\sqrt{1-656/\eta}},\qquad
 6<\theta_-<12.
 \label{eq:v4-time-lower-root}
\end{equation}
At this point $g_{0,\theta}=(\theta-12)/\theta^2<0$.
Both exact scaled times increase with $\theta$ nearby, so their target
face is active. Solving the exact constraint for that face gives
negative derivatives in both core weights, forcing $\eta_L=\eta_R=\eta$.
The constrained imbalance curvature is positive, since
$g_{0,pp}=656\theta/(3\eta)>0$ and
$\partial_p^2\theta_{\rm face}=-g_{0,pp}/g_{0,\theta}>0$.
Uniform $C^2$ convergence keeps these signs strict. Reversal makes the
unique imbalance minimum exactly $p=1/2$. Global localization and
convergence exclude other minima. This proves the exact lower-root prescription uniformly away from the
limiting fold. The simple reflected root obeys
$q_-=\theta_-(\eta)/\epsilon+O_\eta(1)$, hence the separate global
time minima are $\theta_-(\eta)/\epsilon+O_\eta(1)$ and
$3\theta_-(\eta)/\epsilon+O_\eta(1)$, respectively.

\subsubsection{A Lagrangian sufficiency proof through the exact fold}
Near $H_*(\epsilon)$ use the exact-centered coordinates in
Eq.~\eqref{eq:v3-main-error-coordinates}. Set $g_\epsilon=(R-1)/\epsilon-1$ and
$p_\epsilon=-\partial_{\eta_L}g_\epsilon|_*>0$, where
$\eta_L=\epsilon^2Q_L$ and the other three scaled contacts are fixed.
The star denotes the exact reflected critical design. Then
$\mathscr C_{\rm fold}=g_\epsilon/(2p_\epsilon)$ is the normalized constraint expanded in
Eq.~\eqref{v3-eq-full-tolerance-jet}, and feasibility is $\mathscr C_{\rm fold}\le0$.
Its pin Hessian is positive and its core-deficit derivatives are
positive on a fixed small convex box. For either objective
$f=\epsilon\, T_{\rm abs}$ or $\epsilon \,T_{\rm fix}$, its first two derivatives are
bounded and $f_s=\partial_{\xi_s}f$ stays bounded away from zero.

For $d>0$ let $z_-$ be the exact reflected lower-root point, with
$\xi_s=\epsilon(q_--q_*)$ and the other three errors zero.
There $\mathscr C_{{\rm fold},s}=\partial_{\xi_s}\mathscr C_{\rm fold}<0$ with magnitude comparable to $\sqrt d$. Set
\begin{equation}
 \Lambda_{\rm Lag}=-\frac{f_s(z_-)}{\mathscr C_{{\rm fold},s}(z_-)}>0,\quad
 \mathscr L_{\rm Lag}=f+\Lambda_{\rm Lag} \mathscr C_{\rm fold}.
 \label{eq:v4-fold-lagrangian}
\end{equation}
Since $\Lambda_{\rm Lag}\gtrsim d^{-1/2}$, for sufficiently small $d$ the
pin Hessian and the core-deficit derivatives of $\mathscr L_{\rm Lag}$ are
positive throughout the box. Reducing the core deficits first lowers
$\mathscr L_{\rm Lag}$; at zero deficits strict pin convexity gives a unique
minimum at $z_-$. Its symmetric derivative vanishes by construction,
and its antisymmetric derivative vanishes by reversal. Every feasible
point consequently obeys
\begin{equation}
 f=\mathscr L_{\rm Lag}-\Lambda_{\rm Lag} \mathscr C_{\rm fold}\ge\mathscr L_{\rm Lag}
 \ge\mathscr L_{\rm Lag}(z_-)=f(z_-).
 \label{eq:v4-fold-time-bound}
\end{equation}
Equality forces $z=z_-$. Global feasible-set exhaustion puts all
competitors in this box as $(\epsilon,d)\to(0,0)$.
At $d=0$ the exact critical design is the only feasible one. Combining
this neighborhood with compact budget intervals away from the fold
proves the bounded-cap part of Proposition~\ref{prop:v10-allbudget} for every fixed finite
$M_{\rm b}$. For $H>H_*$, strict feasible points approach the lower boundary,
so the same values are infima for the strict target.

The shared optimum applies to small weak-selection targets, bounded
$\epsilon^2H$, neutral times and uniform singleton introduction.
The next argument extends this compact chart to small positive selection.
Appendix~\ref{app:v10-allbudget} treats all larger neutral budgets, where
the optimum can lose reflection. The fixed-$\omega_{\rm lay}$ expansion
in Eq.~\eqref{eq:finite-budget-time-layer-v2}
follows from the exact lower root, which also determines the optimal times when $\omega_{\rm lay}=O(\epsilon)$ and at the fold itself.

\subsubsection{Stability of the optimizer under selection}
\label{app:v7-selected-time}
For an introduced type $\sigma\in\{C,D\}$, let
$L_{\rm int}^{\sigma}$ be its killed interval operator and
$h_{\sigma}$ its probability of fixation. The boundary rates now
include the actual source fecundity. For example, expansion across
the left boundary occurs at
$F_lP_{l,l-1}/\sum_iF_i$, while contraction occurs at
$F_{l-1}P_{l-1,l}/\sum_iF_i$. Payoffs and fecundities are evaluated
in that interval configuration, with the appropriate introduced
trait. With $\mu$ the uniform singleton row, the exact physical
means are
\begin{equation}
 \begin{aligned}
 T_{\rm abs}^{\sigma}&=\mu(L_{\rm int}^{\sigma})^{-1}\mathbf1,\\
 T_{\rm suc}^{\sigma}
 &=\frac{\mu(L_{\rm int}^{\sigma})^{-1}h_{\sigma}}
          {\mu h_{\sigma}}.
 \end{aligned}
 \label{eq:v7-selected-interval-times}
\end{equation}
In particular $T_{\rm abs}^C=T_{\rm abs}$ and $T_{\rm suc}^C=T_{\rm fix}$ in the notation
of the main text. The committor identity preceding
Eq.~\eqref{eq:v4-interval-times} proves the conditional expression
without a time change.

For bounded $\epsilon^2H$, the forest bound in
Eq.~\eqref{v3-eq-selected-global-forest-bound} and the global weak
inequality confine every selected-feasible design to a compact
positive version of the chart in Eq.~\eqref{eq:v4-time-chart}.
The central bridge is then its unique minimum edge. The eighteen
fast transient intervals have a uniformly bounded inverse on a
compact positive-fecundity set, and the two slow factors are exactly
of order $\epsilon$. At $\epsilon=0$, $b=c$ and both aligned
domains have zero payoff for every fixed admissible $\kappa$.
The limiting pin establishment probabilities are one. Block
inversion therefore makes all four scaled objectives analytic,
with two uniformly controlled derivatives and limits
\begin{equation}
 \epsilon T_{\rm abs}^C,\epsilon T_{\rm abs}^D\longrightarrow\theta,
 \qquad
 \epsilon T_{\rm suc}^C,\epsilon T_{\rm suc}^D
 \longrightarrow3\theta.
 \label{eq:v7-selected-time-limits}
\end{equation}
The divided selected constraint is a uniformly smooth perturbation
of $g_0$ in Eq.~\eqref{eq:v4-time-constraint} as $\kappa\to0$.

Away from the fold, each objective increases in $\theta$, so its
constraint face is active. The negative constrained core derivatives
and positive constrained imbalance curvature in the neutral proof
remain strict for sufficiently small $\epsilon,\kappa$.
Thus both cores saturate, and reversal forces the unique pin minimum
to have $p=1/2$. Compactness and the strict limiting minimum exclude
other minimizers. These statements hold uniformly on compact scaled
budgets separated from the limiting fold.

To include arbitrarily small surplus, center on the exact selected
critical design, whose global uniqueness, core saturation and
reflection follow from
Appendix~\ref{v3-app-selected-global-localization}.
Normalize the selected constraint as $\mathscr C_{\rm fold}\le0$. Its pin Hessian and
inward core-deficit derivatives are positive on a fixed small box.
For any of the four scaled objectives $f$, the symmetric pin
derivative $f_s$ stays positive and its first two derivatives remain
bounded. These are the hypotheses used in the constrained-fold argument following Eq.~\eqref{eq:v4-fold-lagrangian}. For the selected surplus $d=\epsilon^2[H-H_*(\epsilon,\kappa)]$, the constraint derivative at the lower root again obeys $-\mathscr C_{{\rm fold},s}(z_-)\asymp\sqrt d$. Its positive multiplier therefore grows as $d^{-1/2}$. Equation~\eqref{eq:v4-fold-time-bound} places the unique minimum at that root for each of the four objectives. Global localization excludes competitors outside the box.
At zero surplus the critical design is the sole feasible point.
Combining this neighborhood with the compact intervals above it
proves Proposition~\ref{prop:v7-selected-time}.

Finally, substituting $q=\theta/\epsilon$ and
$Q=\eta/\epsilon^2$ into Eq.~\eqref{eq:selected-boundary-v2}
gives the leading constraint
$1=A_\kappa/\theta+B_\kappa\theta/\eta$.
Its lower root is Eq.~\eqref{eq:v7-selected-time-frontier}.
Uniform differentiability gives the stated $O(1)$ time errors away
from its fold.

\subsection{Neutral time optimization at every feasible finite budget}
\label{app:v10-allbudget}
\label{app:v10-all-budget-time}
The initial singleton is uniform, the passage times are neutral, and
$R\le1+\epsilon$ is a weak-response design constraint. Time is measured by
the rate-one Poisson process of global copying attempts, including null
attempts. We prove Proposition~\ref{prop:v10-allbudget} under these conventions.

\begin{proof}
Use contacts $(Q_L,q_L,1,q_R,Q_R)$. Retain the endpoint-uniform coordinates $a_L,a_R$ of Eq.~\eqref{eq:asym-coordinates-v2} and retain their harmonic pin-rate sum $\varrho$. Thus
\begin{equation}
\begin{gathered}
\varrho=q_L^{-1}+q_R^{-1},\qquad p=q_R/(q_L+q_R),\\
g_p=p(1-p),\qquad a_L=(q_L+q_R)/Q_L,\\ a_R=(q_L+q_R)/Q_R.
\end{gathered}
\end{equation}
Then $q_L=1/(\varrho p)$, $q_R=1/[\varrho(1-p)]$ and
$Q_L=1/(\varrho g_p a_L)$, $Q_R=1/(\varrho g_p a_R)$.
The global response inequality forces $m=\varrho+a_L+a_R=O(\epsilon)$ uniformly
in $0<p<1$. In these coordinates
\begin{equation}
\begin{gathered}
\mathcal R=R-1=3\varrho+c_L(p)a_L+c_R(p)a_R+O(m^2),\\
c_L=(53-24p)/6,\qquad c_R=(29+24p)/6.
\end{gathered}
\label{eq:v10-time-response-chart}
\end{equation}
The polynomial pullback in Eq.~\eqref{eq:III-endpoint-pullback} gives this response expansion and its derivatives uniformly through $p=0,1$. The amplitude calculation in Appendix~\ref{app:asymmetric-entry-v2} also removes the common factor $p(1-p)$ before taking a ratio. For the time equations, eliminating the 18
fast proper intervals and dividing each of the two aligned-interval rows
by $\varrho$ leaves the limiting positive killed matrix $I/6$.
Consequently the two means have the analytic form
\begin{equation}
\begin{gathered}
T_{\rm abs}=f_{\rm abs}(\varrho,a_L,a_R,p)/\varrho,\quad f_{\rm abs}(0,0,0,p)=2,\\
T_{\rm fix}=f_{\rm fix}(\varrho,a_L,a_R,p)/\varrho,\quad f_{\rm fix}(0,0,0,p)=6.
\end{gathered}
\label{eq:v10-time-analytic}
\end{equation}
These extensions and their derivatives hold on one neighborhood of the
entire closed interval $0\le p\le1$. To check conditional normalization,
the exact reproductive sum satisfies
\begin{align}
Z/\varrho={}&\frac p{1+\varrho p}+\frac{1-p}{1+\varrho(1-p)}\notag\\
&+p(1-p)a_L\left[1+\frac1{1+(1-p)a_L}\right]\notag\\
&+p(1-p)a_R\left[1+\frac1{1+p a_R}\right],
\end{align}
which equals one at the corner. The uniform expansions rest on the divided inverse and this
 normalization.

At fixed $a_L,a_R,p$, increasing $\varrho$ reduces the contacts, increases the
threshold and decreases either time. Thus a time optimum has active target.
For each objective separately, solve the target as $\varrho=\varrho_\epsilon(a_L,a_R,p)$ and denote the restricted time by $\mathscr T_\epsilon(a_L,a_R,p)$. In the derivative formula below, $f$ is the corresponding numerator $f_{\rm abs}$ or $f_{\rm fix}$ from Eq.~\eqref{eq:v10-time-analytic}. Direct differentiation gives
\begin{equation}
\begin{gathered}
(\mathscr T_\epsilon)_{a_L}=
\frac{\varrho f_{a_L}+(f-\varrho f_\varrho)\mathcal R_{a_L}/\mathcal R_\varrho}{\varrho^2}>0,\\
(\mathscr T_\epsilon)_{a_R}>0.
\end{gathered}
\label{eq:v10-time-leakage}
\end{equation}
The numerator has a strictly positive uniform limit. At ideal cores,
$a_L=a_R=0$, the divided pair and interval equations give
\begin{align}
\mathcal R(\varrho,0,0,p)
 &=3\varrho+(1+6g_p)\varrho^2\notag\\
 &\quad+(3+7g_p)\varrho^3+O(\varrho^4),\notag\\
T_{\rm abs}
 &=\frac2\varrho+\frac{44}{3}-4g_p\notag\\
 &\quad+\left(-\frac{37}{18}+6g_p+8g_p^2\right)\varrho+O(\varrho^2),\notag\\
T_{\rm fix}
 &=\frac6\varrho+24-12g_p\notag\\
 &\quad+(21+18g_p+24g_p^2)\varrho+O(\varrho^2).
\label{eq:v10-ideal-core-jets}
\end{align}
Solving the first line for the exact target and substituting in the last
two gives the restricted functions $F_\epsilon(p)=\mathscr T_\epsilon(0,0,p)$,
\begin{align}
F_{\rm abs}&=6/\epsilon+46/3+
\epsilon[-5/54+(8/3)g_p]+O(\epsilon^2),\\
F_{\rm fix}&=18/\epsilon+26+
\epsilon[79/9+8g_p]+O(\epsilon^2),
\label{eq:v10-ideal-time}
\end{align}
uniformly with derivatives. The coefficients follow by the finite pair and
interval recursions after the divisions above.

Set $\eta_H=(H\epsilon^2)^{-1}$. This is the reciprocal of the
scaled cap $\eta=H\epsilon^2$; with $\Lambda=H\epsilon^4$ it obeys
$\eta_H/\epsilon^2=\Lambda^{-1}$. First take
$0<\eta_H\le\eta_0$, with $\eta_0$ fixed and small.
A reflected exact-target trial with $Q_L=Q_R=H$ differs from its ideal time by
$O(\eta_H/\epsilon)$. Any competitive graph first has
$\varrho\asymp\epsilon$ by Eq.~\eqref{eq:v10-time-analytic}.
Integrating Eq.~\eqref{eq:v10-time-leakage} to ideal cores gives
\begin{equation}
\mathscr T_\epsilon(a_L,a_R,p)\ge F_\epsilon(p)+c(a_L+a_R)/\epsilon^2.
\end{equation}
Comparison with the trial and Eq.~\eqref{eq:v10-ideal-time} then forces
\begin{equation}
a_L+a_R\le C\epsilon(\eta_H+\epsilon^2),\qquad
\varrho=\epsilon/3+O(\epsilon\eta_H+\epsilon^2).
\label{eq:v10-time-localization}
\end{equation}
The estimates remain uniform for $0<p<1$, including approach to either endpoint.

Both outer contacts must saturate. If $Q_L<H$, increase $Q_L$ while keeping
$Q_R,p$ fixed and maintaining the exact target. Since
$\varrho\mathcal R_\varrho-a_L\mathcal R_{a_L}-a_R\mathcal R_{a_R}>0$ on the localized set,
\begin{equation}
\begin{gathered}
\varrho_{Q_L}=\frac{(a_L/Q_L)\mathcal R_{a_L}}
{\mathcal R_\varrho-(a_L\mathcal R_{a_L}+a_R\mathcal R_{a_R})/\varrho}>0,\\
(a_L)_{Q_L}=-a_L/Q_L-a_L\varrho_{Q_L}/\varrho<0,\\ (a_R)_{Q_L}=-a_R\varrho_{Q_L}/\varrho<0.
\end{gathered}
\end{equation}
Equation~\eqref{eq:v10-time-leakage} makes this a strict time improvement.
The pins decrease and the other core stays fixed, so the cap is respected.
Reflection gives the same argument for $Q_R$. Thus $Q_L=Q_R=H$ exactly.

After saturation put $z=\eta_H/g_p$, $\varrho=\epsilon \widehat\varrho$ and
$a_L=a_R=\epsilon z/\widehat\varrho$. Every global optimum has
$z=O(\eta_H+\epsilon^2)$ by Eq.~\eqref{eq:v10-time-localization}.
The divided target
\begin{equation}
\epsilon^{-1}\mathcal R(\epsilon \widehat\varrho,\epsilon z/\widehat\varrho,\epsilon z/\widehat\varrho,p)=1
\end{equation}
has a unique analytic root near $\widehat\varrho=1/3$.
At $\epsilon=0$ its competitive branch is
\begin{equation}
3\widehat\varrho+41z/(3\widehat\varrho)=1,\qquad \widehat\varrho_0(z)=\frac{1+\sqrt{1-164z}}6.
\end{equation}
Reflection invariance descends analytically to $g_p=p(1-p)$, including
both endpoints and $g_p=1/4$ by the even series in $p-1/2$.
Hence the exact rescaled restricted time is an analytic function
$\Theta(\epsilon,z,g_p)=\epsilon T$ on one neighborhood of
$\epsilon=z=0$, $0\le g_p\le1/4$. At $\epsilon=0$,
$\Theta_{\rm abs}=2/\widehat\varrho_0(z)$ and $\Theta_{\rm fix}=6/\widehat\varrho_0(z)$, independent of $g_p$.
In particular, $\Theta_{{\rm abs},z}(0,0,g_p)=246$ and $\Theta_{{\rm fix},z}(0,0,g_p)=738$ for every $0\le g_p\le1/4$.
Equation~\eqref{eq:v10-ideal-time} and analytic division give the exact form
\begin{equation}
\begin{gathered}
\Theta_g=\epsilon^2a_{\rm time}(\epsilon,g)+\epsilon zc(\epsilon,z,g),\\
a_{{\rm time},{\rm abs}}=8/3+O(\epsilon),\\
a_{{\rm time},{\rm fix}}=8+O(\epsilon),\\
\partial_g a_{\rm time}=O(\epsilon).
\end{gathered}
\end{equation}
Here $g$ abbreviates $g_p$ only in the derivative formulas. Set
$d_{\rm time}=\Theta_z-\epsilon gc$. Its absorption and successful-fixation values are $246+O(\epsilon+z)$ and $738+O(\epsilon+z)$, respectively, and hence positive. Its derivatives satisfy $(d_{\rm time})_g=O(\epsilon)$ and $(d_{\rm time})_z=O(1)$.

The exact reduced objective is
$L_{\epsilon,\eta_H}(g)=\Theta(\epsilon,\eta_H/g,g)$, whose stationary
condition is
\begin{equation}
\begin{gathered}
L'=\epsilon^2a_{\rm time}-\frac{\eta_H}{g^2}d_{\rm time}=0,\\
\frac{\eta_H}{\epsilon^2}
=\Xi_{\epsilon,\eta_H}(g)=\frac{g^2a_{\rm time}}{d_{\rm time}(\epsilon,\eta_H/g,g)}.
\end{gathered}
\label{eq:v10-time-invariant-proof}
\end{equation}
Its decisive derivative is
\begin{equation}
\begin{aligned}
\frac{g\Xi'}\Xi
&=2+\frac{g\partial_g a_{\rm time}}{a_{\rm time}}-\frac{g(d_{\rm time})_g}{d_{\rm time}}\\
&\quad+\frac{z(d_{\rm time})_z}{d_{\rm time}}
=2+O(\epsilon+z)>0.
\end{aligned}
\label{eq:v10-time-unique-branch}
\end{equation}
Choose a fixed small $z_0$ and then $\eta_0$ sufficiently small relative to $z_0$ and the constant in the localization estimate.
Localization puts every optimum in $z<z_0/2$ after decreasing the common
$\epsilon$ bound. The entire interval from such an optimum to $g=1/4$
stays in the proved chart. It is feasible because the ratios of its pins to the cap are $\epsilon z(1-p)/\widehat\varrho$ and $\epsilon zp/\widehat\varrho$, both below one.
Equation~\eqref{eq:v10-time-unique-branch} permits at most one interior
stationary $g$, where $L'$ changes from negative to positive.
Thus the global optimum is either the reflected endpoint or that unique
interior value of $g$, corresponding to two reversed designs.

The endpoint crossing satisfies
\begin{equation}
\eta_{H,c}(\epsilon)=
\epsilon^2\frac{a_{\rm time}(\epsilon,1/4)}{16d_{\rm time}(\epsilon,4\eta_{H,c},1/4)}
=\epsilon^2/1476+O(\epsilon^3).
\end{equation}
Its defining equation has derivative $1+O(\epsilon^2)$ in $\eta_H$,
so there is no other crossing. Above $\eta_{H,c}$ the reflected endpoint
minimizes the time; below it the unique interior root does.
At the crossing $L''(1/4)>0$. Since $g_p=(1-y^2)/4$ for
$y=(q_L-q_R)/(q_L+q_R)$, the quartic coefficient in $y$ is positive.
The endpoint derivative crosses transversely, proving a supercritical
pitchfork. Increasing the cap decreases $\eta_H$.
The two critical caps therefore satisfy
\begin{equation}
H_{{\rm c},\rm abs}=\epsilon^{-4}[1476+O(\epsilon)],\quad
H_{{\rm c},\rm fix}=\epsilon^{-4}[1476+O(\epsilon)] .
\end{equation}
The leading values agree because $a_{\rm time}$ and $d_{\rm time}$ for fixation are three times their absorption values at $\epsilon=z=0$; the $O(\epsilon)$ corrections depend on the two sets of functions separately.

Finally join this uniform large-budget argument to
the bounded-budget result proved in Appendix~\ref{app:v4-exact-time}.
Fixing $M=1/\eta_0$, that result covers every
feasible $H\epsilon^2\le M$, including its fold, and gives the reflected
optimizer with saturated cores. The present proof covers all
$H\epsilon^2\ge M$. Taking the smaller of their two fixed $\epsilon$
bounds proves the assertion for every feasible finite cap with a single
$\epsilon_0$. For the strict target $R<1+\epsilon$, the values are
infima approached from the admissible side whenever the cap exceeds
feasibility; at the feasibility cap that strict target is empty.
\end{proof}

\subsection{Uncapped infima and the extreme-budget asymptotics}
\label{app:v10-uncapped}
\label{app:v10-time-endpoint}
For neutral copying on $P_6$, uniform singleton introduction and the
target $R\le1+\epsilon$, the uncapped time infima at sufficiently small
positive $\epsilon$ are the exact expressions in
Eqs.~\eqref{eq:v10-uncapped-abs} and \eqref{eq:v10-uncapped-exact}. Positive finite contacts approach but
do not attain these values. At large finite cap the two reversal-related
optimizers approach the ideal-core endpoints, with the uniform
two-parameter time corrections derived below.

\begin{proof}
Without a cap, Eq.~\eqref{eq:v10-time-leakage} strictly improves the time
by lowering either leakage to zero. The ideal derivative satisfies
\begin{equation}
\frac{F_{\rm abs}'(p)}{1-2p}=\tfrac83\epsilon+O(\epsilon^2)>0,
\quad
\frac{F_{\rm fix}'(p)}{1-2p}=8\epsilon+O(\epsilon^2)>0,
\end{equation}
with the quotient continued analytically at $p=1/2$.
Thus the ideal endpoints $p=0,1$ are the exact infima at each sufficiently
small positive target gap. Denote these uncapped infima of $T_{\rm abs}$ and $T_{\rm fix}$ by $I_{\rm abs}(\epsilon)$ and $I_{\rm fix}(\epsilon)$, respectively. Their values are rational functions of one uniquely chosen algebraic root. Parameterize the boundary contact by $\ell$ and define
\begin{equation}
\mathscr R_\partial(\ell)=
\frac{2\ell^5+20\ell^4+74\ell^3+126\ell^2+100\ell+30}
{2\ell^5+14\ell^4+30\ell^3+16\ell^2-12\ell-9},
\end{equation}
and let $\ell_\epsilon$ solve $\mathscr R_\partial(\ell)=1+\epsilon$ on the
branch $\ell_\epsilon=3/\epsilon+O(1)$. The inverse coordinate
$\varrho=1/\ell$ is fixed by $\varrho(0)=0$, $\varrho'(0)=1/3$.
With
\begin{align}
\mathcal P_{\rm abs}(\ell)={}&648\ell^8+11124\ell^7+69228\ell^6\notag\\
&+209469\ell^5+338632\ell^4+303009\ell^3\notag\\
&+149283\ell^2+37923\ell+3960,\\
\mathcal P_{\rm fix}(\ell)={}&36\ell^5+294\ell^4+863\ell^3\notag\\
&+880\ell^2+373\ell+66,
\end{align}
the exact infima are
\begin{align}
I_{\rm abs}(\epsilon)&=
\frac{\mathcal P_{\rm abs}(\ell_\epsilon)}{3\ell_\epsilon(3\ell_\epsilon+2)(\ell_\epsilon+3)}\notag\\
&\quad\times\frac{1}{(\ell_\epsilon^2+5\ell_\epsilon+5)(36\ell_\epsilon^2+42\ell_\epsilon+11)},
\label{eq:v10-uncapped-abs}\\
I_{\rm fix}(\epsilon)&=
\frac{6\mathcal P_{\rm fix}(\ell_\epsilon)}{\ell_\epsilon(\ell_\epsilon+3)
(36\ell_\epsilon^2+42\ell_\epsilon+11)}.
\label{eq:v10-uncapped-exact}
\end{align}
Their leading values are $I_{\rm abs}=6/\epsilon+O(1)$ and
$I_{\rm fix}=18/\epsilon+O(1)$, since $\ell_\epsilon=3/\epsilon+O(1)$.
They are limits of actual positive connected graphs. At fixed small
$\epsilon$, take $p=a_L=a_R=t>0$ and solve the exact target for $\varrho(t)$;
then let $t\downarrow0$. Strict leakage improvement excludes finite
attainment. Canceling the common vanishing factors in the response ratio gives the
well-defined endpoint limit that determines this experiment.

The all-budget theorem controls the number of optimizer branches even at
extreme caps. Their approach to Eqs.~\eqref{eq:v10-uncapped-abs} and \eqref{eq:v10-uncapped-exact} follows
from a separate uniform chart. Put $\Lambda=H\epsilon^4$,
$s=\Lambda^{-1/2}$ and choose the branch $p\le1/2$.
From the exact ideal slope and positive leakage derivatives,
\begin{equation}
T-I(\epsilon)\ge c\epsilon p(1-p)+c(a_L+a_R)/\epsilon^2.
\end{equation}
A trial with $p=sz$ at fixed $z>0$, saturated cores and active target has excess
$O(\epsilon s)$. Every optimizer therefore has $p\le Cs$ and
$a_L+a_R\le C\epsilon^3s$. The cap bounds
$a_L,a_R\ge\epsilon^4s^2/[\varrho p(1-p)]$ give the converse $p\ge cs$.
Write $p=sz$, $Q_L=aH$, $Q_R=eH$. Then $z,a,e$ lie in fixed positive
compact intervals, with $a,e\le1$.

After solving the target for $\varrho=\epsilon\widehat\varrho$, the differences
\begin{equation}
\Psi_{\rm abs}=(T_{\rm abs}-I_{\rm abs})/(\epsilon s),\quad
\Psi_{\rm fix}=(T_{\rm fix}-I_{\rm fix})/(\epsilon s)
\end{equation}
extend analytically to $\epsilon=s=0$. To see the division directly, split the restricted time into its ideal-core value $F_\epsilon(p)$ and its finite-leakage correction. Equation~\eqref{eq:v10-ideal-time}, analyticity and reflection imply
$F_\epsilon(p)-F_\epsilon(0)=\epsilon g_p b_{\rm time}(\epsilon,g_p)$ with analytic $b_{\rm time}$.
In the present chart $g_p=sz(1-sz)$ and
\begin{equation}
 a_L=\frac{\epsilon^3s}{a\widehat\varrho z(1-sz)},\qquad
 a_R=\frac{\epsilon^3s}{e\widehat\varrho z(1-sz)}.
\end{equation}
After this substitution, Eq.~\eqref{eq:v10-time-leakage} makes $\epsilon^2\partial_{a_L}\mathscr T_\epsilon$ and $\epsilon^2\partial_{a_R}\mathscr T_\epsilon$ analytic on the exact target, also along the segment from zero leakage. Integration from zero leakage therefore makes the second difference $\epsilon s$ times an analytic function as well. Both differences have this factor, proving the claimed extension of $\Psi$.
Their leading values are
\begin{equation}
\Psi_{\rm abs}(0,0,z,a,e)=\tfrac83z+\frac{159}{az}+\frac{87}{ez},
\quad \Psi_{\rm fix}=3\Psi_{\rm abs}.
\end{equation}
Both outer derivatives are negative. At $a=e=1$ the remaining function
$8z/3+246/z$ is strictly convex, with minimum at $z_0=\sqrt{369}/2$
and value $8\sqrt{41}$. The persistence argument of
Appendix~\ref{app:v10-persistence}, applied on the chosen $p\le1/2$
branch, and the uniform differentiated remainders give
\begin{align}
p_*&=\frac{\sqrt{369}}{2\sqrt\Lambda}
[1+O(\epsilon)+O(\Lambda^{-1/2})],\\
\mathcal T_{\rm abs}&=I_{\rm abs}(\epsilon)+\frac\epsilon{\sqrt\Lambda}
[8\sqrt{41}+O(\epsilon)+O(\Lambda^{-1/2})],\\
\mathcal T_{\rm fix}&=I_{\rm fix}(\epsilon)+\frac\epsilon{\sqrt\Lambda}
[24\sqrt{41}+O(\epsilon)+O(\Lambda^{-1/2})].
\end{align}
The errors are uniform in the two-parameter corner, including $p_*\ll\epsilon$.
Reflection gives the other optimum. Appendix~\ref{app:v7-selected-time} treats sufficiently small selection at bounded $\epsilon^2H$, where every selected-feasible design lies in a compact positive chart with a uniformly bounded fast inverse. In the present corner $p_*\to0$, so extending the extreme-budget result to selection requires a selected interval reduction uniform as the pin fraction tends to zero.
\end{proof}

\subsection{The five-cycle joint frontier through the chain boundary}
\label{app:III-c5-frontier}
The two endpoints of the five-cycle frontier have different limiting
contact ratios. The global response comparison applies along the whole frontier. Minimizing the time also uses a time estimate that stays uniform as the chain ratio tends to zero or one; Eq.~\eqref{eq:III-c5-uniform-times} below gives it.

We prove the capped frontier and conditional-distribution claims of Proposition~\ref{prop:III-c5-waiting}. Throughout this proof $\eta>185625/512$ is fixed, the target is $J>0$, $R\le3/2+\epsilon$, and contrast is at most $H=\eta\epsilon^{-3}$.
\begin{proof}
Rotate a smallest contact to $1$ and write the remaining cyclic
contacts as $(B,C,D,E)$. Set $S=C+D$ and $\sigma=D/S$.
Equation~\eqref{eq:c5-global}, with
$a(t)=(23+6t-4t^2)/8\ge23/8$ for $0\le t\le1$, implies
$B,E>23/(8\epsilon)$ and $BE<(4/11)\epsilon S$.
The cap gives $S\le2\eta\epsilon^{-3}$, so each pin also has an
upper bound of order $\epsilon^{-1}$. The same comparison bounds
$S$ from below. Equation~\eqref{eq:v10-cycle-temperature-time}
then bounds $B/(B+C)$ and $E/(D+E)$ by $8\epsilon/23$.
Thus every feasible graph satisfies
\begin{equation}
 B,E\asymp_\eta\epsilon^{-1},\qquad
 S\asymp_\eta\epsilon^{-3},\qquad
 C,D\gtrsim_\eta\epsilon^{-2}.
 \label{eq:III-c5-time-localization}
\end{equation}
The pins remain comparable at fixed $\eta$, while $\sigma$ may tend
to zero or one.

Put $\upsilon_1=B^{-1}$, $\upsilon_2=E^{-1}$, $\ell_3=B/(B+C)$,
$\ell_5=E/(D+E)$ and $\varrho_{\rm pin}=\upsilon_1+\upsilon_2$. These four small coordinates
are $O_\eta(\epsilon)$ and $\varrho_{\rm pin}\asymp_\eta\epsilon$.
Their physical relations also give
\begin{equation}
 (1-\sigma)\ell_3\le B/S=O_\eta(\epsilon^2),\quad
 \sigma \ell_5\le E/S=O_\eta(\epsilon^2).
 \label{eq:III-c5-return-products}
\end{equation}
These products control excursions that leave a pin and reach the
opposite pin.

Consider the ancestral subset process killed at a singleton.
At $\upsilon_1=\upsilon_2=\ell_3=\ell_5=0$, the subset $\{1,2\}$ is its sole transient
closed class. The sequence of arrows
$3\to2$, $5\to1$, $4\to3$, $3\to2$ moves every other subset
into the pins. Each arrow has limiting rate $1/5$, independently
of $\sigma\in[0,1]$. A fixed time window therefore has a uniformly
positive probability of reaching $\{1,2\}$ or a singleton.
The other 25 transient rows have a uniformly bounded inverse;
continuity retains this bound for small positive coordinates.

For one fast fringe ancestor, let $h_i$ be the probability of
reaching pin 1 before pin 2. Its Dirichlet equations are
\begin{gather}
 (2-\sigma)h_3-(1-\sigma)h_4=0,\notag\\
 -h_3+2h_4-h_5=0,\qquad
 -\sigma h_4+(1+\sigma)h_5=1.
 \label{eq:III-c5-fringe-hitting}
\end{gather}
Their determinant is $3$ and their solution is
$h_3=(1-\sigma)/3$, $h_4=(2-\sigma)/3$,
$h_5=1-\sigma/3$, including both chain endpoints.
The Schur complement on $\{1,2\}$ consequently has killing coefficient
\begin{align}
 \lambda_{\rm slow}
 &=\frac{\upsilon_1+\upsilon_2}{5}
 +\frac{(1-\sigma)\ell_3+\sigma \ell_5}{15}
 +O_\eta(\epsilon^2)\notag\\
 &=\frac{\varrho_{\rm pin}}5+O_\eta(\epsilon^2).
 \label{eq:III-c5-slow-rate}
\end{align}
The first term is direct pin coalescence; the second comes from
pin excursions and their fast hitting probabilities.
The uniform fast inverse controls the quadratic remainder.
Starting from all five labels, entrance into the pin pair has
probability $1+O_\eta(\epsilon)$ and bounded mean entrance time.
The exact graphical identity in
Eq.~\eqref{eq:v10-success-coalescence}, proved in Appendix~\ref{app:v10-c5-time}, then gives
$T_{\rm fix}=5/\varrho_{\rm pin}+O_\eta(1)$.
For the unconditional mean the separate graphical identity gives
\begin{equation}
 T_{\rm abs}=\frac25T_{\rm fix}
 +\frac15\mathbb E\int_0^{\tau_{\rm coal}}
 (|\mathscr L_t|-2)_+\,dt.
 \label{eq:III-c5-time-difference}
\end{equation}
The integral has a uniformly bounded expectation by the same fast
hitting estimate. Hence, throughout the entire feasible set,
\begin{equation}
 T_{\rm abs}=\frac2{\varrho_{\rm pin}}+O_\eta(1),\qquad
 T_{\rm fix}=\frac5{\varrho_{\rm pin}}+O_\eta(1).
 \label{eq:III-c5-uniform-times}
\end{equation}

\paragraph*{The global variational envelope.}
Write $B=x/\epsilon$, $E=y/\epsilon$ and
$\theta=2xy/(x+y)$, $p=y/(x+y)$. Then
$x=\theta/(2p)$, $y=\theta/[2(1-p)]$ and
$\epsilon T_{\rm abs}=\theta+O_\eta(\epsilon)$.
Since $S\le H/\max(\sigma,1-\sigma)$, the global response comparison
implies
\begin{gather}
 \eta>\frac{11\theta^3}{16\mathscr D(p,\sigma;\theta)}, \label{eq:III-c5-reduced-bound}\\
 \mathscr D(p,\sigma;\theta)=
 \frac{p(1-p)[\theta-2pa(\sigma)-2(1-p)a(1-\sigma)]}
 {\max(\sigma,1-\sigma)}.\notag
\end{gather}
The bracket must be positive, which forces $\theta>23/4$.
Reflection permits $\sigma\le1/2$. With $z=1/2-\sigma$ and
$\mathfrak t_\theta=\theta-25/4$, the factor depending on $z$ is
\begin{equation}
 \frac{\mathfrak t_\theta+z(p-1/2)+z^2}{z+1/2}
 =z+p-1+\frac{\mathfrak t_\theta-p/2+1/2}{z+1/2}.
 \label{eq:III-c5-chain-extrema}
\end{equation}
It is convex when the numerator of the reciprocal term is
nonnegative and strictly increasing when that numerator is negative.
Its maximum is therefore at $z=0$ or $z=1/2$.
At the balanced chain, maximizing over $p$ gives $\mathfrak t_\theta/2$ whenever
this is positive. At the chain endpoint, the maximum of
$p(1-p)(\mathfrak t_\theta+p/2)$ occurs at $p=p_{\rm cyc}(\theta)$ in
Eq.~\eqref{eq:III-c5-frontier}. A negative balanced value cannot
compete with this positive endpoint value.

The two maxima agree once. If $p_c$ is the unique root of
$2p_c^3-4p_c^2+5p_c-2=0$, then
\begin{equation}
 \theta_c=\frac{25}{4}
 +\frac{p_c^2(1-p_c)}{1-2p_c+2p_c^2},\qquad
 \eta_c=\frac{11\theta_c^3}{8(\theta_c-25/4)}.
 \label{eq:III-c5-switch}
\end{equation}
The polynomial derivative $6p^2-8p+5$ is strictly positive.
On the endpoint branch, the derivative of the required cap has
the sign of $2\theta-75/4+3p_{\rm cyc}/2$, which is negative for
$23/4<\theta\le\theta_c$.
On the balanced branch its sign is that of $2\theta-75/4$.
Thus the required cap decreases strictly from infinity to
$185625/512$ as $\theta$ increases from $23/4$ to $75/8$.
Equations~\eqref{eq:III-c5-reduced-bound} and
\eqref{eq:III-c5-uniform-times} prove the lower limiting times.

\paragraph*{Attainment as a limit of positive contacts.}
For matching contacts, fix $x,y>0$ and $0<\sigma<1$, and take
\begin{equation}
 B=x/\epsilon,\quad E=y/\epsilon,\quad
 (C,D)=\frac{H}{\max(\sigma,1-\sigma)}(1-\sigma,\sigma).
 \label{eq:III-c5-frontier-trial}
\end{equation}
The divided ten-pair equations give the regular expansion
\begin{align}
 R-\frac32
 &=\epsilon\left[
 \frac{a(\sigma)}x+\frac{a(1-\sigma)}y
 +\frac{11\max(\sigma,1-\sigma)xy}{4\eta}\right]\notag\\
 &\quad+O_{x,y,\sigma,\eta}(\epsilon^2).
 \label{eq:III-c5-unequal-response}
\end{align}
After dividing its pin-pair row by $\epsilon$, the contractions
have leading terms $K=2\epsilon/(x+y)+O(\epsilon^2)$ and
$J=4\epsilon/[3(x+y)]+O(\epsilon^2)$. The next-order ratio has
coefficient $a(\sigma)/x+a(1-\sigma)/y+
11xy\max(\sigma,1-\sigma)/(4\eta)$, proving
Eq.~\eqref{eq:III-c5-unequal-response} for arbitrary positive
pin ratio and fixed interior chain ratio.

Choose $\theta>\theta_5(\eta)$ sufficiently close to it that
the reduced required cap is strictly below $\eta$.
On the balanced branch set $p=\sigma=1/2$.
On the chain-boundary branch choose a sufficiently small fixed
$\sigma>0$, depending on $\theta$ and $\eta$, and
$p=p_{\rm cyc}(\theta)$, preserving that strict margin.
With $x=\theta/(2p)$ and $y=\theta/[2(1-p)]$,
Eq.~\eqref{eq:III-c5-unequal-response} makes the positive trial
feasible for all sufficiently small $\epsilon$.
Equation~\eqref{eq:III-c5-uniform-times} gives its limiting times.
Letting $\theta\downarrow\theta_5(\eta)$ proves the upper limits.
Equation~\eqref{eq:III-c5-unequal-response} is proved at fixed interior chain ratio, so the construction first takes $\epsilon\to0$ at fixed positive $\sigma$ and then approaches the chain endpoint through these interior trials.

The lower and upper limits agree at every fixed scaled cap above feasibility. Their equality cases also determine the limiting contact geometry. Equation~\eqref{eq:III-c5-time-localization} confines $\epsilon B,\epsilon E,\epsilon^3S$ to fixed positive compact intervals, while $\sigma\in[0,1]$. Every asymptotically minimizing sequence therefore has a convergent subsequence. Saturation of Eq.~\eqref{eq:III-c5-reduced-bound} forces its limiting $(p,\sigma)$ to maximize $\mathscr D$ and saturates the cap inequality. Below $\eta_c$ the maximizer is $(1/2,1/2)$; above it the maximizer is $(p_{\rm cyc}(\theta_5),0)$ after reflection. At $\eta=\eta_c$, either limiting geometry is possible. The strict endpoint maximization in Eq.~\eqref{eq:III-c5-chain-extrema} and the unique access maximizer exclude other limits.
\paragraph*{Conditional distributions and all positive moments.}
By Eq.~\eqref{eq:III-c5-uniform-times}, the condition
$\epsilon T_{\rm abs}\to\bar\theta$ in
Proposition~\ref{prop:III-c5-waiting} is equivalent to
$2\epsilon/\varrho_{\rm pin}\to\bar\theta$.
Along such a sequence, taking the Laplace transform adds $s\epsilon I$ to the positive
Dirichlet matrix of the killed ancestral problem and changes its
slow Schur coefficient to
$\varrho_{\rm pin}/5+s\epsilon+O_\eta(\epsilon^2)$.
The entrance factor is $1+O_\eta(\epsilon)$. Thus the Laplace transform
of $\epsilon\tau$ conditioned on successful fixation tends to
$[1+(5\bar\theta/2)s]^{-1}$. This is the exponential distribution of
mean $5\bar\theta/2$. The limit controls moments as well as weak
convergence. Under the graphical coupling, coalescence from the full
site set dominates the coalescence time from any subset. The uniform
mean estimate therefore bounds every initial mean by
$C_\eta/\epsilon$. Markov's inequality and the Markov property in
windows $2C_\eta/\epsilon$ give
\begin{equation}
 \Pr(\epsilon\tau>2C_\eta n\mid C\text{ fixes})\le2^{-n},
 \quad n=0,1,2,\ldots .
\end{equation}
This common exponential tail ensures uniform integrability of every fixed positive power. Applying the same lower and upper frontier arguments then proves the moment infimum in Eq.~\eqref{eq:III-c5-moment-frontier}, over the positive-weight feasible set at fixed $\eta$.
\end{proof}

The proof above controls the response~\eqref{eq:III-c5-unequal-response} at fixed interior $\sigma$ and the times~\eqref{eq:III-c5-uniform-times} to $O_\eta(1)$. Determining the finite-gap geometry, and the shift of the branch change at $\eta_c$ in Eq.~\eqref{eq:III-c5-switch}, where the two branches have the same leading time, requires higher-order response and time estimates uniform as the chain ratio vanishes. The time estimates rest on neutral graphical identities such as Eq.~\eqref{eq:v10-success-coalescence}, so finite-selection timing also requires retaining fecundity in the physical-time inverse, as in Appendix~\ref{app:v7-selected-time}.

\subsection{Full ancestral coalescence and five-cycle fixation}
\label{app:v10-c5-time}
\label{app:v10-c5-conditional}
For neutral copying on any finite connected support and uniform singleton
introduction, the absorption time conditioned on fixation has the
distribution of full ancestral coalescence, Eq.~\eqref{eq:v10-success-coalescence},
and the means obey $T_{\rm abs}\le T_{\rm fix}\le NT_{\rm abs}/2$.
For the five-cycle claims of Proposition~\ref{prop:III-c5-waiting}, this identity gives the finite-weight survival and moment bounds. We also construct their matching uncapped family and prove convergence of its conditional distribution and all positive moments.
Discrete global attempts have the corresponding
geometric survival comparison.

\begin{proof}
For neutral copying, draw the Poisson arrows first and choose the initial
mutant site independently and uniformly. Let $\mathscr L_t$ be the set
of ancestors of all current sites after tracing the arrows backwards
through a window of length $t$, and let $\tau_{\rm coal}$ be their full coalescence time.
Conditioned on the arrows, all current sites are mutant with probability
$1/N$ when $|\mathscr L_t|=1$, and with probability zero otherwise.
Since the eventual fixation probability is $1/N$,
\begin{equation}
\Pr(\tau\le t\mid\text{fixation})=\Pr(\tau_{\rm coal}\le t).
\label{eq:v10-success-coalescence}
\end{equation}
This graphical identity rests on the distributional reversal of the finite
Poisson-arrow window and does not require detailed balance of the absorbing configuration
chain. The unconditional identity is
\begin{equation}
\Pr(\tau>t)=N^{-1}\mathbb E[|\mathscr L_t|;\ |\mathscr L_t|\ge2],
\end{equation}
and gives $T_{\rm abs}\le T_{\rm fix}\le NT_{\rm abs}/2$.
The lower bound uses $|\mathscr L_t|\le N$ and the upper bound uses $|\mathscr L_t|\ge2$ before coalescence. A sharp conditional lower bound
follows from the ancestral meeting-time comparison below.

For the five-cycle choose the endpoints of its smallest contact as
sites 1 and 2, and put $\epsilon_5=R-3/2$. The global temperature
comparison in Eq.~\eqref{eq:v10-cycle-temperature-time} now relates the
threshold gap to their ancestral jump rate.

Trace the two labels initially at the endpoints of the smallest edge
within the same graphical construction. Full coalescence cannot precede
their meeting, and their meeting cannot precede their first jump.
The latter waiting time is exponential of rate
$(\vartheta_1+\vartheta_2)/5$. Equations
\eqref{eq:v10-success-coalescence} and
\eqref{eq:v10-cycle-temperature-time} therefore give, for
$0<\epsilon_5\le3/2$ and $t>0$,
\begin{equation}
\begin{gathered}
\Pr(\tau>t\mid\text{fixation})>
e^{-8\epsilon_5 t/115},\\
\mathbb E[\tau^m\mid\text{fixation}]>
\Gamma(m+1)\left(\frac{115}{8\epsilon_5}\right)^m\quad(m>0).
\end{gathered}
\label{eq:v10-c5-survival-bound}
\end{equation}
The second inequality follows by integrating
$mt^{m-1}\Pr(\tau>t)$, and both sides of the survival comparison equal
one at $t=0$.

Sharpness is achieved asymptotically by contacts
$(\varepsilon_{\rm cyc}^6,\varepsilon_{\rm cyc}^5,1,\varepsilon_{\rm cyc}^2,\varepsilon_{\rm cyc}^4)$, $\varepsilon_{\rm cyc}\downarrow0$.
These are the family in Appendix~\ref{app:c5-global} multiplied by
$\varepsilon_{\rm cyc}^6$, which leaves the copying process unchanged.
Write $\epsilon_5(\varepsilon_{\rm cyc})=R(\varepsilon_{\rm cyc})-3/2$.
Its expansion at the end of Appendix~\ref{app:c5-global} has positive
leading derivative $23/8$. It therefore selects an analytic
$\varepsilon_{\rm cyc}(\epsilon)$ with
$\epsilon_5(\varepsilon_{\rm cyc}(\epsilon))=\epsilon$. In the ancestral subset process at
$\varepsilon_{\rm cyc}=0$, sites 1 and 2 are absorbing roots. The other directions are
$3\to2$ or $4$, $4\to3$ or $5$, and $5\to1$. The only slow transient
ancestral subset is $\{1,2\}$, reached with unit probability from all
sites. Its direct coalescence rate is $\varepsilon_{\rm cyc}/5+O(\varepsilon_{\rm cyc}^2)$, while excursions have
rate $O(\varepsilon_{\rm cyc}^2)$. The remaining 25 transient subsets have an invertible
limiting killed matrix. Schur elimination and division of the slow row by
$\varepsilon_{\rm cyc}$ give a resolvent analytic in $(\varepsilon_{\rm cyc},s_{\rm L})$ near zero, with $s_{\rm L}\ge0$ the Laplace parameter and limiting transform
\begin{equation}
\mathbb E[\exp(-s_{\rm L}\varepsilon_{\rm cyc}\tau)\mid\text{fixation}]\longrightarrow(1+5s_{\rm L})^{-1}.
\end{equation}
Its derivatives control every integer moment. For a positive real moment,
a larger bounded scaled integer moment ensures uniform integrability.
Consequently $\epsilon_5(\varepsilon_{\rm cyc})\tau$ conditioned on fixation converges to an
exponential distribution of mean $115/8$, with all positive moments.
The unrestricted time-only bounds are therefore sharp in the form
\begin{equation}
\begin{gathered}
\inf T_{\rm fix}=\frac{115}{8\epsilon}+O(1),\\
\lim_{\epsilon\downarrow0}\epsilon^m
\inf\mathbb E[\tau^m\mid\text{fixation}]
=\Gamma(m+1)(115/8)^m.
\end{gathered}
\end{equation}
The infima range over positive cycles with $R\le3/2+\epsilon$.
The lower bound uses the first-jump waiting time of either ancestral label initially placed at an endpoint of the smallest edge. It matches the trial's mean through the leading $\epsilon^{-1}$ term, leaving a bounded difference. Resolving the finite-gap optima requires the full ancestral coalescence resolvent over positive cycles.

If time counts discrete global attempts, the first jump is geometric
on $\{1,2,\ldots\}$. With $p_{\epsilon_5}=8\epsilon_5/115$,
\begin{equation}
\Pr(\tau_{\rm steps}>n\mid\text{fixation})>(1-p_{\epsilon_5})^n,
\qquad n\ge1.
\end{equation}
The corresponding real positive moment lower bound is
$\sum_{n\ge0}[(n+1)^m-n^m](1-p_{\epsilon_5})^n$.
The mean agrees with the Poisson convention, while ordinary higher moments
differ at finite gap. Their leading scaled limits agree.
\end{proof}
\section{Selection, tolerance and the observation of an advantage}\label{app:v10-selection}
This appendix develops the tolerance, finite-selection and observation results of Secs.~\ref{sec:finite-budget-fold-v2} and \ref{sec:v3-selection-window}. The first subsection derives the exact feasible-error geometry and its volume. The positive-forest bound~\eqref{v3-eq-selected-global-forest-bound} then places selected promoters in the protected hierarchy at sufficiently small gap and fixed admissible selection. There, the establishment calculation determines $A_\kappa,B_\kappa$. The final subsection combines these inputs to control the selection window, experimental tolerance and terminal observation cost. It also treats finite surplus, preparation costs, rare introductions and the convergence radius. The tolerance and joint expansions use the exact critical branch as their center.
\subsection{Full parameter tolerance and its volume}
\label{v3-app-full-tolerance}
This subsection constructs the corner chart of Eq.~\eqref{eq:v4-morse-corner} and derives the feasible volume of Eq.~\eqref{eq:v3-main-volume}. Center on the exact critical branch, with $\theta_*=\epsilon q_*$ and $\eta_*=\epsilon^2H_*$. Use the surplus, nonnegative core deficits and pin coordinates of Eq.~\eqref{eq:v3-main-error-coordinates}, together with
$\theta_L=\epsilon q_L$, $\theta_R=\epsilon q_R$,
$\eta_L=\epsilon^2Q_L$, $\eta_R=\epsilon^2Q_R$. Let
$g_\epsilon=(R-1)/\epsilon-1$ and
$p_\epsilon=-\partial_{\eta_L}g_\epsilon|_*>0$,
with the other three scaled weights fixed in that derivative.
Exact differentiation gives
\begin{equation}
 \begin{aligned}
 \frac{g_\epsilon}{2p_\epsilon}
 &=-d+\frac{\alpha_L+\alpha_R}{2}+c_s\xi_s^2+c_a\xi_a^2\\
 &\quad+O(m_{\rm c}^2+m_{\rm c}m_{\rm p}+m_{\rm p}^3),\\
 c_s&=\frac{41}{9}-\frac{2104}{1107}\epsilon+O(\epsilon^2),\\
 c_a&=\frac{41}{9}-\frac{9605}{6642}\epsilon+O(\epsilon^2),
 \end{aligned}
 \label{v3-eq-full-tolerance-jet}
\end{equation}
where $m_{\rm c}=|d|+\alpha_L+\alpha_R$ and $m_{\rm p}=|\xi_s|+|\xi_a|$ are the core and pin scales.
At the reflected critical point write
$g_{LL}=\partial_{\theta_L}^2g_\epsilon$ and
$g_{LR}=\partial_{\theta_L}\partial_{\theta_R}g_\epsilon$.
Then $c_s=(g_{LL}+g_{LR})/(2p_\epsilon)$ and
$c_a=(g_{LL}-g_{LR})/(2p_\epsilon)$.
The derivatives follow the moving exact critical center. The pin Hessian is
positive, while the two inward core derivatives equal $1/2$ in this
normalization. For nearby fixed core parameters minimize over pins, giving
an analytic minimum $m(d,\alpha_L,\alpha_R)$. Taylor's integral Hessian
and its positive analytic square root put the pin part exactly in the form
$X_s^2+X_a^2$. The core contribution is
$m-m(d,0,0)=\alpha_L\mathfrak c_L(d,\alpha_L,\alpha_R)+\alpha_R\mathfrak c_R(d,\alpha_R)$,
where $\mathfrak c_L,\mathfrak c_R>0$ follow by integrating the respective core derivatives.
The changes $\widetilde\alpha_L=\alpha_L\mathfrak c_L$, $\widetilde\alpha_R=\alpha_R\mathfrak c_R$ are invertible and
preserve both positive faces. The exact local feasible set is consequently
\begin{equation}
 \begin{gathered}
 X_s^2+X_a^2+\widetilde\alpha_L+\widetilde\alpha_R\le r_{\rm tol,\epsilon}(d),\\
 \widetilde\alpha_L,\widetilde\alpha_R\ge0,\qquad r_{\rm tol,\epsilon}(0)=0,\quad r_{\rm tol,\epsilon}'(0)>0.
 \end{gathered}
 \label{v3-eq-exact-morse-corner}
\end{equation}
Global localization (Appendix~\ref{v3-app-global-p6}) and the unique limiting optimum ensure that every weak
feasible design at a sufficiently near-critical cap lies in this chart.

For the bridge-fixed logarithmic measure $dV_{\log}$ of Sec.~\ref{sec:finite-budget-fold-v2}, put
$\mathfrak r=(H-H_*)/H_*$. Integration of the leading pin ellipse and core
triangle gives
\begin{equation}
 \begin{aligned}
 \mathcal V_{\log}
 &=\frac{4\pi H_*}{3q_*^2\sqrt{c_sc_a}}\mathfrak r^3[1+O(\mathfrak r)]\\
 &=\frac{4\pi}{3}\left[1+\frac{16057}{60516}\epsilon+O(\epsilon^2)\right]
       \mathfrak r^3[1+O(\mathfrak r)].
 \end{aligned}
 \label{v3-eq-full-tolerance-volume}
\end{equation}
The leading physical volume is $4\pi(H-H_*)^3/(3\sqrt{c_sc_a})$; the
logarithmic density divides by $H_*^2q_*^2$. Odd pin terms in the analytic
Jacobian integrate to zero, giving relative $O(\mathfrak r)$ error.
The resulting volume measures tolerance in the four contact coordinates
under the specified logarithmic measure.

\subsection{Positive forests globalize the selected neighborhood}
\label{v3-app-selected-global-localization}
This subsection proves the positive-reward decomposition of Eq.~\eqref{eq:v4-positive-rewards} and the forest comparison that, at small $\epsilon$ and $\kappa$, confines every competitive selected design to one analytic chart. For $b/c=1+\epsilon$ and $\kappa=\delta c>0$, with $\kappa<1$
for linear fecundity, use the unnormalized backward harmonic
generator $\widetilde{\mathsf A}_\kappa=\sum_{ij}F_iP_{ij}(\mathrm{copy}_{ij}-I)$.
Here $(\mathrm{copy}_{ij}g)(n)=g(n^{j\leftarrow i})$.
On its 62 transient states set $L_\kappa=-\widetilde{\mathsf A}_{\kappa,\mathrm{mix}}$.
Let $\delta_x$ denote the unit point mass at configuration $x$ and set
$\omega_{\rm U}=\frac16\sum_i(\delta_{e_i}+\delta_{\boldsymbol1-e_i})$.
This paired introduction measure has total mass two. With the neutral harmonic function $h_0=\sum_i\pi_i n_i$, including consensus boundary values gives $\Delta=\omega_{\rm U} L_\kappa^{-1}\widetilde{\mathsf A}_\kappa h_0$.
Orient each discordant edge from its cooperative endpoint $i$ to its defective endpoint $j$, so $n_i=1,n_j=0$, and put $\Xi_{ij}=1-(Pn)_i+(Pn)_j>0$. Write $F_C=F_i$, $F_D=F_j$ and likewise for the payoffs. Reciprocity gives
\begin{equation}
 \begin{gathered}\widetilde{\mathsf A}_\kappa h_0=\frac1Z\sum_{\rm discordant}\gamma_{ij}(F_C-F_D),\\
 (f_C-f_D)/c=\epsilon-(1+\epsilon)\Xi_{ij}.\end{gathered}
\end{equation}
For linear fecundity define $\mathfrak u=Z^{-1}\sum \gamma_{ij}$ and
$\mathfrak v=Z^{-1}\sum \gamma_{ij}\Xi_{ij}$ on discordant edges. Both rewards are positive,
and exactly
\begin{equation}
 \begin{gathered}
 \Delta/\kappa=\epsilon\,\mathcal U_\kappa-(1+\epsilon)\mathcal V_\kappa,\\
 \mathcal U_\kappa=\omega_{\rm U} L_\kappa^{-1}\mathfrak u,\qquad
 \mathcal V_\kappa=\omega_{\rm U} L_\kappa^{-1}\mathfrak v,\\
 \mathcal V_0/\mathcal U_0=1-J/K.
 \end{gathered}
\end{equation}
The last equality follows from
$\mathcal U_0=2K/(6Z)$, $\mathcal V_0=2(K-J)/(6Z)$.
For exponential fecundity insert the positive mean-value secant factor
$\Omega_{ij}=\int_0^1e^{\kappa[(1-t)f_D+t f_C]/c}dt$ into both rewards.

The tree and forest edges below are transitions between population
configurations. Identify the two absorbing consensuses as one boundary
vertex. Determinant terms are directed spanning trees rooted there;
inverse cofactors have one fewer transition edge and an additional
marked transient root. If unnormalized transition-rate ratios lie
in $[m,M]$, positive forest
formulas~\cite{AGNamGunawardena2025} make every inverse cofactor
homogeneous of degree 61. The common
degree 62 tree determinant cancels in $\mathcal V_\kappa/\mathcal U_\kappa$.
Positive reward and introduction weights therefore give a lower ratio bound
$(m/M)^{61}\mathcal V_0/\mathcal U_0$ for linear fecundity. Exponential
secant rewards lie between $m$ and $M$ times neutral rewards and add one
power. Using $-1\leq f_i/c\leq1+\epsilon$, selected cooperation requires
\begin{equation}
 \begin{gathered}
 \Gamma_{\rm lin}=\left[\frac{1+\kappa(1+\epsilon)}{1-\kappa}\right]^{61},
 \qquad \Gamma_{\exp}=e^{62\kappa(2+\epsilon)},\\
 1-J/K<\frac{\Gamma\epsilon}{1+\epsilon},\\
 R-1<\frac{\Gamma\epsilon}{1-(\Gamma-1)\epsilon}
 \quad\text{if }\Gamma\epsilon<1+\epsilon.
 \end{gathered}
 \label{v3-eq-selected-global-forest-bound}
\end{equation}
The side condition $\Gamma\epsilon<1+\epsilon$ makes the bound on $1-J/K$ smaller than one and therefore gives $J>0$. Zero advantage uses non-strict
inequalities. These bounds concern fixation probabilities, which are unchanged by the
row rescaling. Mean passage times retain the total-fecundity denominator
in their generator.

On fixed compact admissible selection intervals, this gives $R-1=O(\epsilon)$
and forces the complete weak protected hierarchy globally. For small
$\kappa$, it sharpens to $R-1\leq\epsilon[1+O(\kappa)]$. A reflected
selected trial has $\epsilon^2H_{\rm trial}=656+O(\kappa^2+\epsilon)$.
Every selected-feasible competitor with no greater contrast lies in a compact
positive scaling set and converges to $(12,12,656,656)$ as
$(\epsilon,\kappa)\to(0,0)$. Consequently all such competitors lie in one
fixed local analytic chart for $0<\epsilon<\epsilon_0$,
$0<\kappa<\kappa_0$, for some positive bounds.

The comparison can retain complement parity as well. With centered
fecundity, $\widehat f_i=f_i/c-\epsilon/2=\bar f_i/c$ satisfies
$|\widehat f_i|\leq L=1+\epsilon/2$.
The introduction measure and the neutral rewards are complement invariant, including
$\Xi_{ij}\mapsto\Xi_{ij}$. Pairing each degree-61 forest term with its complement
replaces its linear factors by
$\frac12[\prod_j(1+\lambda\widehat f_j)+\prod_j(1-\lambda\widehat f_j)]$.
Here $j$ labels the 61 forest transition factors, with the reproducing
source site's payoff evaluated in the transition's starting configuration.
Its multi-affine extrema on $|\widehat f_j|\leq L$ occur at box vertices. Balanced
31/30 signs give the minimum $(1-t^2)^{30}$, while equal signs give
$E_{61}(t)=[(1+t)^{61}+(1-t)^{61}]/2$, where $t=L|\lambda|<1$.
For the exponential map, the 61 centered forest payoffs and one secant payoff sum to a
quantity of absolute value at most $62L$; complement pairing gives a
hyperbolic cosine between one and $\cosh(62L\lambda)$. Thus the same inequality
holds with the sharper factors
\begin{equation}
 \widehat\Gamma_{\rm lin}
 =\frac{E_{61}(L\lambda)}{[1-(L\lambda)^2]^{30}},\qquad
 \widehat\Gamma_{\exp}=\cosh(62L\lambda).
 \label{v3-eq-selected-parity-bound}
\end{equation}
Here $\lambda=\kappa/(1+\epsilon\kappa/2)$ for linear fecundity and
$\lambda=\kappa$ for exponential fecundity. Both factors equal
$1+O(\kappa^2)$ uniformly in the weights, so selected success requires
$R-1\leq\epsilon[1+O(\kappa^2)]$ at small selection.

Selected analyticity in this full chart follows from 60
fast states and two slow mixed states. At $\epsilon=0$ each triple aligns
with its frozen pin; the fast killed inverse is bounded. The killed slow Schur
matrix divided by $\epsilon$ tends to $(1/\theta_L+1/\theta_R)I_2$, where
$\theta_L=\epsilon q_L$ and $\theta_R=\epsilon q_R$. Uniform singleton
entrance gives $\rho_C=\rho_D=1/6$, so $\Delta$ is divisible by $\epsilon$.
Complement parity also divides it by centered selection
$\lambda=\kappa/(1+\epsilon\kappa/2)$ for linear fecundity, or
$\lambda=\kappa$ for exponential fecundity.

At finite parameters, continuity on the trial-bounded compact weight
domain gives a minimum for the closed target $\Delta\ge0$; strict
advantage has the same critical contrast as an infimum. At a nearby fixed
cap admitting a nonnegative candidate, every global signal maximizer has
$\Delta\ge0$ and lies in the compact chart just proved. The regular signal
$6\Delta/(\epsilon\lambda)$ has positive core derivatives and a negative
pin Hessian near its unique weak maximum. Increasing a core to its cap
remains admissible. These checks verify the hypotheses of
Appendix~\ref{app:v10-persistence}, applied to the negative signal.
Both cores saturate, and uniqueness with reversal symmetry forces
$Q_L=Q_R=H$ and $q_L=q_R$. The local maximizing branch has a nonzero cap
derivative at its critical zero. That zero is the least selected cap,
because every feasible competitor at any no-larger trial-bounded cap
is already confined to the same chart. A positive candidate at a
controlled nearby cap similarly gives the global positive signal maximum.
Compactness and the strict derivative signs give positive
$\epsilon_0,\kappa_0$. At larger fixed admissible selection, the forest comparison still forces entry into the hierarchy. The saturation and reflection steps above use the positive core derivatives and negative pin Hessian of the regular signal near its unique weak maximum, at small $\epsilon$ and $\kappa$; extending those steps to larger fixed selection needs the same two signs for the selected objective there.

\subsection{The selected establishment correction}
\label{app:selected-boundary-v2}

This appendix derives the local finite-selection boundary in
Eq.~\eqref{eq:selected-boundary-v2}. We use the consecutive
path labeling of the main text, so the pins are sites 3 and 4.
Set \(x=r-1\), \(r=b/c\), \(\kappa=\delta c\),
\(u=1/q\), and \(v=q/Q\). Linear fecundity requires
\(0<\kappa<1\); exponential fecundity allows any fixed
finite \(\kappa>0\). Error estimates are uniform on compact
intervals inside these domains.

At \(x=u=v=0\), every aligned-state payoff vanishes.
The leading interdomain committor is neutral, although
the transient establishment dynamics is still selected.
On all configurations its harmonic extension is
\begin{equation}
 g_0(n)=\frac{n_3+n_4}{2}.
 \label{eq:app-boundary-gzero-v2}
\end{equation}
The limiting kernel \(P_0\) has unit entries in the
directions \(1\to2\), \(2\to1\), \(3\to2\),
\(4\to5\), \(5\to6\), and \(6\to5\).
Every fast event preserves the two pins, and the
other 60 states form four transient blocks of
size 15. Write $S=\{000000,111000,000111,111111\}$ for the
aligned configurations and $F$ for their complement. The two-bit
labels $00,10,01,11$ specify the left and right domain traits.

\subsubsection{Fast Poisson equations and initial establishment}

Let \(F_i^0(n)=\phi[\kappa(-n_i+(P_0n)_i)]\).
For the harmonic calculation it is convenient to
multiply each backward-generator row by its total
fecundity. The resulting fast operator is
\begin{equation}
 (\widetilde{\mathsf A}_0 f)(n)=
 \sum_{ij}(P_0)_{ij}F_i^0(n)
       [f(n^{j\leftarrow i})-f(n)].
 \label{eq:app-boundary-rescaled-generator-v2}
\end{equation}
The positive row factor preserves harmonic functions while changing the
waiting rate in each configuration. Physical times therefore use the
normalized generator below.

Let \(w_u,w_v\) vanish on the aligned sector and
describe the first derivative of the fast harmonic
extension at fixed aligned boundary values.
They solve
\begin{equation}
 \widetilde{\mathsf A}_{0,FF}w_\xi=-b_\xi,\qquad
 w_\xi|_S=0,\qquad \xi=u,v,
 \label{eq:app-boundary-fast-poisson-v2}
\end{equation}
with explicit sources
\begin{align}
 b_u(n)&=\tfrac12[F_3^0(n)-F_4^0(n)](n_3-n_4),\notag\\
 b_v(n)&=\tfrac12\{F_2^0(n)(n_2-n_3)
                         +F_5^0(n)(n_5-n_4)\}.
 \label{eq:app-boundary-sources-v2}
\end{align}
Only a pin-changing first-order event contributes.
A derivative of fecundity or of the global
normalization multiplies the zero increment of
\(g_0\) for each leading fast event and therefore
drops out of these sources. The equations can
be solved separately in the four fast blocks
using exact rational arithmetic.

The aligned rows of the physical normalized generator $\mathcal A$,
at \(x=0\), have the expansion
\(\mathcal A_{S,*}=u\mathsf R_0+u^2\mathsf R_u+O(u^3)\).
They are independent of \(v\), because internal
copies are silent in an aligned triple.
For either fecundity map, their two
configuration-changing rates are
\begin{equation}
 \begin{gathered}
 k_C=\frac u6-\frac{1+\kappa}{6}u^2+O(u^3),
 \\
 k_D=\frac u6+\frac{\kappa-1}{6}u^2+O(u^3).
 \end{gathered}
 \label{eq:app-boundary-cross-expansion-v2}
\end{equation}
Their exact expressions differ.
For linear fecundity they are
\(u[1\mp\kappa u/(1+u)]/[6(1+u)]\).
For exponential fecundity they are
\(u e^{\mp\zeta_{\rm br}}/[(1+u)(4+2\cosh\zeta_{\rm br})]\),
where \(\zeta_{\rm br}=\kappa u/(1+u)\).

After division by $u$, the leading killed mixed-state
physical generator is $-I/3$. Expanding its harmonic equation
therefore gives
\begin{equation}
 h_{M,u}=3(\mathsf R_ug_0+\mathsf R_0w_u)_M,\quad
 h_{M,v}=3(\mathsf R_0w_v)_M ,
 \label{eq:app-boundary-slow-derivatives-v2}
\end{equation}
where \(M=\{10,01\}\) denotes the mixed aligned states, and the derivatives
vanish at both consensus states. Let \(\mathsf E\)
be the limiting entrance matrix selecting the
initial pin assignment. The full derivative
is \(w_\xi+\mathsf E h_{S,\xi}\), so both the initial
transient and the changed competition
probability enter the observable.

To keep the introduction normalization explicit,
recall the paired measure
\begin{equation}
 \omega_{\rm U}=\frac16\sum_{i=1}^6
       (\delta_{e_i}+\delta_{\one-e_i}),\qquad
 \Delta=\omega_{\rm U}h_C-1.
 \label{eq:app-boundary-preparation-v2}
\end{equation}
This measure has total mass two; it combines
the two experiments used in the difference
of fixation probabilities.
In aligned-state order \(00,10,01,11\),
\(\omega_{\rm U}\mathsf E=(2/3,1/3,1/3,2/3)\).
Consequently
\(\partial_\xi\Delta=\omega_{\rm U}w_\xi+
\omega_{\rm U}\mathsf E h_{S,\xi}\).

\subsubsection{Coefficients and the local zero}

For linear fecundity, solving
Eq.~\eqref{eq:app-boundary-fast-poisson-v2}
over \(\mathbb Q(\kappa)\) separates the
contributions to \(-\partial_\xi\Delta\) as
\begin{align}
 c_{u,\mathrm{est}}
 &=\frac{2\kappa(1+\kappa^2)}{3(1-\kappa^2)^2},
 &c_{u,\mathrm{slow}}&=\frac{\kappa}{3}, \label{eq:app-boundary-decomposition-v2}\\
 c_{v,\mathrm{est}}
 &=\frac{\kappa(\kappa^4+2\kappa^2+29)}
 {(9-\kappa^2)(1-\kappa^2)^2},
 &c_{v,\mathrm{slow}}&=\frac{4\kappa}{3(1-\kappa^2)^2}. \notag
\end{align}
Multiplication of each summed coefficient by \(6/\kappa\) gives the linear-fecundity expressions in Eq.~\eqref{eq:selected-linear-coefficients-v2}.
Differentiating the local domain advantage in Eq.~\eqref{b:eq:p6-domain-observables} at $x=r-1=0$ gives $\partial_x\Delta_\infty=\kappa/6$, where $\Delta_\infty$ denotes that hierarchy-limit advantage for both fecundity maps. This fixes the establishment-response normalization.

For exponential fecundity, the fast dimensionless payoffs
at \(r=1\) are \(-1,0,1\), so the calculation
is rational in \(z=e^\kappa\), with the
slow-rate derivative also depending on
\(\kappa\). Writing \(C_\kappa=\cosh\kappa\),
the result is
\begin{equation}
\begin{aligned}
 A_\kappa^{\exp}
 &=2+\frac{2\sinh(2\kappa)(2C_\kappa-1)}{\kappa},
 \\
 B_\kappa^{\exp}
 &=\frac{2\sinh\kappa}{\kappa}
 \frac{(4C_\kappa-1)(32C_\kappa^3+8C_\kappa^2+1)}
      {8C_\kappa^2+1}.
\end{aligned}
 \label{eq:app-exponential-coefficients-v2}
\end{equation}
In particular,
\(c_{u,\mathrm{est}}^{\exp}
=\sinh(2\kappa)(2\cosh\kappa-1)/3\),
while \(c_{u,\mathrm{slow}}^{\exp}=\kappa/3\).
The establishment contribution is the source
of the large selected penalty. Both maps
have removable zero-selection limits
\(A_0=6\) and \(B_0=82/3\), in agreement
with the independent weak-selection
expansion in Eq.~\eqref{eq:p6-expansion}.

The fast inverse is analytic near the corner,
and the exact factor \(u\) in the aligned
rows leaves an invertible mixed-state
operator after division by \(u\).
The response therefore has a regular
expansion in \(x,u,v\). Its nonzero
\(x\) derivative \(\kappa/6\) gives a unique
local zero by the implicit-function
argument, proving
Eq.~\eqref{eq:selected-boundary-v2}.
The response is positive just above this
zero. The first-order boundary determines the sign when the separation from it
exceeds its quadratic error. Closer to that boundary, higher-order terms
or the full finite-state committor resolve the sign.

The dependence on selection also
delimits the uniformity. For the linear
map, \(A_\kappa\sim2/(1-\kappa)^2\)
and \(B_\kappa\sim8/(1-\kappa)^2\)
as \(\kappa\to1^-\).
For the exponential map,
\(A_\kappa^{\exp}\sim e^{3\kappa}/\kappa\)
and \(B_\kappa^{\exp}\sim4e^{3\kappa}/\kappa\)
as \(\kappa\to\infty\).
The remainder bounds are uniform on compact admissible selection intervals. Thus the coefficient limits above are taken after the hierarchy limit at fixed $\kappa$. A simultaneous hierarchy and selection-endpoint limit requires bounds on how the fast inverse and the remainders depend on $\kappa$ as $\kappa\to1^-$ for linear fecundity or $\kappa\to\infty$ for exponential fecundity.

For fixed sufficiently small $\kappa$ and large contrast $H$, put
$q=\widehat q\sqrt H$ and $Q=h_HH$, with $\widehat q,h_H$ in a compact positive set
and $h_H\le1$. The leading gap is
$H^{-1/2}(A_\kappa/\widehat q+B_\kappa\widehat q/h_H)$.
It is minimized at $h_H=1$ and $\widehat q=\sqrt{A_\kappa/B_\kappa}$.
Thus the hierarchy-sector minimum satisfies
\begin{equation}
 \begin{gathered}
 q_{H,\kappa}\sim\sqrt{A_\kappa/B_\kappa}\sqrt H,\\
 r_{\rm crit}-1=2\sqrt{A_\kappa B_\kappa}\,H^{-1/2}+O(H^{-1}).
 \end{gathered}
 \label{eq:selected-design-v2}
\end{equation}
The positive-forest comparison in
Appendix~\ref{v3-app-selected-global-localization} proves global
optimality for sufficiently small positive target gap and selection
intensity, without requiring a relation between their rates of approach
to zero.

The generic fast-block equations and their Poisson residuals were evaluated
exactly for both maps. An independent construction in consecutive path order
solves the full killed 62-state committor at 60 decimal digits, including
parameters on both sides of the leading boundary. These finite checks
corroborate the coefficients and clock conventions; local control follows
from the analytic inverse argument above.

\subsection{Joint selection, tolerance and observation}
\label{app:v3-joint-selection}

We derive the joint analytic expansion near the critical budget and
use it to determine the selection window, signal and observation cost.
Appendix~\ref{v3-app-selected-global-localization} places every globally
competitive design in this analytic neighborhood for sufficiently small
positive gap and selection intensity.

\subsubsection{Centered selection and complement symmetry}
\label{par:v3-joint-selection-coordinates}
Set $b/c=1+\epsilon$, $\kappa=\delta c$, and use the normalized
weights $(Q_L,q_L,1,q_R,Q_R)$. On the reflected branch write
$Q=\eta/\epsilon^2$, $q=\theta/\epsilon$. For linear and exponential
fecundity, use the centered intensity and inverse map of
Eq.~\eqref{eq:v3-main-centered-selection}.
\labelalias{eq:v3-joint-centered-selection}{eq:v3-main-centered-selection}
The centered dimensionless payoff
$\widehat f_i=f_i/c-\epsilon/2=\bar f_i/c$ changes sign under $n\mapsto\mathbf1-n$.
The linear fecundity is a common positive factor times
$1+\lambda\widehat f_i$; this factor cancels in the normalized rates.
For exponential fecundity the common exponential factor cancels in
the same way. Complementing the two introduction experiments therefore
maps $\lambda$ to $-\lambda$ and makes $\Delta$ exactly odd in
$\lambda$. Positive linear fecundity is required in every microscopic
configuration. The local regime stays strictly inside $0\le\kappa<1$.

Let $H_*(\epsilon),q_*(\epsilon)$ denote the exact weak critical
branch, with no truncation in $\epsilon$. Its scaled coordinates are
\begin{equation}
 \theta_*=12+\frac{26348}{5043}\epsilon+O(\epsilon^2),
 \quad
 \eta_*=656+\frac{62032}{123}\epsilon+O(\epsilon^2).
 \label{eq:v3-joint-selected-reference}
\end{equation}
We use $d=\epsilon^2(H-H_*)$, and the pin deviations and core deficits in Eq.~\eqref{eq:v3-main-error-coordinates}. Exact centering is needed whenever the surplus is smaller
than a neglected correction to $H_*$.

\subsubsection{Analyticity at the hierarchy corner}
\label{par:v3-joint-selected-analyticity}
The full-space argument in Appendix~\ref{v3-app-selected-global-localization}
makes
\begin{equation}
 G=\frac{6\Delta}{\epsilon\lambda}
 \label{eq:v3-joint-normalized-signal}
\end{equation}
analytic and even in $\lambda$ on the asymmetric compact chart.
For coefficient evaluation, restrict to the twenty transient cooperative
intervals and their eleven reflection orbits. Defector singletons follow by complementation, $h_C(\mathbf1-n;\lambda)=1-h_C(n;-\lambda)$. Ten orbits are fast and one
mixed aligned orbit is slow. In this reduced space, let $F$ index the ten
fast orbits and $S$ the single mixed aligned orbit. Use centered
fecundities $\widehat F_i=1+\lambda\widehat f_i$ or
$\widehat F_i=\exp(\lambda\widehat f_i)$ and define the harmonic generator
$\widetilde{\mathsf A}^{\rm ctr}=\sum_{ij}\widehat F_iP_{ij}(\mathrm{copy}_{ij}-I)$.
The raw fecundities are $(1+\epsilon\kappa/2)\widehat F_i$ and
$e^{\epsilon\kappa/2}\widehat F_i$, respectively. These common positive
factors cancel in the normalized rates. The harmonic operator keeps
the generator sign, with nonpositive diagonal on physical parameters.
Eliminating its fast block gives the scalar slow coefficient
\begin{equation}
 \widetilde{\mathsf A}^{\rm ctr}_{\rm eff}=-\frac{2\epsilon}{\theta}+O(\epsilon^2).
 \label{eq:v3-joint-slow-divisor}
\end{equation}
This coefficient belongs to the denominator-removed harmonic operator.
Physical waiting times use the normalized generator. The total centered
fecundity of an aligned state tends to six at the corner, so the leading
physical coefficient is $-\epsilon/(3\theta)=-u/3$.

For a reproducible coefficient construction, expand
$\widetilde{\mathsf A}^{\rm ctr}=\sum_j\epsilon^j\mathsf A_j(\lambda)$.
Write $\mathfrak h_j$ for the coefficients of the committor in
$\epsilon$ at fixed $\lambda$, distinct from the selection
coefficients $h_m$ of Sec.~\ref{sec:v4-observable-algebra}. On the
reflected interval space $\mathfrak h_0=(n_3+n_4)/2$. Let
$\mathsf R_F=\mathsf A_{0,FF}^{-1}$ and let $H_F$
be the fast harmonic extension of the slow basis vector, extended
by one on that slow state and zero on the absorbing boundaries.
The coefficient recursion is
\begin{equation}
 \begin{aligned}
 W_{1,F}&=-\mathsf R_F(\mathsf A_1\mathfrak h_0)_F,\qquad W_{1,S}=0,\\
 a_1&=-\frac{(\mathsf A_1W_1+\mathsf A_2\mathfrak h_0)_S}{-2/\theta},\\
 \mathfrak h_1&=W_1+H_Fa_1,\\
 W_{2,F}&=-\mathsf R_F(\mathsf A_1\mathfrak h_1+\mathsf A_2\mathfrak h_0)_F,\quad W_{2,S}=0,\\
 a_2&=-\frac{(\mathsf A_1W_2+\mathsf A_2\mathfrak h_1+\mathsf A_3\mathfrak h_0)_S}{-2/\theta},\\
 \mathfrak h_2&=W_2+H_Fa_2.
 \end{aligned}
 \label{eq:v3-joint-selected-recursion}
\end{equation}
The additional slow solvability equation fixes the coefficient left
undetermined by the singular leading matrix. Taylor expansion to the required order in $\lambda$ gives the
coefficient identities below in the Laurent polynomial ring
$\mathbb Q[\theta^{\pm1},\eta^{\pm1}]$, which allows negative integer powers.

\subsubsection{Mixed hierarchy and selection coefficients}
\label{par:v3-joint-selected-jet}

Writing $G=g_{00}+\epsilon g_{10}
+\lambda^2(g_{02}+\epsilon g_{12})+O(\epsilon^2)+O(\lambda^4)$,
the common weak coefficients are
\begin{equation}
 \begin{aligned}
 g_{00}&=1-\frac6\theta-\frac{82\theta}{3\eta},\\
 g_{10}&=-\frac6\theta+\frac{26}{\theta^2}-\frac{8\theta}{\eta}
             +\frac{308}{9\eta}+\frac{262\theta^2}{3\eta^2}.
 \end{aligned}
 \label{eq:v3-joint-weak-jet}
\end{equation}
Table~\ref{tab:III-selected-response-coefficients} compares the map-dependent coefficients at orders $\lambda^2$ and $\epsilon\lambda^2$.
\begin{table*}[!htbp]
\caption{Map-dependent coefficients in $G=g_{00}+\epsilon g_{10}+\lambda^2(g_{02}+\epsilon g_{12})+O(\epsilon^2)+O(\lambda^4)$ for uniform introduction on the reflected six-site path. Here $\theta=\epsilon q$, $\eta=\epsilon^2Q$ and $\lambda$ is the centered selection variable; $Q=H$ on the core-saturated face. The common weak coefficients are in Eq.~\eqref{eq:v3-joint-weak-jet}; the recursion~\eqref{eq:v3-joint-selected-recursion} determines all four entries.}
\label{tab:III-selected-response-coefficients}
\centering\small
\begin{ruledtabular}
\begin{tabular}{lll}
Fecundity & Coefficient & Expression\\\hline\noalign{\vskip4pt}
Linear & $g_{02}^{\rm lin}$ & $\displaystyle -\frac{12}\theta-\frac{1570\theta}{27\eta}$\\[8pt]
Linear & $g_{12}^{\rm lin}$ & $\displaystyle -\frac{20}\theta+\frac{168}{\theta^2}
 -\frac{1699\theta}{27\eta}+\frac{8402}{27\eta}+\frac{12610\theta^2}{27\eta^2}$\\[8pt]
Exponential & $g_{02}^{\exp}$ & $\displaystyle -\frac{20}{3\theta}-\frac{967\theta}{27\eta}$\\[8pt]
Exponential & $g_{12}^{\exp}$ & $\displaystyle -\frac{12}\theta+\frac{358}{3\theta^2}
 -\frac{1733\theta}{54\eta}+\frac{5725}{27\eta}+\frac{8725\theta^2}{27\eta^2}$
\end{tabular}
\end{ruledtabular}
\end{table*}
The $g_{02}$ terms follow from the fixed-selection
boundary coefficients. The mixed terms $g_{12}$ add finite-hierarchy
control to the joint limit. These are exact coefficient identities obtained from
Eq.~\eqref{eq:v3-joint-selected-recursion}.

The two fold equations $G=0$ and
$\partial_\theta G=0$ have a nonzero Jacobian in $\eta,\theta$ at
$(\epsilon,\lambda^2,\theta,\eta)=(0,0,12,656)$. Their solution is analytic,
with selected critical budget
\begin{equation}
 \eta_{*,\rm sel}=\eta_*+
       [C_0+C_1\epsilon+O(\epsilon^2)]\lambda^2
       +[D_0+O(\epsilon)]\lambda^4+O(\lambda^6).
 \label{eq:v3-joint-selected-critical}
\end{equation}
Table~\ref{tab:III-selected-critical-coefficients} gives the coefficients for the two fecundity maps.

\begin{table}[tbp]
\caption{Exact coefficients in the selected critical-budget expansion~\eqref{eq:v3-joint-selected-critical}, for uniform introduction on the reflected six-site path. The centered selection variable is $\lambda$, and $\epsilon=b/c-1$ is the gap above the threshold floor $R_*=1$.}
\label{tab:III-selected-critical-coefficients}
\begin{ruledtabular}
\begin{tabular}{lccc}
Fecundity&$C_0$&$C_1$&$D_0$\\
\hline
Linear&$24368/9$&$421156528/136161$&$578192/81$\\
Exponential&$14296/9$&$66372124/45387$&$140542/81$
\end{tabular}
\end{ruledtabular}
\end{table}

The $D_0$ terms come from expanding the leading fixed-selection
budget $4A_\kappa B_\kappa$ at $\epsilon=0$, where $\kappa=\lambda$
for both fecundity maps. To obtain $C_1$ directly from the response coefficients, set
$D_g=\partial_\eta g_{00}$ and
$\mathcal V=(26348/5043)\partial_\theta+(62032/123)\partial_\eta$.
With all entries evaluated at $(12,656)$, the coefficients are 
\begin{equation}
 C_0=-\frac{g_{02}}{D_g},\quad
 C_1=-\frac{g_{12}+\mathcal Vg_{02}}{D_g}
 +\frac{g_{02}(\partial_\eta g_{10}+\mathcal VD_g)}{D_g^2}.
 \label{eq:v3-joint-mixed-coefficient-extraction}
\end{equation}

\subsubsection{The joint tolerance and selected boundary}
\label{par:v3-joint-joint-tolerance}
Retain the pin coordinates $\xi_s,\xi_a$ and core deficits
$\alpha_L,\alpha_R$ of Eq.~\eqref{eq:v3-main-error-coordinates}.
The local signal is
\begin{equation}
 \begin{aligned}
 \Delta=\frac{\epsilon\lambda}{7872}
 \biggl[&d-\frac{\alpha_L+\alpha_R}{2}-\frac{41}{9}(\xi_s^2+\xi_a^2)\\
 &-C_0\lambda^2+O(\epsilon\upsilon^2+\upsilon^3)\biggr],
 \end{aligned}
 \label{eq:v3-joint-full-selected-normal-form}
\end{equation}
where $d,\alpha_L,\alpha_R=O(\upsilon^2)$ and
$\xi_s,\xi_a,\lambda=O(\upsilon)$.
When the leading bracket is comparable to or smaller than the remainder bound,
the exact analytic fold and parameter-dependent Morse chart determine
the boundary. Complement symmetry forbids
a term linear in $\lambda$ in $G$, including a quadratic mixed
term $\lambda\xi_s$. The selected localization and optimization in
Appendix~\ref{v3-app-selected-global-localization} already prove
core saturation and exact reflection in this small-parameter window.
The present expansion quantifies the remaining signal and tolerance
around that optimizer.

The leading window edge, Eq.~\eqref{eq:v3-main-window}, is the zero of the contact-optimized signal derived in Appendix~\ref{par:v3-joint-signal-maximum}, whose remainder gives the relative error $O(\epsilon+d)$. For the subleading excess $H-H_*=\omega_{\rm lay}/\epsilon$, fixed $\omega_{\rm lay}>0$,
\begin{equation}
 \lambda_{\rm edge}^2=\frac{\omega_{\rm lay}}{C_0}\epsilon
 -\left(\frac{C_1\omega_{\rm lay}}{C_0^2}
       +\frac{D_0\omega_{\rm lay}^2}{C_0^3}\right)\epsilon^2
 +O_{\omega_{\rm lay}}(\epsilon^3).
 \label{eq:v3-joint-window-two-term}
\end{equation}
Recover $\kappa$ by the exact map in
Eq.~\eqref{eq:v3-joint-centered-selection} for linear fecundity.
After choosing $\lambda_0$ sufficiently small,
$\partial_{\lambda^2}\eta_{*,\rm sel}=C_0+O(\epsilon+\lambda^2)>0$
throughout $0\le\lambda<\lambda_0$. The selected critical budget
therefore crosses $\eta_*+d$ exactly once in that interval for
sufficiently small $d>0$, fixing the positive edge. At fixed leading budget,
selection also shifts the optimal pin coordinate by
$-94\lambda^2/123$ for linear and $-49\lambda^2/41$ for exponential
fecundity, up to $O(\epsilon\lambda^2+\lambda^4)$.
Along the moving selected critical budget the shifts are
$24\lambda^2$ and $40\lambda^2/3$. Since the weak optimum is
stationary, the improvement from retuning first affects the budget
at order $\lambda^4$.

The full design volume acquires the same selected penalty. With the
logarithmic design measure already specified and relative surplus
$\mathfrak r=(H-H_*)/H_*$, its leading fixed-selection value is
\begin{equation}
 \begin{gathered}
 \mathcal V_{\log}(\lambda)=\mathfrak r^3
 \left\{\frac{4\pi}{3}\left[1-\frac{C_0z_\lambda^2}{656}\right]_+^3+o(1)\right\},\\
 z_\lambda=\lambda/\sqrt{\mathfrak r}.
 \end{gathered}
 \label{eq:v3-joint-selected-volume}
\end{equation}
Here $[x]_+=\max(x,0)$. The additive limit is uniform for bounded $z_\lambda$ as
$\epsilon,\mathfrak r\to0$. Choose a fixed $\lambda_0>0$
inside the proved small-selection rectangle. The selected critical
budget increases with $\lambda^2$ there, so the feasible measure
vanishes beyond the shrinking local window. Integrating with
Lebesgue measure on $[0,\lambda_0]$ gives
\begin{equation}
 \int_0^{\lambda_0} \mathcal V_{\log}(\lambda)\,d\lambda
 \sim\frac{64\pi}{105}\sqrt{\frac{656}{C_0}}
                     \mathfrak r^{7/2}.
 \label{eq:v3-joint-selection-volume}
\end{equation}
Equations~\eqref{eq:v3-joint-selected-volume} and \eqref{eq:v3-joint-selection-volume} are joint leading limits as $\epsilon,\mathfrak r\to0$ in the local chart. Finite-$\epsilon$ prefactors depend on the exact local density and curvatures. At $\lambda=0$, the leading small-$\mathfrak r$ prefactor depends on the curvatures $c_s,c_a$ of Eq.~\eqref{v3-eq-full-tolerance-jet}. Its finite-$\epsilon$ correction factor is $1+16057\epsilon/60516+O(\epsilon^2)$ in Eq.~\eqref{v3-eq-full-tolerance-volume}. The extra half-power comes from integrating over selection
as one additional quadratic coordinate of the design measure.

\subsubsection{Contact errors and the probability of promotion}
\label{app:v7-gaussian-tolerance}
A geometric feasible volume becomes an experimental probability
once a distribution of contact errors is specified. Keep the two
cores exactly at the contrast cap and perturb the two optimized
pin contacts by independent, mean-zero Gaussian logarithmic errors
of variance $s_{\rm e}^2$. Since $\epsilon q_*\to12$, the coordinates
of Eq.~\eqref{eq:v3-main-error-coordinates} obey, to leading order,
\begin{equation}
 \xi_s,\xi_a\ \hbox{independent},\qquad
 \operatorname{Var}\xi_s=\operatorname{Var}\xi_a=72s_{\rm e}^2.
 \label{eq:v7-gaussian-pin-variance}
\end{equation}
The selected shift of the pin center is $O(\lambda^2+d)$, smaller
than the error width when $s_{\rm e}^2\asymp d$ and $\lambda^2=O(d)$.
For positive centered intensity $\lambda$, the leading favorable
condition is
$41(\xi_s^2+\xi_a^2)/9<d-C_0\lambda^2$.
The sum of the two squared normalized Gaussian coordinates has
tail $e^{-x/2}$, so
\begin{equation}
 \Pr(\Delta>0)
 =1-\exp\!\left[-\frac{d-C_0\lambda^2}{656s_{\rm e}^2}\right]+o(1)
 \label{eq:v7-gaussian-success}
\end{equation}
when $d-C_0\lambda^2>0$; the limiting probability is zero at a
nonpositive margin. This is an additive limit as
$\epsilon,d,s_{\rm e},\lambda\to0$, with $s_{\rm e}^2/d$ in a compact positive
interval and $\lambda^2/d$ bounded. The remainder in
Eq.~\eqref{eq:v3-joint-full-selected-normal-form}, divided by $d$,
tends to zero in probability, while the Gaussian tails outside the
local chart vanish. Thus the sharp leading sign test is legitimate
also at the zero-margin limit.

For the weak target use the divided response at $\lambda=0$.
With $\mathfrak r=(H-H_*)/H_*$ and $d=656\mathfrak r[1+o(1)]$,
\begin{equation}
 \Pr(R\le1+\epsilon)=1-e^{-\mathfrak r/s_{\rm e}^2}+o(1).
 \label{eq:v7-gaussian-weak-success}
\end{equation}
Solving Eq.~\eqref{eq:v7-gaussian-success} for fixed confidence $1-p_{\rm fail}$, with $d=656\mathfrak r[1+o(1)]$, gives Eq.~\eqref{eq:v7-gaussian-confidence}; its $\lambda=0$ case follows equally from Eq.~\eqref{eq:v7-gaussian-weak-success}.
This distribution keeps both cores at the cap and samples the two quadratic pin directions. Random core deficits enter the leading bracket of Eq.~\eqref{eq:v3-joint-full-selected-normal-form} linearly through $-(\alpha_L+\alpha_R)/2$. Their inclusion requires averaging the conditional promotion probability over a specified joint distribution supported on nonnegative deficits.

\subsubsection{Signal maximum and terminal observation cost}
\label{par:v3-joint-signal-maximum}
\label{par:v3-joint-detection}\label{app:v7-terminal}

After optimization of geometry in the controlled neighborhood, the leading signal is the cubic
$\epsilon\lambda(d-C_0\lambda^2)/7872$. Its derivative vanishes
at $\lambda^2=d/(3C_0)$, whereas the window edge has zero signal.
The optimized remainder is sharper than the remainder for a generic
perturbed contact geometry. The normalized signal $G$ is analytic and
even in $\lambda$. At the exact weak center its pin gradient vanishes
and its pin Hessian is negative definite. The implicit-function
theorem therefore makes the maximizing pin displacement analytic in
$(\epsilon,d,\lambda^2)$ and of order $d+\lambda^2$. After substituting
that displacement,
\begin{equation}
 \max_{\rm contacts}G
 =\frac{d-C_0\lambda^2}{1312}
  +O\!\left(\epsilon(d+\lambda^2)+(d+\lambda^2)^2\right),
\end{equation}
with the corresponding differentiated estimates in the controlled
neighborhood. On writing $\lambda=\sqrt d\,\ell$, the scaled advantage
$\Delta/(\epsilon d^{3/2})$ has a nondegenerate interior maximum with perturbation
$O(\epsilon+d)$. Consequently, with these relative errors,
\begin{equation}
 \begin{aligned}
 \lambda_{\rm peak}&=\sqrt{\frac{d}{3C_0}}[1+O(\epsilon+d)],\\
 \Delta_{\rm peak}&=\frac{\epsilon d^{3/2}}{11808\sqrt{3C_0}}
                       [1+O(\epsilon+d)].
 \end{aligned}
 \label{eq:v3-joint-signal-peak}
\end{equation}
At $d=\omega_{\rm lay}\epsilon$, the best signal in the controlled window is of order
$\epsilon^{5/2}$. Its coefficients are fixed by Eq.~\eqref{eq:v3-joint-selected-recursion}.

The peak fixes the advantage that the terminal experiments must resolve.
Let $h_i^+$ and $h_i^-$ be the all-cooperator absorption
probabilities after, respectively, a cooperator and a defector
singleton at site $i$. Then
\begin{equation}
 \Delta=\frac1N\sum_i(h_i^++h_i^-)-1.
 \label{eq:v7-stratified-estimand}
\end{equation}
Treat the $2N$ probabilities as independently unknown parameters.
For $m_i^{\pm}$ independent completed runs in each class, the
sample means define the unbiased estimate
$\widehat\Delta_{\rm str}=N^{-1}\sum_i
(\widehat h_i^++\widehat h_i^-)-1$.
Writing $\varsigma_{i,\pm}^2=h_i^{\pm}(1-h_i^{\pm})$ and denoting the
corresponding expected run durations by $t_i^{\pm}$, its exact
variance and expected aggregate duration are
\begin{equation}
 \operatorname{Var}(\widehat\Delta_{\rm str})
 =\frac1{N^2}\sum_{i,\pm}\frac{\varsigma_{i,\pm}^2}{m_i^{\pm}},
 \qquad
 \mathcal T=\sum_{i,\pm}m_i^{\pm}t_i^{\pm}.
 \label{eq:v7-stratified-variance}
\end{equation}
Applying Cauchy--Schwarz to the two sums gives Eq.~\eqref{eq:v7-stratified-time-bound} for general $N$, with equality at the continuous allocation $m_i^\pm\propto\sqrt{\varsigma_{i,\pm}^2/t_i^\pm}$. The diagonal Bernoulli
Fisher matrix has entries $m_i^{\pm}/\varsigma_{i,\pm}^2$, and the gradient
of the estimand has every entry $1/N$. Its inverse gives the same
bound for locally unbiased estimation; regular large-sample
estimators obey the corresponding local asymptotic bound. For fixed
class probabilities, regularity means that the limiting rescaled estimation
error is unchanged by parameter shifts on the inverse-square-root sample
scale. For a joint small-$\epsilon$ limit, take sequences on which all four pin class counts diverge and require regularity under local perturbations of each pin probability on its own inverse-square-root count scale. The pin probabilities tend to $1/2$, so their Bernoulli variances remain uniformly positive. Expanding each binomial log likelihood on that scale gives its quadratic Fisher-information term; the bounded individual scores satisfy the Lindeberg condition, and the cubic remainder is $O((m_i^\pm)^{-1/2})$. The product of the four pin experiments therefore has the Gaussian local limit along these sequences. Fixing the follower probabilities defines a submodel of the full unknown-probability experiment, so regularity of the full estimator enforces the corresponding pin bound. Cauchy--Schwarz and $\mathcal T\ge\sum_{\rm pins}m_i^\pm t_i^\pm$ give the leading variance--duration lower coefficient $N^{-2}(\sum_{\rm pins}\sqrt{\varsigma_{i,\pm}^2t_i^{\pm}})^2\sim4/\epsilon$. Equal class sampling attains it through the exact variance calculation below, without a Gaussian approximation for follower outcomes.
A pilot estimate can determine the allocation. Integer rounding
has no leading effect when the required class counts grow.
The cost uses expected completed-run durations, and each run enters the estimate only through its terminal absorbing state.

In the six-site layer $d=\omega_{\rm lay}\epsilon$ at the signal optimum,
the four pin-and-type classes have
$\varsigma_{i,\pm}^2=1/4+o(1)$ and $t_i^{\pm}=36/\epsilon+o(\epsilon^{-1})$.
A follower introduction reaches a mixed aligned state with
probability $O(v)$. At $v=0$ it cannot change either pin, and
the fast inverse remains bounded. The harmonic and physical-time
reconstructions identified in Sec.~\ref{b:sec:domain-input}
therefore give, for either introduced trait,
\begin{equation}
 \begin{gathered}
 \Pr_i(\text{mutant fixes})=O(v),\quad \mathbb E_i\tau=O(1+v/u),\\
 i\in\{1,2,5,6\}.
 \end{gathered}
 \label{b:eq:follower-observation-input}
\end{equation}
The first estimate retains the small entrance probability;
the second multiplies it by the $O(u^{-1})$ slow residence
time and adds the bounded establishment time. Here $u,v=O(\epsilon)$
with bounded $v/u$, so these classes have
$\varsigma_{i,\pm}^2=O(\epsilon)$ and $t_i^\pm=O(1)$.
 Taking the infimum over allocations and locally unbiased or regular
asymptotic terminal estimators in this experiment therefore gives
\begin{equation}
 \inf\operatorname{Var}(\widehat\Delta)\mathcal T
 =\frac4\epsilon[1+o(1)].
 \label{eq:v7-terminal-optimal-coefficient}
\end{equation}
Equal counts in all twelve classes already attain this leading
coefficient, giving Eq.~\eqref{eq:v3-main-variance}.
For comparison, independently redrawing a uniform site for each run
and discarding its label gives exact variance
\begin{equation}
 \frac{2\rho_C(1-\rho_C)+2\rho_D(1-\rho_D)}M
 \sim\frac5{9M}.
 \label{eq:v3-joint-estimator-variance}
\end{equation}
Thus balancing or stratifying the introduction sites reduces the leading
variance--time coefficient from $20/(3\epsilon)$ to $4/\epsilon$.
With equal fixed counts in every type-and-site class, pooling afterward
is algebraically the same estimator and retains the smaller variance.
For independently randomized sites, stratifying by the observed class
counts attains the smaller coefficient asymptotically.

For target signal-to-noise ratio $\mathfrak s$, the peak signal in
Eq.~\eqref{eq:v3-joint-signal-peak} and equal class counts give
\begin{align}
 M_{\rm str}&\sim\mathfrak s^2\,11808^2C_0\,
                 \omega_{\rm lay}^{-3}\epsilon^{-5},\notag\\
 \E[T_{{\rm total},{\rm str}}]
 &\sim12\mathfrak s^2\,11808^2C_0\,
                 \omega_{\rm lay}^{-3}\epsilon^{-6}.
 \label{eq:v3-joint-detection-cost}
\end{align}
These costs assume independent completed runs and negligible reset time. They count aggregate elapsed run time; parallel experimentation has a different wall-clock objective. A known class-dependent reset cost
enters Eq.~\eqref{eq:v7-stratified-time-bound} by replacing
$t_i^{\pm}$ with the sum of the run and reset durations.
The lower bound treats the site-specific terminal probabilities as
independently unknown. With the game and graph known, estimating the
microscopic intensity uses their shared dependence on that intensity and
therefore a different likelihood.

\subsubsection{Finite-surplus signal and observation design}
\label{app:III-finite-surplus}
The cubic signal resolves a shrinking neighborhood of the least contrast.
At fixed scaled contrast, the selected establishment coefficients retain
the finite-surplus dependence. They also distinguish the design with the
largest advantage from the design that estimates an advantage in the least
aggregate time.

Use uniform singleton introduction on the reflected path
$(H,q,1,q,H)$ and independent completed terminal trials, with the
site-specific absorption probabilities treated as unknown.
Put $H=\eta/\epsilon^2$, $q=\theta/\epsilon$ and $b/c=1+\epsilon$.
Fix an admissible positive $\kappa$, with $\kappa<1$ for linear
fecundity, and suppose $\eta>4A_\kappa B_\kappa$.
The coefficients are those of Eqs.~\eqref{eq:selected-linear-coefficients-v2}
and \eqref{eq:app-exponential-coefficients-v2}. Define
\begin{equation}
 G_\kappa^{(0)}(\theta,\eta)
 =1-\frac{A_\kappa}{\theta}-\frac{B_\kappa\theta}{\eta}.
 \label{eq:III-finite-surplus-profile}
\end{equation}
As $\epsilon\downarrow0$,
\begin{equation}
 \begin{aligned}
 \Delta&=\frac{\epsilon\kappa}{6}G_\kappa^{(0)}+O(\epsilon^2),\\
 \inf_{\{m_i^\pm\}}
 \operatorname{Var}(\widehat\Delta_{\rm str})\mathcal T
 &=\frac{\theta}{3\epsilon}[1+O(\epsilon)].
 \end{aligned}
 \label{eq:III-finite-surplus-signal-time}
\end{equation}
The infimum uses continuous class counts; integer counts attain the
leading limit when the counts grow. The variance bound also applies to
locally unbiased or regular terminal estimators under the hypotheses
already stated for Eq.~\eqref{eq:v7-stratified-time-bound}.
For a fixed target signal-to-noise ratio $\mathfrak s$, the leading
aggregate duration is therefore
\begin{equation}
 \mathcal T_{\mathfrak s}
 =\frac{12\mathfrak s^2}{\epsilon^3}
 \frac{\theta}{\kappa^2[G_\kappa^{(0)}(\theta,\eta)]^2}
 [1+O(\epsilon)].
 \label{eq:III-finite-surplus-observation-time}
\end{equation}
These remainders are uniform on compact admissible parameter sets with
$G_\kappa^{(0)}$ bounded away from zero.
At fixed $(\eta,\kappa)$, the leading signal is largest at
$\theta_{\rm sig}=\sqrt{A_\kappa\eta/B_\kappa}$, whereas the leading
observation duration has the unique minimum
\begin{equation}
 \theta_{\rm obs}
 =\frac{6A_\kappa}
 {1+\sqrt{1+12A_\kappa B_\kappa/\eta}}.
 \label{eq:III-observation-optimal-pin}
\end{equation}
Writing $\theta_-^{\rm sel}$ for the lower positive zero of
$G_\kappa^{(0)}$, their order is
$\theta_-^{\rm sel}<\theta_{\rm obs}<\theta_{\rm sig}$.
These optima compare reflected designs $(H,q,1,q,H)$ with both cores at the cap. An unrestricted comparison at fixed selection must control independent variations of both pins and both cores in the signal and, for observation cost, in the terminal variances and mean durations.
\begin{proof}
In Eq.~\eqref{eq:selected-boundary-v2}, substitute
$r-1=\epsilon$, $u=\epsilon/\theta$ and
$v=\epsilon\theta/\eta$. Analyticity of the fast inverse and the
divided slow operator makes the resulting $O(\epsilon^2)$ error uniform
on the indicated compact sets. This proves the first line of
Eq.~\eqref{eq:III-finite-surplus-signal-time} without expanding in
$\eta-656$ or in $\kappa$.

At the hierarchy corner a pin introduction fixes its own domain during
the fast stage. The two mixed-domain exits each have physical rate
$\epsilon/(6\theta)+O(\epsilon^2)$, so each of the four pin-and-type
classes has fixation probability $1/2+O(\epsilon)$ and mean duration
$3\theta/\epsilon+O(1)$. A follower introduction has fixation
probability $O(v)=O(\epsilon)$ and mean duration
$O(1+v/u)=O(1)$. These estimates are uniform for either trait on the
same compact sets. In Eq.~\eqref{eq:v7-stratified-time-bound}, the pin
contribution is consequently
\begin{equation}
 \frac1{36}
 \left(4\sqrt{\frac{3\theta}{4\epsilon}}
       [1+O(\epsilon)]+O(\sqrt\epsilon)\right)^2
 =\frac{\theta}{3\epsilon}[1+O(\epsilon)].
\end{equation}
The eight follower classes account for the $O(\sqrt\epsilon)$ term.
The cost-weighted allocation attains the exact continuous-count bound.
Equal counts already attain its leading term, since their variance is
$(3M)^{-1}[1+O(\epsilon)]$ and their expected aggregate duration is
$M\theta/\epsilon[1+O(\epsilon)]$.

The constraint $G_\kappa^{(0)}>0$ is a bounded open interval in
$\theta$ when $\eta>4A_\kappa B_\kappa$.
The derivative of $\theta/[G_\kappa^{(0)}]^2$ has the sign of
\begin{equation}
 G_\kappa^{(0)}-2\theta\partial_\theta G_\kappa^{(0)}
 =1-\frac{3A_\kappa}{\theta}+\frac{B_\kappa\theta}{\eta}.
\end{equation}
This expression increases strictly with $\theta$. It is negative at
the lower feasible endpoint and positive at $\theta_{\rm sig}$,
where it equals $1-2\sqrt{A_\kappa B_\kappa/\eta}$.
Its unique zero is Eq.~\eqref{eq:III-observation-optimal-pin}.
The duration diverges at either feasible endpoint, proving the stated
minimum and ordering.
\end{proof}

Within this hierarchy the exact finite-surplus leading peak over pins is
\begin{equation}
 \max_\theta\Delta
 =\frac{\epsilon\kappa}{6}
 \left[1-2\sqrt{\frac{A_\kappa B_\kappa}{\eta}}\right]
 +O(\epsilon^2).
 \label{eq:III-finite-surplus-peak-profile}
\end{equation}
The profile is positive when $4A_\kappa B_\kappa<\eta$.
Expanding the limiting profile at $\eta=656+d$,
$\kappa=O(\sqrt d)$ gives the cubic coefficient of
Eq.~\eqref{eq:v3-joint-signal-peak}. Equation~\eqref{eq:III-finite-surplus-peak-profile} takes $\epsilon\to0$ at fixed $(\eta,\kappa)$. When the surplus shrinks together with $\epsilon$, the peak after optimizing selection is given by Eq.~\eqref{eq:v3-joint-signal-peak}, centered on the exact weak critical branch $H_*(\epsilon)$ and controlled by the joint remainder in Appendix~\ref{par:v3-joint-signal-maximum}. At finite scaled surplus the profile retains the higher-order dependence on surplus and selection omitted by the cubic expansion. Table~\ref{tab:III-observation-scales} compares this profile with finite-state evaluations at its signal-maximizing designs.
The observation optimum uses a smaller pin contact because its gain
in completed-run speed outweighs the accompanying reduction in signal.
At the fold the two pin choices coincide. For each fixed feasible
surplus their observation durations retain the $\epsilon^{-3}$ power,
with different finite coefficients.

\paragraph*{Uniqueness after optimizing selection.}
For either fecundity map, the leading reflected observation objective
has a unique joint optimum in $(\kappa,\theta)$ at each fixed
$\eta>656$. The leading signal also has a unique joint maximum.
Equation~\eqref{eq:III-finite-surplus-observation-time} controls their
finite-process approximation on compact admissible parameter sets with
$G_\kappa^{(0)}$ bounded away from zero.

\begin{proof}
Put $x=\log\theta$. Both
$A_\kappa$ and $B_\kappa$ are positive and strictly log-convex
functions of $\kappa$ in their admissible positive intervals.
Consequently
\begin{equation}
 G_\kappa^{(0)}=1-A_\kappa e^{-x}-B_\kappa e^x/\eta
\end{equation}
is strictly concave in $(\kappa,x)$, and its positive domain is convex.
The logarithm of the signal divided by the square root of its
variance--duration coefficient is, up to constants,
\begin{equation}
 \log\kappa-\frac{x}{2}+\log G_\kappa^{(0)}.
\end{equation}
It is strictly concave. The unlogged quantity $\kappa e^{-x/2}G_\kappa^{(0)}$ tends to zero at the boundary
of its feasible domain, including $\kappa=0$, so its positive interior maximum exists
and is unique. The two stationary equations are
Eq.~\eqref{eq:III-observation-optimal-pin} and
\begin{equation}
 1-\frac{A_\kappa+\kappa A_\kappa'}{\theta}
   -\frac{(B_\kappa+\kappa B_\kappa')\theta}{\eta}=0.
 \label{eq:III-observation-optimal-selection}
\end{equation}
The same argument without $-x/2$ proves uniqueness of the leading
signal maximum. Define $\mathfrak h(\kappa)=2\sqrt{A_\kappa B_\kappa}$.
Its selection coordinate obeys
\begin{equation}
 \mathfrak h(\kappa)+\kappa\mathfrak h'(\kappa)=\sqrt\eta.
 \label{eq:III-signal-optimal-selection}
\end{equation}

The log-convexity used above follows from the factors
of the coefficient functions. For linear fecundity put
$p_A=\kappa^4+3$, $p_B=3\kappa^4+2\kappa^2+123$ and $t=\kappa^2$.
Then
\begin{align}
 p_Ap_A''-(p_A')^2&=4\kappa^2(9-\kappa^4)\ge0,\\
 p_Bp_B''-(p_B')^2
 &=4(123+1105t-3t^2-9t^3)\\
 &\ge4(123+1093t)>0
 \qquad(0\le t\le1).\notag
\end{align}
Thus both positive numerator factors are log-convex.
The reciprocal quadratic factors are strictly log-convex since
$\partial_\kappa^2[-\log(a-\kappa^2)]
=2(a+\kappa^2)/(a-\kappa^2)^2>0$ for $a>\kappa^2$.
The products defining $A_\kappa$ and $B_\kappa$ are consequently
strictly log-convex throughout $0\le\kappa<1$.

For exponential fecundity,
\begin{equation}
 A_\kappa=2+2\int_0^1\cosh(s\kappa)\,ds
               +2\int_2^3\cosh(s\kappa)\,ds
\end{equation}
is the moment-generating function of a finite positive symmetric
measure, so its logarithm is convex, strictly so because the measure
is not concentrated at one point.
Also $B_\kappa=2(\sinh\kappa/\kappa)F(\cosh\kappa)$, where
\begin{equation}
 F(C)=\frac{(4C-1)(32C^3+8C^2+1)}{8C^2+1}.
\end{equation}
The first factor has the same positive-integral argument.
For the second, the sign of $\partial_\kappa^2\log F(\cosh\kappa)$
is the sign of
$CF F'+(C^2-1)[FF''-(F')^2]$, with primes here taken in $C$.
Its denominator is positive and its numerator is
\begin{align}
 P(C)={}&1310720C^{10}+294912C^9+1310720C^8
 \notag\\
 &-223232C^7+102400C^6-6400C^5
 \notag\\
 &+14848C^4-2912C^3
 \notag\\
 &+2048C^2-196C+16.
\end{align}
For $C\ge1$, pair each negative term with the preceding positive
term to obtain
\begin{align}
 P(C)\ge{}&1310720C^{10}+294912C^9+1087488C^7
 \\
 &+96000C^5+11936C^3+1852C+16>0. \notag
\end{align}
This proves strict log-convexity. The removable values at zero are
$A_0=6$, $B_0=82/3$. Both functions increase away from zero and
diverge at the far endpoint of their admissible selection interval.
Thus $4A_\kappa B_\kappa<\eta$ defines a bounded feasible selection
interval whenever $\eta>656$.
\end{proof}

\paragraph*{Common preparation cost.}\label{app:III-preparation-cost}
Take a common known preparation duration $r_{\rm prep}\ge0$, independent
of the contacts and selection intensity, and set
$\mathcal R_{\rm prep}=\epsilon r_{\rm prep}$. In the same reflected
hierarchy, Eq.~\eqref{eq:III-preparation-cost} is the leading aggregate
cost at fixed $\mathcal R_{\rm prep}$, uniformly on compact admissible
sets with positive leading signal. At fixed $(\eta,\kappa)$ its minimizing
pin coordinate increases from $\theta_{\rm obs}$ toward
$\theta_{\rm sig}$ as $\mathcal R_{\rm prep}$ increases from zero to
infinity. For each $\eta>656$ and fixed $\mathcal R_{\rm prep}\ge0$,
the joint optimum over selection and pin contact is unique.

\begin{proof}
Let $\mathcal T^{\rm prep}=\sum_{i,\pm}m_i^\pm(t_i^\pm+r_{\rm prep})$ be the aggregate run-plus-preparation duration. Replace each $t_i^\pm$ in the exact allocation bound by $t_i^\pm+r_{\rm prep}$. The four pin terms dominate the sum of square roots,
while the follower terms have relative size at most $O(\sqrt\epsilon)$.
Consequently
\begin{equation}
 \inf_{\{m_i^\pm\}}\operatorname{Var}(\widehat\Delta_{\rm str})
       \mathcal T^{\rm prep}
 =\frac{3\theta/\epsilon+r_{\rm prep}}9
       [1+O(\sqrt\epsilon)].
\end{equation}
The error improves to $O(\epsilon)$ for bounded preparation duration.
Together with Eq.~\eqref{eq:III-finite-surplus-signal-time}, this proves
Eq.~\eqref{eq:III-preparation-cost}. The minimizing $\theta$ is the unique
feasible root of
\begin{equation}
 1-\frac{3A_\kappa}{\theta}
 +\frac{B_\kappa\theta}{\eta}
 -\frac{2\mathcal R_{\rm prep}A_\kappa}{3\theta^2}
 +\frac{2\mathcal R_{\rm prep}B_\kappa}{3\eta}=0.
\end{equation}
The derivative of its left-hand side is
$3A_\kappa/\theta^2+B_\kappa/\eta
+4\mathcal R_{\rm prep}A_\kappa/(3\theta^3)>0$.
It is negative at the lower feasible endpoint and positive at
$\theta_{\rm sig}$. At positive $\mathcal R_{\rm prep}$ it is also negative
at $\theta_{\rm obs}$. Implicit differentiation proves monotone motion
toward $\theta_{\rm sig}$, with that limit as
$\mathcal R_{\rm prep}\to\infty$.

The joint optimum over selection remains unique. In $x=\log\theta$,
the relevant logarithmic objective is
$\log\kappa+\log G_\kappa^{(0)}
-\tfrac12\log(3e^x+\mathcal R_{\rm prep})$.
The first two terms are strictly concave by the coefficient
argument above, and the last is concave. The unlogged quantity
$\kappa G_\kappa^{(0)}/\sqrt{3\theta+\mathcal R_{\rm prep}}$ vanishes at the
feasible boundary. If $r_{\rm prep}\gg\epsilon^{-1}$, the leading cost is
preparation duration times the least required trial count. Optimal count
allocation has variance--count coefficient $1/9+O(\sqrt\epsilon)$,
compared with $1/3+O(\epsilon)$ for equal class counts. Signal maximization
therefore minimizes the leading preparation-dominated cost.
\end{proof}

\paragraph*{Correlated terminal pairs.}\label{app:III-correlated-pairs}
Let $X_i,Y_i$ be the all-cooperator terminal indicators in any coupled pair
with the correct singleton marginals, and put $Z_i=X_i+Y_i-1$.
Sharing the neutral copying arrows between complementary singleton
introductions gives $Z_i=0$ almost surely. At fixed positive contacts,
maximal coupling of the selected source choices gives
$\operatorname{Var}Z_i=O(|\kappa|)$ near neutrality.

\begin{proof}
Exactly,
\begin{equation}
 \begin{aligned}
 \mathbb E Z_i&=h_i^++h_i^--1,\\
 \operatorname{Var}Z_i&=\Pr(X_i=Y_i)-(h_i^++h_i^--1)^2.
 \end{aligned}
\end{equation}
Every neutral copying map commutes with type complementation. Sharing all
neutral directed copying events therefore gives $Y_i=1-X_i$ pathwise.
Here both indicators score all-cooperator consensus. If each run
scores fixation of its introduced trait, the two neutral indicators agree.
For fixed positive contacts, a maximal coupling of the selected source
choices fails before absorption with probability at most
$C|\kappa|\mathbb E\tau_\kappa$. Here $\tau_\kappa$ is the absorption time of one selected marginal. Its mean stays bounded
near neutrality at those fixed contacts, proving the $O(|\kappa|)$
variance bound. Along the compact reflected hierarchy, a pin introduction has mean absorption time $3\theta/\epsilon+O(1)$, as shown in the proof of Eq.~\eqref{eq:III-finite-surplus-signal-time}. Thus the factor $\mathbb E\tau_\kappa$ in the fixed-contact coupling bound grows as the gap closes. A joint small-gap variance exponent requires controlling the coupling-failure probability together with this growing absorption time.
\end{proof}

\subsubsection{Rare introductions on a shrinking signal scale}
\label{app:v7-rare-signal}

To specify the introduction clock away from consensus, add independent
symmetric flips at each site at rate $\nu/6$ to the copying
process. The total spontaneous-flip rate is $\nu$, and its site at either consensus is uniform. Define the cooperative fraction $X(n)=N^{-1}\sum_i n_i$ and let $\langle X\rangle_\nu$ be its stationary mean in this mutation-regularized process.
At fixed positive weights, the leading rates between the two consensus states are $\nu\rho_C$ from all defectors to all cooperators and $\nu\rho_D$ in the reverse direction. Their stationary fraction gives $X_C^{\rm rare}=\rho_C/(\rho_C+\rho_D)$ as $\nu\to0$, by the reduced-chain argument~\cite[Sec.~4, Theorem~2]{FudenbergImhof2006}. The structured-population result for mutation at replacement has appearance rates weighted by turnover~\cite[Theorem~6]{AllenTarnita2014}; the independent flips here fix a uniform introduction rule at consensus.

Errors in rare-mutation reductions have been quantified for well-mixed evolutionary processes~\cite{WuEtAl2012RareMutation}. Here the approximation error must also be small relative to a vanishing invasion advantage.

In the compact reflected optimization chart, put
$\eta_{\rm mut}=\nu q$ and use the centered intensity $\lambda$.
The stationary mean is analytic through $\eta_{\rm mut}=0$ after two
Schur eliminations. First eliminate fast configurations using their
bounded inverse. The zero-frequency Schur complement of the physical
generator retains the four aligned domain states and their effective
rates. Divide this reduced generator by the common factor $q^{-1}$,
which leaves its invariant measure unchanged, and then eliminate the
two mixed domain states in this slow clock. Each departure from consensus
contains a mutation factor, so the final two-state off-diagonal
rates contain a common factor $\eta_{\rm mut}$. After canceling it, the two remaining slow-time rates and their
normalization are positive and analytic at the corner. Reconstruction of the eliminated probabilities has the
same regularity. At $\eta_{\rm mut}=0$ this ratio is the exact
rare-introduction ratio at the same finite weights.

The difference between these two abundances vanishes at
$\eta_{\rm mut}=0$ and is odd in $\lambda$ by type complementation.
At $\epsilon=0$ the aligned domain payoffs vanish, so both limiting
means equal one half for every small $\lambda,\eta_{\rm mut}$.
Analytic division by these three zeros gives
\begin{equation}
 \langle X\rangle_\nu-X_C^{\rm rare}
 =\epsilon\lambda\eta_{\rm mut}\,
 \mathcal R_{\rm mut}(\epsilon,\lambda^2,\eta_{\rm mut},\text{scaled weights}),
 \label{eq:v7-rare-signal-error}
\end{equation}
where $\mathcal R_{\rm mut}$ is bounded on a sufficiently small compact chart.
At the local signal optimum with $d>0$, $\lambda_{\rm peak}\asymp\sqrt d$ and
$X_C^{\rm rare}-1/2\asymp\Delta_{\rm peak}
\asymp\epsilon d^{3/2}$ because $\rho_C+\rho_D\to1/3$.
Equation~\eqref{eq:v7-rare-signal-error} bounds the relative correction by
$O(\eta_{\rm mut}/d)$. Thus $\eta_{\rm mut}=o(d)$ makes it smaller
than the positive bias, or equivalently
\begin{equation}
 \nu=o(\epsilon d),\qquad q\asymp\epsilon^{-1}.
 \label{eq:v7-rare-signal-condition}
\end{equation}
For $d=\omega_{\rm lay}\epsilon$ at fixed $\omega_{\rm lay}>0$ this becomes
$\nu=o(\epsilon^2)$; at fixed sufficiently small positive $d$,
$\nu=o(\epsilon)$ suffices. Both rate conditions are sufficient, since they come from the bound on $\mathcal R_{\rm mut}$ in Eq.~\eqref{eq:v7-rare-signal-error}, and they refer to the symmetric spontaneous-flip model specified above.
\subsubsection{The reflected response radius for linear fecundity}
\label{par:v3-joint-radius-window}
For linear fecundity the reflected interval fast determinant at the
hierarchy corner is
\begin{equation}
 \det\widetilde{\mathsf A}^{\rm ctr}_{FF}(0,\lambda)
 =-4(\lambda-3)(\lambda-1)^6(\lambda+1)^2(\lambda+3).
 \label{eq:v3-joint-fast-determinant}
\end{equation}
Together with the nonzero divided slow factor, this proves uniform
analyticity on every closed disk $|\lambda|\le r_0<1$, for sufficiently
small $\epsilon$, uniformly on compact positive $(\theta,\eta)$ sets.
At $\epsilon=0$ the centered intensity obeys $\kappa=\lambda$.
Writing $A_\lambda=A_\kappa|_{\kappa=\lambda}$ and
$B_\lambda=B_\kappa|_{\kappa=\lambda}$ for the linear-map coefficients,
the limiting normalized response is
$G(0,\theta,\eta,\lambda)=1-A_\lambda/\theta-B_\lambda\theta/\eta$,
and its pole at $\lambda=1$ is nonremovable, since 
\begin{equation}
 \lim_{\lambda\to1}(1-\lambda)^2G
 =-\frac2\theta-\frac{8\theta}\eta<0.
 \label{eq:v3-joint-limiting-pole}
\end{equation}
The reflected interval chain has eleven transient states, and its linear-fecundity harmonic matrix is affine in $\lambda$. Cramer's rule bounds the denominator degree of singleton fixation by eleven in $\lambda$. Pairing its values at $\lambda$ and $-\lambda$ bounds the denominator degree of $\Delta/\lambda$ by eleven in $\lambda^2$. This restriction for singleton introduction sharpens the full observable-space bound of fifteen from Eq.~\eqref{eq:III-parity-resolvent}. Both constructions bound the degree uniformly as $\epsilon\to0$. If a subsequence had a pole-free
disk of radius $1+\varepsilon_{\rm rad}$ for some fixed
$\varepsilon_{\rm rad}>0$, normalize its reduced denominator at zero.
Its bounded degree and roots outside that disk bound its coefficients.
Local convergence then bounds the numerator coefficients. A convergent
subsequence would represent an analytic limit across $\lambda=1$,
contradicting Eq.~\eqref{eq:v3-joint-limiting-pole}. Consequently the
centered Taylor radius tends to one. The fast determinant~\eqref{eq:v3-joint-fast-determinant}, the degree bound and the limiting pole~\eqref{eq:v3-joint-limiting-pole} use the reflected interval chain, so this full-disk result holds on the reflected family. For asymmetric contacts in the compact local chart, Appendix~\ref{par:v3-joint-selected-analyticity} proves analyticity on a common neighborhood of $\lambda=0$.

In the layer $d=\omega_{\rm lay}\epsilon$ with fixed $\omega_{\rm lay}>0$, $\lambda_{\rm edge}=O(\sqrt\epsilon)$ lies well inside the
reflected linear-response convergence disk, whose radius tends to one.
The adverse nonlinear response reverses the sign within a convergent
series. At fixed finite size, rationality gives coefficient growth
governed by poles. 

\labelalias{b:eq:input-floors}{tab:III-input-floors}
\labelalias{eq:v7-birth-c4-chambers}{tab:III-birth-c4-chambers}
\labelalias{eq:v3-joint-selected-critical-coefficients}{tab:III-selected-critical-coefficients}
\labelalias{app:v7-fixed-mixture}{sec:v10-crossover}

\labelalias{eq:v4-dual-constants}{tab:III-positive-balances}
\labelalias{eq:joint-frontiers}{tab:III-time-frontiers}
\labelalias{eq:III-p6-birth-onset-array}{tab:III-p6-birth-onset}
\labelalias{eq:app-p6-A}{tab:III-p6-response-coefficients}
\labelalias{eq:app-p6-B}{tab:III-p6-response-coefficients}
\labelalias{eq:v3-joint-linear-jet}{tab:III-selected-response-coefficients}
\labelalias{eq:v3-joint-exp-jet}{tab:III-selected-response-coefficients}
\bibliography{references}
\end{document}